\documentclass[11pt]{article}
\pdfoutput=1
\usepackage{jheppub}
\usepackage{upgreek}
\usepackage{float, extarrows, tikz-cd}
\usepackage{graphicx}
\usepackage{slashed}
\usepackage{tabularx,ragged2e}
\usepackage{amssymb}
\usepackage{subcaption}
\usepackage{amsmath,amssymb}
\usepackage{slashed}
\usepackage{caption}
\usepackage{xcolor}
\usepackage{dsfont}
\usepackage{verbatim}
\usepackage{mathtools, xcolor,ytableau, amsfonts,tikz}
\usepackage{graphicx}
\usepackage{physics}
\usepackage{simpler-wick}
\usepackage{microtype}
\usepackage[nobottomtitles*]{titlesec}
\newcolumntype{C}{>{\Centering\arraybackslash}X}

\numberwithin{equation}{section}

\newcommand{\mL}{\mathcal{L}}
\newcommand{\mO}{\mathcal{O}}

\newcommand{\pd}{\partial}

\newcommand{\nt}{\tilde{\nu}}

\newcommand{\bz}{\bar{z}}

\def\le{\left}
\def\ri{\right}
\newcommand\ov{\over}
\newcommand\p{\ensuremath{\partial}}
\newcommand{\es}[2] {\begin{equation} \label{#1} \begin{split} #2 \end{split} \end{equation}}

\def\<{\langle}
\def\>{\rangle}
\newcommand\al{{\alpha}}

\newcommand\lam{\lambda}
\newcommand\Lam{\Lambda}
\newcommand\om{\omega}
\newcommand\Om{\Omega}
\newcommand\ga{{\ensuremath{{\gamma}}}}

\newcommand\de{{\ensuremath{{\delta}}}}
\newcommand\De{{\ensuremath{{\Delta}}}}

\DeclareMathOperator*{\SumInt}{%
\mathchoice%
  {\ooalign{$\displaystyle\sum$\cr\hidewidth$\displaystyle\int$\hidewidth\cr}}
  {\ooalign{\raisebox{.14\height}{\scalebox{.7}{$\textstyle\sum$}}\cr\hidewidth$\textstyle\int$\hidewidth\cr}}
  {\ooalign{\raisebox{.2\height}{\scalebox{.6}{$\scriptstyle\sum$}}\cr$\scriptstyle\int$\cr}}
  {\ooalign{\raisebox{.2\height}{\scalebox{.6}{$\scriptstyle\sum$}}\cr$\scriptstyle\int$\cr}}
}

\author[a,b]{Ofer Aharony,}
\author[c,d,e,f]{Gabriel Cuomo,}   
\author[e,f]{Zohar Komargodski,}        
\author[g,e]{M\'ark Mezei}                         
\author[e]{and Avia Raviv-Moshe}   
              
                            \affiliation[a]{Department of Particle Physics and Astrophysics, Weizmann Institute of Science, Rehovot, Israel}                                                              
                            \affiliation[b]{School of Natural Sciences, Institute for Advanced Study, Princeton, NJ 08540, USA}
              \affiliation[c]{Center for Cosmology and Particle Physics, Department of Physics, New York University, New York, NY 10003, USA}
              \affiliation[d]{Department of Physics, Princeton University, Princeton, NJ  08544, USA}
               \affiliation[e]{Simons Center for Geometry and Physics, SUNY, Stony Brook, NY 11794, USA}      
              \affiliation[f]{C. N. Yang Institute for Theoretical Physics, Stony Brook University, Stony Brook, NY 11794, USA}
               \affiliation[g]{Mathematical Institute, University of Oxford, Woodstock Road, Oxford, OX2 6GG, United Kingdom}  

\emailAdd{ofer.aharony@weizmann.ac.il}                                            
\emailAdd{gc6696@princeton.edu}				
\emailAdd{zkomargodski@scgp.stonybrook.edu}       
\emailAdd{mezei@maths.ox.ac.uk}   
\emailAdd{araviv-moshe@scgp.stonybrook.edu}

\begin{document}

\title{Phases of Wilson Lines: Conformality and Screening}

\abstract{We study the rich dynamics resulting from introducing static charged particles (Wilson lines) in 2+1 and 3+1 dimensional gauge theories. Depending on the charges of the external particles, there may be multiple defect fixed points with interesting renormalization group flows connecting them, or an exponentially large screening cloud can develop (defining a new emergent length scale), screening the bare charge entirely or partially. We investigate several examples where the dynamics can be solved in various weak coupling or double scaling limits. Sometimes even the elementary Wilson lines, corresponding to the lowest nontrivial charge, are screened. We consider Wilson lines in 3+1 dimensional gauge theories including massless scalar and fermionic QED$_4$, and also in the ${\mathcal N}=4$ supersymmetric Yang-Mills theory. 
We also consider Wilson lines in 2+1 dimensional conformal gauge theories such as QED$_3$ with bosons or fermions, Chern-Simons-Matter theories, and the effective theory of graphene. Our results in 2+1 dimensions have potential implications for graphene, second-order superconducting phase transitions, etc. 
Finally, we comment on magnetic line operators in 3+1 dimensions ('t Hooft lines) and argue that our results for the infrared dynamics of electric and magnetic lines are consistent with non-Abelian electric-magnetic duality.}

\maketitle

\section{Introduction and summary}

Conformal Field Theory is by now a mature subject in some ways. A great deal is understood about the space of local operators and their correlation functions, see \cite{Poland:2018epd} for a review. 

Yet, relatively little is understood about extended operators. The simplest class of extended operators are line operators.  For a line operator that is stretching in the time direction at a point $\vec{x}=0$ in space, we say that the line operator is conformal if it preserves an 
\begin{equation}SL(2,\mathbb{R})\end{equation}
subgroup of the conformal group. The $SL(2,\mathbb{R})$ subgroup acts at $\vec{x}=0$ by $t\to {\alpha t+\beta\over \gamma t +\delta}$, with $\alpha\delta-\beta\gamma=1$. A conformal line operator admits local defect operators $\hat O_i(t)$ transforming under $SL(2,\mathbb{R})$. The defect operators have scaling dimensions $\hat \Delta_i\geq 0$ and one can perform an Operator Product Expansion (OPE) among them. Very importantly, bulk local operators can be expanded as $|\vec{x}|\to 0$ in terms of defect operators in the following schematic form 
\begin{equation}\label{bbope}O(x,t) = \sum a_i \, x^{-\Delta_O+\hat \Delta_i}\hat O_i(t)~.\end{equation}
In particular, in the presence of a line operator, bulk operators can have a nonzero one-point function due to the unit operator appearing on the right hand side of~\eqref{bbope}.  The $a_i$ on the right hand side of~\eqref{bbope}, along with the $\hat \Delta_i$ and also the defect OPE coefficients, are observables associated to conformal line defects. 
This whole structure is referred to as a Defect Conformal Field Theory (DCFT), see~\cite{Billo:2016cpy} for a review. A special defect operator which always appears in nontrivial DCFTs is the displacement operator~\cite{McAvity:1993ue,Jensen:2015swa}, whose scaling dimension is $\hat \Delta=2$. Integrating this operator on the defect is equivalent to changing the shape of the defect.

When one defines line operators in a conformal field theory, it is not guaranteed that they are conformal line operators preserving  $SL(2,\mathbb{R})$. While the bulk theory remains conformal, there can be renormalization group flows between line operators, and one typically expects the deep infrared limit of line operators to be a DCFT. The space of distinct DCFTs in a given CFT is far from understood. Line operators are of great interest also in the context of condensed matter, since they represent localized impurities/defects. The long distance limit of various defects in condensed matter is a subject that goes back to the Kondo problem \cite{Affleck:1995ge}. 

We would like to make three general remarks on the general theory of line defects in CFTs:

\begin{itemize}
\item Renormalization group (RG) flows on line defects are constrained by the so-called defect entropy \cite{Cuomo:2021rkm}. For the connection between the defect entropy and entanglement entropy see~\cite{Lewkowycz:2013laa,Casini:2022bsu,Casini:2023kyj}. This allows one to make consistency checks on various proposals, and sometimes to prove that a DCFT cannot be trivial (screened). 

\item  Local bulk operators transform in a representation of the ordinary (0-form) symmetry of the CFT, $G$. The interplay of line operators with the 0-form symmetry of the theory is more complicated. First of all, it is possible for a conformal line operator to break the symmetry $G$ altogether. That means that the intersections of the line with the co-dimension 1 $G$-surfaces are not topological. If we wrap the line operator with the co-dimension 1 $G$-surfaces, we must obtain a new line operator in which the bulk VEVs of $G$-charged operators change accordingly. For a continuous symmetry $G$ this means that there are tilt operators on the defect which have $\hat\Delta=1$ exactly. These are analogous to the displacement operator. See~\cite{Cuomo:2021cnb,Drukker:2022pxk} for a review and some examples of tilt operators. 

If the line operator preserves $G$,\footnote{This means that symmetry defects admit a topological intersection with the line operator. For invertible symmetries, this also implies that the corresponding defect state is an eigenstate of the symmetry operator; this is not necessarily true for non-invertible symmetries \cite{Choi:2023xjw}.} defect operators that appear on the right hand side of~\eqref{bbope} are in representations of $G$.
However, defect-changing operators, and in particular, end-point operators, do not have to be in representations of $G$.
Physically, line operators can be viewed as capturing the response of the CFT to heavy/external objects. The full symmetry in the presence of the heavy objects could be an extension of $G$. Some simple examples (without gauge fields) where the end-point operators indeed only transform in projective representations of $G$ were studied in~\cite{Liu:2021nck,Cuomo:2022xgw,Nahum:2022fqw,Weber:2022ada} along with many examples in 1+1 dimensions in the context of the Kondo defect (see~\cite{affleck2010kondo} for a review with an emphasis on the screening cloud, a theme we will discuss below in higher dimensions). We will see examples of this phenomenon in gauge theories in this paper.

If the end-point operators transform in a projective representation of $G$, this has consequences for RG flows on such lines, since they cannot become trivial lines with no degeneracy (i.e. the unit operator). Indeed, if they were completely trivial lines with no degeneracy in the infrared, we would not be able to attach operators in projective representations to them. This argument is presented in more detail in the body of the paper. 

\item If the theory admits a one-form symmetry $\Gamma$ then the line operators transform under $\Gamma$ \cite{Gaiotto:2014kfa}. For a line operator to furnish a nontrivial DCFT it is not necessary for it to be charged under $\Gamma$. However, under additional assumptions that we will discuss later, it is possible to prove that a line defect transforming under $\Gamma$ must have a nonzero displacement operator at long distances.

\end{itemize} 

A special class of line operators, that exist in any gauge theory in any dimension, are Wilson lines
\es{Wilson}{
W_R(\gamma) = {\rm tr}_R \left( P\exp (i \int_{\gamma} A_{\mu} dx^{\mu}) \right),
}
labelled by a representation $R$ of the gauge group and by a closed, or infinite, contour $\gamma$. 

Historically, Wilson lines have been introduced to diagnose the confinement/deconfinement transition in gapped theories. Then, the interesting Wilson lines are those charged under the one-form symmetry $\Gamma$, since such Wilson lines serve as order parameters for the confinement/deconfinement transition.

Here our interest is in conformal field theories. Then, as discussed above, Wilson lines are interesting observables whether or not the Wilson line transforms nontrivially under $\Gamma$. In fact, Wilson lines are interesting line operators even in theories with trivial $\Gamma$. 

A peculiarity of~\eqref{Wilson} that makes them into intriguing line operators is that there is no free continuous parameter in the definition~\eqref{Wilson}.
As we will see, that does not mean that no RG flow takes place!

This paper is an extended version of~\cite{Aharony:2022ntz}.
The central goal of this paper is to determine the long distance limit of \eqref{Wilson} as a function of the representation $R$. 
We will investigate this question in various examples of conformal gauge theories in four and three space-time dimensions. There are two complementary ways to analyze this question. 

\begin{itemize} 
\item In the ``bulk approach'', we view the insertion of the Wilson line \eqref{Wilson} as setting a boundary condition for the dynamical fields of the gauge theory at $\vec{x}=0$, which includes an electric field emanating from there; as usual we need to regularize this by putting the boundary and the boundary condition at some $|\vec{x}|=r_0$, and asking what the theory behaves like for $|\vec{x}| \gg r_0$. This corresponds to the infrared limit of the defect. The simplest possible answer, is that the lowest energy state just involves the electric field going as $1/|\vec{x}|^2$ (recall that in Abelian gauge theories the dimension of $F_{\mu \nu}$ is always $\Delta=2$). In other cases we will find that the dynamical fields react non-trivially to the Wilson line source, and screen it, either partially or completely. The infrared can then be trivial or partially screened. Another important comment is that specifying the electric field at $|\vec{x}|=r_0$ is not sufficient. The boundary conditions (and possible boundary interactions) of the charged bulk fields at the insertion should be specified too, and this leads to new coupling constants that must be added to~\eqref{Wilson} to define the problem properly. These coupling constants inevitably run and lead to much of the rich dynamics that we will encounter here and review soon. 

\item In the ``defect approach'', we view the insertion of the Wilson line \eqref{Wilson} as modifying the action of the gauge theory by some extra terms that are localized on a ``defect'' at $\vec{x}=0$. One can then discuss a renormalization group flow of the conformal gauge theory in the presence of these extra defect terms (with an ultraviolet cutoff $\mu$, which is inversely related to $r_0$ discussed above); for simplicity we can assume that the bulk theory has already flowed to its low-energy fixed point, and then the non-trivial flow involves only the action of the defect. This approach is convenient since it utilizes the ideas behind the renormalization group more directly. While the charge of the Wilson line is quantized and does not flow under the renormalization group, we will see that in many cases other couplings (localized on the defect) related to the additional fields in the theory do flow non-trivially (and, as we remarked above, ignoring them is inconsistent), reproducing in a different language the bulk physics discussed above. \end{itemize}

In the rest of this section we briefly summarize our results.

\subsection{Scalar and fermionic \texorpdfstring{QED$_4$}{QED4}}

Scalar and Fermionic massless QED$_4$ are not strictly conformal theories (due to the Landau pole) and hence ideas of DCFT do not rigorously apply. However, at weak enough gauge coupling the running coupling constant is an insignificant perturbation (and furthermore there is a double scaling limit in which it is truly a subleading effect, as will be explained in what follows), and there the physics of these models does lend itself to the language of DCFT. Also, understanding these examples will be a valuable springboard towards more complicated 3+1 dimensional gauge theories which are truly conformal. Needless to say, understanding Wilson lines in QED is of great interest in and of itself. It is perhaps surprising that there is much new to say on this subject.

We will consider Wilson lines of charge $q$ in either scalar or fermion QED$_4$. These are given by the ``naive'' expression:
\es{WilsonAb}{
W_q(\gamma) =   \exp (iq \int_{\gamma} A_{\mu} dx^{\mu})~,
}
and we take the contour $\gamma$ to be localized at $\vec{x}=0$.

For concreteness let us start from scalar QED$_4$ with a single complex charge $1$ scalar field $\phi$. If one tries to interpret~\eqref{WilsonAb} as a conformal defect, one can compute the scaling dimension of the gauge-invariant bilinear $\phi^\dagger \phi$. In the bulk at sufficiently weak coupling the scaling dimension is of course $\Delta=2$. But we can also ask about the scaling dimension of $\phi^\dagger \phi$ as a defect operator. As we will show in section \ref{sec_ScalarQED},
one finds 
\begin{equation}\label{sqf}\hat \Delta_{\phi^\dagger \phi} =1+\sqrt{1-{e^4q^2\over 4\pi^2}}~. \end{equation}
This formula is reliable as long as the fine structure constant is small enough (more precisely, it is {\it exact} in the limit of $e\to 0$ and $e^2|q|$ fixed). A consistency check is that for $q=0$ the bulk and defect scaling dimensions coincide.

Interestingly,~\eqref{sqf} implies that for ${e^2|q|\over 2\pi}=1$ the operator becomes marginal on the defect, while for ${e^2|q|\over 2\pi}>1$ there is a disease with our defect. Since the operator is marginal at ${e^2|q|\over 2\pi}=1$ and slightly irrelevant as we approach ${e^2|q|\over 2\pi}=1$ from below, we learn that ignoring it in RG flows is inconsistent, and hence we must study the more general line defect
\es{WilsonAb2}{
W_q(\gamma) =   \exp (iq \int_{\gamma} A_{\mu} dx^{\mu}-g \int \phi^\dagger \phi \,|dx|)~.
}
The parameter $g$ has a nontrivial beta function with the following properties: 
\begin{itemize}
\item For ${e^2|q|\over 2\pi}<1$ there are two fixed points. One corresponds to a stable DCFT (with no relevant operators) and the other to an unstable DCFT (with one relevant operator, $\phi^\dagger\phi$). 
\item For ${e^2|q|\over 2\pi}=1$ the two fixed points merge and the operator $\phi^\dagger \phi$ is marginal. 
\item For ${e^2|q|\over 2\pi}>1$ the coupling $g$ flows to $-\infty$.
\end{itemize} 
This behavior is reminiscent of how conformality is (presumably) lost in QCD (a.k.a. Miransky scaling~\cite{Kaplan:2009kr}). Here, conformality is lost at ${e^2|q|\over 2\pi}=1$ in the sense that no DCFTs with finite $g$ exist for ${e^2|q|\over 2\pi}>1$. However, the flow $g\to -\infty$ must still be analyzed to determine the infrared behavior of Wilson lines with sufficiently large charge. 

Again, analogously to QCD, one finds that an exponentially low energy scale is generated when ${e^2|q|\over 2\pi}$ is slightly larger than $1$ (dimensional transmutation). The dynamics is that of an exponentially large cloud of bosons surrounding the Wilson line. We present the properties of the cloud, which is essentially a new soliton, and argue that it screens the charge of the Wilson line entirely. 
We find that the defects with ${e^2|q|\over 2\pi}>1$ are trivial DCFTs in the infrared for a generic bulk scalar potential.

While we find two fixed points for ${e^2|q|\over 2\pi}<1$, we do not claim that our analysis of that region is complete. Indeed, for sufficiently small ${e^2|q|\over 2\pi}$ the quartic $|\phi|^4$ defect operator becomes relevant in the unstable defect fixed point, and the dynamics must be re-analyzed. We leave this for the future.

For fermionic QED$_4$ the story, which we analyze in section \ref{sec_Fer}, is conceptually similar, except that the instability occurs at ${e^2|q|\over 4\pi}=1$, which in nature corresponds to $|q|\sim 137$. The fact that nuclei with charge $\sim 137$ lead to difficulties with the Dirac equation was observed already decades ago~\cite{pomeranchuk1945energy}. Another difference from the scalar theory is that for ${e^2|q|\over 4\pi}>1$ we find an exponentially large charged condensate of fermions that only screens the line down to ${e^2|q|\over 4\pi}\simeq 1$, and not to a trivial line. 

In summary, unlike in pure Maxwell theory, in QED Wilson lines with sufficiently large $|q|$ are screened, i.e. do not lead to new interesting DCFTs. 
The transition between the two regimes involves the annihilation of two fixed points and dimensional transmutation due to the running coupling $g$ on the Wilson line. Furthermore, for small $|q|$, there are multiple fixed points, not all of which we have analyzed. 

We also consider two interesting variations on the above themes. 
The first variation is multi-flavor scalar QED. Our soliton that screens the Wilson line for ${e^2|q|\over 2\pi}>1$ then transforms nontrivially under an internal symmetry and there is thus some sort of symmetry breaking -- a zero mode of the soliton. We show that this zero mode must be integrated over, and the true screening cloud does not lead to symmetry breaking. The Wilson line leaves no measurable trace of its existence at distances much larger than the cloud. It is therefore completely transparent to all bulk observables. However, on the Wilson line itself, depending on its charge, there is some degeneracy of states (a 0+1d TQFT stacked on the trivial, screened line), which follows from symmetry considerations alone. This happens precisely because of a fact we already mentioned above: the end-point operators in the multi-flavor model are in a projective representation of the symmetry group, and this constrains the infrared limit of the line defect.

The second variation on the above themes is to consider QED$_4$ with a scalar of charge $q_\phi > 1$ but no scalars of charge $1$. Now the theory has a $\mathbb{Z}_{q_\phi}$ electric one-form symmetry, and hence Wilson lines of $q\neq 0\ {\rm mod} \ q_\phi$ cannot flow to trivial DCFTs, no matter how large $|q|$ is. Similarly, in the $\mathcal{N}=4$ supersymmetric Yang-Mills (SYM) theory with, say, gauge group $SU(2)$, a Wilson line in a large half-integer spin representation cannot be completely screened due to the $\mathbb{Z}_2$ one-form symmetry. The question of what precisely is the infrared limit in these two cases goes beyond the leading order we analyze here. We speculate about the possible infrared phases that are consistent with the one-form symmetry in both cases.

\subsection{Non-Abelian gauge theories and \texorpdfstring{${\cal N}=4$}{N=4} SYM}

Much of what we have found for Abelian theories carries over to essentially all weakly coupled conformal gauge theories in four dimensions. We discuss non-Abelian theories in section \ref{secNonAbelian}. Let us consider for concreteness the Wilson lines in the ${\cal N} = 4$ SYM theory with gauge group $SU(2)$.

There are multiple possible Wilson lines in this theory, including supersymmetric versions of the Wilson line which include also scalar fields from the vector multiplet, and which preserve some fraction of the supersymmetry. These Wilson lines were the subject of many investigations in the last decades, see e.g.~\cite{Drukker:2007qr}. All the supersymmetric Wilson lines break the $SO(6)_R$ global symmetry of ${\cal N} = 4$ SYM to a subgroup. Here we are interested in the $SO(6)_R$-invariant Wilson line~\eqref{Wilson}, which breaks all of the supersymmetry.
As in the Abelian case, one must not ignore scalar bilinears, which turn out to be important defect operators. We find again that for Wilson lines in the spin $s$ representation of $SU(2)$, when $g_{YM}^2s\sim 1$ the Wilson line is screened (for large half-integral $s$ the one-form symmetry prevents the line from being completely screened). Therefore, as the coupling constant is increased, fewer Wilson lines survive as nontrivial DCFTs with $SO(6)_R$ symmetry. 
Since the theory admits electric-magnetic duality, this suggests that $SO(6)_R$-invariant 't Hooft lines have interesting dynamics already at weak coupling. In other words, there should be very few nontrivial $SO(6)_R$-invariant 't Hooft lines at weak coupling. We will see that this is indeed the case!

\subsection{2+1 dimensional critical points}

Wilson lines in conformal 2+1 dimensional theories, which we analyze in section \ref{Sec3dCFTs}, are interesting both theoretically but also because they correspond to charged impurities in 2+1 dimensional second order quantum phase transitions, and hence the predictions we make may be testable (in addition, there are recent numerical techniques which appear very promising~\cite{Hu:2023ghk} as well as advances on bootstrapping defects, e.g.~\cite{Collier:2021ngi,GimenezGrau:2022izf,Barrat:2022psm,Gimenez-Grau:2022czc,Bianchi:2022sbz,Ghosh:2023lwe}). We analyze 2+1 dimensional scalar and fermionic QED$_3$, with and without a Chern-Simons term. In these theories again Wilson lines of small enough charge flow to nontrivial DCFTs, while the others are screened. 

We consider both the tricritical and the ordinary scalar QED$_3$, which are related to second order superconducting transitions. In the tricritical point, all the Wilson lines that survive in the scaling limit we study are trivial (what that means precisely is that the number of nontrivial Wilson lines is much smaller than $N_f$ for large $N_f$), while for the ordinary scalar QED$_3$ we expect the number of nontrivial Wilson lines to scale with $N_f$ for large $N_f$ (but we do not determine the value of the critical charge in this theory). 
Extrapolating to small $N_f$ this has repercussions for charged impurities in the superconducting phase transition, and there could also be implications for 3d dualities.
For the fermionic QED$_3$ we find that the number of nontrivial Wilson lines scales as $\sim N_f$ and we determine in detail the precise bound on the charge of conformal line operators. We do not analyze explicitly the fate of Wilson lines with super-critical charge in any of these 2+1 dimensional examples, i.e. we do not compute in detail the screening cloud solitons, and we leave it for future work as well.

Finally, we study the 2+1 dimensional theory of graphene. This has four 2+1 dimensional fermions coupled non-relativistically to the electric field in 3+1 dimensions. Charged impurities of relatively low charge are screened and a cloud develops \cite{shytov2007atomic}. For better analytic control, we consider a generalization of graphene with $2N_f$ fermions, and compute the critical charge in the limit of large $N_f$ and compare with the experimental result \cite{wang2013observing}. Charged impurities with charge smaller than the critical one admit conformal line phases and interesting RG flows that have not been observed yet. 

\subsection{Comments on 't Hooft lines}

 We end this paper in section \ref{sectHooft} with a few  comments on 't Hooft lines in Abelian and non-Abelian gauge theories. We emphasize the properties of 't Hooft lines as DCFTs.
We compute the anomalous dimension of $\phi^\dagger\phi$ as a defect operator in scalar QED$_4$. Unlike the situation with Wilson lines, it always remains  irrelevant, and in fact, picks up a large positive anomalous dimension as the charge of the monopole or of the scalar field grows. 

For fermionic QED$_4$, it is well known that the lowest angular momentum modes of the fermion can penetrate the centrifugal barrier and the fermions should then be  treated carefully in the background of a monopole. We reinterpret these statements in terms of the spectrum of the defect. We show that there exists a marginal operator at tree level which is a fermion bilinear. Therefore, at tree level there is a continuous conformal manifold of possible 't Hooft lines in fermionic QED$_4$, corresponding to different boundary conditions for the fermions at the defect. This manifold is lifted at one loop and only one fixed point remains (see~\cite{vanBeest:2023dbu} and references therein).

 We note that internal symmetries that participate in a nontrivial two-group structure with the magnetic one-form symmetry are necessarily broken by the 't Hooft loops. This leads to tilt operators with $\hat \Delta =1 $ exactly. (In gauge theories with $\sum_i q_i\neq 0$, but where $\sum_i q^3_i\neq 0$, no spherically symmetric stationary 't Hooft loops exist. This can be interpreted as due to a two-group involving the Lorentz symmetry.) 

In the ${\mathcal{N}}=4$ SYM theory, we consider 't Hooft loops which are $SO(6)$-invariant. These are non-BPS 't Hooft loops which we can study at weak coupling. We argue that with gauge group $SU(N)$, they are all screened -- there is an instability towards condensing vector bosons which presumably form a screening cloud, canceling the bare magnetic field at the core altogether. 
With non-simple gauge groups, such as $PSU(N)= SU(N)/\mathbb{Z}_N$, we argue that there exist nontrivial 't Hooft line DCFTs, corresponding to anti-symmetric representations. This is consistent with the $\mathbb{Z}_N$ magnetic 1-form symmetry.
This picture is nicely consistent with $S$-duality which exchanges $PSU(N)$ and $SU(N)$ gauge groups~\cite{Aharony:2013hda}. We expect that all Wilson lines are screened as we increase the coupling constant in the $PSU(N)$ gauge theory. Wilson lines in the $SU(N)$ gauge theory cannot completely disappear though at strong coupling due to the electric $\mathbb{Z}_N$ one-form symmetry. The minimal conjecture, that only $N$ Wilson lines survive at strong coupling, is compatible with the fact that we have exactly $N$ nontrivial 't Hooft lines at weak coupling in the $PSU(N)$ gauge theory.

\section{Scalar \texorpdfstring{QED$_4$}{QED4}}\label{sec_ScalarQED}

\subsection{Two DCFT fixed points}\label{subsec_2DCFTs}
\label{subsec_TwoDCFTFixedPoints}

We first consider scalar QED$_4$ in (mostly minus) Minkowski signature with a charge $q$ Wilson line extending in the time direction:
\es{ScalarQED}{
\expval{W_q\, \mO_1 \dots}&=\int D\phi \, DA_\mu\,  \exp\le[i\int d^4x\, \le(\mL[A,\phi]-q \,\delta^3(\vec{x})A_0(x) \ri)\ri]\, \mO_1 \dots\,,\\
\mL[A,\phi]&=-{1\ov 4 e^2} F_{\mu\nu}^2+\abs{D_\mu\phi}^2-{\lam\ov 2 } \abs{\phi}^4\,.
}
The setup~\eqref{ScalarQED} defines a so-called defect QFT (DQFT). In the following we will study the properties of such DQFTs as a function of the charge of the Wilson line $q$ at weak coupling $e^2\sim\lambda\ll 1$. In most of the section we will assume that the scalar mass is tuned to zero, in order to get interesting low-energy physics.

By rescaling $\phi=\Phi/e$, we can see that taking the scaling limit 
\es{Scaling}{
e\to 0\,, \quad \lam \to 0 \,, \quad q\to \infty\,,\\
{\lam\ov e^2}=\text{fixed} \,, \quad {q\, e^2}=\text{fixed}\,,
}
leads to a problem that can be treated in the saddle point or semiclassical approximation, i.e.~by solving classical differential equations.
The saddle point equations are 
\es{SPeqs}{
\p_\mu F^{\mu\nu}+J^\nu&=e^2 q \, \de^\nu_t \,\delta^3(\vec{x})\,,\\
D_\mu D^\mu \Phi+{\lam\ov e^2} \abs{\Phi}^2 \Phi&=0\,,
}
where the expression of the current is $J_\mu={i\ov e}(\Phi^\dagger \p_\mu \Phi-\p_\mu \Phi^\dagger \Phi-2 i e \abs{\Phi}^2)$. In this section we look for interesting classical solutions of this system, which would be related to various infrared phases of the line operator. 

The first, obvious classical solution is
\es{SPsol}{
A_t&={e^2 q\ov 4\pi r}\,, \qquad \Phi=0\,.
}
This is the intuitive solution corresponding to the Coulomb field of a point charge and vanishing scalar field. This solution automatically obeys the $SL(2,\mathbb{R})$ symmetry and hence leads to a DCFT. However we will find that, depending on the parameters, the resulting DCFT may be sick and hence a different classical solution would have to be identified.

To investigate the DCFT associated to the saddle point~\eqref{SPsol} we consider fluctuations around the saddle point. We focus on scalar fluctuations, and we find that close to the Wilson loop they behave as
 \es{PhiFluct}{
 \Phi(x)&=\SumInt_{\om,\ell,m} \Phi_{\om \ell m}(r) \,e^{-i\om\tau}\, Y_{\ell m}(\Om) \,,\\
  \Phi_{A}(r) &=\al_A \, r^{-\nu_\ell-1/2}\le(1+{q\,\om \,r\ov \nu_\ell-1/2}+\dots\ri)+\beta_A\, r^{\nu_\ell-1/2}\le(1-{q\,\om\, r\ov \nu_\ell+1/2}+\dots\ri)\,,\\
    \nu_\ell&\equiv\sqrt{\frac14+\ell(\ell+1)-g^2}\,,\qquad g\equiv {e^2 q\ov 4\pi}\,,
}
where $A\equiv \om \ell m$ is a superindex. To make the setup well-defined we have to choose boundary conditions that will fix $\beta_A=F(\al_A)$; we have more to say on this below.
As long as $\nu_\ell$ is real, with appropriate boundary conditions, the respective modes lead to sensible creation operators. However, as we increase $g$, starting from the $\ell=0$ mode, we encounter imaginary $\nu_\ell$. This leads to an instability as we will soon explain. 
The  ${\ell=0}$ mode will be in our focus in the following, and we repeat the expression for $\nu\equiv \nu_{\ell=0}$ here:  
\es{nu_highlight}{
    \nu&\equiv\sqrt{\frac14-g^2}\,,\qquad g\equiv {e^2 q\ov 4\pi}\,.
}

Before we plunge into the details let us summarize what we do below: 
\begin{itemize} 
\item For $g^2<1/4$ we have $\nu>0$ and two different possible power law falloffs, as in~\eqref{PhiFluct}. We will find  two conformal boundary conditions, and hence two different versions of the DCFT with the one-point functions~\eqref{SPsol}. These are two different conformal Wilson lines between which there is an RG flow (one of the conformal Wilson lines is RG stable), and we will analyze the spectrum of defect operators in both. We will also see that something interesting happens for $\nu=1/4$ to the unstable conformal boundary conditions.
\item We get a critical line for $g^2=1/4$ for which the two power laws degenerate.  The DCFT has a marginally irrelevant operator at this point. 
\item For $g^2>1/4$ we see that $\nu$ is purely imaginary and $\Phi$ fluctuations exhibit oscillations in the radial direction. We will argue that this signals an instability, in part, because defect operators cannot have imaginary scaling dimensions. In this regime the saddle point~\eqref{SPsol} needs to be replaced by a different one. Importantly, we find a new saddle point which we call the ``scalar cloud'', and we show that it leads to a sensible physical picture for $g^2>1/4$, where very far away from the scalar cloud the initial charge is completely screened. For $\lambda=0$ the scalar operator admits a nontrivial one-point function at long distances, while for a nontrivial bulk scalar potential all the one-point functions vanish. For $g^2$ slightly larger than $1/4$ the scalar cloud is exponentially large and generates a new scale in the system. 
\end{itemize}

\begin{figure}[h!]
\centering
\includegraphics[width=77mm]{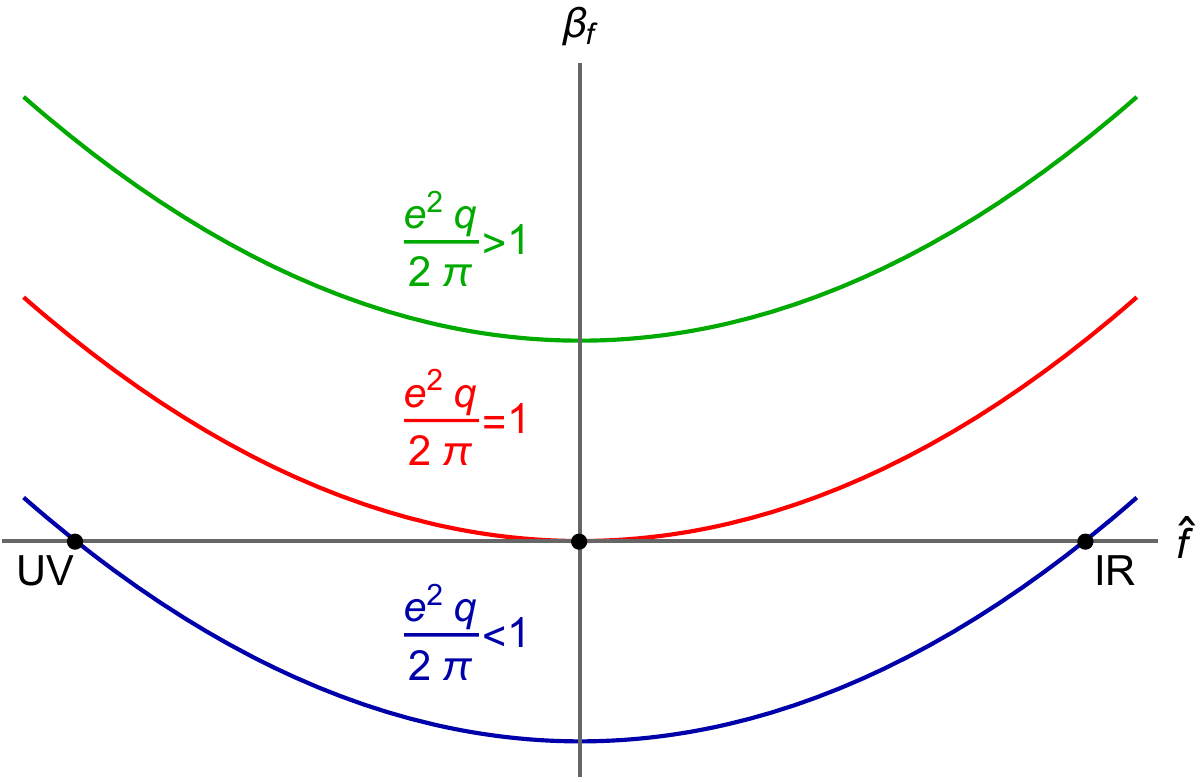}
\caption{An illustration of the $\beta$-function associated with the parameter $\hat f$ in equation \eqref{eq_beta_final}.   \label{fig_betaFunction}}
\end{figure}

It will be technically advantageous to exploit the fact that massless scalar QED$_4$ is classically conformally invariant, and perform the Weyl transformation\footnote{Whether the theory is a DCFT at the classical level depends on the boundary conditions we choose. However, even if the boundary conditions break (boundary) conformal invariance, we can perform the Weyl transformation, since the bulk remains conformal.} 
\es{WeylTF}{
ds^2&=r^2\le[{-dt^2+dr^2\ov r^2}+ds_{S^2}^2\ri]\\
&\equiv r^2 d\tilde s^2_\text{AdS$_2\times S^2$}
}
which maps the flat space problem to a problem in AdS$_2\times S^2$, with the defect now at the asymptotic boundary of AdS$_2$. The scalar fluctuations in AdS are related to those in~\eqref{PhiFluct} through $\Phi_A={1\ov r} \tilde\Phi_A$ and the gauge field background from~\eqref{SPsol} is unchanged. Through this mapping, we can readily borrow results from the AdS/CFT literature about boundary conditions on scalar fields; we give a quick overview below, with the final result obtained in  \eqref{eq_bdry_action_final} and figure~\ref{fig_betaFunction}.
Specializing to the $\ell=0$ mode, the near-boundary behavior is: 
\es{special_PhiFluct}{
\tilde\Phi_{\om}(r) &=\al_\om \, r^{1/2-\nu}\le(1+{q\,\om \,r\ov \nu-1/2}+\dots\ri)+\beta_\om\, r^{1/2+\nu}\le(1-{q\,\om\, r\ov \nu+1/2}+\dots\ri)\,.
}
We will drop the tilde from $\Phi$ from here onwards.

It will be useful for our purposes below to introduce a small radial cutoff at $r=r_0$ and never remove it throughout the computation. First, we return to the question of boundary conditions. These are determined from the variation of the action and keeping track of boundary terms. Varying the action $S_\text{bulk}=\int_\text{AdS$_2\times S^2$}\sqrt{-g}\, \mL[A,\phi]$ from \eqref{ScalarQED} and imposing the bulk equations of motions gives the boundary term (for $\nu > 0$)
\es{BulkVar}{
\de S_\text{bulk}=r_0^{-2\nu}\,{1-2\nu\ov 2}\int d\om \le(\al_\om^\dagger \de\al_\om+\text{c.c.}\ri)+\text{(subleading)}\,.
}
The subleading terms will be important at the next step, where we will write them out explicitly. 
The leading term vanishes if
\es{StandBC}{
\al_\om=\al_\om^\dagger=0\,,
}
whereas $\beta_\om$ is a fluctuating degree of freedom. These boundary conditions are analogous to the Dirichlet boundary conditions, since the more singular mode in~\eqref{special_PhiFluct} is set to zero. Since the boundary terms have to vanish identically on the equations of motion, at finite $r_0$ we need to slightly correct the boundary conditions, or add additional boundary terms, to cancel the subleading terms. We will only write out the most important such terms. The same comment applies below. 

We can read out the scaling dimension of the defect operators for the conformal boundary condition~\eqref{StandBC}. We see that the $\beta_\om$ correspond to operators of dimension  $1/2+\nu$. Remember that we are studying a gauge theory and hence only gauge-invariant operators should be considered, and thus the scaling dimension of the bilinear $\beta^\dagger\beta$ is \begin{equation}\hat \Delta(\beta^\dagger\beta) =1+\sqrt{1-4g^2} ~.\end{equation}
 For $g\rightarrow 0$ (i.e. the trivial defect $q=0$) the scaling dimension becomes $\hat\Delta=2$, coinciding with the dimension of the bulk operator $\Phi^\dagger\Phi$. Therefore the boundary condition \eqref{StandBC} defines the usual Wilson line operator, in the sense that if we set $q=0$ this boundary condition means that there is no defect at all.\footnote{From the perspective of standard perturbation theory (where $q$ is usually taken to be $O(1)$), the result for $\nu$ in \eqref{PhiFluct} re-sums the contribution of infinitely many Feynman diagrams to the anomalous dimensions of defect operators of the form  $\pd^\#\Phi^\dagger\pd^\#\Phi$.  This is analogous to what happens in other semiclassical limits, see e.g. \cite{Beccaria:2022bcr,Cuomo:2022xgw,Rodriguez-Gomez:2022gbz,Rodriguez-Gomez:2022xwm} in a similar context. We checked explicitly the agreement of the semiclassical result \eqref{PhiFluct} with a one-loop diagrammatic calculation of the anomalous dimension of $\Phi^\dagger\Phi$ on the defect.}

There is a twist in the story: the boundary condition~\eqref{StandBC} is not unique. 
We can add boundary terms and they can change the boundary conditions and the boundary operator spectrum~\cite{Klebanov:1999tb,Witten:2001ua,Gubser:2002vv,Faulkner:2009wj,Aharony:2015afa}. Let us add the following boundary term:
\begin{equation}\label{eq_action_bdry1}
S_\text{bdy}^{(1)}=-\frac{1-2\nu}{2}\int_{r=r_0} dt\,\sqrt{-\hat{g}}\,|\Phi|^2\,,
\qquad d\hat{s}^2=\frac{-dt^2}{r_0^2}\,,
\end{equation} 
which is carefully tailored to cancel the leading term in the variation~\eqref{BulkVar}. Combined with $S_\text{bulk}$ we now have the variation:
\begin{equation}\label{BulkBdyVar}
\begin{split}
\de \le(S_\text{bulk}+S_\text{bdy}^{(1)}\ri)=\int d\om&\le[2\nu\le(\beta_\om^\dagger \de\al_\om+\text{c.c.}\ri)+2\nu \,r_0^{2\nu}\le(\beta_\om^\dagger \de\beta_\om+\text{c.c.}\ri)
\right.\\ &\left.
+{2q\,\om \,r_0^{1-2\nu}\ov 2\nu-1}\le(\al_\om^\dagger \de\al_\om+\text{c.c.}\ri)+\text{(subleading)}\ri]\,.
\end{split}
\end{equation}
Since for $g^2<1/4$ we have $0< \nu< 1/2$, the first term is dominant and permits the choice of boundary condition\footnote{Note that the boundary condition in~\eqref{StandBC} remains viable. With that choice the boundary term we added evaluates to $0$ (to the order in $r_0$ we are working).}
\es{AltBC}{
\beta_\om=\beta_\om^\dagger=0\,.
}
$\al_\om$ becomes a boundary degree of freedom with dimension $1/2-\nu$. 
Again, only bilinears are gauge-invariant.  In the AdS/CFT literature this is known as the `alternative quantization' \cite{Klebanov:1999tb} of the scalar $\Phi$.  The boundary condition~\eqref{AltBC} describes a new DCFT, with a different operator spectrum from the usual Wilson line defined by~\eqref{StandBC}. In particular, the lowest dimensional gauge-invariant operator is the bilinear $\alpha^\dagger\alpha$, which has scaling dimension $\hat \Delta = 1-2\nu<1$ and it is therefore relevant. This will be important below, as adding this operator to the action leads to an RG flow. 

A remark is in order. Our analysis of the linearized fluctuations suggests that alternative quantization defines a unitary DCFT for arbitrary $0<\nu<1/2$, i.e. for arbitrary $|q|<2\pi^2/e^2$. This would mean that even low-charge Wilson lines have two different possible fixed points. 
As we will explain in the next section, this conclusion is not entirely correct due to nonlinear interactions between the fluctuations. Eventually, we will only claim that alternative quantization defines a healthy DCFT in the window $0<\nu<1/4$, which means that there are two fixed points for the Wilson line starting from charge $|q|={ \sqrt 3 \pi\over e^2}$   up to the unitarity bound $|q|= {2\pi \over e^2}$.

In our problem we have $1/2>\nu>0$, which in the infrared fixed point~\eqref{StandBC} means that the lowest lying bilinear operators cover the range from $\hat\De=2 $ with no impurity to $\hat\De=1$ at the bound $\nu=0$,  while in the alternative quantization fixed point~\eqref{AltBC} we cover the range from $\hat \De=0$ when there is no impurity to $\hat \De=1$ at the bound $\nu=0$. These are consistent with the unitarity bound $\De\geq\max\le({d-2\ov2},0\ri)$, which for $d=1$ is $\hat\De\geq0$.\footnote{The alternative quantization window commonly quoted in the literature is $0<\nu<1$. However, the range $1/2<\nu<1$ (which is not realized in our problem) clearly would give rise to a non-unitary alternatively quantization DCFT since the scaling dimension would be negative. The scalar theory in $AdS_2$ in this mass range therefore develops a sickness with alternative boundary conditions earlier than in higher dimensions. See \cite{Iqbal:2011aj} for related comments.}

There is a way to interpolate between alternative and standard quantization: they are connected by an RG flow (referred to as the double-trace flow in AdS/CFT) that is triggered by adding the relevant operator $\abs{\al}^2$ to the alternative quantization DCFT action. This is implemented by the additional boundary term:
\begin{equation}\label{eq_action_bdry2}
S_\text{bdy}^{(2)}=-f_0\int_{r=r_0} dt\,\sqrt{-\hat{g}}\,r_0^{2\nu}|\Phi|^2\,.
\end{equation} 
This term is chosen so that it reduces to $\abs{\al}^2$ in the limit $r_0\rightarrow 0$.
Upon imposing \linebreak
\mbox{$\de \le(S_\text{bulk}+S_\text{bdy}^{(1)}+S_\text{bdy}^{(2)}\ri)=0$} and defining the dimensionless coupling constant $f$ through $f_0=f r_0^{-2\nu}$, we obtain  the boundary condition
\es{eq_deltaS}{
 \frac{\beta_\om}{\alpha_\om}&=\frac{f_0}{2\nu -f_0 r^{2\nu}_0}\\
&={f \ov 2\nu-f}\,r_0^{-2\nu} \,,
}
where we took $r_0 \omega\ll 1$ and only kept the leading term.\footnote{While we may contemplate whether $f_0$ needs to be a lot smaller than the cutoff scale $r_0^{-2\nu}$, and hence $f$ needs to be infinitesimal, this result is trustworthy for $f=O(1)$.}
For $\nu > 0$ in the limit $r_0\rightarrow 0$ this is the well known result for the double-trace deformation~\cite{Witten:2001ua,Gubser:2002vv,Faulkner:2009wj,Aharony:2015afa}.  To extract the beta function of $f$ for arbitrary $\nu$ we demand that the boundary conditions \eqref{eq_deltaS} (and hence the physical theory) are left invariant by a simultaneous rescaling of the cutoff and the coupling. This Callan-Symanzik style argument~\cite{Kaplan:2009kr} 
implies that \mbox{$r_0\frac{\pd (\beta_\om/\alpha_\om)}{\pd r_0}-\beta_f \frac{\pd (\beta_\om/\alpha_\om)}{\pd f}=0$}, which leads to the beta function
\begin{equation}\label{eq_beta_epsBig}
\beta_f=-2\nu f+f^2\,.
\end{equation}
For real $\nu$, we find two fixed points: $f=0$ is the UV (alternative quantization) and  $f=2\nu$ the IR (standard quantization) DCFT limit of the resulting RG flow. The RG flow corresponds to interpolating between the boundary conditions in~\eqref{AltBC} and \eqref{StandBC}. For $f<0$ we have $\beta_f>0$ leading to the runaway behavior $f\to  -\infty$ in the IR. In other words, in the alternative quantization fixed point, which exists for positive $\nu$ at $f=0$, with positive sign for $f$ the deformation~\eqref{eq_action_bdry2} leads to the standard Wilson line fixed point, while with a negative sign for $f$, a long flow towards infinitely negative $f$ ensues and the dynamics has to be understood. 
We will provide a physical interpretation of this runaway in the next section.

Let us consider in some detail the special case when the coupling $f$ is marginal, $\nu=0$, and only exhibits logarithmic running. Then the two falloffs in \eqref{PhiFluct} degenerate and we have to implement the change of basis
\begin{equation}\label{eq_critical_falloffs_trick}
\alpha_\om=\frac12 \left(-\frac{a_\om}{\nu}+b_\om\right)r_0^{\nu}\qquad\text{and}\qquad
\beta_\om = \frac12\left(\frac{b_\om}{\nu}+a_\om\right)r_0^{-\nu}\,,
\end{equation} 
for some arbitrary cutoff radius $r_0$. Then we can take the limit $\nu\rightarrow 0$:
\begin{equation}\label{eq_critical_falloffs}
\Phi\sim b_\om\sqrt{r}\log( r/r_0)+a_\om\sqrt{r}\,,\qquad
\nu=0\,.
\end{equation}
We can directly read off the evolution of the coupling constant $f$: 
\es{flog}{
f_*\log(r/r_*)+1&=f(r_0)\log(r/r_0)+1 \quad\implies\quad f(r_0)={f_*\ov 1- f_* \log(r_0/r_*)},
}
where $r_*$ is some reference scale, or equivalently 
\es{flog2}{
\beta_f=f^2\,,\qquad
\nu=0\,.
}
This agrees with \cite{Aharony:2015afa} for $\nu=0$. Note that while for $\nu>0$ we could have taken the $r_0\to 0$ limit from early in the calculation, it is essential to keep $r_0$ finite to make sense of the marginal case with $\nu=0$ that corresponds to the critical Wilson loop.
The cutoff is also necessary to study the supercritical case, where $\nu$ is imaginary and no real DCFT exists. We will study the supercritical case below.

Since our theory is naturally equipped with a cutoff $r_0$, the beta function \eqref{eq_beta_epsBig} should make sense also in the region $g^2>1/4$, where  the coupling $f$ however cannot be thought as a perturbation of a (unitary) DCFT.  There is a trick to rewrite the beta function and the coupling in a way that would make sense with real couplings for both $g^2>1/4$ and $g^2<1/4$. To see this, note that in terms of the dimensionless coupling $f=f_0 r_0^{2\nu}$, the boundary action with the double-trace deformation can be written as:
\begin{equation}\label{eq_bdry_action_final}
S_\text{bdy}^{(1)}+S_\text{bdy}^{(2)}=-\frac{1+2\hat{f}}{2}\int_{r=r_0} dt \sqrt{-\hat{g}} \,|\Phi|^2\,,\qquad
\hat{f}\equiv f-\nu\,.
\end{equation}
We can make sense of this in the region where $\nu^2<0$ by choosing a complex $f$, so that the coupling $\hat{f}$ in~\eqref{eq_bdry_action_final} is real. In terms of $\hat f$ the beta function \eqref{eq_beta_epsBig} reads:
\begin{equation}\label{eq_beta_final}
\beta_{\hat{f}}=-\nu^2+\hat{f}^2\,.
\end{equation}
This result holds both for $\nu^2\geq 0$ and $\nu^2<0$. Crucially,  the coupling $\hat{f}$ is real in both regions, making sure that the theory stays unitary.

For $\nu^2<0$ the beta function  \eqref{eq_beta_final} does not admit fixed points at finite coupling. Instead, it is associated with a dimensional transmutation phenomenon usually referred to as ``walking" behavior \cite{Kaplan:2009kr,Gorbenko:2018ncu}.  Suppose that the RG flow starts from a {small} initial value for the coupling $\hat{f}(\mu_{UV})$. The coupling constantly  decreases along the RG, but the {rate} is slower near $\hat{f}=0$ where the beta function is small. Eventually $-\hat{f}(\mu)$ blows up at a scale $\mu_{IR}$ given by
\begin{equation}\label{eq_walking}
\mu_{IR}=\mu_{UV}e^{-\frac{\pi}{|\nu|}}\,.
\end{equation}
We see that the dynamically generated scale is exponentially separated from the UV one for small $\nu^2<0$. There is therefore dimensional transmutation on the line defect. We will discuss the physical implications of this RG flow in section~\ref{subsec_scalar_supercritical}.

In summary, we learn that there are two DCFT fixed points for subcritical Wilson lines in scalar QED in the double scaling limit. They correspond to different boundary conditions for the field $\phi$ and are connected by an RG flow. The flow is triggered by the gauge-invariant relevant defect operator $\abs{\phi}^2$ that has dimension $1-2\nu$ in the UV DCFT (alternative quantization) and that becomes irrelevant with dimension $1+2\nu$ in the IR DCFT. For $\nu=0$, corresponding to the critical Wilson line with $q=2\pi/e^2=1/(2\al_\text{QED})$, the two fixed points merge. For $q>2\pi/e^2$ they annihilate and there is no DCFT. Instead there is a runaway towards large negative $\hat f$.

The fixed point annihilation corresponds to the supercritical regime of Wilson lines, whose physics we explore in section \ref{subsec_scalar_supercritical}. This problem is closely related to another one we encountered above:
If we deform the subcritical UV DCFT (alternative quantization) with $f<0$, we encounter a runaway behavior. Finally, in the following we also analyze the stability of the alternative quantization DCFT in the subcritical regime at the nonlinear level.

\subsection{Stability in the subcritical regime}\label{subsec_scalar_stability}

The abstract DQFT viewpoint was very useful to  interpret the $f>0$ regime of the phase diagram, where we found two DCFTs. In particular, standard quantization (with $f=2\nu$) is stable for small deformations. From the beta function in~\eqref{eq_beta_epsBig} we also found a runaway behavior for $f<0$. It is not uncommon that such a behavior indicates instability. We will show below that this is indeed the case. We will find that for $\nu>1/4$ this instability leads to the formation of a  classical soliton that we construct numerically. Perhaps more surprisingly, we will find hints that for $\nu>1/4$ the alternative quantization fixed-point admits solitons with arbitrary negative energy for any value of the deformation $f$. We will explain why $\nu=1/4$ is special.

To perform the analysis, we view our setup as a problem in differential equations. By changing $f$ we are changing the boundary conditions for the  equations of motion (EOM) \eqref{SPeqs}. Since these are nonlinear, it is possible that there is an interesting phase diagram as we change~$f$. 

The profile $\Phi=0,\, A_t=g/r$ is always a solution of the EOM. The RG analysis predicts that for $f>0$ this solution is stable. For $f<0$ we will find that it becomes unstable, and a new soliton solution takes over. The instability can be diagnosed in two different ways. First, in appendix \ref{app_KG_prop} we compute the $\expval{\al_\om\al^\dagger_\om}$ retarded two-point function and show that for $f<0$ it has a tachyon pole in the upper half plane, the telltale sign of a dynamical instability. The endpoint of the instability is the soliton that we construct below. Second, we show that the soliton has lower energy than the $\Phi=0,\, A_t=g/r$ solution, which is a ``thermodynamical'' demonstration of instability.

A simple argument establishes that the soliton cannot end up partially screening the Wilson line: the screening has to be complete.\footnote{That is to leading order in the double scaling limit. Later we discuss situations in which total screening is in tension with symmetry considerations at subleading order in $e^2$.} Let us assume that partial screening was possible; in the IR we have $A_t=g'/r$. If $\Phi=$const, the Maxwell equation \eqref{SPeqs} is not satisfied due to the $A_t \abs{\Phi}^2$ term in the gauge current. So we have to assume that $\Phi$ is small. We know the possible small $\Phi$ behavior in a Coulomb background from \eqref{PhiFluct}, $\Phi\sim r^{1/2\pm \nu}$ with $0<\nu<1/2$. Hence $\Phi$ is growing (instead of decaying), and it starts backreacting on $A_t$, ruining the assumed Coulomb behavior. We have thus reached a contradiction, and the only possible way out is to have complete screening (or $\Phi=0$ throughout the bulk). Next we construct the explicit new soliton corresponding to screened Wilson loops.

To perform the computation, first, we forget about the boundary conditions at $r=r_0$ and construct solutions to the EOMs of $\Phi$ and $A_t$ which are regular as $r\to\infty$. Since the equations are second order in derivatives and regularity provides two conditions, the resulting solution depends on two constants. One is simply a length scale $\xi$, while the other is a dimensionless parameter  that we denote by $c$.  These parametrize the asymptotic form of the solution as $r\to\infty$.  Explicitly, the asymptotics are different depending on the value of $\bar\lam\equiv\lam/e^2$,  and are given by (we have obtained a couple of more orders of the asymptotic expansions that we suppress here to avoid clutter):
\es{ThreeCases}{
\lam=0: \qquad&\begin{cases}
\Phi={c\ov\sqrt 2}+\dots\\
r\, eA_t=(r/\xi)^{-\frac12-\sqrt{\frac14+c^2}}+\dots\end{cases}\\[8pt]
\bar\lam<2: \qquad&\begin{cases}
\Phi={1\ov \sqrt{2\bar\lam\,\log(r/\xi)}}+\dots\\
r \,eA_t=c\le[\log(r/\xi)\ri]^{-1/\bar\lam}+\dots\end{cases}\\[8pt]
\bar\lam>2: \qquad&\begin{cases}
\Phi={1\ov 2\sqrt{\log(r/\xi)}}\le[1+\dots+c \le[\log(r/\xi)\ri]^{1-{\bar\lam\ov2}}+\dots\ri]\\
r \,eA_t={\sqrt{\bar\lam-2}\ov 2\sqrt{\log(r/\xi)}}\le[1+\dots+{2c\ov\bar\lam-2} \le[\log(r/\xi)\ri]^{1-{\bar\lam\ov2}}+\dots\ri]\end{cases}
}
Note that $c$ denotes different things in the different cases and  for $\bar\lam>2$ it is hiding at a subleading order (as a noninteger power term). Also note that we set the AdS radius equal to one, which makes up for the missing dimensions in the above equations.

Given a solution with the above asymptotics as $r\to\infty$, we may integrate the EOMs towards smaller $r$. 
We may then obtain the near-defect boundary conditions, namely the charge of the Wilson line $g$ and the value of the double-trace coefficient $f$, that correspond to a given choice of the parameters $\xi$ and $c$. 

In practice, we can only solve the EOMs numerically. To this aim, we set $\xi=1$ and use the asymptotics as initial data for the numerical integration of the EOM starting at some $r=r_c$ and integrating towards smaller $r$ (up to some small $r_0$). Let us denote the resulting solution by $\{\varphi^{(c)}(r),\,{\cal A}^{(c)}_t(r)\}$. We can reinstate $\xi$ by a simple rescaling, thereby obtaining a two-parameter family of solutions
\es{twopar}{
\le\{\varphi^{(c)}(r/\xi),\,{{\cal A}^{(c)}_t(r/\xi)\ov\xi}\ri\}\,.
}
If we denote the near-defect asymptotic data corresponding to the solution $\{\varphi^{(c)}(r),\,{\cal A}^{(c)}_t(r)\}$ by $\{\al^{(c)},\,\beta^{(c)},\, g^{(c)}\}$, as in \eqref{special_PhiFluct}:\footnote{The subleading behavior of the gauge field provides an additional boundary datum. We omit it below, as it does not play a role in our discussion.}
\es{solitonAsym}{
\varphi^{(c)}(r) &=\al^{(c)} \, r^{1/2-\nu}+\beta^{(c)}\, r^{1/2+\nu}\,,\\
 {\cal A}^{(c)}_t&={g^{(c)}\ov r}  \,,
}
 then the two-parameter family of solutions in \eqref{twopar} has asymptotic data
\es{twoparAsymp}{
\le\{\al={\al^{(c)}\ov \xi^{1/2-\nu}},\,\beta_\text{sol}(\al)={\beta^{(c)}\ov \xi^{1/2+\nu}},\, g^{(c)}\ri\}\,,
}
where by writing $\beta_\text{sol}(\al)$ we emphasize that the family of solitons characterized by {\it fixed $c$ and varying $\xi$} gives a curve in the $(\al,\beta)$ plane.

From~\eqref{twoparAsymp}, we learn that we can use $c$ as a proxy for $q$ (or $g$). Then for fixed $q$ (or equivalently $c$) we can use $\xi$ to tune the absolute value of the ratio of $\al/\beta_\text{sol}(\al)$. The sign of the ratio $\beta^{(c)}/\al^{(c)} $ determines the sign of the coupling $f$ corresponding to the so-constructed soliton.  More in detail, combining \eqref{eq_deltaS} with \eqref{twoparAsymp},  for infinitesimal $f$ we have
\es{fxi}{
{\beta^{(c)} \ov \al^{(c)} }={f \ov 2\nu}\, (\xi/r_0)^{2\nu}\,.
}
Since without loss of generality we choose $\Phi$ to be positive near the boundary giving $\al>0$, it is the sign of $\beta^{(c)}$ that correlates with that of $f$.  Naively, one might expect all the solitons that can be constructed in this way to correspond to a negative value of the double-trace coefficient $f$ (since this is the region where an instability exists for positive $\nu$, while for positive $f$ we expect the dominant saddle point to be~\eqref{SPsol}). Rather surprisingly, we find the following intriguing pattern of signs as a function of $g=e^2 q/(4\pi)$ with blue negative and red positive:
\begin{figure}[h!]
\centering
\includegraphics[scale=0.7]{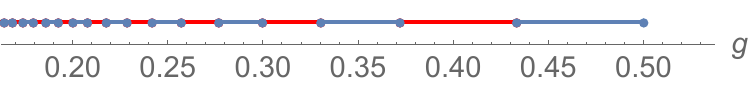}
\caption{Sign of $f$ corresponding to the numerical soliton as a function of $g$.}
\end{figure}

We will explain this pattern below analytically, showing that the red regions are associated with the existence of dangerous configurations whose energy seems unbounded from below. These have a simple interpretation in terms of the renormalization group, as we will see.

From \eqref{twoparAsymp} we observe that from the combination
\es{sDef}{
s(g)\equiv - {\beta_\text{sol}(\al)\ov \al^{1/2+\nu\ov1/2-\nu}}=- {\beta^{(c(g))}\ov \le[\al^{(c(g))}\ri]^{1/2+\nu\ov1/2-\nu}}
} 
$\xi$ drops out, hence $s(g)$ is a useful characterization of the nonlinear response of our system to turning on $\Phi$. The sign of $s(g)$ is then anti-correlated with the sign of $f$ that is required to get a solitonic solution. A useful construction borrowed from the AdS/CFT literature is as follows \cite{Faulkner:2010fh,Faulkner:2010gj}. To avoid clutter let us use the notation $A=\sqrt{\al^\dagger \al},\, B=\sqrt{\beta^\dagger \beta}$ and specify the general non-linear boundary condition
\es{GeneralBC}{
B={\cal W}'(A)\,.
}
For example the interpolating flow between alternative and standard boundary conditions corresponds to ${\cal W}(A)=f A^2/2$. We also define 
\es{WVdef}{
{\cal W}_0(A)&\equiv\le(\frac12-\nu\ri)s(g)\,A^{1\ov 1/2-\nu}\,,\\
{\cal V}(A)&\equiv 4\nu\le[{\cal W}(A)+{\cal W}_0(A)\ri] \,,
}
where ${\cal V}(A)$  is the effective potential. That is,  ${\cal V}(A)$ 
is genuinely the (leading order) quantum effective potential, i.e.~(minus) the 1PI effective action evaluated for constant $\expval{A}$; see Appendix~\ref{app:EffectivePot} for a derivation similar to~\cite{Faulkner:2010fh,Faulkner:2010gj}.   One can verify that the solitonic solution that satisfies the boundary condition \eqref{GeneralBC} is a critical point of ${\cal V}(A)$; this is a consistency check of the formalism.  (Recall that ${\cal W}'_0(A)=s(g) \, A^{1/2+\nu\ov1/2-\nu}=-B_\text{sol}(A)$, where we used \eqref{sDef}.)  Since the value of the effective potential is zero for the naive saddle $A=0$ and negative for the soliton critical point, we conclude that it is the energetically favored configuration, hence establishing thermodynamic stability.

Next we ask if we can provide an analytic understanding of the sign structure of $s(g)$ (previewed below \eqref{fxi}). We will first explain what happens to the bulk scalar profile at the special points where the sign of $s(g)$ flips, and then we interpret the values of $g$ where the sign flips take place from the point of view of the defect renormalization group.

\begin{figure}[!h]
\centering
\includegraphics[scale=0.8]{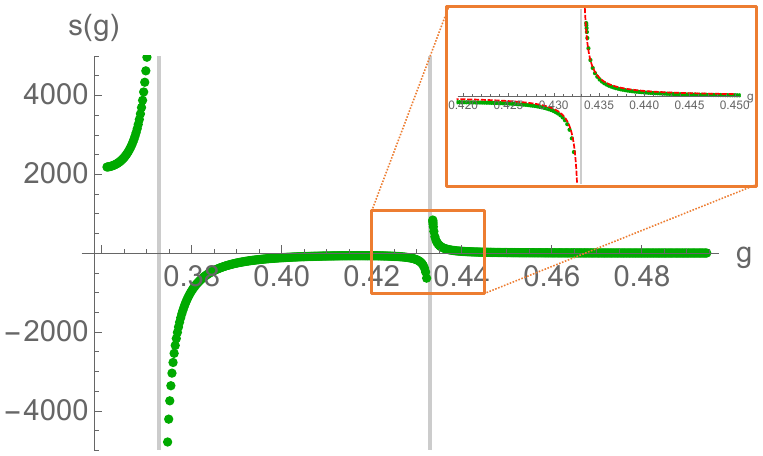}
\caption{Plot of $s(g)$ for $\lam=0$ obtained from numerics. In the inset the numerical data are plotted together with the analytic formula \eqref{sgnu} (red dashed line) describing the behavior of $s(g)$ near the singularity at $g=\sqrt{3}/4$; we get perfect agreement.}\label{fig:sgplot}
\end{figure}

The near boundary analysis of the equations (at $\om=0$, but going beyond the terms displayed in \eqref{PhiFluct}) gives for $\Phi$:
\es{PhiNB}{
\Phi=&\al\, r^{1/2-\nu}\le[1+ \al^2\, {1+2(1+\bar \lam)\nu\ov 4\nu(1-2\nu)(1-4\nu)}\, r^{1-2\nu}+\dots \ri]+\beta  \, r^{1/2+\nu}\le[1+\dots \ri]\\
&+\text{(cross terms between $\al$ and $\beta$)}\,.
}
Note that the exponent in the $\al^3$ term coincides with that of the $\beta$ term for $\nu=\frac14$. Exactly at this point, the coefficient of the $\al^3$ term diverges.
Using the relation between $\beta$ and $\al$ from \eqref{sDef}, we see that the only way for us to get a regular scalar profile $\Phi$ (which we expect, since nothing drastic happens in the bulk), is to have:
\es{sgnu}{
s(g)=-{(3+\bar \lam)/4\ov \nu(g) -1/4}+\dots\,, \qquad \text{(for $\nu\to 1/4$).}
}
Since $\nu=1/4$ corresponds to $g=\sqrt{3}/4=0.43$, we have successfully explained the first sign change (counting from $g=1/2$) of $s(g)$ that we see on the diagram below \eqref{PhiFluct}, see also figure~\ref{fig:sgplot}.
 It turns out that all the sign changes are explained by an $\al^{2k+1}$ term colliding with the $\beta$ term, giving 
$\nu_k={k\ov 2(k+1)}$ for the $k$'th sign change point. 

From the defect point of view we have already mentioned that something special happens at $\nu=1/4$, now we can explain why. 
At these points the operators $\le(\al^\dagger \al\ri)^{k+1}$ become marginal. On the other side of this transition point, we have a new relevant operator, which can be added to the action (included in ${\cal W}$) without spoiling the UV alternative quantization DCFT.\footnote{We also need new boundary terms for each new relevant operator that supplements \eqref{eq_action_bdry1}. A worked out example can be found in Appendix D of \cite{Faulkner:2010gj}.} This can be understood straightforwardly also from~\eqref{AltBC}, where we have found that in alternative quantization the scaling dimension of the charged boson is $(1-2\nu)/2$, and hence, at $\nu=1/4$ the quartic operator becomes relevant in the alternative quantization fixed point.

Irrespective of what coefficients we choose for the relevant terms, the asymptotics of the effective potential is always determined by the conformal term $s(g) A^{1\ov1/2-\nu}$, hence when $s(g)<0$ the effective potential is unbounded from below.\footnote{In itself this does not in itself signal pathology as evidenced by the negative effective potential for the Hubbard-Stratonovich field in the large $N$ critical $O(N)$ model \cite{Coleman:1974jh}, which is a perfectly healthy theory.} We speculate that in our case this indicates an instability of the system, as it seems to allow for the construction of a configuration with arbitrarily negative energy.\footnote{A circumstantial evidence is that the analogous transition in holography from $s>0$ to $s<0$ is accompanied by the loss of the positive energy theorem \cite{Faulkner:2010fh}.} However, we have not constructed such a configuration explicitly and we have not found a defect QFT explanation for the potential loss of stability of alternative quantization. 
One scenario is that since the quartic operator becomes relevant, the alternative quantization fixed point is lost, due to the joint beta function of the bilinear and quartic operator not having mutual zeroes. What we are sure about is that alternative quantization is a healthy DCFT in the regime $\sqrt{3}/4<g<1/2$. We leave the interesting problem of understanding the regime $g<\sqrt{3}/4$ for future work. 

\begin{figure}[!h]
\centering
\includegraphics[width=48mm]{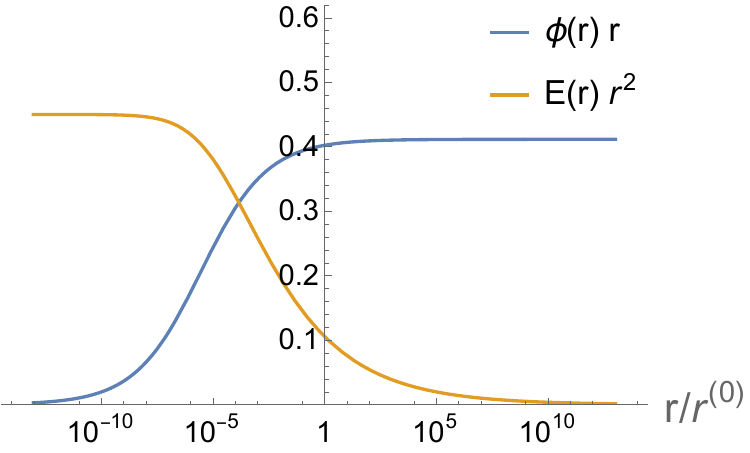}\hspace{0.2cm}
\includegraphics[width=48mm]{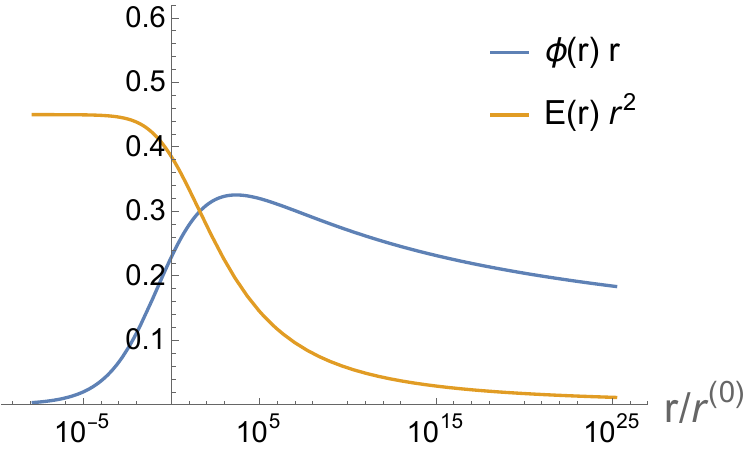}\hspace{0.2cm}
\includegraphics[width=48mm]{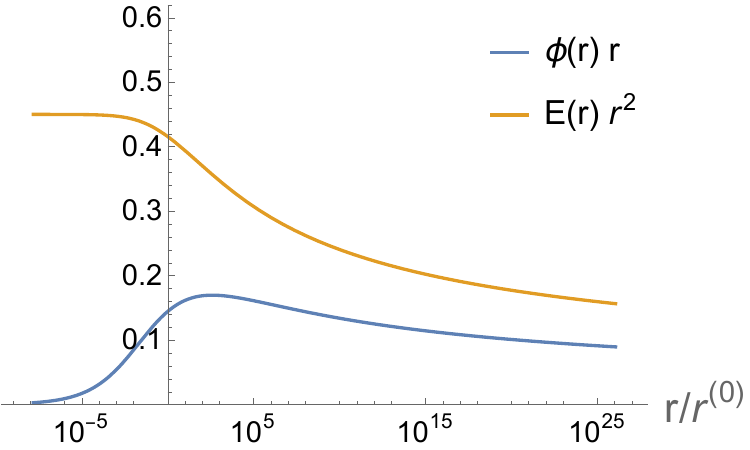}
\caption{Plots of the scalar profile (blue) and the electric field $E=F_{tr}$ (orange) as functions of the distance from the probe charge, all normalized to be dimensionless.  
The analysis was carried out in the subcritical regime for $\frac{e^2q}{2\pi}=0.9$ corresponding to $g=0.45$. The left plot is for $\lam=0$, the middle is for $\bar \lambda=\lambda/e^2=\frac12$, and the right one is for $\bar \lambda=8$. These curves were obtained numerically by solving the saddle point equations of motion \eqref{SPeqs} following the three step procedure explained in the beginning of this section. Note that on the left plot the scalar (multiplied by $r$) goes to a constant in the IR, in the middle one it decays slower than the electric field (multiplied by $r^2$), while in the right plot, their decay rate is the same ($\sim 1/\sqrt{\log (r)}$).
\label{fig:subcrit}}
\end{figure}

We end this subsection with three examples of soliton solutions in the regime $\sqrt{3}/4<g<1/2$ for the three cases of $\bar\lam=\lam/e^2$ considered above, see figure~\ref{fig:subcrit}. These are the lowest energy states when $f<0$ and the runaway behavior of the RG equation~\eqref{eq_beta_epsBig} leads to the physical interpretation of complete screening of the Wilson line. For these plots, we determine the value of $r_0$ from the dimensionful coupling $f_0$ as $r_0=(f/f_0)^{1/(2\nu)}$. By \eqref{fxi}, this equals\footnote{That is, we set the dimensionless coupling $f=-2\nu/100$, so that $\abs{f}\ll 2\nu$ and that $f$ is negative.} 
\es{manipulater0}{
r_0=\le[{2\nu\ov f}\, {\beta^{(c)} \ov \al^{(c)} }\ri]^{1/(2\nu)} \xi\,.
}
We expect $r_0$ to be the scale where nonlinearities operate, and the core of the screening cloud should be localized on this scale. Indeed, the three examples shown in figure~\ref{fig:subcrit} confirm this expectation. The core region is followed by an extended tail region, as explored in \eqref{ThreeCases} from the point of view of differential equations. A complementary IR DQFT perspective on these tails is given in section \ref{subsec_IR_line}, while in 
 this section we provided a UV perspective on screening.

\subsection{Screening in the supercritical regime}\label{subsec_scalar_supercritical}

The bulk physics of the supercritical regime resembles that of the subcritical regime with the boundary conditions triggering an instability to forming a screening scalar cloud. There are two distinctions: the screening cloud forms for any boundary condition (i.e. for either sign of $f$, as is evident from the beta function~\eqref{eq_beta_final} which leads to negative infinite $f$ regardless of the initial conditions for imaginary $\nu$) and the cloud slightly above criticality is generically exponentially large, $\exp\le(\pi/\abs{\nu}\ri)$. These differences originate from the near defect dynamics as we explain below.

Let us recall from section \ref{subsec_2DCFTs} that in the supercritical regime there is no genuine DCFT and we need to keep the cutoff $r_0$ finite. It is natural to take the boundary condition at the cutoff surface to be
\es{SupCritBC}{
0=\le(\hat f+\frac12\ri)\Phi+\p_n \Phi\big\vert_{r_0}\,,
}
where we take $n$ to point towards the origin, we converted the boundary condition \eqref{eq_deltaS} into one given in terms of $\Phi$,  and we look for a real scalar profile. We use the obvious boundary conditions $F_{0r}\vert_{r_0}=g/r_0^2$ for the gauge field. The profiles $\Phi=0$ and $A_t=g/r$ will always be a solution, but for any $\hat{f}$ we will always find (infinitely many) scalar solitons. The reason for the existence of an infinite family of solutions is that the solution of the  linearized equations has a discrete scale symmetry which can be used to generate new solutions from existing ones. The same phenomenon was discussed in \cite{Kaplan:2009kr,Iqbal:2010eh,Iqbal:2011aj}.

In more detail, the solutions for the linearized equations are $r^{1/2+i \abs{\nu}}$, so that we can write:
\es{PhiAsymSupCrit}{
\Phi=C \sqrt{r}\, \cos\le(\abs{\nu}\log\le({r\ov r_0}\ri)-\ga\ri)\,,
}
where $C$ is fixed by bulk regularity of the full nonlinear problem and 
 $\ga$ is fixed by the boundary condition \eqref{SupCritBC}  to be:
\es{PhiAsymSupCritBC}{
\ga=\arctan\le({\hat f /  \abs{\nu}}\ri)\,.
}
 Under discrete scale transformations the profile \eqref{PhiAsymSupCrit} transforms by scaling and hence also satisfies the boundary condition \eqref{SupCritBC} and gives rise to a new soliton 
\es{DiscreteScale}{
r&\to \Lam^n\, r\,, \qquad \Phi(r) \to (-1)^n \Lam^{n/2} \, \Phi(r)\,, \qquad \Lam\equiv \exp\le(-{\pi\ov \abs{\nu}}\ri)\,.
}
Since the envelope of $\Phi$ in \eqref{PhiAsymSupCrit} grows, the solution eventually exits the linear regime and stops oscillating. The discrete scale invariance is broken by nonlinear effects.
Let $C=C_0$ give rise to a regular solution of the equations with zero nodes of the $\Phi$ profile. Then by the discrete scale invariance \eqref{DiscreteScale}, the amplitude $C_n\approx C_0 \,\Lam^{n/2}$ with $n\in \mathbb{Z}_+$ will also give rise to a regular scalar profile with $n$ nodes.\footnote{We write approximately equal, since some nonlinear effects correct the profile \eqref{PhiAsymSupCrit}. For increasing $n$ these corrections are decreasing in importance. We note that we may regard the $\Phi=0$ solution as corresponding to $C_\infty$.} 

From the infinitely many potential solitons characterized by $C_n$, we have to choose the one that is physically realized: this can be done by comparing the energies of field configurations or by dynamical stability analysis.
 In Appendix~\ref{app_KG_tachyons} we determine the spectrum of fluctuations around the $\Phi=0$ background and we find infinitely many tachyon modes with sizes $R_k \simeq \Lam^{-k}\,r_0$ with $k\geq 1$. Since we can treat the $n$th soliton (with parameter $C_n$) as consisting of a linearized oscillating region of size $R_{\text{lin},n} \simeq \Lam^{-n}\,r_0$ followed by a nonlinear region, we can fit tachyons with $k\leq n$ (a total of $n$ of them) into the linearized regions, and we find that the all  $C_{n>0}$ solitons are unstable. Hence we conclude that only the $C_0$ soliton is stable, since it lacks a large linearized region, where tachyons could reside.

\begin{figure}[!h]
\centering
\includegraphics[width=77mm]{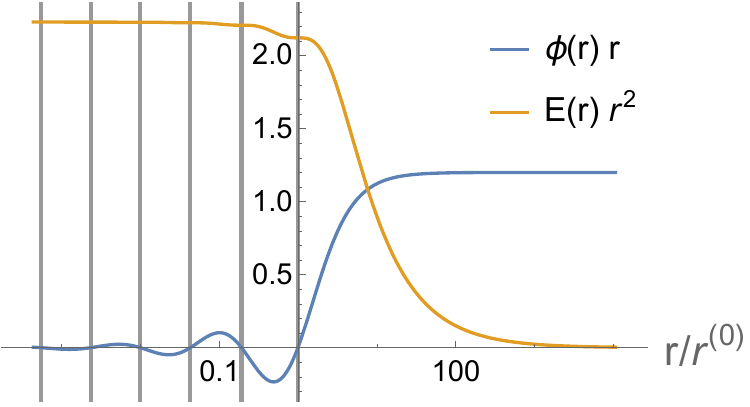}
\caption{Plot of the screening cloud for the supercritical case with $g\sim 2$, $\lam=0$, and Dirichlet boundary condition for the scalar. We cut off part of this solution with $r<r_0$ thereby obtaining the physical profiles.
If we set $r_0=r^{(0)}$ (rightmost gray line) we get the stable scalar soliton with Dirichlet boundary condition. If we set $r_0=r^{(n)}$ (the $(n+1)$th gray line), we get the scalar soliton with $n$ nodes (and $n$ tachyons). A subtlety is that in the oscillating region $g$ changes slightly, so we have to make slight adjustments in parameter space to keep $g$ constant as we increase the number of nodes.
\label{fig:oscillations}}
\end{figure}

As in the subcritical case discussed in section~\ref{subsec_scalar_stability}, there are three possible IR asymptotics of the scalar soliton depending on the value of $\bar\lam=\lam/e^2$ as listed in \eqref{ThreeCases}. Starting from these and setting $\xi=1$ we obtain a scalar soliton. We then reinstate $\xi$ as in \eqref{twopar}. We choose $\xi$ such that we satisfy the boundary condition \eqref{SupCritBC} at $r=r_0$. An illustrative example is given in figure~\ref{fig:oscillations}, where we chose $\xi=\xi_0$ such that Dirichlet boundary conditions are obeyed.
Since the soliton is oscillating in the small $\Phi$ region as in \eqref{PhiAsymSupCrit}, it is always possible to satisfy any boundary conditions from the class \eqref{SupCritBC}. This is unlike the subcritical case, where the sign of $f$ decided if we have a soliton solution. In fact with the choice $\xi_n=\Lam^{-n} \, \xi_0$ we again obtain a soliton that obeys the same boundary conditions. The corresponding asymptotic amplitude is $C_n$. This is clearly demonstrated in figure~\ref{fig:oscillations}.

\begin{figure}[!h]
\centering
\includegraphics[width=77mm]{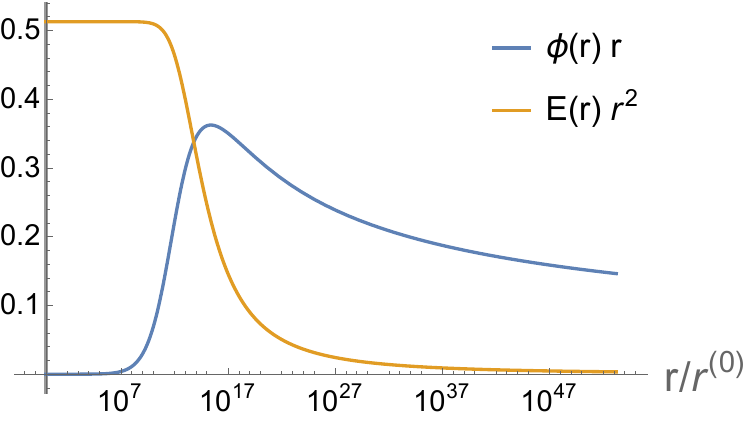}
\caption{Plot of the screening cloud for the supercritical case with $g=0.51$. The striking feature of this plot is the large cloud size for which the theoretical estimate $R_\text{cloud}/r_0\approx 4 \cdot 10^{13}$ is consistent with the plotted numerical result. We have chosen $\bar\lam=1/2$  and the tail region is identical to the (middle) $\bar\lam=1/2$ plot in figure~\ref{fig:subcrit}. We chose Dirichlet boundary condition $\Phi(r_0)=0$ for the scalar field (corresponding to $\hat f \to \infty$ in \eqref{SupCritBC}). Different boundary conditions for the scalar field lead to a qualitatively similar plot.
\label{fig:supcrit}}
\end{figure}

The most striking feature of supercritical clouds is that for small $\abs{\nu}$ the core of the soliton is huge, of order $R_\text{cloud}\simeq {r_0\ov \Lam}=r_0\exp\le(\eta /\abs{\nu}\ri)$, where the constant $\eta=O(1)$ is determined by nonlinear physics. In contrast, in the subcritical case the soliton has a natural size, $R_\text{cloud}\simeq r_0$, with $r_0$ fixed by the dimensionful coupling constant $f_0\sim r_0^{-2\nu}$ as demonstrated in figure~\ref{fig:subcrit}. Since the tail of the cloud is identical to what we have already shown for the subcritical case in figure~\ref{fig:subcrit}, in figure~\ref{fig:supcrit} we only show a $\bar\lam=1/2$ cloud.

We conclude this section by providing an intuitive RG explanation of the $C_n$ solitons that we constructed numerically. The exponential size for the cloud $R_\text{cloud}\simeq r_0\exp\le(\pi/\abs{\nu}\ri)$ associated with the ground state solution $C_0$ is clearly a consequence of the ``walking" behavior discussed around~\eqref{eq_walking}. To understand the solutions with $n\geq 1$ nodes,  it is convenient to re-express the boundary condition~\eqref{SupCritBC} in terms of the angle $\gamma$ parametrizing the linear solution~\eqref{PhiAsymSupCrit}:
\begin{equation}
e^{i\gamma}=\frac{|\nu|+i f}{\sqrt{f^2+|\nu|^2}}\equiv z\,.
\end{equation}
In terms of the phase $z$, the solution to the beta-function~\eqref{eq_beta_final} for $\nu^2<0$ reads
\begin{equation}\label{eq_cyclic}
z(\mu)=z(\mu_0)\left(\frac{\mu}{\mu_0}\right)^{i|\nu|}\,,
\end{equation}
where $\mu$ is the running scale and $\mu_0$ some reference initial scale.~\eqref{eq_cyclic} describes a cyclic RG flow with period $\Lambda^{-2}$. In practice, the beta-function~\eqref{eq_beta_final} describes only the linear regime, and the nonlinearities drive the RG flow away from the cyclic regime. 

Similarly to the Efimov effect \cite{Efimov:1970zz}, the discrete scale invariance \eqref{DiscreteScale} is a consequence of the approximate cyclic RG flow~\eqref{eq_cyclic}. The $C_n$ solitons are then simply interpreted as RG flows in which the fields linger in the linear regime for $\sim n/2$ cycles before entering the nonlinear regime and screening the Wilson line. We remark however that the ground state solution always exits the linear regime before performing a full cycle.

\subsection{Effective defect field theory for screened Wilson lines}\label{subsec_IR_line}

Remarkably, the numerical analysis in the previous sections provides an exact solution for the defect RG flow triggered by unstable Wilson lines, both in the case of negative double-trace deformations and supercritical charges. In this section we complement that analysis by interpreting the long distance tail of the screening cloud in terms of an \emph{effective defect field theory} description of the final stages of the flow.

Let us consider first the theory with no quartic coupling, $\lambda=0$. In this case the long distance limit of the screening cloud analyzed in the previous section admits a non-trivial one-point function for the scalar field with conformal scaling:
\begin{equation}\label{eq_eft_phi2}
\langle|\Phi|^2(r)\rangle=\frac{e^4 v^2}{2(4\pi r)^2}\,,
\end{equation}
where $v\sim 1/e^2$ is a dimensionless number which depends upon the initial charge and the boundary condition. 

In the absence of gauge fields, conformal defects sourcing the scalar operator can be constructed by straightforwardly integrating the fundamental field along the line contour, see e.g.  \cite{Allais:2014fqa,Cuomo:2021kfm,Popov:2022nfq,Pannell:2023pwz}. An equally explicit construction is not available in a gauge theory.
Since all gauge-invariant operators have engineering dimension larger or equal than $2$, the effective defect field theory corresponding to \eqref{eq_eft_phi2} cannot be obtained by deforming the trivial line defect with a local operator. Rather, it should be understood in terms of boundary conditions for the scalar field at $r\rightarrow 0$.  To write the corresponding defect explicitly, we notice that \eqref{eq_eft_phi2} is equivalent to a constant profile for the AdS$_2$ rescaled field; in other words,   \eqref{eq_eft_phi2} describes a \emph{Higgs phase} on AdS$_2$. It is therefore natural to decompose the scalar field into a radial and a Goldstone component
\begin{equation}
\Phi(x)=\frac{1}{\sqrt{2}}h(x)e^{i\pi(x)}\,,\qquad
\Phi^\dagger(x)=\frac{1}{\sqrt{2}}h(x)e^{-i\pi(x)}\,,
\end{equation}
so that the action reads
\begin{equation}\label{eq_eft_bulk_free}
S=\frac{1}{e^2}\int d^4x\left[-\frac{1}{4 }F_{\mu\nu}^2+
\frac{1}{2 }(\pd h)^2+\frac{ 1}{2}h^2(\pd_\mu\pi- A_\mu)^2\right]\,.
\end{equation}
To obtain the profile \eqref{eq_eft_phi2} we then simply introduce a source in the Higgs equations of motion
\begin{equation}\label{eq_eft_Heom}
-\pd^2 h+A^2_\mu h= -e^2v \,\delta^{3}(x_\bot)\,,\qquad
\pd_\mu F^{\mu\nu}+h^2(\pd^\nu\pi-A^\nu)=\pd_\mu \left[h^2(\pd^\mu\pi-A^\mu)\right]=0\,,
\end{equation}
with the solution (up to gauge transformations)
\begin{equation}\label{eq_eft_saddle_free}
A_\mu=\pi=0\,,\qquad
h=\frac{e^2 v}{4\pi r}\equiv h_s(r)\,.
\end{equation}
The source in \eqref{eq_eft_Heom} can be formally represented with a term localized at $r=0$
\begin{equation}\label{eq_eft_free}
S_D=v\int_{r=0} dt \, h=v\int_{r=0} dt \sqrt{2|\Phi|^2} \,.
\end{equation}
We stress that, despite its formal representation \eqref{eq_eft_free}, this defect cannot be understood as a perturbation of the trivial defect by a local operator, but it is rather thought as setting a boundary condition for the scalar field, somewhat similarly to a 't Hooft line. As in that case, the corresponding defect is perfectly local. To appreciate this point further, one could imagine obtaining such a line operator by starting from an interface separating the theory \eqref{eq_eft_bulk_free} from a deformed model with a bulk potential $V(|\Phi|^2)$ which Higgses the gauge group. The defect \eqref{eq_eft_free} is then obtained upon deforming the interface into a cylinder of radius $r_0$ along the time direction, and taking the limit $r_0\rightarrow 0$ while simultaneously scaling the coefficients of the potential $V(|\Phi|^2)$ with inverse powers of $r_0$ according to their dimension. 

In practice, to concretely study the model given by \eqref{eq_eft_free} and \eqref{eq_eft_bulk_free} we simply need to expand the fields around the saddle-point \eqref{eq_eft_saddle_free}.  We consider a gauge-fixing inspired by the usual Feynman - ’t Hooft choice
\begin{equation}\label{eq_eft_gf}
S_{g.f.}=-\frac{1}{2 e^2}\int d^4x\left(\pd_\mu A^\mu+ h^2_s\pi\right)^2\,.
\end{equation}
Upon rescaling fluctuations with a factor of $e$, the quadratic action reads
\begin{equation}
S+S_{g.f.}\simeq\int d^4x\left\{-\frac{1}{2}(\pd_\mu A_\nu)^2
+\frac12 h^2_sA_\mu^2
+\frac12(\pd \delta h)^2+h^2_s\left[\frac12(\pd\pi)^2-\frac12 h^2_s\pi^2\right]
\right\}\,,
\end{equation}
where $\delta h=h-h_s$. Clearly $\delta h$ behaves as a free field in the absence of a defect. By studying the propagators of $A_\mu$ and $\pi$, we find that the lowest dimensional operator in the bulk-to-defect OPE of the $U(1)$ current $j_\mu\simeq h_s^2(\pd_\mu \pi-A_\mu)$ has dimension
\begin{equation}
\delta=\frac12+\sqrt{\frac14+\frac{e^4 v^2}{(4\pi)^2}}\,.
\end{equation}
In particular, there is a (defect) scalar operator with dimension $\delta$ corresponding to the $r\rightarrow 0$ limit of $j_0$. The corresponding deformation of the defect action \eqref{eq_eft_free} can be written as
\begin{equation}\label{eq_eft_free_deformation}
\delta S_D=- \tilde{q}\int_{r=0} dt(A_0-\dot{\pi})\,.
\end{equation}
Note that because of the nontrivial profile of the Higgs field $h\sim 1/r$ we can write gauge-invariant defect operators using both the gauge field and the Goldstone mode. The corresponding coupling $\tilde{q}$ in \eqref{eq_eft_free_deformation} is thus not quantized, and it is in fact irrelevant since $\delta>1$. Analyzing perturbatively the deformation \eqref{eq_eft_free_deformation},  we find the following one-point function for the gauge field\footnote{This is derived from the propagator of $A_0$, whose zero-mode, with the chosen gauge-fixing, behaves analogously to an AdS$_2$ scalar field with dimension $\delta$.}
\begin{equation}\label{eq_eft_free_F_tail}
\langle F_{0i}\rangle\propto x^i\frac{e^2 \tilde{q}}{4\pi r^{2+\delta}}\,.
\end{equation}
\eqref{eq_eft_free_F_tail} agrees with the functional form for the screening tail of the gauge field previously derived from the equations of motion in \eqref{ThreeCases} (setting $c={e^2 v\ov 4\pi}$ in \eqref{ThreeCases}).  Further subleading corrections to the screening cloud are reproduced by other irrelevant deformations of the defect \eqref{eq_eft_free}.

Let us now discuss the theory with a quartic coupling
\begin{equation}\label{eq_eft_bulk}
S=\frac{1}{e^2}\int d^4x\left[-\frac{1}{4 }F_{\mu\nu}^2+
\frac{1}{2 }(\pd h)^2-\frac{\lambda}{8e^2}h^4+\frac{ 1}{2}h^2(\pd_\mu\pi- A_\mu)^2\right]\,.
\end{equation}
Inspired by the previous analysis, we consider the following defect deformation
\begin{equation}\label{eq_eft_2}
S_D=\int_{r=0} dt \left[v\,h-\tilde{q}(A_0-\dot{\pi})\right]\,.
\end{equation}
We focus on the double-scaling limit
\begin{equation}\label{eq_eft_double_scaling}
e^2\sim \lambda\rightarrow 0\,,\quad
v\sim \tilde{q}\rightarrow\infty\quad\text{with}\quad
e^2v\sim e^2\tilde{q}=\text{fixed}\,.
\end{equation}
In this limit the Goldstone mode can be neglected.  Including the gauge-fixing \eqref{eq_eft_gf} we thus consider
\begin{equation}
S+S_{g.f.}+S_D=
\frac{1}{e^2}\int d^4x\left[-\frac{1}{2 }(\pd_\mu A_\nu)^2
+\frac{ 1}{2}h^2 A_\mu^2
+\frac{1}{2 }(\pd h)^2-\frac{\lambda}{8e^2}h^4\right]+\int_{r=0} dt
\left(v\,h-\tilde{q}\,A_0\right)\,.
\end{equation}

\begin{figure}[t]
\centering
\subcaptionbox{ \label{fig:leading_order_fig1a}}
{\includegraphics[scale=0.2]{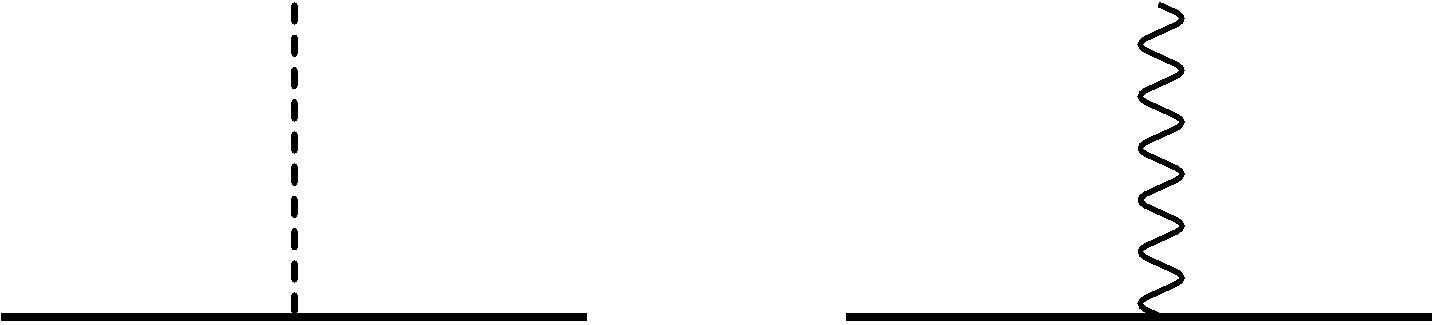}}\\[1em]
\subcaptionbox{\label{fig:subleading_fig2}}
{\includegraphics[scale=0.18]{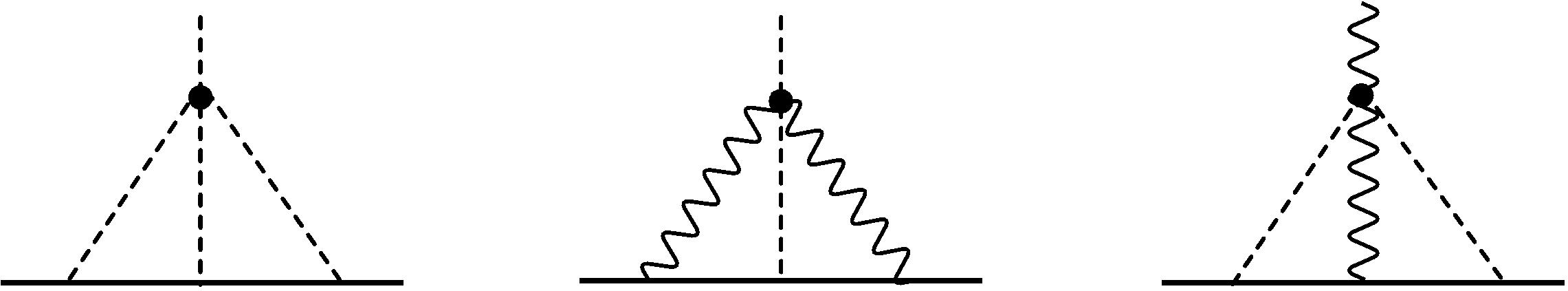}}
\caption{Diagrams contributing to the scalar and gauge field one-point functions. Dashed lines denote scalar fields while wiggly lines stand for gauge fields. The solid line represents the defect and dots stand for bulk couplings. . }
\end{figure}

In what follows, we self-consistently focus on the regime $e^2 v\sim e^2\tilde{q} \ll 1$. In this regime the one-point functions for the scalar and gauge field admit the following expansion 
\begin{align}\label{eq_eft_h1pt}
\langle h(r)\rangle &=\frac{e^2v}{4\pi r}\left[F_0\left(\frac{\tilde{q}}{v},\frac{\lambda}{e^2},r\right)+\frac{e^4 v^2}{(4\pi)^2} F_1\left(\frac{\tilde{q}}{v},\frac{\lambda}{e^2},r\right)+O\left(\frac{e^8 v^4}{(4\pi)^4}\right)\right]\,,\\
\label{eq_eft_A1pt}
\langle A_0(r)\rangle &=\frac{e^2v}{4\pi r}\left[G_0\left(\frac{\tilde{q}}{v},\frac{\lambda}{e^2},r\right)+\frac{e^4 v^2}{(4\pi)^2} G_1\left(\frac{\tilde{q}}{v},\frac{\lambda}{e^2},r\right)+O\left(\frac{e^8 v^4}{(4\pi)^4}\right)\right]\,.
\end{align}
The leading order terms $F_0$ and $G_0$ are determined from the linearized equations of motion
\begin{equation}
\pd^2 h=e^2 v\delta^3(x_{\bot})\,,\qquad
\pd^2A_\mu=e^2\tilde{q}\delta^3(x_{\bot})\,,
\end{equation}
from which we obtain
\begin{equation}
F_0\left(\frac{\tilde{q}}{v},\frac{\lambda}{e^2},r\right)=1\,,\qquad
G_0\left(\frac{\tilde{q}}{v},\frac{\lambda}{e^2},r\right)=\frac{\tilde{q}}{v}\,.
\end{equation}
Diagrammatically, the leading order result is associated with a single insertion of the defect couplings and no insertion of the bulk vertices, as in figure \ref{fig:leading_order_fig1a}.

The subleading contributions arise from the diagrams in figure \ref{fig:subleading_fig2}. The resulting integrals are divergent; as usual in QFT, this signals a nontrivial RG flow for the defect couplings $v$ and $\tilde{q}$. To extract the corresponding beta-functions, we evaluate the divergent parts of $F_1$ and $G_1$ in dimensional regularization:
\begin{equation}
\begin{aligned}
F_1\left(\frac{\tilde{q}}{v},\frac{\lambda}{e^2},r\right)&=
\left(\frac{\tilde{q}^2}{v^2}-\frac{\lambda}{2e^2}\right)\times\frac{1}{2\varepsilon}+\text{finite}
\,,\\
G_1\left(\frac{\tilde{q}}{v},\frac{\lambda}{e^2},r\right)&=-
\frac{\tilde{q}}{v}\times\frac{1}{2\varepsilon}+\text{finite}\,,
\end{aligned}
\end{equation}
where $\varepsilon=4-d$.
For the physical one-point functions \eqref{eq_eft_h1pt} and \eqref{eq_eft_A1pt} to be finite, we need to rewrite the defect couplings in terms of bare ones. Working in the minimal subtraction scheme, we find
\begin{equation}
\begin{aligned}
v&\rightarrow  v_0=vM^{\varepsilon/2}\left[1+
\frac{e^4v^2}{(4\pi)^2}\left(\frac{\lambda}{2e^2}-\frac{\tilde{q}^2}{v^2}\right)\frac{1}{2\varepsilon}+\ldots
\right]\,,\\
\tilde{q}&\rightarrow \tilde{q}_0=\tilde{q}M^{\varepsilon/2}
\left[1+
\frac{e^4v^2}{(4\pi)^2}
\frac{1}{2\varepsilon}+\ldots
\right]\,,
\end{aligned}
\end{equation}
where $M$ is the sliding scale. Demanding that the bare couplings $v_0$ and $\tilde{q}_0$ be independent of $M$, with textbook manipulations \cite{Weinberg:1996kr} we obtain the beta-functions for the physical couplings $v$ and $\tilde{q}$:
\begin{equation}\label{eq_eft_beta}
\begin{aligned}
\beta_v&=\frac{\pd v}{\pd\log (M)}=
v\left[\left(\frac{\lambda e^2 v^2}{2}-e^4\tilde{q}^2\right)
+O\left( e^8v^4\right)\right]
\,,
 \\
\beta_{\tilde{q}}&=\frac{\pd \tilde{q}}{\pd\log (M)}=
\tilde{q}\left[ e^4v^2
+O\left( e^8v^4\right)\right]
\,.
\end{aligned}
\end{equation}

The equations \eqref{eq_eft_beta} imply that both $v$ and $\tilde{q}$ run logarithmically to zero in the IR ($M\rightarrow 0$). We therefore conclude that the defect \eqref{eq_eft_2}, describing a fully screened Wilson line, flows to a trivial defect in the IR.\footnote{For $\tilde{q}=0$, the model effectively reduces to the pinning field defect in $d=4$, which was studied in \cite{Allais:2014fqa,Cuomo:2021kfm} and also flows to a trivial defect.} Hence the description \eqref{eq_eft_2} of screened Wilson lines as a scalar line is useful as an {\it intermediate energy description}. In the following we show how to reproduce the tail of the screening cloud previously derived from the classical equations of motions.

We consider the one-point functions for the scalar and the gauge field in the long distance limit
\begin{equation}\label{eq_eft_1ptf}
\langle h(r)\rangle\stackrel{r\rightarrow\infty}{=}\frac{e^2 v(1/r)}{4\pi r}\,,\qquad
\langle A_0(r)\rangle\stackrel{r\rightarrow\infty}{=}\frac{e^2 \tilde{q}(1/r)}{4\pi r}\,,
\end{equation}
where the couplings are expressed at the scale $1/r$. For sufficiently large $r$, the coupling can be written using the asymptotic solution to \eqref{eq_eft_beta} for $t=-\log (M/M_0)\gg 1/(e^2 v)$, where $M_0$ is the scale at which the initial conditions for the couplings are specified; physically, $M_0$ represents the cut-off of the effective description \eqref{eq_eft_2}. The explicit result depends on the ratio $\lambda/e^2\equiv \bar{\lambda}$. For $\bar{\lambda}<2$, we find
\begin{equation}\label{eq_eft_asymp1}
\begin{aligned}
e^2v(M_0e^{-t})&\stackrel{t\rightarrow\infty}=
\frac{t^{-1/2}}{\sqrt{\bar{\lambda} }}+b^2\frac{\sqrt{\bar{\lambda}}}{2(\bar{\lambda}-1)}t^{1/2-2/\bar{\lambda}}+\ldots\,,\\
e^2\tilde{q}(M_0e^{-t})&\stackrel{t\rightarrow\infty}=b\, t^{-1/\bar{\lambda}}+\ldots\,,
\end{aligned}
\end{equation}
while for $\bar{\lambda}>2$ the asymptotic solution reads
\begin{equation}\label{eq_eft_asymp2}
\begin{aligned}
e^2v(M_0 e^{-t})&\stackrel{t\rightarrow\infty}=
\frac{t^{-1/2}}{\sqrt{2 }}+b\frac{\bar{\lambda}-2}{2\sqrt{2}}t^{1/2-\bar{\lambda}/2}+\ldots\,,\\
e^2\tilde{q}(M_0 e^{-t})&\stackrel{t\rightarrow\infty}= 
\frac{\sqrt{\bar{\lambda}-2}}{2}t^{-1/2}
+b
\frac{\sqrt{\bar{\lambda}-2}}{2}t^{1/2-\bar{\lambda}/2}
\ldots\,.
\end{aligned}
\end{equation}
The parameter $b$ in \eqref{eq_eft_asymp1} and \eqref{eq_eft_asymp2} depends upon the initial condition for the coupling constants.\footnote{An additional free parameter shows up in the asymptotic solutions \eqref{eq_eft_asymp1} and \eqref{eq_eft_asymp2} in the subleading orders; we did not report additional corrections to \eqref{eq_eft_asymp1} and \eqref{eq_eft_asymp2} since these depend upon the higher order terms neglected in the beta functions \eqref{eq_eft_beta}.} Unsurprisingly, using \eqref{eq_eft_asymp1} and \eqref{eq_eft_asymp2} in the one-point functions \eqref{eq_eft_1ptf} we recover the form for the tail of the screening cloud \eqref{ThreeCases} previously derived in section \ref{subsec_scalar_stability}.

\subsection{Constraints from 0-Form symmetry and multi-flavor scalar \texorpdfstring{QED$_4$}{QED4}}\label{subsec_multiflavor}

In this section we briefly discuss the generalization of our results to multi-flavor QED$_4$.  We consider the action
\begin{equation}\label{eq_MultiFlavorAction}
S=\frac{1}{e^2}\int d^4x\left[\abs{D_\mu\Phi_a}^2-\frac{\lam}{2e^2} (\abs{\Phi_a}^2)^2 -\frac{1}{4}F_{\mu\nu}^2\right]-q\int dt A_0\,,
\end{equation}
where $a=1,2,\ldots,N$. The theory is invariant under the action of the internal symmetry group $PSU(N)=SU(N)/\mathbb{Z}_N$ which rotates among the scalars.\footnote{The symmetry group is $PSU(N)$ and not $SU(N)$ because all gauge-invariant operators transform in representations whose Young diagram consists of $p=0\mod N$ boxes.}  
Consider inserting a Wilson line of charge $q$. This represents the wordline of a massive external particle of charge $q$. If we represent the external particle by a heavy massive field $\Psi$ with no $PSU(N)$ quantum numbers, then the total global symmetry of the system is now $(U(1)\times SU(N))/\mathbb{Z_{N}}$ where the $U(1)$ factor is particle number, normalized such that $\Psi$ carries charge 1, and $\mathbb{Z}_N$ is generated by a rotation of $\Psi$ by angle $q/N$ accompanied by a transformation in $SU(N)$ given by the matrix ${\rm diag}(\exp(2\pi i/N),...,\exp(2\pi i/N))$. The identification by $\mathbb{Z}_N$ means that in a sector with one $\Psi$ particle the $SU(N)$ representation must have $q \ {\rm mod} \ N$ boxes in the Young diagram. This means that the state in the presence of a Wilson line of charge $q$ must transform under a representation with $q \ {\rm mod} \ N$ boxes, i.e.~the Wilson line can only end on operators transforming in a projective representation of $PSU(N)$. (To make this precise, we can insert the Wilson loop as a localized charge on the sphere.) This does not mean that the infrared cannot be completely screened for $q\neq 0\ {\rm mod} \ N$. It is possible that the infrared theory has a decoupled representation on the line and all the bulk Green functions coincide with those without a defect. In this situation the infrared $g$ function is the dimension of the representation. The line operator is simply the trivial line defect stacked with a quantum mechanical system with vacuum degeneracy. 
Similar comments apply whenever we are dealing with symmetric defects in a system whose global symmetry $G$ can be nontrivially centrally extended (i.e. whenever $H^2(G,U(1))$ is nontrivial). In some situations this can lead to interesting constraints related to the $g$ theorem, since the infrared $g$ function is given by the dimension of the representation if the line is otherwise screened.

Let us now analyze in detail the Wilson line in the theory~\eqref{eq_MultiFlavorAction}.

For $0<\abs{q}\leq\frac{2\pi}{e^2}$ each of the $\ell=0$ modes of the scalars $\Phi_a$ admit either standard or alternate boundary conditions on the defect. This leads overall to $2^N$ fixed points, many of which partially break the internal symmetry group.  To analyze the defect RG flows in this setup, consider the fixed points where all fields are in alternate quantization.  As in the discussion around \eqref{eq_action_bdry2}, this is achieved by supplementing the action \eqref{eq_MultiFlavorAction} with the following defect term
\begin{equation}
S_{bdry}^{(1)}=-\frac{1-2\nu}{2}
\int_{r=r_0\hspace{-1em}} dt\,\sqrt{\hat{g}}\abs{\Phi_a}^2\,.
\end{equation}
We now deform this theory with a relevant double-trace deformation as in section \ref{subsec_2DCFTs}. The most general bilinear is parametrized by a Hermitian matrix $f_0^{ab}=(f_0^{ba})^*$:
\begin{equation}\label{eq_multi_DT}
S_{bdry}^{(2)}=-\int_{r=r_0\hspace{-1em}} dt\sqrt{\hat{g}} r_0^{2\nu}\sum_{a,b}\bar{\Phi}_a f_0^{ab}\Phi_b\,.
\end{equation}
\eqref{eq_multi_DT} imposes the following mixed boundary conditions among the modes
\begin{equation}\label{eq_multi_C}
\beta_a=  C_{ab}\alpha_b , \qquad C=  f_0\cdot\left[2\nu\mathds{1} - f_0 r_0^{2\nu}\right]^{-1}.
\end{equation}
Proceeding as in the single flavor case, we obtain the beta function of the dimensionless coupling $f^{ab}=r_0^{2\nu}f_0^{ab}$ by applying the Callan-Symanzik equation to the expression \eqref{eq_multi_C}. The result is formally identical to \eqref{eq_beta_final}:
\begin{equation}\label{eq_beta_multiflavor}
\beta_f^{ab}=\mu\frac{\pd f^{ab}}{\pd\mu}=-2\nu f^{ab}+(f^2)^{ab}\,,
\end{equation}
where $(f^2)^{ab}=\sum_c f^{ac} f^{cb}$.

To illustrate the result consider $N=2$. We call $\sigma^4=\mathds{1}$ and denote the Pauli matrices with $\sigma^i$, $i=1,2,3$. It is then convenient to decompose the coupling as (in matrix notation):
\begin{equation}
f=\sum_{a=1}^4 \alpha_a\sigma^a\,,\quad\implies\quad
\alpha_a=\frac{1}{2}\text{Tr}\left[\sigma^a f\right]\,.
\end{equation}
The result \eqref{eq_beta_multiflavor} can be written in terms of the beta functions of the components $\beta_a=\frac{1}{2}\text{Tr}\left[\sigma^a \beta_f\right]$:
\begin{align}
\beta_i&=-2\nu \alpha_i+2 \alpha_4 \alpha_i\,,
\quad
\beta_4=-2\nu \alpha_4+\sum_{a=1}^4\alpha_a\alpha_a\,. \label{eq_beta_multiflavor2}
\end{align}
\eqref{eq_beta_multiflavor} admits the following fixed points:
\begin{enumerate}
\item $\alpha_4=\alpha_i=0$: this an unstable fixed point corresponding to alternate boundary conditions for all the fields:
\item $\alpha_4=2\nu$ and $\alpha_i=0$: this is a stable fixed point corresponding to standard boundary conditions for all the fields.
\item $\alpha_4=\nu$ and $\alpha_i^2=\nu$: this is a manifold of unstable fixed points corresponding to standard boundary conditions in one field direction, and alternate in the orthogonal one.  In this fixed point the internal symmetry group $SU(2)$ is broken to $U(1)$.
\end{enumerate}
In the first two cases the line preserves $PSU(N)$, while in the last case it is explicitly broken and therefore there are protected tilt operators.

We end this section with some comments on the case with $e^2|q|>2\pi$ where 
the Wilson lines are supercritical and expected to be screened. At a classical level, the analysis proceeds along the lines of section \ref{subsec_scalar_supercritical}; in particular the classical scalar profile \emph{spontaneously} breaks the internal symmetry.  However,  quantum-mechanically we have to integrate over the zero modes of the screening saddle-point.\footnote{In more detail,  it was recently argued, under general assumptions, that a line defect that spontaneously  breaks a continuous internal symmetry can only flow to a decoupled one-dimensional sector on the line, tensored with a DCFT which does not break the symmetry \cite{Cuomo:2023qvp}. Therefore at large distances only singlet operators are allowed to acquire a VEV.} Therefore only flavor singlets acquire an expectation value in the screening cloud, e.g.
\begin{equation}
\langle \Phi_a^*\Phi_b(r)\rangle\propto \delta_{ab}\,.
\end{equation}
Considering the equivalent theory on AdS$_2\times S^2$, the long distance limit of the screened line is well approximated by a defect setting Dirichlet boundary conditions for the (AdS$_2$ rescaled) radial mode, as in section \ref{subsec_IR_line},  with Neumann boundary conditions imposed on the Goldstone modes.

This shows that sufficiently far from the line, i.e. much farther than the screening cloud, the bulk expectation values and Green functions are those of the theory without the defect, i.e. there is screening in this sense. However, as we argued above, the symmetries of the system force the line defect to carry a representation with $q \ {\rm mod} \ N$ boxes under $SU(N)$. In the language of defect QFT, this implies that supercritical Wilson lines with charge $q\neq 0 \mod N$ do not furnish simple line defects, and they are completely screened in the bulk. In particular, supercritical Wilson lines with $q\neq 0 \mod N$ admit a nontrivial $g$-function in the deep infrared. We will see another example in~\ref{ScalarQEDTwoPlusOne}.

\subsection{Constraints from 1-form symmetry and \texorpdfstring{QED$_4$}{QED4} with charge \texorpdfstring{$q_\phi$}{q} particles}\label{subsec_1form}

In the previous section we discussed the dynamics of line defects when there are interesting constraints from 0-form symmetry, i.e. when the infrared symmetry can be extended by the external (heavy) particles. 
Here we discuss the constraints imposed by 1-form symmetry. To motivate the discussion consider a charge $q$
Wilson line in a theory of a charge $q_{\phi}>1$ scalar field:
\begin{equation}\label{eq_ChargeGen}
S=\frac{1}{e^2}\int d^4x\left[\abs{D_\mu\Phi}^2-\frac{\lam}{2e^2} (\abs{\Phi}^2)^2 -\frac{1}{4}F_{\mu\nu}^2\right]-q\int dt \,A_0\,,
\end{equation}
with  $D=\partial-iq_\phi A$.

One might expect that in such a theory Wilson lines with charge $q\mod q_{\phi}\neq 0$ cannot be fully screened by the scalar particles. Let us make this precise. Wilson lines with charge $q\mod q_{\phi}\neq 0$
are charged under the electric $\mathbb{Z}_{q_\phi}$ one-form symmetry of the theory.  We are therefore led to the following question: what can be inferred about line defects charged under a one-form symmetry? This question is very general and arises in several different contexts; we will encounter it again  in section~\ref{sectHooft} in the analysis of 't Hooft lines. In the following we thus discuss this problem in full generality.  At the end of this subsection we will discuss the implications of our findings for charge $q_{\phi}$ scalar QED.

It is useful to introduce some terminology that we will use below: 
\begin{itemize} 
\item A line defect is a nontrivial DCFT if and only if the displacement operator is nonzero: $D_\perp \neq 0$.
\item A line defect is said to be topological if the displacement operator vanishes $D_\perp=0$, i.e. it is trivial as a DCFT, but the line defect can braid nontrivially with co-dimension 2 surfaces. 
\item A line defect $L$ is said to be completely trivial if none of the two definitions above apply, i.e. if it is  completely transparent (that is trivial as a DCFT and also transparent to co-dimension 2 surfaces).
\end{itemize} 
In this language, the lines which are stacked with a 0+1 dimensional TQFT that we have encountered in the previous subsection are completely trivial (but not simple). 

We will now argue that with some additional conditions the existence of a one-form symmetry implies that the charged Wilson line must necessarily define a {\it nontrivial} DCFT. 

\begin{figure}[t]
\centering
\includegraphics[scale=0.3,trim={0em 5em 2em 18em}, clip]{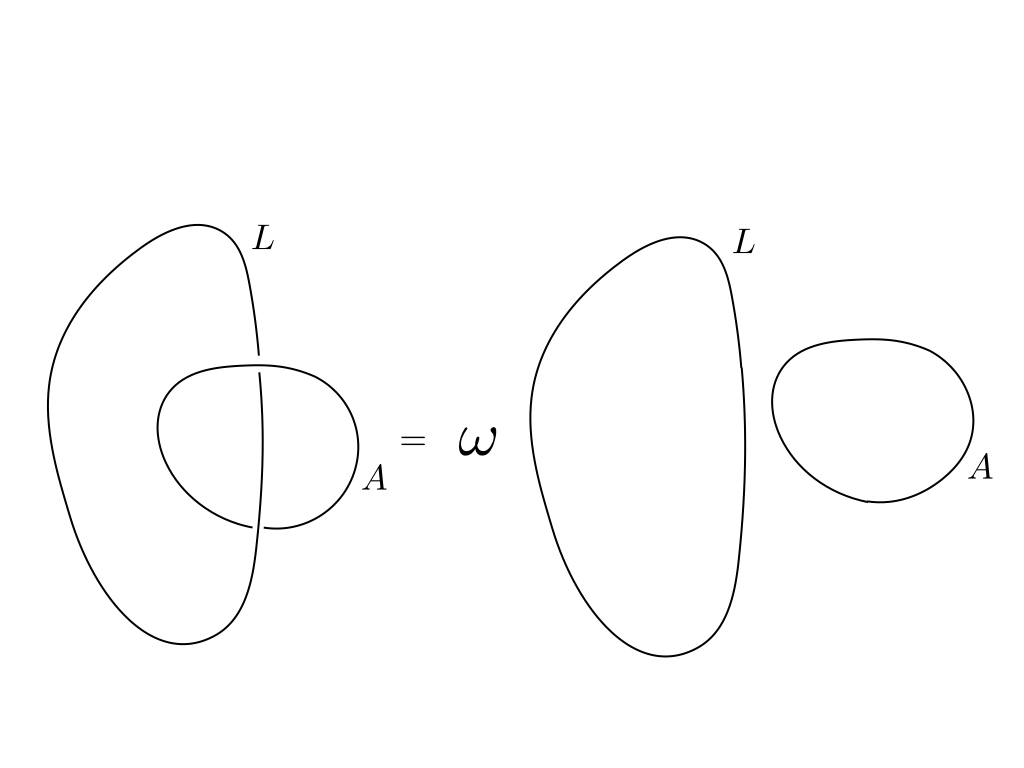}
\caption{A line defect charged under a one-form symmetry in 2+1 dimensions. }
\label{FigLA}
\end{figure}

A line $L$ is charged under a one-form symmetry if and only if it braids nontrivially with a topological co-dimension $2$ operator. Figure~\ref{FigLA} represents a line defect that is charged under a one-form symmetry in 2+1 dimensions. In the figure $L$ is a line defect, $A$ is a one-form symmetry charge and $\omega\neq 1$ is a root of unity.  Note that this immediately implies that the line $L$ cannot be completely trivial. The interesting question, that we address below, is under which conditions the one-form symmetry forces the displacement operator to be nontrivial.

\begin{figure}[h!]
\centering
\includegraphics[scale=0.3,trim= 0 18em 0em 7em, clip]{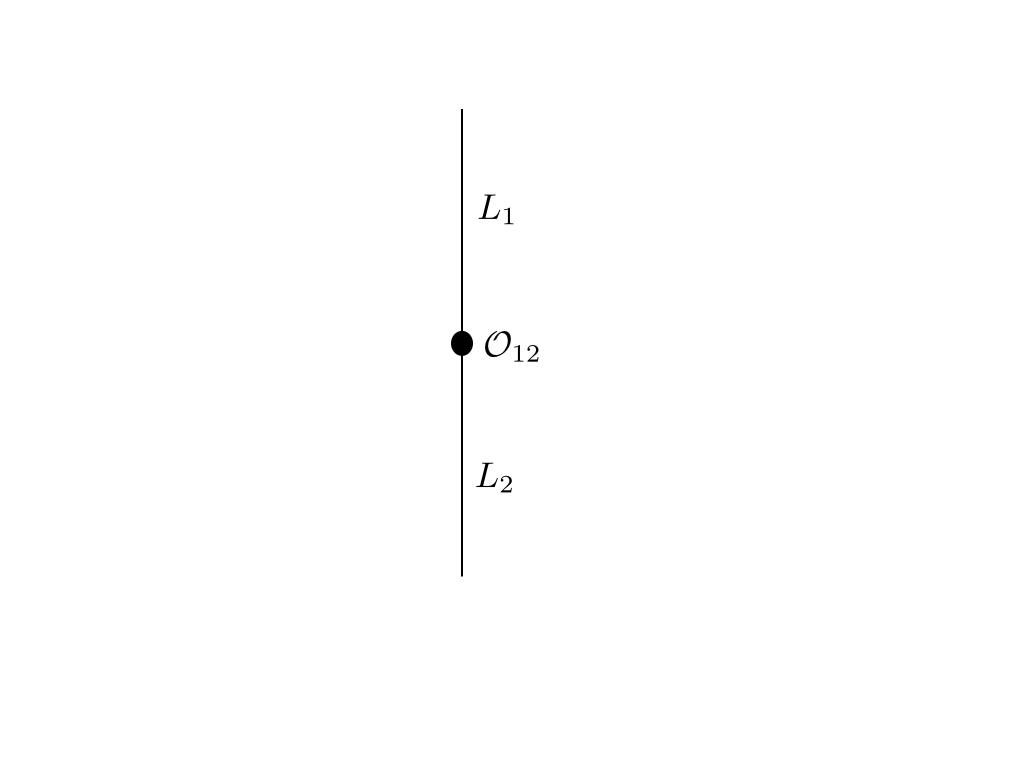}
\caption{An intertwining operator $\mO_{12}$ between the lines $L_1$ and $L_2$.}
\label{FigLoL}
\end{figure}

Note that if there is an intertwining operator between the lines $L_1$ and $L_2$,  such as in figure~\ref{FigLoL}, then the two lines carry the same charge under the one-form symmetry. In particular, if a line can end (meaning that either of $L_1$ or $L_2$ is trivial), then the line is not charged under the one-form symmetry. Importantly, this does not mean that the line furnishes a trivial DCFT in general. The example of line defects in 2+1 dimensional TQFT (which are trivial, topological, but not transparent) that cannot end demonstrates that just because a line cannot end, it is not necessary a nontrivial DCFT.

\begin{figure}[h!]
\centering
\includegraphics[scale=0.3,trim={0em 12em 0em 6cm}, clip]{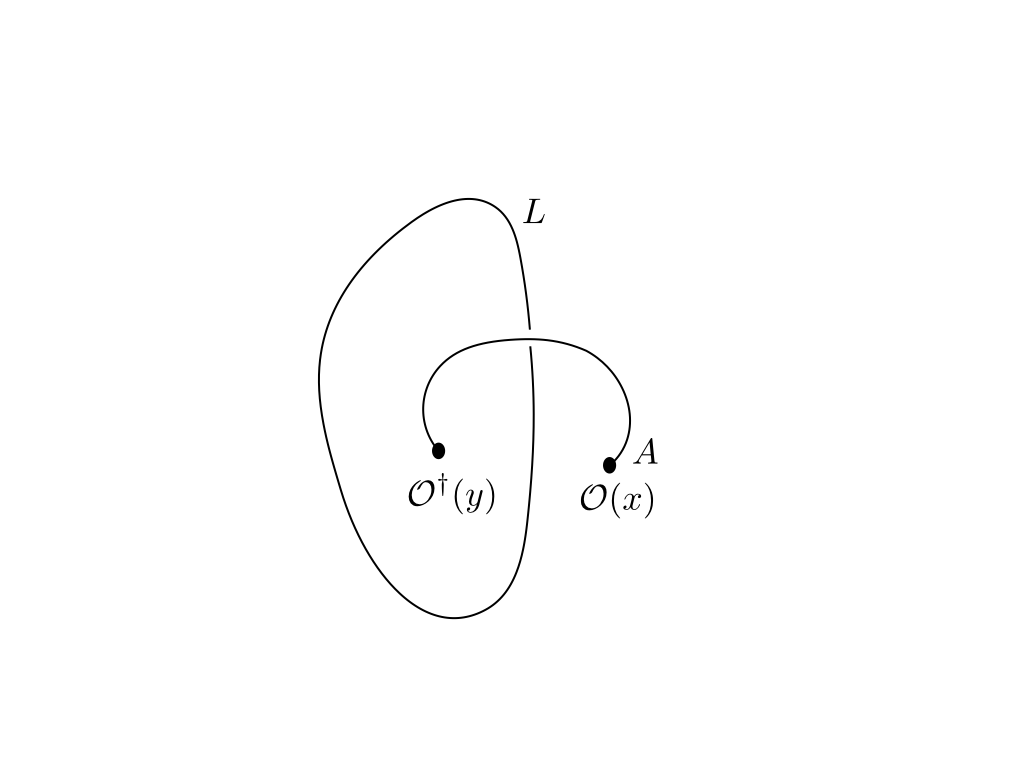}
\caption{A one-form symmetry generator cut open in 2+1 dimensions.}
\label{FigCut}
\end{figure}

Let us assume that the 1-form symmetry charge $A$ can be cut open.  This means that the one-form symmetry charge can be terminated on codimension $3$ (twist) operators, as illustrated in figure~\ref{FigCut} in 2+1 dimensions. The end points of $A$ are not topological in general.

In this case, it is evident that $L$ cannot be topological. This is because if we move $L$ through $A$ we get a phase $\omega$, but if we move $L$ to the same final location without crossing $A$ then we do not get a phase.
Therefore we have shown that if the one-form symmetry charge can be cut open, the Wilson lines charged under it must have a nontrivial displacement operator $D_\perp\neq0$.

In summary, if the one-form symmetry charge $A$ can be terminated on codimension $3$ operators, then charged lines cannot furnish a trivial DCFT.
The central question is therefore when can the one-form symmetry charges be cut open. Here we will make two comments about it. 

In $d=3$ both the one-form symmetry surface and the Wilson line are 1 dimensional defects. If the one-form symmetry has an anomaly then the one-form symmetry lines certainly cannot be cut open because they are charged under themselves. One can find many conformal gauge theories with vanishing one-form symmetry anomaly. For instance, this is the case in ABJM theory. 

In 4d gauge theories, the electric 1-form symmetry surfaces can in many cases terminate on improperly quantized 't Hooft lines. Hence Wilson lines charged under the 1-form symmetry cannot be topological, and must furnish a nontrivial DCFT in such theories.  This certainly applies in pure Yang-Mills theory with gauge group $SU(N)$, QED with charge $q_\phi$ particles, and in $\mathcal{N}=4$ SYM theory with gauge group $SU(N)$.

Let us now return to~\eqref{eq_ChargeGen} where lines with charge $q\mod q_{\phi}\neq 0$ are charged under the one-form symmetry and should flow to nontrivial DCFTs at large distances. Note that this argument does not specify which properties this DCFT should have; in particular it does not imply that the electric field should be non-zero at large distances. All the argument says is that there should be some remaining response to displacing the line defect.

It is instructive to discuss this prediction within the formalism of section~\ref{subsec_IR_line}, where we described the effective defect field theory describing the long distance limit of a screened Wilson line.\footnote{Note that in this model all lines with charge $q>2\pi/(q_{\phi}e^2)$ are unstable. } In that setup we can model a Wilson of charge $q\mod q_{\phi}=\delta q\neq 0$ as a small perturbation of a neutral (under the one-form symmetry) line of charge $q-\delta q$ (for $\delta q\sim q_{\phi}\sim O(1)\ll q$).  To this aim we simply add to the EFT defect action~\eqref{eq_eft_2} the following perturbation
\begin{equation}\label{eq_eft_d_q}
\delta S_D=-\delta q\int_{r=0} dt \,A_0=
-\delta q \int_{r=0} dt (A_0-\dot{\pi}/q_{\phi})-\frac{\delta q}{q_{\phi}}\int_{r=0}dt\, \dot{\pi}\,.
\end{equation}
The rest of the analysis proceeds as in the case $q_{\phi}=1$. In particular in perturbation theory we can neglect the total derivative in \eqref{eq_eft_d_q} and proceed similarly to what we did below \eqref{eq_eft_free_deformation} and \eqref{eq_eft_2}, where we showed that the operator $A_0-\dot{\pi}/q_{\phi}$ is irrelevant (both for $\lambda=0$ and $\lambda>0$).  Note however that the Goldstone field is $2\pi$-periodic and thus for $\delta q\neq 0$ the last term in \eqref{eq_eft_d_q}, while it has no effect in perturbation theory,  implies a nontrivial braiding between the defect and the one-form symmetry surface operator, as expected. 

We may contrast these findings with the argument above, namely that Wilson lines of charge $\de q\neq 0$ cannot furnish trivial DCFTs.  For zero quartic coupling, $\lambda=0$, the scalar admits a nontrivial conformal one-point function \eqref{eq_eft_saddle_free} and the line defines thus a nontrivial DCFT.  It is perhaps still surprising that we do not measure any Coulomb field at large distances. Physically, this is because in a massless theory the $\Phi$-particles may have arbitrarily delocalized wave-function; it is therefore possible to store fractional units of charge at $r\rightarrow \infty$.\footnote{This is similar to the fate of Wilson lines in the Schwinger model (QED$_2$) with fermions of charge $q_{\psi}>1$. As in our setup, there the electric flux of Wilson lines with charge $q\neq 0\mod q_{\psi} $ is not fully screened by massive fermions, while it is in the massless limit \cite{Gross:1995bp}. A similar behavior appears also in QCD$_2$ with massless adjoint fermions, where fundamental Wilson lines are screened. In all cases, the IR limit of charged line defects remains nontrivial and it is given by a topological line.  Indeed, with massless fermions, the infrared theory has multiple vacua and the Wilson line flows to the defect interpolating between these degenerate vacua. This also means that far away from the Wilson line there is no electric field but there is a scalar VEV due to an order parameter distinguishing these vacua. Note that the IR limit being topological does not contradict the general theorem, since the one-form symmetry charge is a local operator in a $2d$ QFT and thus cannot be cut open. It would be interesting to study $1+1$d theories with charged massless scalars, to see if these can also screen fractional charges.} 

To substantiate this interpretation, in appendix \ref{app_quantization} we study the screening cloud for a scalar of small mass squared $m^2>0$, such that $m^{-1}\gg R_{cloud}$, where $R_{cloud}$ is the scale over which the massless soliton we have found is localized.  In that case, due to the mass term, the scalar profile decays exponentially at distances of order of the Compton wavelength $r\sim m^{-1}$.   We find that Wilson lines charged under the electric one-form symmetry retain an $O(1)$ amount of charge.  Namely, we show that the flux of the electric field $~ r^2 F_{tr}$ decreases until it reaches a minimum at distances $r\sim 1/m$. After that the flux increases again and eventually settles into a constant $O(1)$ value.  Therefore the Wilson line is nontrivial and, even at distances such that the scalar profile has decayed completely, $r\gg 1/m$, there is an $O(1)$ remnant Coulomb field.

 For a small negative mass $m^2<0$ instead the bulk theory flows to a Higgs phase, described by a $\mathbb{Z}_{q_{\phi}}$ gauge theory. In $\mathbb{Z}_{q_{\phi}}$ gauge theory the one-form symmetry surface operator cannot be cut-open because of the emergent two-form symmetry, and thus there is no obstruction for a charged line to be topological in the IR; this is obviously the fate of Wilson lines with charge $\de q\neq 0$.\footnote{This may be seen e.g. from the comments below \eqref{eq_eft_d_q} and the fact that the scalar field one-point function \eqref{eq_eft_saddle_free} is modified so that it decays exponentially to a constant value for $r\gg|m|^{-1}$.}

The situation is more puzzling for the massless theory with a nonzero quartic coupling $\lambda>0$. Indeed, the beta functions~\eqref{eq_eft_beta} imply that the defect approaches logarithmically a trivial DCFT in the infrared, irrespectively of the charge of the scalar field. This is tension with the conclusion that Wilson lines charged under the electric one-form symmetry should furnish nontrivial DCFTs.  The resolution of this apparent paradox might require analyzing the fate of Wilson lines beyond the double-scaling limit~\eqref{eq_eft_double_scaling} in which we worked so far. We leave the investigation of this fascinating issue for future work.

We conclude this section by noticing that in scalar QED there is a $U(1)$ magnetic one-form symmetry, whose topological charge can be terminated on improperly quantized Wilson lines. This implies that all 't Hooft lines, which are charged under the magnetic one form symmetry, furnish nontrivial DCFTs. Physically,  this is because there are no monopoles to screen them.  We will analyze 't Hooft lines in greater detail in section~\ref{sectHooft}.

\section{Fermionic \texorpdfstring{QED$_4$}{QED4}}\label{sec_Fer}

In this section we consider fermionic QED in $d=4$ dimensions in the presence of a Wilson line of charge $q>0$, extending in the time direction. The action is given by:
\begin{equation} \label{eq_Action_fer_main4d}
\begin{aligned}
& S = S_{\psi,A}-q\int dt\, A_0\,,\\
& S_{\psi,A} = \frac{1}{e^2}\int d^4x \left[-\frac{1}{4}F_{\mu\nu}^2+i\bar{\Psi}_D\slashed{D}\Psi_D  \right],  \\
\end{aligned}
\end{equation}
where $\Psi_D$ is a massless Dirac spinor in four dimensions that carries a charge $1$ under the $U(1)$ gauge group, $F_{\mu\nu}$ is the electromagnetic field tensor,  $\slashed{D} \equiv \Gamma^\mu D_\mu$ where $D_\mu=\pd_\mu-i A_\mu$ denotes the gauge covariant derivative, and $\Gamma^\mu$ are the Dirac Gamma matrices in $d=4$, satisfying $\{\Gamma^\mu,\Gamma^\nu\}=2\eta^{\mu\nu}$.

As in the previous section, we tune the fermion mass to zero and work in the semiclassical limit specified by the following double-scaling limit :
\begin{equation}\label{eq_fer_double}
\begin{aligned}
&e \to 0, \quad q\to \infty,\\
&e^2 q = \text{fixed}.
\end{aligned}
\end{equation} 
In this limit the generated mass scale associated with QED becomes infinite and thus we can ignore any RG flow in the bulk.  
In this limit we expand the fields around the classical saddle point $A_0 = \frac{e^2q}{4\pi r}\equiv \frac{g}{r}$, $\psi_D=0$.  In the rest of this section we analyze the fluctuations of the Dirac field in the Coulomb background profile.

\subsection{Dirac fermion on \texorpdfstring{AdS$_2\times S^{d-2}$}{Ad2.Sdminus2}}
\label{subsec_fermions_general_con}

When studying scalar QED in section  \ref{subsec_TwoDCFTFixedPoints}, it was convenient to map the theory to AdS$_2\times S^2$ and perform a Kaluza-Klein (KK) decomposition over the sphere. We shall adopt a similar strategy for the model \eqref{eq_Action_fer_main4d}. In this section we thus describe the general KK decomposition for a Dirac field. For future purposes we consider aribtrary spacetime dimensions $d$, specializing to $d=4$ later.

We consider a $d$-dimensional Dirac fermion, which consists of $2^{\lfloor d/2\rfloor} $ complex components, coupled to an external gauge field.  The action is:
\begin{equation} \label{eq_Dirac_action_gauge_coupling}
S=\int d^d x \,i\bar{\Psi}_D\left( \slashed{\partial}-i\slashed{A}\right)\Psi_D\,.
\end{equation}
Just like in \eqref{WeylTF}, we can map the theory from $\mathbb{R}^d$ to $\text{AdS}_2\times S^{d-2}$   via a Weyl rescaling
\begin{equation}\label{eq_metric_general_d}
 ds^2=r^2\left[ \frac{dt^2-dr^2}{r^2}-d\Omega^2_{d-2}\right] 
= r^2 d\tilde{s}^2_{\text{AdS}_2\times S^{d-2}}\,.
\end{equation}
In a factorized geometry of the form $\mathcal{M}_2\times S^{d-2}$, such as that in \eqref{eq_metric_general_d}, there exists a convenient decomposition of the Dirac field and the associated Clifford algebra  \cite{Lopez-Ortega:2009flo}.
The fermionic field is written in terms of the following expansion:
\begin{equation} \label{eq_fermionic_decomposition}
\Psi_D = \frac{1}{r^{\frac{d-1}{2}}}\sum_{\ell,s} \sum_{\delta=+,-}\psi^{(\delta)}_{\ell s}(t,r)\otimes \chi^{(\delta)}_{\ell s}(\hat{n})\,,
\end{equation}
where $\psi^{(\delta)}_{\ell s}(t,r)$ are AdS$_2$ Dirac spinors with two complex components, while the $\chi^{(\delta)}_{\ell s}(\hat{n})$ are the spinor harmonics on $S^{d-2}$ with $2^{\lfloor d/2\rfloor-1} $ components.
The summation over $\ell$ runs over the non-negative integers, $\ell=0,1,2,...$; for every $\ell$ the $\chi_{\ell s}$ form a representation of spin $j=|\ell|+1/2$ of the $\text{Spin}(d-1)$ group,  with the index $s$ running over the components.\footnote{For general $\ell$ and $d$ the multiplicity of $s$ is given by $\frac{2^{\lfloor\frac{d-2}{2}\rfloor}(d-3+\ell)!}{\ell!(d-3)!}$ \cite{Camporesi:1995fb}.}
The spinor harmonics $\chi^{(\delta)}_{\ell\,s}(\hat{n})$ satisfy the following equation \cite{Lopez-Ortega:2009flo,Camporesi:1995fb}:
\begin{equation}\label{eq_angular_eq}
\slashed{\nabla}_{S^{d-2}}\chi_{\ell s}^{(\pm)}(\hat{n})=\pm i \left(\ell+\frac{d-2}{2}\right)\chi_{\ell s}^{(\pm)}(\hat{n})\,,
\end{equation}
where $\slashed{\nabla}_{S^{d-2}}$ is the Dirac operator on $S^{d-2}$,
as well as the orthogonality relation:
\begin{equation}
\int d\Omega_{d-2} \, \chi^{\dagger\, (\delta)}_{\ell s}(\hat{n}) \chi^{\, (\delta')}_{\ell' s'}(\hat{n}) = \delta_{\ell \ell'} \delta_{ss'}\delta^{\delta\delta'}\,.
\end{equation}
We can similarly decompose the $d$-dimensional gamma matrices $\Gamma$.  We denote by $\gamma^0$ and $\gamma^1$ the two-dimensional gamma matrices in Lorentzian signature, satisfying $\{\gamma^a,\gamma^b\}=2\eta^{ab}=2\, \text{diag}(1,-1)$ ($a,b=0,1$).  We additionally introduce a $2^{\lfloor(d-2)/2\rfloor}\times 2^{\lfloor(d-2)/2\rfloor}$ dimensional representation of the Euclidean Clifford algebra $\hat{\gamma}_E^i$, $i=1,2,\ldots,d-2$, which satisfies $\{\hat{\gamma}_E^i,\hat{\gamma}_E^j\}=2\delta^{ij}$. Then we have the following decomposition:
\begin{equation}\label{eq_gen_construction_gamma_mat}
\begin{aligned}
&\Gamma^0= \gamma^0 \otimes \hat{\mathds{1}},\\
&\Gamma^1= \gamma^1\otimes \hat{\mathds{1}},\\
&\Gamma^2 = i\gamma^3 \otimes  \hat{\gamma}^1_E,\\
&\Gamma^3 = i\gamma^3\otimes \hat{\gamma}^2_E,\\
&\vdots \\
&\Gamma^{d-1} = i\gamma^3\otimes \hat{\gamma}^{d-2}_E,
\end{aligned}
\end{equation}
where $\gamma^3$ is the $2\times 2$ AdS$_2$ chirality matrix defined by
\begin{equation}\label{eq_gamma3def}
\gamma^3=\gamma^0\gamma^1,
\end{equation} 
and $\hat{\mathds{1}}$ is the identity matrix of dimension $2^{\lfloor(d-2)/2\rfloor}\times 2^{\lfloor(d-2)/2\rfloor}$.\footnote{Note that in $d=3$ under the decomposition on $\text{AdS}_2\times S^1$, $\hat{\gamma}^1_E=\hat{\mathds{1}}=1$ are just numbers.}

Using \eqref{eq_fermionic_decomposition} and \eqref{eq_gen_construction_gamma_mat} we can write the action \eqref{eq_Dirac_action_gauge_coupling} as a sum over the AdS$_2$ spinors:
\begin{equation} \label{eq_action_fer1}
S = \sum_{\ell,s} \,\sum_{\delta=+,-} \int_{\text{AdS}_2} d^2x \sqrt{g}\, \bar{\psi}_{\ell s}^{(\delta)} \left[i\left( \slashed{\nabla}_{\text{AdS}_2}-i\slashed{A}\right)-\delta i\gamma^3 m_\ell \right]\psi^{(\delta)}_{\ell s}\,,
\end{equation}
where $\bar{\psi}_{\ell s}^{(\delta)} = ({\psi}_{\ell s}^{(\delta)})^\dagger\gamma^0$ and the masses $m_\ell$ are given by
\begin{equation} \label{eq_fer_mass_Ads2}
m_\ell = \ell+\frac{d-2}{2}\,.
\end{equation}
In \eqref{eq_action_fer1}, $\slashed{\nabla}_{\text{AdS}_2}=\gamma^a e_a^\mu\nabla_\mu$ is the AdS$_2$ Dirac operator,  with $e_a^\mu$ the vielbeins (which we take to be in the diagonal convention), and $\nabla_\mu$ the covariant derivative; similarly $\slashed{A}=\gamma^a e_a^\mu A_\mu$. We may bring the action \eqref{eq_action_fer1} to a more symmetric form by performing the following axial transformation:\footnote{Note that this transformation is not anomalous since we rotate the fields $\psi^{(+)}_{\ell s}$ and $\psi^{(-)}_{\ell s}$ by opposite angles.} 
\es{eq_rotated_fields}{
& \psi_{\ell s}^{(\pm)} \to e^{\mp i\frac{\pi}{4}\gamma^3}\psi_{\ell s}^{(\pm)}, \\ 
& \bar{\psi}_{\ell s}^{(\pm)} \to (\psi_{\ell s}^{(\pm)})^\dagger e^{\pm i \frac{\pi}{4}\gamma^3}\gamma^0. 
}
In this basis, the action \eqref{eq_action_fer1} takes a simple form:
\begin{equation} \label{eq_action_fer2}
S = \sum_{\ell,s}\, \sum_{\delta=+,-} \int_{\text{AdS}_2} d^2x \sqrt{g}\, \bar{\psi}_{\ell s}^{(\delta)} \left[i\left( \slashed{\nabla}_{\text{AdS}_2}-i\slashed{A}\right)-m_\ell \right]{\psi}_{\ell s}^{(\delta)}.
\end{equation}

The action \eqref{eq_action_fer1},  in addition to the internal $Spin(d-1)$ symmetry and the $U(1)$ gauge transformations,  is clearly invariant under $SO(2)$ rotations acting on $\{\psi_{\ell s}^{(+)},\psi_{\ell s}^{(-)}\}$, i.e.
\begin{equation}\label{eq_axial3}
 \begin{aligned}
 & \psi_{\ell s}^{(\pm)} \to \cos(\theta)\psi_{\ell s}^{(\pm)}\mp \sin(\theta)\psi_{\ell s}^{(\mp)}\,.
 \end{aligned}
 \end{equation} 
In even dimensions, this global symmetry is the AdS$_2$ avatar of the axial symmetry of the massless Dirac action. To see this in $d=4$, we recall the standard definition of $\Gamma^5$:
\begin{equation}
\Gamma^5=i\Gamma^0\Gamma^1\Gamma^2\Gamma^3.
\end{equation}
The axial transformation of the Dirac spinor in flat space reads:
 \begin{equation}\label{eq_axial_trans_4d}
 \Psi_D\to e^{i\Gamma^5\theta}\Psi_D, \qquad \bar{\Psi}_D \to \bar{\Psi}_D e^{i\Gamma^5\theta},
 \end{equation}
which is a symmetry of the theory \eqref{eq_Action_fer_main4d} (at a classical level). Using the decompositions \eqref{eq_fermionic_decomposition}, \eqref{eq_gen_construction_gamma_mat},  and recalling the field redefinitions \eqref{eq_rotated_fields}, \eqref{eq_axial_trans_4d} is easily seen to be equivalent to \eqref{eq_axial3}.  

The action \eqref{eq_Action_fer_main4d} for the Dirac field in flat $d=4$ space is also invariant under the following discrete parity transformation  
\begin{equation}
\begin{aligned}
&\Psi_D(t,r,\theta,\phi) \to \Gamma^5 \Gamma^2\Psi_D(t,r,\pi-\theta,\phi+\pi),\\
&\bar{\Psi}_D(t,r,\theta,\phi) \to \bar{\Psi}_D(t,r,\pi-\theta,\phi+\pi)\Gamma^5\Gamma^2,
\end{aligned}
\end{equation}
where $\theta$ and $\phi$ are the  azimuthal angles in spherical coordinate system. The above translates (up to an overall real factor) into the following transformation for the reduced fields on $AdS_2$:
\begin{equation}\label{eq_parity_red}
\psi_{\ell s}^{(\pm)}\to \pm \psi_{\ell s}^{(\pm)}, \qquad \bar{\psi}_{\ell s}^{(\pm)}\to \pm \bar{\psi}_{\ell s}^{(\pm)},
\end{equation}
which clearly leaves the action \eqref{eq_action_fer1} invariant.  Note that the transformations \eqref{eq_axial3} and \eqref{eq_parity_red} form an $O(2)$ group. 
The transformation rule \eqref{eq_axial3} of the fermions under axial symmetry as well as the discrete symmetry \eqref{eq_parity_red} will be useful to classify defect operators made of fermion bilinears in section  \ref{subsec_fer4}.

In conclusion, the action for a $d$-dimensional massless Dirac field in the presence of an Abelian gauge field can be decomposed into a sum over KK modes with angular momentum $j=|\ell|+1/2=1/2,3/2,\ldots$, each corresponding to a Dirac field in AdS$_2$. The result is compactly given in \eqref{eq_action_fer2}. For what follows it is important to note that the $\ell=0$ modes in the decomposition \eqref{eq_fermionic_decomposition} have the lowest mass, see \eqref{eq_fer_mass_Ads2}. Their degeneracy is  $2\times 2^{\lfloor\frac{d-2}{2}\rfloor} $, where the first  factor of $2$ arises from the index $\delta=+,-$, while the second factor is related to the spin degeneracy associated with $s$.

\subsection{Conformal Wilson lines: old and new fixed points} 
\label{subsec_TwoConformalLines_Fermions}

\subsubsection{Dirac fermion in \texorpdfstring{AdS$_2$}{ADS2} and boundary RG flows}\label{subsec_fer_toy}

Motivated by the decomposition that led to \eqref{eq_action_fer2}, in this section we study a single Dirac fermion in AdS$_2$ in the presence of a Coulomb field $A_0=g/r$. The action is:\footnote{Here we used $ \bar{\psi}\overset{\leftrightarrow}{\slashed{\nabla}}\psi=\frac12\bar{\psi}\gamma^a\nabla_a\psi-\frac12\left(\nabla_a\bar{\psi}\right)\gamma^a\psi$, which ensures that the action is exactly Hermitian (and not just up to boundary terms). }
\begin{equation}\label{eq_fer_toy_action}
S=\int_{\text{AdS}_2} d^2x\sqrt{g}\,\bar{\psi} \left[
i\left( \overset{\leftrightarrow}{\slashed{\nabla}}_{\text{AdS}_2}-i\slashed{A}\right)-m\right]\psi\,.
\end{equation}
By restoring the proper indices $\psi\to \psi_{\ell s}^{(\delta)}$ and setting the mass $m\to m_{\ell}$ as in \eqref{eq_fer_mass_Ads2}, we recover \eqref{eq_action_fer2}.  In the following we consider arbitrary $m>0$ and $g>0$. We choose the following representation for the gamma matrices
\begin{equation}\label{eq_gamma_matrices_2d}
\gamma^0 =\sigma_1= \begin{pmatrix} 0 & 1\\
1 & 0 
\end{pmatrix}, \qquad 
\gamma^1 =i\sigma_3= \begin{pmatrix} i & 0\\
0 & -i
\end{pmatrix}.
\end{equation}

Following the analysis in section \ref{subsec_TwoDCFTFixedPoints}, we can extract the scaling dimension of defect fermionic operators by studying the equations of motion of the Dirac field for $r\to 0$. Neglecting the time dependence, the equations of motion associated with the action \eqref{eq_fer_toy_action} for the Dirac fermion on $\text{AdS}_2$ coupled to a Coulomb field $A_0 =  g/r$ are given by: 
\begin{equation}\label{eq_fer_EOMr0}
\left[ i\left(r\gamma^1\partial_r-\frac{1}{2}\gamma^1-i g\gamma^0\right)-m\right]\psi = 0\,.
\end{equation}
We decompose the field explicitly in its components as
\begin{equation} \label{fer_two_comp_def}
\psi \equiv 
\begin{pmatrix}
\chi \\
\xi
\end{pmatrix},
\end{equation}
where $\chi$ and $\xi$ are single-component complex Grassmannian fields, in terms of which \eqref{eq_fer_EOMr0} reads:
\es{eq_fer_EOMs_comp}{
& \left( r\partial_r -\frac{1}{2}+m\right)\chi-g\xi =0\,,\\
& \left(r\partial_r-\frac{1}{2}-m\right)\xi+g\chi=0\,.
}
To leading order near the line defect,  the dependence of the modes in the radial coordinate $r$ is of the form $\sim r^\Delta$, for both $\chi$ and  $\xi$. Substituting such a dependence into the equations above yields a quadratic equation for the scaling dimension $\Delta$ of the (non-gauge-invariant) boundary operators associated with $\psi$. This results in the following:
\begin{itemize}
\item For $m^2  > g^2$: there are two real solutions to the quadratic equation for the scaling dimensions,  given by $\Delta_{\pm} = \frac{1}{2}\pm \sqrt{m^2-g^2}$.  They correspond to the two possible conformal boundary conditions for the fermionic modes, as will be detailed below.  
\item For  $m^2  <g^2$ there are no real solutions for the scaling dimensions $\Delta$. 
\end{itemize}

Thus when $m^2 =g^2$,  the parameter $g$ (which is related to the charge $q>0$ of the Wilson line) is at a critical value $g_c$. For $g<g_c$ there are two unitary conformal boundary conditions, while for $g>g_c$ there are no real solutions for the scaling dimension of the fermionic mode. We will see that this implies an instability of the vacuum for $g>g_c$. Both boundary conditions are normalizable in the window $0<\sqrt{m^2-g^2}<1/2$, where the upper limit arises from the unitarity bound $\Delta>0$.
 This behavior is analogous to the one which was observed for scalar QED in the previous section.  In $d=4$,  the mass of the lowest $\ell=0$ mode is $m=1$ and criticality is achieved for $g=g_c=1$. Using $g=\frac{e^2 q}{4\pi}$ we obtain the critical value $q_c=4\pi/e^2$,  in agreement with classic results in the literature \cite{pomeranchuk1945energy}.
Note that this value of $q_c$ differs by a factor $1/2$ from the one obtained for a scalar. In $d=3$, $m=\frac{1}{2}$, and criticality implies $g_c=\frac{1}{2}$, in agreement with the previously known results (see e.g. \cite{Pereira_2007} and references therein).  

Using the real world value for the electromagnetic coupling, the previous analysis gives a critical charge for point-like nuclei $q_c\approx 137$. In practice to estimate the real value of the critical charge one needs to account for both the size of the nucleus $r_0$ and the mass of the electron $m_e$,  and the critical charge is much larger, $q_c\approx 173$ \cite{pomeranchuk1945energy,Greiner:1985ce}. The huge discrepancy between the real world instability and the massless result might be surprising given the smallness of the dimensionless product $r_0 m_e\approx 10^{-3}$. As we will explain in section~\ref{subsec_fermions_supercritical},  this discrepancy is  naturally expained as a consequence of dimensional transmutation, and is similar to the explanation of the proton mass in QCD.

In the rest of this section we analyze the subcritical regime $m^2>g^2$, postponing a discussion of the supercritical instability to section  \ref{subsec_fermions_supercritical}. In particular we will show that, analogously to what we found in scalar QED,  the two conformal boundary conditions are related by RG flow (when both are allowed).

It is convenient to define the following dimensionless parameter: 
\begin{equation}
\nu \equiv  \sqrt{m^2-g^2}.
\end{equation}
In the subcritical regime,  the parameter $\nu$ is real and positive, and when it is also within the range $\nu <\frac{1}{2}$, both boundary conditions discussed above result in normalizable modes for the fermion.\footnote{Of course, for $g=0$ this regime corresponds to the usual double quantization window in AdS, see e.g. \cite{Iqbal:2009fd}.} 
The leading order physical solution near the boundary at $r\to 0$ explicitly reads:
\begin{equation}\label{eq_fer_modes_basic}
\begin{aligned}
&\chi = \alpha r^{\frac{1}{2}-\nu}+\frac{g}{m+\nu}\beta \,r^{\frac{1}{2}+\nu},\\
&\xi = \beta r^{\frac{1}{2}+\nu}+\frac{g}{m+\nu}\alpha \,r^{\frac{1}{2}-\nu},
\end{aligned}
\end{equation}
where $\alpha$ and $\beta$ are two independent Grassmann modes that depend only on the line coordinate $t$ and where we have omitted subleading terms whose coefficients are fixed by $\al,\beta$; see \eqref{special_PhiFluct} for such terms in the scalar case. 
The mode expansion \eqref{eq_fer_modes_basic} formally describes also the critical case $m^2=g^2$ by setting
\begin{equation}\label{eq_fer_modes}
\alpha=\left(\frac{a}{4}-\frac{bg}{2\nu}\right)r_0^{\nu}, \qquad 
\beta=\left(\frac{a}{4}+\frac{bg}{2\nu}\right)r_0^{-\nu},
\end{equation}
where $a$ and $b$ are complex Grassmann fields, and $r_0$ is an arbitrary cutoff radius. In the critical limit $\nu\to 0$ the fermionic components read:
\begin{equation}\label{eq_fer_modes_inAB}
\begin{aligned}
 &\chi \to  \frac{\sqrt{r}}{2}\left(a-b\right)+g\sqrt{r}\,b\log\left({\frac{r}{r_0}}\right)\,,\\
& \xi \to \frac{\sqrt{r}}{2}\left(a+b\right)+g\sqrt{r}\,b\log\left({\frac{r}{r_0}}\right)\, .
\end{aligned}
\end{equation}

We now continue the discussion in the spirit of the analysis presented in subsection \ref{subsec_TwoDCFTFixedPoints}.  In particular we want to construct the appropriate boundary terms corresponding to the two possible conformal boundary conditions in the window $0<\nu<1/2$, for which both modes in \eqref{eq_fer_modes_basic} are normalizable.

Note first that, unlike the scalar case,  the on-shell action vanishes for arbitrary boundary conditions, since it is linear in derivatives. The variation of the action \eqref{eq_fer_toy_action} for configurations which satisfy the bulk equations of motion is written purely in terms of the boundary modes as follows
\begin{equation}
\begin{aligned}
\delta S&= -\frac{i}{2}\int_{r=r_0}dt\, \sqrt{g g^{rr}} \left(\bar{\psi}\gamma^1\delta\psi -\delta\bar{\psi}\gamma^1\psi\right)\\
& =  \frac{1}{2}\int_{r=r_0}dt \,\sqrt{g g^{rr}} \left( \xi^\dagger\delta\chi-\chi^\dagger\delta\xi+\delta\chi^\dagger\xi-\delta\xi^\dagger\chi  \right)\\
&= \frac{\nu}{m+\nu}\int_{r=r_0}dt\, \left( \bar{\beta}\delta\alpha-\bar{\alpha}\delta\beta+\delta\bar\alpha\beta-\delta\bar{\beta}\alpha\right),
\end{aligned}
\end{equation}
where we use $\bar{\alpha}$, $\bar{\beta}$ to denote $\alpha^\dagger$, $\beta^\dagger$, respectively, and $r_0$ is a small cutoff radius.  As in subsection \ref{subsec_TwoDCFTFixedPoints},  we do not impose Dirichlet boundary conditions, but leave boundary modes free to fluctuate. Thus we are faced again with the question of adding boundary terms such that the variation of the action vanishes for either $\alpha=0$ or $\beta=0$, while leaving the other mode free to fluctuate.

At this stage, it is technically convenient to notice that we can use an arbitrary linear combination of the following four bilinears: $\bar{\psi}\psi$, $\bar{\psi}\gamma^0\psi$, $i\bar{\psi}\gamma^1\psi$ and $i\bar{\psi}\gamma^3\psi$. For infinitesimal $r_0$, such that we can neglect subleading terms in  \eqref{eq_fer_modes_basic}, this is equivalent to considering the most general linear combination of bilinears in the boundary modes: $\bar{\beta}\beta r_0^{2\nu}$, $\bar{\beta}\alpha$, $\bar{\alpha}\beta$ and $\bar{\alpha}\alpha r_0^{-2\nu}$. In the following it will be simpler to write operators directly in terms of the boundary modes.

One possible choice of a boundary term $S_{bdy}^{(1)}$ would be:
\begin{equation}\label{eq_bdry_fer1}
S_{bdy}^{(1)} = \frac{\nu}{m+\nu}\int_{r=r_0}dt \left( \bar{\beta}\alpha+\bar{\alpha}\beta +2\bar{\beta}\beta r_0^{2\nu}\right).
\end{equation}
This term is chosen so that it admits a smooth limit for $\nu\to 0$, for which it reduces to $S_{bdy}^{(1)}=\frac{g}{4m}\left(\bar{a}b+\bar{b} a\right)$ using  \eqref{eq_fer_modes_inAB}.
The total variation $\delta S+\delta S_{bdy}^{(1)} $ then reads:
\begin{equation}\label{eq_bdry_fer1b}
\delta S+\delta S_{bdy}^{(1)} = \frac{2\nu}{m+\nu}\int_{r=r_0}dt \left[ \bar{\beta}\delta\alpha+\delta\bar{\alpha}\beta+
\left(\delta\bar{\beta}\beta+\bar{\beta}\delta\beta\right)r_0^{2\nu}\right],
\end{equation}
which vanishes for $\beta=\bar{\beta}=0$, and corresponds to alternate quantization, where the most singular falloff in \eqref{eq_fer_modes_basic} is allowed to fluctuate; in the limit $\nu\rightarrow 0$ \eqref{eq_bdry_fer1b} reduces to $\delta S+\delta S_{bdy}^{(1)} =\frac{g}{4m}\left(\delta\bar{a}b+\bar{b} \delta a\right)$, and thus sets the logarithmic mode to zero, $\bar{b}=b=0$ in \eqref{eq_fer_modes_inAB}.

The other fixed point is obtained considering the following boundary term:
\begin{equation}\label{eq_bdry_fer2}
S_{bdy}^{(2)} = -\frac{\nu}{m+\nu}\int_{r=r_0}dt \left( \bar{\beta}\alpha+\bar{\alpha}\beta+2\bar{\alpha}\alpha r_0^{-2\nu}\right)\,.
\end{equation}
It can be checked that the total variation $\delta S+\delta S_{bdy}^{(2)} $ vanishes for $\bar{\alpha}=\alpha=0$ and thus corresponds to standard quantization, in which the less singular term in \eqref{eq_fer_modes_basic} is allowed to fluctuate.  We also note that the boundary term \eqref{eq_bdry_fer2} coincides with \eqref{eq_bdry_fer1} in the limit $\nu\rightarrow 0$.

The two fixed points are related by RG flow. As in the scalar case, this is triggered by a double-trace relevant perturbation $\bar{\alpha}\alpha$ of the alternate quantization boundary fixed point.  In practice,  it is convenient to keep the cutoff radius $r_0$ finite and consider the following deformation of the theory specified by \eqref{eq_bdry_fer1}
\begin{equation}\label{eq_fer_DTD}
S_{bdy}^{DTD}=-{2f_0}\int_{r=r_0} dt \,r_0^{2\nu} \left(\bar{\beta}\beta r_0^{2\nu}+\bar{\beta}\alpha+\bar{\alpha}\beta+\bar{\alpha}\alpha r_0^{-2\nu} \right)
\end{equation}
where $f_0$ is a dimensionful (bare) coupling.  The deformation \eqref{eq_fer_DTD} is in general fully equivalent to a standard double-trace $\sim\bar{\alpha}\alpha$ when considered as a perturbation of the UV DCFT corresponding to $\beta=\bar{\beta}=0$. In the limit $r_0\rightarrow 0$ it reduces explicitly to $-2f_0\bar{\alpha}\alpha$.  However the combination in \eqref{eq_fer_DTD} is chosen so that in the $\nu \to 0$ limit it becomes $-f_0\bar{a}a/2$, which is the appropriate double-trace deformation for the logarithmic case (see e.g. \cite{Aharony:2015afa}).

We now require that the total variation of the action and boundary terms vanishes: $\delta S+ \delta S^{(1)}_{bdy}+\delta S_{bdy}^{DTD}=0$. The boundary condition fixes the ratio between the modes: 
\begin{equation}\label{eq_fer_bc}
\beta = c\,\alpha\,,\qquad
c = \frac{f_0 (m+\nu)}{\nu-\left(m+\nu\right)f_0 r_0^{2\nu}}\,.
\end{equation}
Note that the limit $r_0\rightarrow 0$ simply yields $\beta=(f_0 (m+\nu)/\nu)\alpha $, while in the limit $\nu\to 0^+$ with finite $r_0$, plugging $\beta=c\,\alpha$  into the modes expansion \eqref{eq_fer_modes_basic} yields $b=f_0\, a$ in terms of the modes in \eqref{eq_fer_modes_inAB}. 

From \eqref{eq_fer_bc} we can compute the beta function associated with the perturbation \eqref{eq_fer_DTD}. To this aim we denote by $f$ the dimensionless coupling, $f=f_0 r_0^{2\nu}$. From the Callan-Symanzik equation, one finds the following beta function
\begin{equation}\label{eq_beta_fer_toy}
\beta_{f} = -2\nu f +2\left(m+\nu\right)f^2.
\end{equation}

The beta function \eqref{eq_beta_fer_toy} is the main result of this section. It has the same physical significance as in the scalar case discussed in section \ref{subsec_TwoDCFTFixedPoints}.  It admits two fixed points: an unstable one at $f=0$, corresponding to alternate boundary conditions $\beta=0$, and a stable one at $f=\nu/(m+\nu)$, corresponding to standard boundary conditions $\alpha=0$.  
For $f>0$ and $\nu>0$ \eqref{eq_beta_fer_toy} thus describes the RG flow from alternate to standard boundary conditions.  At $\nu=0$ the two fixed points merge into a unique one at $f=0$.

For $f<0$ the coupling has a runaway behavior toward $f=-\infty$.  
Analogously to the scalar case, this is associated with an instability of the vacuum. We will analyze this instability in section  \ref{subsec_fer_negative}, where we will argue that as a consequence of the Pauli exclusion principle it leads to the screening of a single unit of charge.

In the limit $g\rightarrow 0 $ the beta function \eqref{eq_beta_fer_toy} describes the well-known double-trace RG flow in AdS$_2$ from alternative to standard quantization.\footnote{It can be checked that \eqref{eq_beta_fer_toy} indeed holds for arbitrary spacetime dimensions $d$ and agrees with the previous result in \cite{Laia:2011wf} up to coupling redefinitions.} In particular the beta function \eqref{eq_beta_fer_toy} vanishes for $m=g=0$, since then $\nu=0$. This can be understood considering the solution \eqref{eq_fer_modes_inAB} in the limit $\nu\to 0$. Indeed from \eqref{eq_fer_modes_inAB} we see that the logarithmic falloffs are proportional to $g$.  Taking the limit $g\rightarrow 0$ in \eqref{eq_fer_modes_inAB} we thus find two independent complex modes proportional to $\sqrt{r}$.  This implies that for $\nu=g=0$ there is a marginal operator which rotates between the possible conformal boundary conditions.  Correspondingly, the beta function \eqref{eq_beta_fer_toy} vanishes for $m=\nu=0$.  In practice this regime is not relevant for our discussion of Wilson lines, since $m_0=\frac{d-2}{2}>0$ for $d>2$.

As a final comment, we note that, irrespectively of the value of $\nu$, the alternate fixed point does not admit relevant perturbations other than the one we considered, $\bar{\alpha}\alpha$. Indeed higher trace deformations of the form $(\bar{\alpha}\alpha)^n$ vanish for $n>1$ since the modes are Grassmanian, and all other defect operators involve derivatives and are thus irrelevant.  This is to be contrasted with the scalar setup, where higher trace deformations could become relevant at the alternate fixed point, as discussed in section \ref{subsec_scalar_stability}.

\subsubsection{Defect fixed points in four dimensions}\label{subsec_fer4}

We now apply the analysis of the previous section to the case of a Dirac fermion in $d=4$. We thus consider the action \eqref{eq_action_fer2} and focus on the $\ell=0$ modes of the decomposition \eqref{eq_fermionic_decomposition}, as these have the lowest mass in AdS$_2$. As explained in section \ref{subsec_fermions_general_con}, there are four $\ell=0$ modes. We denote them simply by $\psi^{(\delta)}_{s}$, where $\delta=+,-$ and $s=+\frac{1}{2},-\frac{1}{2}$ such that $\psi^{(+)}_{s}$, and $\psi^{(-)}_{s}$ transform both as doublets under the rotation group $SU(2)$.  The vector $\{\psi^{(+)}_{s},\psi^{(-)}_{s}\}$ rotates under the action of the $O(2)$ group associated with the axial transformations and parity.

To extend the analysis of subsection \ref{subsec_fer_toy} to $d=4$ we thus promote $\psi \to \psi_{s}^{(\delta)}$, $\chi \to \chi_{s}^{(\delta)}$ and $\xi \to \xi_{s}^{(\delta)}$, such that as in  \eqref{fer_two_comp_def}:
\begin{equation}
\psi_{s}^{(\delta)}  \equiv \begin{pmatrix}
&\chi_{s}^{(\delta)} \\
&\xi_{s}^{(\delta)} 
\end{pmatrix}.
\end{equation}
As a result, the modes $\alpha$ and $\beta$ in \eqref{eq_fer_modes_basic} are promoted to $\alpha_{s}^{(\delta)}$, $\beta_{s}^{(\delta)}$ respectively.  We introduce the notation: 
\begin{equation}\label{eq_fer_SU2_not}
\beta^{(\delta)} = \begin{pmatrix}
\beta_{+\frac{1}{2}}^{(\delta)}\\
\beta_{-\frac{1}{2}}^{(\delta)}
\end{pmatrix}\,, \quad 
\alpha^{(\delta)} = \begin{pmatrix}
\alpha_{+\frac{1}{2}}^{(\delta)}\\
\alpha_{-\frac{1}{2}}^{(\delta)}
\end{pmatrix}\,,\quad 
\quad 
\bar{\beta}^{(\delta)} = \left(\beta^{\dagger\, (\delta)}_{\frac{1}{2}}\,,\beta^{\dagger\, (\delta)}_{-\frac{1}{2}}\right), \quad \bar{\alpha}^{(\delta)} = \left(\alpha^{\dagger\, (\delta)}_{\frac{1}{2}},\alpha^{\dagger\, (\delta)}_{-\frac{1}{2}}\right)\,,
\end{equation}
to conveniently denote the $SU(2)$ doublets. 

In $d=4$, $m=m_0=1$ and $g=\frac{e^2q}{4\pi}$, where $q>0$ is the charge of the Wilson line.  According to the discussion below equation \eqref{eq_fer_EOMs_comp}, the fields $\psi^{(\delta)}_s$ admit both standard and alternate boundary conditions for $0<\sqrt{1-e^4q^2/(4\pi)^2}<1/2$. Differently than in the case of scalar $\text{QED}_4$ described in subsection \ref{subsec_TwoDCFTFixedPoints}, this condition provides both a lower and a upper bound on $q$:\footnote{Using the physical value of the QED coupling, in natural units this condition reads $119 <q<137$.}
\begin{equation}\label{eq_fer_windows}
\frac{\sqrt{3}}{2}<\frac{e^2q}{4\pi}<1\,.
\end{equation}
In this window, for each of the modes $\psi^{(\delta)}_s$ there are two conformal boundary conditions, leading to $2^4=16$ defect fixed points overall. We will focus on this window in what follows, and study the corresponding RG flows.

Consider the fixed point where $\beta_s^{(\delta)}=0$ for all modes. This is specified by a defect term of the form \eqref{eq_bdry_fer1} (promoting $\beta\to \beta^{(\delta)}_s$ etc., as discussed above). We are interested in deformations of this fixed point by fermion bilinear operators on the Wilson line.  It is natural to classify the possible bilinears according to their $SU(2)\times SO(2)$ charges, associated with the symmetries discussed in subsection \ref{subsec_fermions_general_con}.  For convenience, we use the notation $\sigma^K \equiv \left( \mathds{1},  \sigma^i\right)$, where $K=0,\cdots,3$,  the matrix $\mathds{1}$ is the $2\times 2$ dimensional identity matrix and $\sigma^i$, $i=1,2,3$ are the Pauli matrices.  Then,
 in the notation \eqref{eq_fer_SU2_not}, the most general gauge-invariant bilinear defect operator without derivatives can be written as a linear combination of the following terms:
\begin{equation}\label{eq_fer_Phi_Pert}
\Phi^{(\delta \gamma) K} \equiv r_0^{2\nu} \left( \bar{\beta}^{(\delta)}\sigma^K\beta^{(\gamma)} r_0^{2\nu} +\bar{\beta}^{(\delta)}\sigma^K\alpha^{(\gamma)} +\bar{\alpha}^{(\delta)} \sigma^K\beta^{(\gamma)}+r_0^{-2\nu}\bar{\alpha}^{(\delta)}\sigma^K\alpha^{(\gamma)} \right). 
\end{equation}
There are $16$ independent real bilinears that are invariant under the gauge symmetry: eight preserve the global $SO(2)$, among which two preserve the $SU(2)$ while the other six break it, and eight that break the global $SO(2)$, among which two are invariant under $SU(2)$ while the remaining six break it.  In addition, eight of the bilinears are invariant under parity $P$ while the other eight break it. There is a single bilinear invariant under all symmetries. 
The classification is summarized in table \ref{table_bilinears}. 
\begin{table}
    \begin{center}
        \begin{tabular}{ |c | c | c | c | c |}
        \hline
           Bilinears  & $SU(2)$ & $SO(2)$ & $P$ & $\#$ \\      \hline 
             $f_0^0\left(\Phi^{++0}+\Phi^{--0} \right)$  & $\checkmark$ & $\checkmark$ & $\checkmark$ & $1$ \\
            $ik_0^0\left(\Phi^{+-0}-\Phi^{-+0} \right) $ & $\checkmark$ & $\checkmark$ & $\cross$ & $1$\\
            $f_0^i\left(\Phi^{++i}+\Phi^{--i}\right)$  & $\cross$ & $\checkmark$ & $\checkmark$& $3$ \\
            $ik_0^i\left(\Phi^{+-i}-\Phi^{-+i}\right) $  & $\cross$ & $\checkmark$ & $\cross$ & $3$ \\
           $h_0^0\left(\Phi^{+-0}+\Phi^{-+0} \right)$  & $\checkmark$ & $\cross$ & $\cross$& $1$\\
            $q_0^0\left(\Phi^{++0}-\Phi^{--0} \right) $  & $\checkmark$ & $\cross$ & $\checkmark$& $1$\\
            $h_0^i\left(\Phi^{+-i}+\Phi^{-+i}\right)$ & $\cross$ & $\cross$ & $\cross$ & $3$\\
            $q_0^i\left(\Phi^{++i}-\Phi^{--i} \right) $ & $\cross$ & $\cross$ & $\checkmark$ & $3$\\
        \hline 
        \end{tabular}  \caption{A classification of the gauge-invariant fermion bilinear line operators.   The last column represents the number of independent bilinears of the specified type.  $f_0^{K}$, $k_0^K$, $h_0^K$ and $q_0^K$ denote the (real) bare coupling constants.} \label{table_bilinears}
    \end{center}
    \end{table}

We may now calculate the beta-functions associated with the fermion bilinears introduced in table \ref{table_bilinears}.  The derivation is analogous to the one discussed in subsection \ref{subsec_multiflavor} for multi-flavor scalar QED$_4$. We perturb the DCFT by adding the most general relevant perturbation on the line.  This can be written as:
\begin{equation} \label{eq_generalDTD_4d_fer}
S_{DTD} = -2 \int dt\, r_0^{2\nu}\left( \bar{\beta} F_0 \beta r_0^{2\nu}+\bar{\beta} F_0\alpha+\bar{\alpha} F_0\beta+r_0^{-2\nu}\bar{\alpha} F_0\alpha\right),
\end{equation}
where $F_0$ is a $4\times 4$ symmetric matrix, which collectively denote all the bare coupling constants associated with the double-trace deformations \eqref{eq_fer_Phi_Pert}.  $\alpha,\beta$ in the above carry four-components each in accordance with all the possible combinations of $(\delta)$ and $s$. 
Explicitly, the coupling constants in table \ref{table_bilinears} are related to the matrix $F_0$ via
\begin{equation} 
\lambda_A = \frac{1}{4}\Tr\left(\Sigma^A F_0\right),
\end{equation}
with $\lambda_A = f^0_0, f^i_0, k^0_0, k^i_0, h^0_0, h^i_0, q^0_0, q^i_0$, and the matrices $\Sigma^A$ are $4\times 4$ matrices given by:
\es{SigmaA}{
& \Sigma^{f^0} = 
\begin{pmatrix}
\sigma^0 & 0\\
0 & \sigma^0
\end{pmatrix}\,, \quad 
\Sigma^{f^i} = \begin{pmatrix}
\sigma^i & 0 \\
0 & \sigma^i
\end{pmatrix}\,, \quad 
 \Sigma^{k^0} = \begin{pmatrix}
0 & i\sigma^0 \\
-i\sigma^0 & 0 
\end{pmatrix}\,, \quad 
 \Sigma^{k^i} = \begin{pmatrix}
0 & i\sigma^i\\
-i\sigma^i &0
\end{pmatrix},\\
& \Sigma^{h^0} = \begin{pmatrix}
0 & \sigma^0 \\
\sigma^0 & 0 
\end{pmatrix}\,, \quad 
\Sigma^{h^i}= \begin{pmatrix}
0 & \sigma^i \\
\sigma^i & 0
\end{pmatrix}\,, \quad 
\Sigma^{q^0} = \begin{pmatrix}
\sigma^0 & 0 \\
0 & -\sigma^0 
\end{pmatrix}\,, \quad 
\Sigma^{q^i} = \begin{pmatrix}
\sigma^i & 0 \\
0 & -\sigma^i
\end{pmatrix}.
}

Similarly to the analysis around \eqref{eq_fer_bc}, by requiring that the total variation of the action and boundary terms vanish we find the following ratios between the modes:
\begin{equation}
\beta=  C\alpha , \qquad C= (m+\nu) F_0\cdot\left[\nu\mathds{1} -(m+\nu)F_0 r_0^{2\nu}\right]^{-1},
\end{equation}
where  $\mathds{1}$ is the $4\times 4$ identity matrix.  Defining the dimensionless coupling as $F=F_0 r^{2\nu}_0$,  from the Callan-Symanzik equation we find the beta-function:
\begin{equation}\label{fer_beta_fab_4d}
\beta_F = -2\nu F +2(m+\nu)F\cdot F \,.
\end{equation}
It follows that the beta-functions for each of the couplings in table \ref{table_bilinears} is given by:
\begin{equation}
\beta_{\lambda^A} = \frac{1}{4}\Tr\left(\Sigma^A\beta_f\right),
\end{equation}
where $\beta_F$ is the $4\times 4$ matrix whose terms are given by \eqref{fer_beta_fab_4d}. 

As an illustration,  we write explicitly the system of beta functions for the $SU(2)$ preserving couplings (i.e. setting $f^i=k^i=h^i=q^i=0)$:
\es{betafns}{
&\beta(f^0) = -2\nu f^0+2(m+\nu)\left[(f^0)^2+(k^0)^2+(h^0)^2+(q^0)^2\right]\,,\\
&\beta(k^0)=-2\nu k^0+4(m+\nu)k^0f^0\,,\\
&\beta(h^0)=-2\nu h^0+4(m+\nu)h^0f^0\,,\\
&\beta(q^0)=-2\nu q^0+4(m+\nu)q^0f^0\,.
}
The fixed points are classified similarly to the analysis in subsection \ref{subsec_multiflavor}, according to: 
\begin{itemize}
\item  $\left(f^0,k^0,h^0,q^0\right)=\left(0,0,0,0\right)$: an unstable fixed point which corresponds to alternate boundary conditions to all modes. The anomalous dimensions read: $\left(\gamma_{f^0},\gamma_{k^0},\gamma_{h^0},\gamma_{q^0}\right) = \left(-2\nu,-2\nu,-2\nu,-2\nu\right)$. 
\item $\left(f^0,k^0,h^0,q^0\right)=\left(\frac{\nu}{m+\nu},0,0,0\right)$: a stable fixed point that corresponds to standard boundary conditions for all modes. The anomalous dimensions read: $\left(\gamma_{f^0},\gamma_{k^0},\gamma_{h^0},\gamma_{q^0}\right) = \left(2\nu,2\nu,2\nu,2\nu\right)$. 
\item $f^0 = \frac{\nu}{2(m+\nu)}$ while $(k^0)^2+(h^0)^2+(q^0)^2=\frac{\nu^2}{4(m+\nu)^2}$: a family of unstable fixed points corresponding to mixed boundary conditions.
For example,  there are two $SO(2)$ preserving fixed points with $q^0=h^0=0$ and  $\left(f^0,k^0\right)=\left( \frac{\nu}{2(m+\nu)},\mp \frac{\nu}{2(m+\nu)}\right)$, with anomalous dimensions $\left(\gamma_{f^0},\gamma_{k^0},\gamma_{h^0},\gamma_{q^0} \right) = \left( \mp2\nu,\pm2\nu,0,0\right)$.
\end{itemize}

 It is worth mentioning that similarly to the analysis of the fixed points structure in the $SU(2)$ preserving case discussed above, one can consider perturbing the UV DCFT with $SU(2)$ breaking deformations as described in table \ref{table_bilinears}, and straightforwardly find fixed points that correspond to DCFTs (invariant under $SL(2, \mathbb{R})$) that break spatial rotation symmetry. 
 
Finally, let us remark that, differently than in the case of a single AdS$_2$ Dirac fermion analyzed in subsection \ref{subsec_fer_toy}, as we lower the charge below criticality there are additional operators that become relevant or marginal on the line at the alternate quantization fixed point.  In fermionic $\text{QED}_4$, $4$-fermion operators become relevant when $\frac{e^2q}{4\pi}<\frac{\sqrt{15}}{4}$.
Because of the fermionic statistics, there is only a finite number of marginal or relevant operators. The term that contains a polynomial of the highest number of fields is an $8$-fermion term and it becomes marginal when $\frac{e^2q}{4\pi}<\frac{\sqrt{55}}{8}$.\footnote{Using the physical value for the fine structure constant,  $4$-fermion defect operators become relevant for $q<132$, and $8$-fermion ones for $q<127$.}

\subsection{Partial charge screening from double-trace perturbation}\label{subsec_fer_negative}

Consider the model \eqref{eq_fer_toy_action} tuned to alternate boundary conditions. In this section we study the effect of a double-trace deformation of the form \eqref{eq_fer_DTD} with negative coefficient, i.e.~we study the model
\begin{equation}\label{eq_fer_toy_negF}
S=S_{alternate}-f\int_{r=0} dt\, \bar{\alpha}\alpha\,,\qquad
f<0\,,
\end{equation}
where $S_{alternate}$ schematically denotes the action for the UV defect fixed point. In the following we always assume $\nu>0$. 

As remarked in section \ref{subsec_fer_toy},  the beta function \eqref{eq_beta_fer_toy} shows that a negative double-trace coupling flows to infinitely large values.   In the scalar case, we found that this kind of RG flow is associated with the existence of an instability of the vacuum. Somewhat similarly, we will argue that a negative double-trace deformation leads to a change in the structure of the vacuum also in the fermionic case.  However, the fate and the signature of this instability are different than in the scalar case as a consequence of fermionic statistics.   In particular a negative double-trace deformation for a single $AdS_2$ Dirac fermion leads to the screening of a single unit of charge.

In this section we will also introduce some tools that will be relevant in section \ref{subsec_fermions_supercritical}, where we will discuss the supercritical Coulomb potential.  Our discussion in this section will largely be inspired by classic results about QED in strong electromagnetic fields \cite{Greiner:1985ce}.  Several technical details are given in appendix \ref{app_DC_problem}.

When we analyzed scalar QED$_4$ with a negative double-trace deformation in section \ref{subsec_scalar_stability}, we found that the retarded Green's function of the defect mode displayed a tachyon pole in the upper half plane. 
This is not the case for the fermion defect propagator. In appendix \ref{app_DC_problem_sub} we compute the propagator $G_\alpha(\omega)$ for the mode $\alpha(\omega)$ in the theory \eqref{eq_fer_toy_negF} and verify explicitly that no tachyon pole exists.\footnote{The absence of a tachyon pole is in fact expected as a consequence of the Fermi statistics as opposed to Bose statistics \cite{Faulkner:2009wj}.} Instead, the Green's function takes qualitatively the same form for both signs of $f$.

To understand the physical implication of the double-trace deformation with $f<0$, it is simplest to momentarily consider a massive theory. Consider the $4d$ model \eqref{eq_Dirac_action_gauge_coupling} with a mass term $\delta S=-\int d^4x M\bar{\Psi}_D\Psi_D$. Upon performing the KK decomposition on AdS$_2\times S^{2}$ explained in section \ref{subsec_fermions_general_con}, we find that this amounts to modifying the action \eqref{eq_action_fer2} by a term
\begin{equation}
\delta S_M= \sum_{\ell,s}\sum_{\delta=\pm}
i\delta\int_{\text{AdS}_2}d^2x\,\sqrt{g}\, r M\bar{\psi}_{\ell s}^{(\delta)}\gamma^3\psi_{\ell s}^{(\delta)}\,,
\end{equation}
where the overall factor of $\delta=\pm$ arises from the redefinition in \eqref{eq_rotated_fields}. Note that the mass term breaks explicitly both the axial symmetry \eqref{eq_axial3} that rotates the $(\pm)$ fields, as it should, and  part of the AdS$_2$ isometries. We are thus led to consider the model \eqref{eq_fer_toy_negF} deformed by the term
\begin{equation}\label{eq_fer_M_def}
\delta S^{(\pm)}_M= \pm
i\int_{\text{AdS}_2}d^2x\sqrt{g}\,r M\bar{\psi}\gamma^3\psi\,,
\end{equation}
where $M>0$ and we will consider both a positive and a negative prefactor for generality. Note the deformation \eqref{eq_fer_M_def} vanishes for $r\rightarrow 0$ and thus does not modify the near defect behavior of the field \eqref{eq_fer_modes_basic}. Therefore the boundary conditions read \begin{equation}\label{eq_fer_AB_bc_M}
\beta /\alpha =\frac{m+\nu}{\nu}f=\text{sgn}(f)\mu^{2\nu}\,,
\end{equation}
where we defined for convenience $\mu\equiv\left(\frac{m+\nu}{\nu}|f|\right)^{1/(2\nu)}$ as the mass scale introduced by the deformation, and momentarily considered both signs for $f$.  The massless limit is recovered for $M/\mu\rightarrow 0$.

As well known, in the presence of a mass gap the spectrum for the Dirac equation in an external potential organizes itself into an infinite number of discrete (bound) states with frequency $-M<\omega<M$ (with an accumulation point for $\omega\rightarrow M$),  a positive energy continuum for $\omega\geq M$ and a negative energy continuum with $\omega\leq -M$. In appendix \ref{app_DC_problem_M_sub}, we study the discrete part of the spectrum and find the quantization condition on the frequencies $\omega_n$ of the discrete bound states at $f=0$ and $f\rightarrow+\infty$, the latter case coinciding with the well known relativistic Hydrogen atom spectrum \cite{Greiner:1985ce}.  As we increase $\mu/M$ while keeping $f$ positive,  all the bound states at $f=0$ increase their energy and smoothly approach the standard quantization energies $\omega_n$ (corresponding to $f\to+\infty$). For negative $f$ instead,  as we increase $\mu$ while keeping $M$ fixed we find that the lowest energy bound state decreases its energy $\omega_0$. Eventually,  $\mu$ reaches a critical value given by
\begin{equation}
\mu_c^{(\pm)}\equiv g M\left[\frac{\pi  \,2^{2\nu }( m\pm\nu )}{g\,\sin\left(2\pi\nu   \right)  \Gamma (2\nu ) \Gamma (1+2\nu )}\right]^{\frac{1}{2\nu}}\,,\
\end{equation}
where the $(\pm)$ distinguishes the two signs in  \eqref{eq_fer_M_def}; in the following we will drop this supscript for notational simplicity. At $\mu=\mu_c$ we find $\omega_0=-M$, and the bound state becomes completely delocalized; for larger values of $\mu$ the state joins the negative energy continuum and the bound state ceases to exist. We will prove at the end of this section that for $\mu\gtrsim\mu_c$ this state still manifests itself as a resonant pole in the second sheet of the retarded Green's function.\footnote{To avoid the mention of resonances one can introduce an IR cutoff $r_{far}\gg 1/M,1/\mu$, so that the full spectrum is discrete, and a resonance simply corresponds to a single state mixing with many nearby (quasi-continuum) states (see e.g. the discussion in section 2.2 of \cite{Giudice:2000av}). The picture of a \emph{diving bound state} then simply follows by continuity of the spectrum and the fact that discrete states cannot disappear as we change continuously the potential.  Similar arguments are at the heart of two related well known classical results: Levinson's theorem in quantum mechanics \cite{osti_4434712,MIT_levinson} and the Friedel sum rule in condensed matter physics \cite{doi:10.1080/14786440208561086,solyom2008fundamentals}.} This phenomenon is referred to as the \emph{dive} of a bound state into the negative energy continuum \cite{Greiner:1985ce}. All other bound states smoothly approach the standard quantization (corresponding to $f\to+\infty$) energies as $f \to -\infty$.

The dive of a bound state implies that the vacuum of the theory acquires one unit of (negative) charge. To see this, we need to properly define the vacuum. While many choices are ultimately equivalent in the $M\rightarrow 0$ limit, a natural one is to define the vacuum as the state which minimizes the following modified Hamiltonian \cite{Greiner:1985ce}:
\begin{equation}\label{eq_fer_Heff}
\hat{H}_{M}=\hat{H}-M \hat{Q}\,,
\end{equation}
where $\hat{H}$ is the Dirac-Coulomb Hamiltonian and $\hat{Q}$ the gauge charge (normalized so that the field has negative unit charge). In old-fashioned language, this means that we consider as filled holes all states with energy less than $-M$, while states with energy larger than $-M$ are particle excitations on top of the vacuum. The definition \eqref{eq_fer_Heff} is natural if we imagine turning on the potential adiabatically starting from the usual vacuum. The term $M\hat{Q}$ is a chemical potential, which accounts for the fact that, by charge conservation, the transition to the new ground state can only happen by creation of an electron-positron pair,  with a positron that escapes far away from the Wilson line. 

According to the Hamiltonian \eqref{eq_fer_Heff}, the dive of the bound state into the low energy continuum at $\omega<-M$ for $\mu>\mu_c$ is thus interpreted as a change in the nature of the state from a particle energy level to a hole in the Dirac sea. Since all holes must be filled in the ground state, this leads to the screening of one unit of charge.  The same remains true in the massless limit.  We can understand this phenomenon physically by interpreting  the double-trace deformation in \eqref{eq_fer_toy_negF} as an attractive potential localized on the defect. For sufficiently large $\mu$, the potential traps an electron energy level close to the defect, similarly to the creation of a bound state by a Dirac delta function potential in quantum mechanics.

We finally discuss how to compute the charge cloud created by this process around the Wilson line.  Note that in the double-scaling limit \eqref{eq_fer_double} we can safely neglect the change in the electromagnetic field induced by this process, differently than in the case of scalar QED$_4$ analyzed in section  \ref{subsec_scalar_stability}, where we needed to account for the backreaction of the large-charge scalar cloud that formed.

Consider the mode decomposition of the Dirac field in terms of the two continuum modes and the discrete states,
\begin{equation}\label{eq_fer_psiDec}
\psi(t,r)=
\int_{M}^{\infty}\frac{d\omega}{2\pi} e^{-i\omega t}\psi_{\omega}(r) b_{\omega}
+\sum_{-M<\omega_n<M}\hspace*{-1em}e^{-i\omega_n t}\psi_n(r) b_n+
\int_{-\infty}^{-M}\frac{d\omega}{2\pi} e^{-i\omega t}\psi_{\omega}(r) d^{\dagger}_\omega\,.
\end{equation}
Different wave functions are orthogonal and we normalized them as
\begin{equation}
\int dr\sqrt{g g^{rr}}\psi^\dagger_\omega(r)\psi_{\omega'}(r)=(2\pi)\delta(\omega-\omega')\,,\qquad
\int dr\sqrt{g g^{rr}}\psi^\dagger_n(r)\psi_{k}(r)=\delta_{nk}\,.
\end{equation}
The canonical commutation relation $\{\psi^\dagger(t,r),\psi(t,r')\}=\delta(r-r')/\sqrt{g g^{rr}}$ implies
\begin{equation}
\{b_{\omega},b^{\dagger}_{\omega'}\}=\{d_{\omega},d^{\dagger}_{\omega'}\}=(2\pi)\delta(\omega-\omega')\,,\qquad
\{b_n,b^{\dagger}_k\}=\delta_{nk}\,.
\end{equation}
According to the discussion around \eqref{eq_fer_Heff},  the vacuum satisfies 
\begin{equation}
b_{\omega}|0\rangle=b_{n}|0\rangle=d_{\omega}|0\rangle=0\,.
\end{equation}
Note that all these equations are true for arbitrary values of $\mu/M$.

We can use \eqref{eq_fer_psiDec} to give an explicit formula for the charge polarization of the vacuum~\cite{Greiner:1985ce}\footnote{We are cavalier about short distance divergences; these can be taken care of by subtracting the charge density in a reference state, such as the usual vacuum.} 
\begin{equation}\label{eq_fer_j01}
\begin{split}
\langle j_0(r)\rangle &=-\lim_{x\rightarrow x'}\frac12\langle[\bar{\psi}(x')\gamma^0,\psi(x)]\rangle\\
&=\frac12\left[
\int_M^{\infty}\frac{d\omega}{2\pi}\psi^\dagger_\omega(r)\psi_\omega(r)+
\sum_{M>\omega_n>-M}\psi^\dagger_n(r)\psi_n(r) -
\int^{-M}_{-\infty}\frac{d\omega}{2\pi}\psi^\dagger_\omega(r)\psi_\omega(r)\right]\,,
\end{split}
\end{equation}
which has the obvious physical interpretation of the particle contribution minus the hole contribution. Note the factor $1/2$ upfront.  It is sometimes convenient to express \eqref{eq_fer_j01} in terms of the retarded Green's function $i S_R(x;x')=\theta(t-t')\langle \left\{\psi(x),\bar{\psi}(x')\right\}\rangle$ as\footnote{Analogous formulas can be written using the advanced and Feynman propagators.}
\begin{equation}\label{eq_fer_j02}
\langle j_0(r)\rangle=-\int_{-M}^{\infty} \frac{d \omega}{2\pi}
\text{Im} (\text{Tr}[\gamma^0 S_R(\omega;r,r)])+\int^{-M}_{-\infty} \frac{d \omega}{2\pi}
\text{Im} (\text{Tr}[\gamma^0 S_R(\omega;r,r)])\,,
\end{equation}
where 
\begin{equation}
S_R(t,r;t',r')=\int\frac{d\omega}{2\pi}e^{-i\omega (t-t')}S_R(\omega;r,r')\,.
\end{equation}
\eqref{eq_fer_j02} simply follows from evaluating the imaginary part of the Green's function using the K{\"a}ll{\'e}n-Lehman representation.
Note that \eqref{eq_fer_j01} and \eqref{eq_fer_j02} hold also in the massless case (in which case the contribution of  bound states is absent). Obviously we recover the total amount of screened charge via 
\begin{equation}\label{eq_fer_Qscreen}
Q_{screen}=\int dr\sqrt{g g^{rr}}\langle j_0(r)\rangle\,.
\end{equation}

We could now use \eqref{eq_fer_j01} and \eqref{eq_fer_j02} to compute the charge density profile. It is however obvious on dimensional grounds that for $\mu\gg M$ the charge is localized at distances $r\sim 1/\mu$ from the defect, and in the massless limit the IR defect simply corresponds to a Wilson line with charge $q-1$ and standard boundary conditions. We instead conclude this section by showing how \eqref{eq_fer_j02} implies a discontinuity for the screened charge as $\mu$ changes from below to above the critical value $\mu_c$ for fixed $M$. Our analysis will also elucidate the aforementioned relation between \emph{diving} bound states and resonances. 

Consider the difference between the charge density for $\mu=\mu_c+\delta\mu $ and $\mu=\mu_c-\delta\mu$ for a positive $\delta\mu/M\ll 1$. From \eqref{eq_fer_j01} and \eqref{eq_fer_j02} we see that the only significant contribution to the charge density in this limit arises from the lowest bound state $\psi_0(r)$ and the negative energy continuum:
\begin{multline}\label{eq_j0_resF1}
\langle j_0(r)\rangle_{\mu=\mu_c+\delta\mu}-\langle j_0(r)\rangle_{\mu=\mu_c-\delta\mu}=-\frac12\psi_0^\dagger(r)\psi_0(r)\\
+\int^{-M}_{-\infty} \frac{d \omega}{2\pi}
\left\{\text{Im} (\text{Tr}[\gamma^0 S_R(\omega;r,r)]_{\mu=\mu_c+\delta\mu})-
\text{Im} (\text{Tr}[\gamma^0 S_R(\omega;r,r)]_{\mu=\mu_c-\delta\mu})\right\}
+O(\delta\mu/M)
\,.
\end{multline}
Standard arguments about mixing of isolated states with a continuum let us express the difference in the second line in terms of the wave-function $\psi_0(r)$ of the \emph{diving} state just below criticality \cite{di1992lezioni,Greiner:1985ce}: 
\begin{multline}\label{eq_j0_resF2}
\text{Im} (\text{Tr}[\gamma^0 S_R(\omega;r,r)]_{\mu=\mu_c+\delta\mu})-
\text{Im} (\text{Tr}[\gamma^0 S_R(\omega;r,r)]_{\mu=\mu_c-\delta\mu})\\
=-\frac{\psi_{0}^\dagger(r)\psi_{0}(r)\Gamma_{res}/2}{(\omega-E_{res})^2+\Gamma_{res}^2/4}+O\left(\frac{\delta\mu}{M}\right)\,,
\end{multline}
where $E_{res}+M=O(\delta\mu)$ (with $E_{res}<-M$) while $\Gamma_{res}=O(\delta\mu^2/M)$. In other words, the \emph{diving} bound state became a resonance in the negative energy continuum.\footnote{It may be argued that the width $\Gamma_{res}$ is associated with the inverse decay time of the \emph{wrong vacuum}, where the hole is not filled \cite{Greiner:1985ce}. \label{footnote_decay_fer}} In practice  \eqref{eq_j0_resF2} only applies \emph{locally} for $r\lesssim 1/M$, since the analytic continuation of the wave-function $\psi_0$ changes its behavior at infinity when $\omega_0$ becomes complex. Using \eqref{eq_j0_resF2} in \eqref{eq_j0_resF1} we conclude 
\begin{equation}\label{eq_j0_resF3}
\begin{split}
\langle j_0(r)\rangle_{\mu=\mu_c+\delta\mu}-\langle j_0(r)\rangle_{\mu=\mu_c-\delta\mu}
=-\psi_{0}^\dagger(r)\psi_{0}(r)+O(\delta\mu/M)\,.
\end{split}
\end{equation}
\eqref{eq_j0_resF3} implies a discontinuity of the Green's function at $\mu=\mu_c$.  Integrating \eqref{eq_j0_resF3} we recover the expected discontinuity for the screening charge:
\begin{equation}
Q_{screen}\vert_{\mu=\mu_c+\delta\mu}-Q_{screen}\vert_{\mu=\mu_c-\delta\mu}=-1\,.
\end{equation}

As explained in section \ref{subsec_fermions_general_con}, in $d=4$ there are $4$ independent $\ell=0$ modes. Thus, negative double-trace deformations of the UV fixed point may lead to up to $4$ units of charge screening.  Intriguingly, this remains true also for massive Dirac fields. It would be interesting to analyze further possible implications of this analysis for real world nuclei (note that the window \eqref{eq_fer_windows} implies $q>119$ for our world, corresponding to theoretically predicted super-heavy elements).

\subsection{Supercritical Wilson lines}\label{subsec_fermions_supercritical}

In this section we address the fate of Wilson lines with supercritical charge, $q>q_c$.  Differently from the scalar setup, we will argue that the charge of the Wilson line is screened only down to $q=\lfloor q_c\rfloor$, as a consequence of the Pauli exclusion principle.  We will be particularly interested in the nearly supercritical regime, where we will show that dimensional transmutation leads to an exponentially large matter cloud screening the Wilson line.  

While our main focus is on $4d$ massless QED, whenever possible we keep the notation general. Indeed our analysis applies almost verbatim to setups where the matter fields live in $d=3$; we discuss some of these in section \ref{Sec3dCFTs}. In particular, our analysis is largely inspired by previous works on charged impurities in two-dimensional graphene sheets \cite{shytov2007atomic,shytov2007vacuum,Nishida:2014zfa}.

We consider the model \eqref{eq_Action_fer_main4d} in the presence of a Wilson line of charge $4\pi/e^2<q<8\pi/e^2$ so that $m_0<g<m_1$ (in the notation of \eqref{eq_fer_mass_Ads2}). The trivial saddle-point $\Psi_D=0$, $A_0=g/r$ corresponds to the supercritical regime for the $\ell=0$ modes of the decomposition \eqref{eq_fermionic_decomposition}, according to the discussion in section \ref{subsec_fer_toy}. 
In this case, the solution of the equations of motion \eqref{eq_fer_EOMs_comp} for $r\rightarrow 0$ is written as\footnote{In this section we will omit the subscript $s$ and the supscript $(\delta)$ from the fields, since they are inessential for our analysis (besides introducing a degeneracy).}
\begin{equation}\label{eq_fer_modes_super}
\begin{aligned}
&\chi = \frac{g}{m+i \nt}\beta r^{\frac{1}{2}+i\nt}+\alpha r^{\frac{1}{2}-i\nt}\,,\\
&\xi = \beta r^{\frac{1}{2}+i\nt}+\frac{g}{m+i \nt} \alpha r^{\frac{1}{2}-i\nt}
\end{aligned}
\end{equation}
where we let $m=m_0$ and we defined
\begin{equation}
\nt=\sqrt{g^2-m^2}\,.
\end{equation}
The nearly supercritical regime we will be interested in corresponds to $\nt\ll 1$.
\eqref{eq_fer_modes_super} shows that there are no unitary conformal boundary conditions for the Dirac field as $r\rightarrow 0$.  

We are thus forced to choose non conformal boundary conditions on the defect.  While our results are ultimately independent of this choice, for the sake of definiteness, we follow \cite{pomeranchuk1945energy} and imagine that the charge of the Wilson line is localized inside a small cutoff surface at $r=r_0$ (modelling the nucleus as a uniformly charged ball). Thus the Wilson line in \eqref{eq_Action_fer_main4d} becomes
\begin{equation}
-q\int dt \,A_0 \;\;\rightarrow\;\;
-\frac{q}{4\pi} \int_{r=r_0} \hspace{-1em}dt \,d\Omega_2 \,A_0\,.
\end{equation}
This implies that for $\Psi_D=0$ the saddle-point profile for the gauge field reads
\begin{equation}\label{eq_fer_A0_super}
A_0=\begin{cases}\displaystyle
\frac{g}{r_0}& \text{for }r<r_0\\[0.7em]
\displaystyle\frac{g}{r} & \text{for }r\geq r_0\,,
\end{cases}
\end{equation} 
so that there is no electric field for $r<r_0$. The $\ell=0$ AdS$_2$ Dirac fields 
now satisfy standard boundary conditions for $r/r_0\rightarrow 0$,
\begin{equation}\label{eq_fer_standard_in}
\psi\sim\left(\begin{array}{c}
0 \\
r^{\frac12+\frac{m}{2}}
\end{array}\right)
\qquad\text{for }r\rightarrow 0\,,
\end{equation}
as well as being continuous at $r=r_0$.  

We now show that in the presence of a supercritical field \eqref{eq_fer_A0_super}, the trivial saddle-point $\Psi_D=0$ admits infinitely many diving states in the massless limit. In the spirit of section \ref{subsec_fer_negative}, we momentarily consider a field with mass $M>0$. In appendix \ref{app_DC_problem_Sup_M}, we find that for $\nt\ll 1$ the condition for having a bound state with energy $\omega=-M$ is\footnote{A solution with $M\sim 1/r_0$, schematically corresponding to $n=0$ in \eqref{eq_DC_sup_M_result}, may exist for different boundary conditions. Such an $n=0$ diving state is somewhat analogous to the one created by a negative double-trace deformation discussed in section \ref{subsec_fer_negative} and does not play any essential role for us.}
\begin{equation}\label{eq_DC_sup_M_result}
\nt\log (2M g r_0)=\nt \eta-\pi n\,,\qquad
n=1,2,\ldots\,,
\end{equation}
where $\eta$ is an $O(1)$ number (which depends on the sign in \eqref{eq_fer_M_def} and which we determine in appendix \ref{app_DC_problem_Sup_M}). \eqref{eq_DC_sup_M_result} admits infinitely many solutions given by
\begin{equation}\label{eq_fer_Mn}
M_n=M_0\, \Lam^n \,,\qquad
M_0=\frac{e^{\eta}}{2 g r_0}\,, \qquad \Lam=e^{-\pi /\nt}\,,
\end{equation}
where $\Lam$ is the same small number we encountered in the discussion of scalar tachyons in subsection~\ref{subsec_scalar_supercritical}. 
Note also that just like in the scalar case there are multiple solutions and they \eqref{eq_fer_Mn} are log-periodic $\log(M_n/M_{n+1})=\pi /\nt$. We comment more on this below.

Imagine now lowering $Mr_0$ for a single AdS$_2$ Dirac fermion.  When $M>M_1$, all bound states have energy $\omega>-M$. For $M_1>M\geq M_2$ there is one diving state, then as we lower to $M_2> M\geq M_3$ we have two diving states, etc..  In general, for $M_n>M\geq M_{n+1}$, $n$ states have dived into the negative energy continuum. Somewhat pathologically, in the massless limit $Mr_0\rightarrow 0$ infinitely many states have joined the negative energy continuum.  Physically this implies that the trivial saddle-point with gauge field \eqref{eq_fer_A0_super} is not a good approximation to the true ground state.  This is reminiscent of the discussion around \eqref{eq_j0_resF2}: in appendix \ref{app_DC_problem_Sup} we show that the diving states are reflected in the existence of infinitely many resonances in the negative energy continuum,\footnote{Somewhat improperly, we call resonances complex poles of the retarded Green's function analytically continued to the second sheet. These are in one-to-one correspondence with solutions of the Dirac-Coulomb equation satisfying outgoing boundary conditions: $\psi_n\sim e^{-i\omega_n t}e^{i\omega_n r}$  with $\Re\omega_n<0$ and $\Im\omega_n<0$.  Hence the $\psi_n$'s decay in time and grow exponentially for $r\rightarrow\infty$. We remark however that the corresponding frequencies have comparable real and imaginary part $\Re\omega_n\sim \Im\omega_n$, signifying that these cannot be understood as usual resonances, which arise due to a weak mixing between a discrete and a continuum spectrum as in the discussion which led to \eqref{eq_j0_resF2}.} whose (complex) frequencies satisfy a logarithmic periodicity property analogous to \eqref{eq_fer_Mn}; this fact was formerly pointed out in \cite{shytov2007atomic}.

The result~\eqref{eq_DC_sup_M_result} in $d=4$ was originally derived by Pomeranchuk and Smorodinsky \cite{pomeranchuk1945energy}, who used it to argue that the critical charge $q_c$ for real world nuclei is in fact mach larger than $137$, the value that is obtained in the massless theory.   In fact,  for realistic values of $r_0$ and $M$, we get $q_c\approx 173\divisionsymbol 175$ \cite{Greiner:1985ce}.  This is because $M_1\ll 1/r_0$ for small $\nt$, as~\eqref{eq_fer_Mn} shows.  In light of the connection with the \emph{walking} behavior associated with the fixed-point merger, as we discussed in section~\ref{subsec_2DCFTs} for scalar QED, the discrepancy between the massless result and the real world is understood as a consequence of dimensional transmutation. Indeed, the mass of the first diving state $M_1$ coincides parametrically with the scale $\mu_{IR}$ at which the double-trace coupling blows up, cf.~\eqref{eq_walking}, given a UV scale $\mu_{UV}\sim 1/r_0$. This is analogous to dimensional transmutation in QCD, where the proton mass parametrically coincides with the strong coupling scale of the one-loop beta function of the gauge coupling.

Another important remark is the following. The log-periodic structure of the solution \eqref{eq_fer_Mn} reflects an approximately cyclic RG, as for the scalar tachyons discussed in section \ref{subsec_scalar_supercritical}. 
Such an RG structure implies that the mass scale $M_n$ at which the $n$th state dives is exponentially larger than $M_{n+1}$ for $\nt\ll 1$.\footnote{This is completely analogous to the phenomenon of Efimov bound states \cite{Efimov:1970zz,Kaplan:2009kr}.} Note that this is also true for the first diving state, since $M_1$ is exponentially smaller than the cutoff scale $1/r_0$ as we commented above.
A finite small mass $M$ provides an IR cutoff to this periodic flow after $\sim \frac{\nt}{\pi}\log(M_1/M)$ cycles. 

In the massless limit, the periodic flow does not persist at arbitrary long distances once we account for the screening cloud created by the matter field and the corresponding backreaction of the gauge field.  In particular, after $q-q_c$ units of charge have been screened, the Coulomb field becomes subcritical. According to the analysis in the previous sections, no further instability can occur beyond this point (up to the one discussed in section \ref{subsec_fer_negative}, which may only change the final charge by an $O(1)$ amount) and the RG flow terminates at the standard quantization fixed point. Nonetheless,  the approximate cyclic flow plays an important role at intermediate scales; we will momentarily use this observation to our advantage to estimate the size of the screening cloud for $\nt\ll 1$.  Note that this is different from scalar QED, for which, as we discussed in section~\ref{subsec_scalar_supercritical}, all the screening solitons corresponding to more than one RG cycle are unstable.

To this aim, let us consider the formula \eqref{eq_fer_j01} expressing the charge density in terms of the single-particle (AdS$_2$) wave-functions:
\begin{equation}\label{eq_fer_charge_sup}
\langle j_0(r)\rangle_{\mathbb{R}^d}=\frac{\kappa_0/2}{\Omega_{d-1}r^{d-1}}
\int_0^{\infty}\frac{d\omega}{2\pi}
\left[\psi^\dagger_\omega(r)\psi_\omega(r)-
\psi^\dagger_{-\omega}(r)\psi_{-\omega}(r)\right]\,,
\end{equation}
where the prefactor arises due to rescaling to flat space and $\kappa_0=2^{\lfloor\frac{d}{2}\rfloor}$($=4$ in $d=4$) is the degeneracy of the $\ell=0$ modes; note the result is spherically symmetric since we are summing over all the spinor harmonics of the degenerate modes. In appendix \ref{app_DC_problem_Sup} we show that the wave-functions satisfy the following property
\begin{equation}\label{eq_fer_wf_prop}
\psi_{\omega}(\Lam^n\,  r )\
\simeq \Lam^{n/2}\,\psi_{ \Lam^n\omega}\, (r)\,, \qquad
n\in \mathds{Z}\,,
\end{equation}
which holds as long as $\omega r_0\ll 1$ and $\Lam^n\omega  r_0\ll 1$. The property \eqref{eq_fer_wf_prop} implies that in the absence of backreaction the charge density at distances $r\gg r_0$ satisfies a log-periodicity property similar to \eqref{eq_fer_Mn}:
\begin{equation}\label{eq_fer_j0_prop}
r^{d-1}\langle j_0(r)\rangle_{\mathbb{R}^d}\simeq
(\Lam^{-n} \,r)^{d-1}\langle j_0(\Lam^{-n}\, r)\rangle_{\mathbb{R}^d}\qquad
\text{for }
n\in \mathbb{Z}\,.
\end{equation}
This property \eqref{eq_fer_j0_prop} was formerly noticed in \cite{Nishida:2014zfa}. It  reflects the aforementioned cyclic RG flow.   In particular, there are $n\kappa_0$ units of screening charge 
between some $r\gg r_0$ and $ \Lam^{-n} \,r$, for every $n$.\footnote{To see this, consider introducing a mass $M$ such that $M_{n+1}\ll M\ll M_n$, for which thus $n\kappa_0$ states dived into the negative energy continuum.  Such a deformation provides an IR cutoff to the radius of the screening cloud at distances $R\sim 1/M$ with $1/M_n\ll R\ll 1/M_{n+1}$. Consistency demands that there are exactly $n\kappa_0$ units of screening charge for $r\lesssim R$. By iteration of this argument for different $n$, we conclude that there must be $\kappa_0$ units of screening charge localized at distances $R\sim  \Lam^{-n} \, r_0 $ for every $n$.}

Let us now define $r_1$ to be the radius of the region inside which $\kappa_0$ units of screening charge are contained; while its precise value  depends on the boundary condition, we generically expect $r_1\sim  r_0/\Lam $. According to the aforementioned periodicity property, it is not until exponentially larger distances $r_2\simeq r_1/\Lam $ that an additional $\kappa_0$ units of charge get screened. In between we may thus safely assume that the Coulomb potential is well approximated by
\begin{equation}\label{eq_fer_gauge_r12}
A_0\simeq\frac{e^2(q-\kappa_0)}{4\pi r}\qquad
\text{for }r_1\ll r \ll r_2\,.
\end{equation}
This implies that in computing the radius $r_2\simeq r_1e^{\pi /\nt}$ we should use the value of $\nt$ corresponding to the backreacted gauge field \eqref{eq_fer_gauge_r12}. This is a small correction to $r_2$ itself in the double-scaling limit \eqref{eq_fer_double}.  We can now repeat this process self-consistently for $r_3$, $r_4$, etc., where $r_n$ denotes the size of the region where $n\kappa_0$ units of charge have been screened. In general, defining
\begin{equation}
\nt(n)=\sqrt{\frac{e^4(q-n \kappa_0)^2}{(4\pi)^2}-m^2}\,,
\end{equation}
this leads to the following the equation
\begin{equation}\label{eq_fer_screening_discrete}
\log(r_n/r_{n-1})= \frac{\pi}{\nt(n)}\,.
\end{equation}
For a sufficiently supercritical charge (but still such that $\nt\ll 1$) we may treat $n$ as a continuous variable and approximate \eqref{eq_fer_screening_discrete} with a differential equation 
\es{nflow}{
\frac{d n}{d\log (r)}\simeq\frac{\nt(n)}{\pi}=
\frac{1}{\pi}\sqrt{\frac{e^4(q-n\kappa_0)^2}{(4\pi)^2}-m^2}
\,,
}
where $-\kappa_0 n(r)$ is the amount of screened charged at distance $r$.
We obtain the ratio $r_n/r_0$ by integrating this equation:
\begin{equation}
\begin{split}
\log (r_n)-\log (r_0) &\simeq
\int_0^{n} d x\frac{\pi}{\sqrt{\frac{e^4(q-\kappa_0 x)^2}{(4\pi)^2}-m^2}} \\
&= \frac{4\pi^2}{e^2\kappa_0}\left[\cosh^{-1}\left(\frac{q}{q_c}\right)-
\cosh^{-1}\left(\frac{(q-\kappa_0  n)}{q_c}\right)\right]\,,
\end{split}
\end{equation}
where we wrote the last line using the value for the critical charge $q_c=4\pi m/e^2$.
In particular we obtain an estimate for the total radius of the screening cloud by setting $q-\kappa_0  n=q_c$:
\begin{equation}\label{eq_fer_cloud}
\begin{split}
R_{cloud}&\simeq r_0\exp\left[\frac{4\pi^2}{e^2\kappa_0}\cosh^{-1}\left(\frac{q}{q_c}\right)\right]
\simeq r_0\exp\left[\frac{8\pi^2}{e^2 \kappa_0}\sqrt{\frac{q-q_c}{2q_c}}\right]
\,,
\end{split}
\end{equation}
where we expanded for $q/q_c-1\ll 1$. \eqref{nflow}-\eqref{eq_fer_cloud} were formerly derived in \cite{shytov2007vacuum} via different, less direct means. \eqref{eq_fer_cloud} predicts an exponentially large cloud in the limit where $e$ is infinitesimal and $\nt=m\sqrt{q^2/q_c^2-1}$ is fixed (and small).  Note that the exponent in \eqref{eq_fer_cloud} is larger by a factor of $2$ than the naive estimate that does not account for backreaction $R_{cloud}^{(naive)}\simeq \exp\left[\frac{q-q_c}{\kappa_0}\frac{\pi}{\nt(0)}\right]\simeq 
\exp\left[\frac{4\pi^2}{e^2 \kappa_0}\sqrt{\frac{q-q_c}{2q_c}}\right]$.

The extrapolation of the first equation of \eqref{eq_fer_cloud} to $q\gg q_c$ predicts a power law increase for the radius of the cloud $R_{cloud}\sim r_0(2q/q_c)^{4\pi^2/(e^2\kappa_0)}$. In the future it would be interesting to compare this behavior with a more accurate analysis of the screened line, beyond the regime $\nt \ll 1$.  The numerical methods previously developed to study Fermi surfaces in AdS/CFT \cite{Sachdev:2011ze,Allais:2012ye,Allais:2013lha} might prove useful in this context.

We finally comment on the generalization to fermions with charge $q_{\psi}>1$. In this case Wilson lines are screened to the largest possible value $q_{IR}\leq q_c$, which is compatibe with the condition that the charge difference $q-q_{IR}$ has to be quantized in units of $q_{\psi}$. In particular, the IR limit of a supercritical Wilson line is always a non-trivial (as well as a non-topological) defect, in agreement with the general constraints discussed in section \ref{subsec_1form}.

\section{Non-Abelian gauge theory}\label{secNonAbelian}

\subsection{Non-Abelian saddle point} 

In this section we discuss the generalization of our analysis to weakly coupled non-Abelian conformal gauge theories in $4d$,\footnote{Former discussions of instabilities for Wilson lines in non-Abelian gauge theories can be found in \cite{Mandula:1976uh,Mandula:1976xf,Shuryak:2003ja,Klebanov:2006jj}.} focusing on the illustrative case of an $SU(2)$ gauge group.  Schematically, the action of the models of interest is given by:
\begin{equation}\label{eq_actionSU2}
\mL_{bulk}=-\frac{1}{4g_{YM}^2}F^a_{\mu\nu}F_a^{\mu\nu}
+\frac{\theta}{32\pi^2}F^a_{\mu\nu}\tilde{F}_a^{\mu\nu}
+\text{matter}\,,
\end{equation}
where $a=1,2,3$ and $F^a_{\mu\nu}=\pd_\mu A_\nu^a-\pd_\nu A^a_\mu+\varepsilon^{abc}A^b_\mu A^c_\nu$.  Relevant examples of such theories include $\mathcal{N}=4$ SYM and the $\mathcal{N}=2$ SCFT with $N_f=4$ hypermulitplets in the fundamental.

Our former analysis of the DCFT fixed points associated with a Wilson line in QED crucially relied on expanding the gauge field around a ``Coulomb"-like fixed point. To do the same in the non-Abelian gauge theory we introduce a convenient representation of the line operator. Consider a Wilson line in the $(2s+1)$-dimensional representation of $SU(2)$
\begin{equation}\label{eq_standard_rep_WL}
W_s=\text{Tr}\left[Pe^{i\int dx^\mu A^a_\mu T^a}\right]\,,
\end{equation}
where $T^a$ form a spin-$s$ representation of the $SU(2)$ algebra.  An equivalent representation of the defect \eqref{eq_standard_rep_WL} can be given in terms of a bosonic $SU(2)$ doublet $z=\{z_1,z_2\}$ on the line, subject to the constraint $\bar{z} z=2s$. In this formulation, the total action (bulk and defect) of the defect quantum field theory (DQFT) reads 
\begin{equation}\label{eq_free_DCFT0}
S=S_{bulk}+\int d\tau\left[i\bz\dot z+\bz\frac{\sigma^a}{2}z\,A^a_\mu\dot{x}^\mu \right]\,,\qquad
\bz z=2s\, ,
\end{equation}
where $\sigma^a$ are Pauli matrices and $x^\mu(\tau)$ is an affine parametrization of the line contour. 
The action~\eqref{eq_free_DCFT0} is invariant under $SU(2)$ gauge transformations, and thanks to the constraint it is also invariant under the $U(1)$ gauge transformations $z\rightarrow e^{i\alpha(\tau)}z$.  We refer the reader to \cite{Clark_1997,Cuomo:2022xgw} for details on the equivalence between \eqref{eq_standard_rep_WL} and \eqref{eq_free_DCFT0}.  In the representation \eqref{eq_free_DCFT0}, the color matrices are given by the following bilinear operator:\footnote{More precisely,  \eqref{eq_T_def} involves a point splitting procedure $T^a=\lim_{\eta\rightarrow 0^+}\bz(\tau+\eta)\frac{\sigma^a}{2}z(\tau)$, see \cite{Cuomo:2022xgw}. This subtlety will not play a role in our analysis.}
\begin{equation}\label{eq_T_def}
T^a(\tau)=\bz(\tau)\frac{\sigma^a}{2}z(\tau).
\end{equation}
Physically, the variable $z$ is a quantum-mechanical representation of the color degrees of freedom of the heavy probe modeled by the Wilson line. 

The representation \eqref{eq_free_DCFT0} makes it straightforward to generalize the analysis of the previous sections to the non-Abelian case. Rescaling $z\rightarrow \sqrt{s}z$ we recast \eqref{eq_free_DCFT0} as
\begin{equation}
S=\frac{1}{g_{YM}^2}\hat{S}_{bulk}+s\int dt \left[i\bz\dot z+\bz\frac{\sigma^a}{2}z\,A^a_0 \right]\,,
\end{equation}
with $\bz z = 2$, where we pulled out explicitly the coupling in front of the bulk action $S_{bulk}=\hat{S}_{bulk}/g_{YM}^2$ and we assumed a straight line at $r=0$.  It is then clear that we can work in the double-scaling limit
\begin{equation}\label{eq_NonAbelian_scaling}
g_{YM}^2\rightarrow 0\,,\quad
s\rightarrow \infty\quad\text{with}\quad g_{YM}^2 s=\text{fixed}\,.
\end{equation}
The saddle-point profile (assuming trivial values for the matter fields) takes the form
\begin{equation}\label{eq_NonAbelianSaddle}
z=z_0=\text{const.}\,,\qquad
A^a_0
=\frac{g_{YM}^2 s}{4\pi r}\,\bz_0\frac{\sigma^a}{2}z_0\,.
\end{equation}
There is a $S^2$ manifold of saddle-points: this is accounted for by the integration over the zero modes which rotate the solution as $z_0\rightarrow U z_0$, where $U$ is an arbitrary element of $SU(2)$, modulo the $U(1)$ gauge transformations.  The integration over the zero modes has a trivial effect on (gauge-invariant) correlation functions. If we take only the first component of $z_0$ to be non-zero we obtain $A_0^3 = g_{YM}^2 s / (4 \pi r )$, as in the Abelian case for charge $q=s$.

On the saddle-point \eqref{eq_NonAbelianSaddle} we may then effectively decompose the matter fields according to their charge under the \emph{unbroken} $U(1)$ generated by the direction $T^a\propto\bz_0\frac{\sigma^a}{2}z_0$. For instance, a field in the fundamental decomposes into components of charge $-1$ and charge $1$, a field in the adjoint has a neutral component and charge $\pm 2$ components, etc.\footnote{We work in conventions such that $U(1)$ charges are quantized in integer units.} The rest of the analysis thus proceeds as in the Abelian case.  In particular we find that
\begin{itemize}
\item A scalar in the $2S+1$ representation of $SU(2)$ becomes tachyonic when $\left|\frac{g_{YM}^2 s S}{2\pi }\right|>1$;
\item A fermion in the $2S+1$ representation of $SU(2)$ leads to an instability for $\left|\frac{g_{YM}^2 s S}{4\pi }\right|>1$;
\item The charged components of the vector bosons also become tachyonic when $\left|\frac{g_{YM}^2 s}{6\pi }\right|>1$ (see appendix \ref{app_vectors} for the derivation).
\end{itemize}
Additionally, depending on the flavor group and the charge $s$ of the line defect, several fixed points may exist; these are connected by RG flows analogous to the ones discussed in the Abelian case (including the one corresponding to screening).

The generalization to arbitrary gauge groups $G$ is straightforward (still working in the semi-classical large-representation limit as above). In some cases, like symmetric and anti-symmetric representations of $SU(N)$, there is a simple worldline action for the Wilson line as in \eqref{eq_free_DCFT0}, while in other cases writing down a worldline action is more complicated (see, for instance, \cite{Gomis:2006sb,Beccaria:2021rmj}). If the Wilson line is in a representation $R$, and $\vec{\rho}_R$ is the highest weight vector of this representation, then without loss of generality we choose  the Wilson line to generate an electric field in the Cartan subalgebra going as $\vec{A_0} = g_{YM}^2 \vec{\rho}_R / 4 \pi r$ (and we will have zero modes that will rotate this inside the group as in the discussion above). Here we normalized the roots to square to one, to agree with the $SU(2)$ case discussed above. If we have in the bulk a scalar field in a representation $r$ with weights $\vec{\mu}_r$, then it obtains an effective mass on $\text{AdS}_2$ proportional to $g_{YM}^2 |\vec{\rho}_R \cdot \vec{\mu}_r|$, and we have an instability whenever for some component of this field this becomes larger than $2\pi$. The analysis for fermions and gauge bosons is similar with instabilities at $4\pi$ and $6\pi$, respectively, as above.

For example, if we consider the $SU(N)$ $\mathcal{N}=4$ SYM theory, where all fields are in the adjoint representation, the first instability arises from the adjoint scalars, and specifically for the scalar with $\vec{\mu} = \vec{\alpha}_1 + \cdots + \vec{\alpha}_{N-1}$.\footnote{
This is the generic case. For special weight vectors of the Wilson line, some other scalars will also become unstable at the same time, but not before.}
If we write the highest weight vector of the Wilson line as $\vec{\rho}_R = \sum_{k=1}^{N-1} \lambda_k \vec{\mu}_k$, where $\lambda_k$ are non-negative integers and $\vec{\mu}_k$ are the fundamental weights of $SU(N)$ (satisfying $\vec{\mu}_k \cdot \vec{\alpha}_j = \delta_{jk}$), then the instability arises when $g_{YM}^2 \sum_{k=1}^{N-1} \lambda_k = 4\pi$. For any fixed non-zero value of $g_{YM}^2$, only a finite number of Wilson line representations lead to stable DCFTs.

The extrapolation\footnote{
Strictly speaking, the semiclassical saddle-point described in this section is only guaranteed to apply to the large representation limit at small coupling $g^2_{YM}$ with fixed $N$ \cite{Beccaria:2022bcr}. We nonetheless expect the qualitative features of our results to survive in the 't Hooft large $N$  limit.} of  our results to the  't Hooft large $N$ limit with fixed $g_{YM}^2 N$ suggests that we must consider representations with weights of order $N$ in order to obtain instabilities. We will discuss the holographic interpretation of this below.

\subsection{Example: \texorpdfstring{$\mathcal{N}=4$}{N=4} SYM}

Let us discuss in more detail the concrete example of the $\mathcal{N}=4$ SYM theory with gauge group $SU(2)$. As well known, the theory consists of $6$ scalars $\Phi_i$ in the $\mathbf{6}$ of the $SO(6)\simeq SU(4)$ $R$-symmetry group,  $4$ Dirac fermions in the $\mathbf{4}$ of $SU(4)$, and the non-Abelian gauge field (which is not charged under the $R$-symmetry). All matter fields are in the adjoint of the gauge group $SU(2)$. 

The $\mathcal{N}=4$ SYM theory has famous half-supersymmetric Wilson lines which involve a coupling to the scalar fields that breaks the $SU(4)$ R-symmetry; these always flow to stable DCFTs, and we will discuss them more below. Here we consider Wilson lines that preserve the $SU(4)$ R-symmetry. This does not allow any coupling to single scalar fields on top of \eqref{eq_standard_rep_WL}, but scalar and fermion bi-linears are allowed.

Let us consider the (non-supersymmetric) Wilson line \eqref{eq_standard_rep_WL}. 
According to the previous discussion, all matter fields are stable around the saddle-point \eqref{eq_NonAbelianSaddle} as long as $g_{YM}^2|s|\leq 2\pi$, above which value the scalars develop an instability.  It is instructive to analyze explicitly defect operators in this setup.  For concreteness, we will focus on scalar bilinears and consider the case where standard boundary conditions are imposed on all fields. We denote the scalars as $\Phi^a_i$, where $a$ is the $SU(2)$ index and $i$ is an $SO(6)$ index. In the formalism of \eqref{eq_free_DCFT0}, we can construct $SU(2)$ invariants by contracting the $SU(2)$ indices with the line color matrix $T^a$, and the most general gauge-invariant defect operators made from two scalars take the form: 
\begin{align}\label{eq_N4_flcut}
&\mO_{ij}^{(1)}=\frac{1}{s^2}\Phi_i ^a  \Phi_j^bT^aT^b=\mO_{ji}^{(1)}\,,\\
&\mO_{ij}^{(2)}=\Phi_i ^a  \Phi_j^a-\frac{1}{s^2}\Phi_i ^a  \Phi_j^bT^aT^b=\mO_{ji}^{(2)}\,,\\
&\mO_{ij}^{(3)}=\frac{1}{s}\varepsilon_{abc}\Phi_i ^a  \Phi_j^bT^c
=-\mO_{ji}^{(3)}\,.
\end{align}
Without loss of generality we can consider a saddle point such that $T^a=s \delta^a_3$, since the zero-modes' integration does not affect gauge-invariant correlators.  Adapting to the $U(1)$ unbroken by the saddle-point \eqref{eq_NonAbelianSaddle}, $\Phi_a^i$ can be written in terms of a neutral component $\phi^3_i\equiv\Phi^3_i$ and a charge $1$ complex field $\phi^{\pm}_i\equiv\frac{1}{\sqrt{2}}\left(\Phi^1_i\pm i\Phi^2_i\right)$.  In terms of this decomposition the quadratic expansion of the operators in \eqref{eq_N4_flcut} is:
\begin{align}\label{eq_N4_flcut2}
&\mO_{ij}^{(1)}= \phi_i^3\phi_j^3+\ldots
\,,\\
&\mO_{ij}^{(2)}= \left(\phi_i^+\phi_j^-+\phi_j^+\phi_i^-\right)+\ldots
\,,\\
&\mO_{ij}^{(3)}= i\left(\phi_i^+\phi_j^--\phi_j^+\phi_i^-\right)
+\ldots\,.
\end{align}
From the analysis of the previous section we thus conclude that to leading order in the double-scaling limit \eqref{eq_NonAbelian_scaling} the dimension of the defect operators is
\begin{equation}
\Delta\left(\mO_{ij}^{(1)}\right)=2\,,\qquad
\Delta\left(\mO_{ij}^{(2)}\right)=\Delta\left(\mO_{ij}^{(3)}\right)=1+\sqrt{1-\frac{g^4_{YM} s^2}{4\pi^2}}\,.
\end{equation}
We can also consider more general fixed points where some operators are instead relevant and the $SO(6)$ group is broken to a subgroup, analogously to the discussion in section  \ref{subsec_multiflavor}.  

Thus, as in the previous subsection, $SO(6)$ preserving Wilson lines become unstable to scalar condensation for $g_{YM}^2|s|>2\pi$. The screening mechanism is completely analogous to the one discussed for scalar QED with no quartic coupling; note in particular that the screening cloud naturally aligns on a flat direction, such that the potential trivializes $V\sim\text{Tr}\{[\Phi_i,\Phi_j]^2\}=0$. Therefore the IR DCFT admits a nontrivial one-point function for the scalar field, with a coefficient that depends on the initial charge and the boundary condition. Such a coefficient therefore represents  a marginal parameter in the double-scaling limit. We expect that quantum corrections will lift this marginal direction.
At a quantum level, the $R$-symmetry group is preserved by the scalar cloud according to the discussion in subsection \ref{subsec_multiflavor}. 

The generalization of our discussion to $SU(N)$ gauge group is straightforward, with \eqref{eq_standard_rep_WL} deformed by $\Phi_i^a T^a \Phi_j^b T^b$ and all the other possible billinears; the only difference is that for $N>2$ there are more than three independent bilinear operators.

We close this section with some comments on a well studied generalization of the standard Wilson line \eqref{eq_standard_rep_WL}, which also includes a coupling to the adjoint scalar:
\begin{equation}\label{eq_zeta_WL}
W^\text{BPS}_s=\text{Tr}_{2s+1}\left[P\exp\left( \int_{\mathcal{C}}dt( i\dot x^\mu  A_\mu^a +\zeta |\dot x|\Phi_1^a)T^a\right)\right]\,.
\end{equation}
The coupling to the scalar (conventionally chosen in the ``$1$" direction) breaks the $R$ symmetry group to $SO(5)$. For $\zeta=1$ the line \eqref{eq_zeta_WL} additionally preserves half of the supersymmetry charges, and many exact results are available about this case \cite{Maldacena:1998im,Rey:1998ik,Erickson:2000af, Drukker:2000rr,Zarembo:2002an,Gomis:2006sb,Pestun:2007rz,Gaiotto:2010be,Correa:2012at,Cordova:2013bza,Lewkowycz:2013laa,Fucito:2015ofa,Fiol:2015spa,Cordova:2016uwk,Giombi:2018qox,Bianchi:2018zpb,Bianchi:2019dlw,Galvagno:2021qyq}.\footnote{For a general approach to supersymmetric line defects in diverse dimensions see~\cite{Agmon:2020pde,Giombi:2021zfb,Penati:2021tfj}.}  The coupling $\zeta$ has a nontrivial beta function to one-loop order in perturbation theory, and there is a nontrivial RG flow from the standard fixed point at $\zeta=0$ to the superconformal one at $\zeta=1$ \cite{Polchinski:2011im} (see also \cite{Beccaria:2021rmj,Beccaria:2017rbe,Beccaria:2018ocq,Beccaria:2022bcr}). 

Let us consider the operator \eqref{eq_zeta_WL} in the double-scaling limit \eqref{eq_NonAbelian_scaling}. The main difference with respect to the previous case is that the saddle-point \eqref{eq_NonAbelianSaddle} now also includes a nontrivial scalar profile proportional to $\zeta$:
\begin{equation}
\Phi_i^a
=\frac{\zeta g_{YM}^2 s}{4\pi r}\bz_0\frac{\sigma^a}{2}z_0\,.
\end{equation}
It is now straightforward to repeat the analysis in section \ref{subsec_2DCFTs} and compute the scaling dimensions of the defect operators \eqref{eq_N4_flcut}. Since to leading order in the double-scaling limit the running of the scalar coupling in \eqref{eq_zeta_WL} is negligible, we can write the result for arbitrary values of $\zeta$.  For $\frac{g_{YM}^4 s^2}{4\pi^2}\left(1-\zeta^2\right)<1$ the corresponding DCFT is unitary and the possible values of the scaling dimensions are given by
\begin{equation}
\Delta\left(\mO_{ij}^{(1)}\right)=2\,,\qquad
\Delta\left(\mO_{ij}^{(2)}\right)=\Delta\left(\mO_{ij}^{(3)}\right)=1\pm \sqrt{1-\frac{g_{YM}^4 s^2}{4\pi^2}\left(1-\zeta^2\right)}\,,
\end{equation}
where the sign in front of the square root for each operator depends on the boundary conditions as before. Importantly, for $\zeta=1$ no value of $s$ leads to an instability: this is a consequence of the supersymmetry preserved by the line, which as well known ensures that the ground state has zero energy.  Additionally, for $\zeta=1$ there is always a unique Wilson line.

The final remark concerns the RG flow from $\zeta=0$ to $\zeta=1$ initially studied in \cite{Polchinski:2011im}.  Our analysis shows that the starting point of such an RG flow is ill-defined for $s>2\pi/g_{YM}^2$. The running of $\zeta$ can nonetheless be analyzed for smaller values of $s$ in the double-scaling limit \eqref{eq_NonAbelian_scaling}.  This in general requires the analysis of one-loop corrections around the saddle-point profile; see \cite{Beccaria:2022bcr} for recent progress in this direction.

\subsection{Holographic description}

The $SU(N)$ $\mathcal{N}=4$ SYM theory is famously dual to type IIB string theory on $AdS_5\times S^5$, with a fixed weakly coupled string theory background appearing in the 't Hooft limit of large $N$ with fixed $g_{YM}^2 N$. Supersymmetric Wilson lines have been extensively discussed in this context, but for the non-supersymmetric Wilson lines \eqref{eq_standard_rep_WL} the discussion has mostly been limited to the fundamental representation. For that representation, as discussed in \cite{Alday:2007he}, the Wilson line \eqref{eq_standard_rep_WL} maps to a string ending on the appropriate contour on the boundary of $AdS_5$, with Neumann boundary conditions for the $S^5$ position of the string.\footnote{As opposed to the supersymmetric Wilson loop that obeys Dirichlet boundary conditions.} This string is stable under $SO(6)$-preserving deformations, consistent with our discussion above.

Our discussion suggests that instabilities should occur for non-supersymetric Wilson lines with weights of order $N$.  (The stability of Wilson lines in small representations  follows simply from the large $N$ factorization of correlation functions.) Supersymmetric WLs with weights of that order may be described by D-branes \cite{Drukker:2005kx,Yamaguchi:2006tq,Gomis:2006sb}; for instance, the supersymmetric WL in the $k$'th anti-symmetric representation is described by a D5-brane wrapping an $S^4 \in S^5$ (and carrying some electric field that gives it the appropriate fundamental string charge). It seems natural to conjecture that non-supersymmetric Wilson lines in representations with weights of order $N$ would be described by non-BPS D-branes wrapping the $S^5$; for instance, the straight anti-symmetric representation Wilson line may be described by a non-BPS D6-brane on $\text{AdS}_2\times S^5$ (with an appropriate electric field on $\text{AdS}_2$). At large 't Hooft coupling where the $S^5$ is weakly curved, any such non-BPS D-brane has a tachyonic instability, and it is tempting to identify this with the instability discussed above; note that for large 't Hooft coupling any WL with weights of order $N$ is expected to be unstable. Condensation of the tachyon in a non-uniform fashion that breaks $SO(6)$ to $SO(5)$ can describe the flow to the supersymmetric WLs, while the end-point of an $SO(6)$-preserving tachyon condensation is less clear. It would be interesting to study further the holographic description of non-supersymmetric WLs in various representations and their instabilities.

\section{2+1 dimensional CFTs}\label{Sec3dCFTs}

In this section we analyze Wilson lines in $2+1$ dimensional CFTs, focusing on Abelian gauge theories.  The main difference with respect to four-dimensional theories is that the standard kinetic term $F_{\mu\nu}^2$ for the gauge field is not conformal.  Nonetheless, Abelian gauge fields lead to interesting conformal fixed points in Chern-Simons theories or when they interact with matter fields in certain strongly coupled models, some of which may be analyzed perturbatively in a large $N_f$ expansion. Additionally, it is possible to couple Abelian gauge fields in four dimensions to matter fields confined on a three-dimensional interface, a setup that for instance describes the long wavelength limit of graphene. We discuss several examples below.

\subsection{Chern-Simons theories with and without matter}\label{sec5CS}

Here we review some of the properties of Chern-Simons theories with gauge group $U(1)$ coupled to a fermion or a scalar, and then study the Wilson line operators of these theories.

The Chern-Simons term at level $k$ is given by 
\begin{equation}\label{CSterm}
kS_{CS}={k\over 4\pi}\int d^3x\epsilon^{\mu\nu\rho} A_\mu \partial_\nu A_\rho~.
\end{equation}
Without additional matter fields it describes a topological theory. Wilson lines with electric charge $q$ harbor magnetic flux $-2\pi q/k$. The magnetic flux is localized to the worldline. This familiar fact can be reproduced by solving the equations of motion in the presence of the line operator $e^{iq \int dt A_0}$, leading to 
\begin{equation}
{k\over 2\pi} F_{xy}=q \delta^{2}(x_\perp)\,,
\end{equation}
where $x_\perp$ stands for the coordinates on the plane $(x,y)$.  In other words, we have a holonomy in polar coordinates: 
\begin{equation}\label{CS_holonomy}
A_\theta ={q\over k}~.
\end{equation}
Since there are no charged matter fields in this theory, Wilson lines with arbitrary $q$ do not lead to instabilities. Note that for $q=n k$, with $n\in\mathds{Z}$ the Wilson line harbors an integer multiple of the $2\pi$ flux unit, which is why such a line is transparent in the language of topological field theory. Such a Wilson line is equivalent to a shift of the gauge field by an integral holonomy $A_\theta\rightarrow A_\theta+ n$. Even in the presence of dynamical matter fields $\Phi_a$ (of any spin)
an integral holonomy $A_\theta=n$ does not have any physical consequence, as it can be eliminated via a field redefinition of the form $\Phi_a\rightarrow e^{-i n\theta}\Phi_a$. (Note that this redefinition preserves the boundary conditions around the defect only for $n\in\mathds{Z}$.)
More generally this implies that Wilson lines with different charges in Chern-Simons matter theories are identified modulo $k$: $q\sim q+k$. 
This observation will be important in section \ref{subsec_U1k_Nf}.

Adding matter fields to~\eqref{CSterm} famously leads to a very rich set of nontrivial conformal field theories in 2+1 dimensions.  These theories can be analyzed perturbatively for 
$k\gg 1$. Let us consider adding $N_f$ scalar fields $\Phi^i$ of charge 1 under the gauge symmetry and $N_f$ Dirac fermions $\psi^i$ of charge 1. 
There are four independent $SU(N_f)$-invariant marginal terms in the bulk:
\begin{equation}\label{fourv}
 \bar \psi_i\psi^i \Phi^\dagger_k\Phi^k\,,\,\bar \psi_i\psi^j \Phi^\dagger_j\Phi^i \,,\,(\bar \psi_i\bar\psi_j \Phi^i\Phi^j+ \psi^i\psi^j \Phi^\dagger_i\Phi^\dagger_j) \,,\,(\Phi^\dagger_i\Phi^i)^3~.
 \end{equation}
The system of beta functions was written in \cite{Avdeev:1992jt}. At large $k$ and fixed $N_f$, there are perturbative fixed points where all the couplings with fermions scale like $1/k$ while the sextic coupling scales like $1/k^2$.
Curiously, without fermions, no perturbative fixed points exist.
 In the following we will not need to know the precise value of the couplings at which the fixed points appear. 
The action reads 
\begin{equation}\label{CSMAbelianPre}
\begin{split}
S=kS_{CS}+\int d^3x&\left[|D\Phi_i|^2+i\bar \psi_i \slashed{D}\psi^i+\frac{\alpha}{k} \bar \psi_i\psi^i \Phi^\dagger_k\Phi^k+\frac{\beta}{k} \bar \psi_i\psi^j \Phi^\dagger_j\Phi^i
\right.\\
&\left.\qquad
+\frac{\gamma}{4 k}(\bar \psi_i\bar\psi_j \Phi^i\Phi^j+ \psi^i\psi^i \Phi^\dagger_i\Phi^\dagger_j)-\frac{h}{6 k^2}(\Phi^\dagger_i\Phi^i)^3\right]\,,
\end{split}
\end{equation}
where we normalized the couplings so that  the coefficients $\alpha$, $\beta$, $\gamma$ and $h$ are all $O(1)$ at the fixed points of interest.

Here we would like to make some observations about Wilson lines in these theories. Let us consider a Wilson line of a charge $q$ particle. 
This amounts to again deforming the action \eqref{CSMAbelianPre} by $-q\int A_0 \delta^{2}(x_\perp)$. It is convenient to normalize the fields $\Phi\to\sqrt k \Phi$ and $\psi\to\sqrt k\psi$. Then all four vertices~\eqref{fourv} become of order $O(k)$ and hence the action admits a double scaling limit for large $k$ and fixed $q/k$: 
\begin{equation}\label{CSMAbelian}
\begin{split}
S=k\Bigg\{S_{CS}+\int d^3x&\left[|D\Phi|^2+i\bar \psi \slashed{D}\psi+\alpha \bar \psi_i\psi^i \Phi^\dagger_k\Phi^k+\beta \bar \psi_i\psi^j \Phi^\dagger_j\Phi^i
\right.\\
&\left.
+\frac14\gamma(\bar \psi_i\bar\psi_j \Phi^i\Phi^j+ \psi^i\psi^i \Phi^\dagger_i\Phi^\dagger_j)-\frac{h}{6}(\Phi^\dagger_i\Phi^i)^3\right]
-{q\over k}\int d^3x A_0 \delta^{2}(x_\perp) \Bigg\}~.
\end{split}
\end{equation}
The parameters $\alpha,\beta,\gamma,h$ are all $O(1)$ in this normalization.

The classical solution we will be expanding about has $\Phi=0$, fermions in their ground state, and the gauge field given by \eqref{CS_holonomy}.
The fluctuations around this background can be analyzed by writing  dropping nonlinear terms in the action as in section \ref{subsec_2DCFTs}. Consider first the scalar field.
We decompose the field in components with different angular momenta as 
\begin{equation}
\Phi=\sum_{\ell=-\infty}^{\infty} {e^{i\ell \theta}\over \sqrt r}R_\ell(t,r)\,,
\end{equation}
where we can interpret the $\{R_{\ell}(t,r)\}$ as the KK modes of the scalar for the theory on $\text{AdS}_2\times S^1$. Going to frequency space $R_{\ell}(t,r)=e^{-i\omega t}R_{\ell}(r)$, the linearized equations of motion read
\begin{equation}\label{eq_fluctu2}
-\partial_r^2R_\ell+{-{1\over 4}+\left(\ell-{q\over k}\right)^2\over r^2}R_\ell=\omega^2R_\ell\,.
\end{equation}
The coefficient of the $1/r^2$ term in \eqref{eq_fluctu2} corresponds to the AdS$_2$ mass of the $\ell$'th KK mode. Since this coefficient is always greater than the BF bound, $-1/4$, for every $\ell$ and $q/k$,  we find there is no perturbative instability for the Wilson line.  Note however that for $q=0$ the mass of the $\ell=0$ KK mode sits exactly at the BF bound, and thus may be destabilized by arbitrary small perturbations. This observation will be important for the theories that we analyze in the next subsections.

There is still more to say since the mode $R_\ell$ admits two possible conformal boundary conditions if $|\ell-{q\over k}|<1$. 
Let us focus first on standard boundary conditions. Proceeding as in the previous sections, from \eqref{eq_fluctu2} we find the following scaling dimensions for (non-gauge-invariant) defect operators:
\begin{equation}\label{CS_scaling}
\Delta(D^{\ell}_z\Phi) = \frac12+\left|\ell-{q\over k}\right|~,\quad \Delta(D^{\ell}_{\bar z}\Phi) = \frac12+\left|-\ell-{q\over k}\right|\,,
\end{equation}
where we denoted the transverse complex coordinate by $z=x-i y$ (and $\bar z=x+iy$).
Note that for small $q/k$ \eqref{CS_scaling} indeed corresponds to small corrections to the classical dimension $\frac12+|\ell|$.

Consider now deforming the action by quadratic terms corresponding to operators with $ 0 < |\ell-{q\over k}| <1/2$ .
Since $\ell$ is integral, there is at most one $\ell$ satisfying $ |\ell-{q\over k}| <1/2$; we denote it by $\ell_0$ and assume $\ell_0>0$ for simplicity.
We can then deform the line operator by
$\delta S\propto\int dt D^{\ell_0}_z\Phi D^{\ell_0}_{\bar z}\Phi^\dagger $.  This deformation is irrelevant, but formally it leads to a UV fixed point, at which the scaling dimension is flipped for $\ell=\ell_0$,  i.e. $\Delta(D^{\ell_0}_z\Phi) = \frac12-|\ell_0-{q\over k}|$, while the scaling dimension of the operators with $\ell\neq \ell_0$ remain the same as in the infrared fixed point. The dimension of the bilinear $D^{\ell_0}_z\Phi D^{\ell_0}_{\bar z}\Phi^\dagger$ in the UV fixed point is  $1-2|\ell_0-{q\over k}|<1$, which is positive only for $|\ell_0-{q\over k}|<1/2$.  Note however that as we change $|q/k|$ quartic operators may become relevant, and the fate of the ultraviolet fixed point has to be re-considered similarly to section \ref{subsec_scalar_stability}.

As in the analysis of 4d QED,  we can ask what happens when we deform the ultraviolet defect fixed point by a negative coupling double-trace deformation. Let us focus on the case where $|\frac{q}{k}|<1/2$, so that the operator with flipped scaling dimension at the UV fixed point is $\Phi^\dagger \Phi$. The perturbation we consider is given by (in Minkowski signature)
\begin{equation}
S_{DCFT_{UV}}\rightarrow S_{DCFT_{UV}}+f^{1- \Delta_{\Phi^\dagger \Phi}}\int_{r=0\hspace*{-0.7em}} dt \,\Phi^\dagger \Phi\,,
\end{equation}
where $f>0$ and $\Delta_{\Phi^\dagger \Phi}=1-2|\frac{q}{k}|$.
As in four dimensions, this deformation leads to a classical instability of the vacuum, and the new ground state is provided by a nontrivial solitonic profile. This profile can be interpreted as an RG flow to a different, screened, defect. Unlike in scalar QED$_4$,  here we do not solve for the RG flow  numerically, and we content ourselves with providing the endpoint of this flow, focusing on the theory with a single scalar and nonzero sextic coupling $\frac{h}{6} \left(\Phi^\dagger\Phi\right)^3$. In this case, a straightforward asymptotic analysis of the equations of motion shows that at large distances from the defect both the electric and magnetic field decay faster than $1/r^2$, where $r$ is the distance from the defect, hence the gauge field is fully screened.  Instead, the scalar decays according to a conformal scaling law with a coefficient which does not depend on $q$:
\begin{equation}\label{eq_k_1pt}
\langle \Phi^\dagger\Phi(r)\rangle=\frac{h^{-1/2}/2}{r} \quad
\text{for }r\gg f^{-1}\,.
\end{equation}
\eqref{eq_k_1pt} represents a nontrivial one-point function for the scalar field, which Higgses the gauge field close to the defect. Note that the coefficient is fixed in terms of the bulk coupling.\footnote{ Interestingly, we can represent the boundary conditions leading to \eqref{eq_k_1pt} in terms of the following defect:
\begin{equation}\label{eq_k_D}
\mathcal{D}=\exp\left[i \pi\int_{r=0\hspace*{-0.7em}} dt \,\Phi^\dagger\Phi\right]\,.
\end{equation}
To prove that the defect \eqref{eq_k_D} indeed leads to \eqref{eq_k_1pt} it is enough to take the variation of the scalar action including the defect term \eqref{eq_k_D}. Upon regularizing the defect by introducing an infinitesimal thickness $r_0$, to be taken to zero at the end of the calculation, it is  easily seen that the variations of both the boundary term and the bulk action vanish on a solution of the form \eqref{eq_k_1pt}.  } The subleading falloff of the scalar and the gauge field depend on the ratio $h/(q/k)$ and we will not discuss them here.

An analogous discussion concerns the fluctuations of the Dirac field.  We find that no value of $q/k$ leads to an instability.  In terms of the decomposition discussed in section  \ref{subsec_fermions_general_con},  we see that the holonomy \eqref{CS_holonomy} simply shifts the AdS$_2$ mass of the $(\pm,\ell)$ mode as $m_{\ell}^{(\pm)}\rightarrow m_{\ell}^{(\pm)}\pm q/k$, where $m_{\ell}^{(\pm)}=\frac12+\ell=|j|$ in terms of the angular momentum $j\in\frac12+\mathbb{Z}$. We conclude that the scaling dimensions of defect operators corresponding to KK modes with different spin are given by
\begin{equation}
\Delta_j=\frac{1}{2}+\left|j +{q\over k}\right|\,,
\end{equation}
which for small $q/k$ is a small perturbation of the free theory result.  As in the scalar case, when $\left|j +{q\over k}\right|<1/2$ for some $j=j_0$ there exists a UV fixed point on the Wilson line, which flows to the infrared one via a double-trace deformation.

\subsection{Large \texorpdfstring{$N_f$}{Nf} critical points}

\subsubsection{\texorpdfstring{QED$_3$}{QED3} with \texorpdfstring{$2N_f$}{2Nf} fermions}\label{subsec_QED3_N}

Abelian gauge fields coupled to matter fields are expected to lead to several interesting conformal fixed points.  Perhaps the most interesting and well studied example is given by QED$_3$ with $2N_f$ charge $1$ Dirac fields (complex fermions with two components):
\begin{equation}\label{eq_QED3}
\mL=i\sum_{a=1}^{2N_f}\bar{\Psi}_a\left(\slashed{\pd}-i\slashed{A}\right)\Psi_a\,,
\end{equation}
where we omitted the kinetic term for the gauge field, since it is irrelevant in the sense of RG. The theory \eqref{eq_QED3} enjoys a $SU(2N_f)$ internal symmetry and it is parity invariant; Chern-Simons terms are therefore disallowed.

The model \eqref{eq_QED3} is believed to flow to an interacting CFT, at least for sufficiently large $N_f$.  The theory \eqref{eq_QED3} can be studied perturbatively in the $\varepsilon$-expansion \cite{DiPietro:2015taa} and in the large $N_f$ limit \cite{Pisarski:1984dj}. We will focus on the latter limit in what follows.

To leading order in $N_f$ the fermions behave as free fields. The gauge field has a more interesting large $N_f$ limit instead. To see this, it is convenient to integrate out the fermions in \eqref{eq_QED3} and write a nonlocal action for the gauge field. In Euclidean signature this reads:
\begin{equation}\label{eq_QED3_A}
\begin{split}
S_{eff}[A]&\equiv-2N_f\text{Tr}\left[\log\left(\slashed{\pd}-i\slashed{A}\right)\right]\\
&=\text{const.}+\frac{N_f}{16}\int \frac{d^3k}{(2\pi)^3}A_\mu(k)
|k|\left(\delta^{\mu\nu}-\frac{k^\mu k^\nu}{k^2}\right)A_\nu(-k)+\ldots\,,
\end{split}
\end{equation}
where in the second line we expanded around $A_\mu=0$ and computed the loop integral with a gauge-invariant regulator.  Therefore, to leading order in $1/N_f$, the gauge field two-point function is given by:
\begin{equation}\label{eq_QED3_2ptA}
\begin{split}
\langle A_\mu(k)A_\nu(-k)\rangle
&=N_f\frac{8}{|k|}\left(\delta_{\mu\nu}-\frac{k_\mu k_\nu}{k^2}\right)
+\text{gauge dependent terms}\,.
\end{split}
\end{equation}
Equivalently, the result \eqref{eq_QED3_2ptA} can be seen as the resummation of infinitely many bubble diagrams with fermion loops.

We now consider the theory \eqref{eq_QED3} in the presence of a Wilson line. Upon integrating out the fermions, the Euclidean action reads:
\begin{equation}
S_q[A]=-2N_f\text{Tr}\left[\log\left(\slashed{\pd}-i\slashed{A}\right)\right]+i q \int d\tau A_0\,,
\end{equation}
where the factor of $i$ in front of $q$ is due to the Euclidean signature of the metric and we assume $q\sim N_f$.  The gauge field sourced by the Wilson line is determined by the following nonlocal equation:
\begin{equation}\label{eq_QED3_SaddleEq}
\frac{\delta \text{Tr}\left[\log\left(\slashed{\pd}-i\slashed{A}\right)\right]}{\delta A_\mu(x)}=i\delta^\mu_0\frac{q}{2N_f}\delta^2(x_\bot)\,.
\end{equation}
Because of conformal invariance, the field which solves \eqref{eq_QED3_SaddleEq} is Coulomb-like
\begin{equation}\label{eq_QED3_Ftr}
F_{\tau i}=iE\frac{x^i}{r^3}\,,
\end{equation}
where $E$ is a nontrivial function of $q/N_f$. For $q/N_f\ll 1$, we can linearize the fluctuation determinant using \eqref{eq_QED3_A} and solve for $E$: 
\begin{equation}\label{eq_QED3_lin}
E=\frac{4q}{\pi N_f}+O\left(\frac{q^2}{N_f^2}\right)\,.
\end{equation}

According to the general analysis in section  \ref{subsec_fer_toy}, in the presence of the electric field \eqref{eq_QED3_Ftr}, the scaling dimensions of the single-trace defect operators with spin $j=\pm\frac12$ is given by
\begin{equation}\label{eq_QED3_Delta}
\Delta=\frac12+\sqrt{\frac{1}{4}-E^2}\,.
\end{equation}
When the electric field becomes as large as $|E|=1/2$ the $j=\pm 1/2$ modes of the Dirac field develop an instability. In the following we would like to determine the critical value of $q/N_f$ for which $|E|=1/2$.  Clearly, the linearized approximation \eqref{eq_QED3_lin} is not enough and we need to solve the saddle-point equation \eqref{eq_QED3_SaddleEq} in the nonlinear regime.

\begin{figure}[t]
\centering
\includegraphics[scale=0.6]{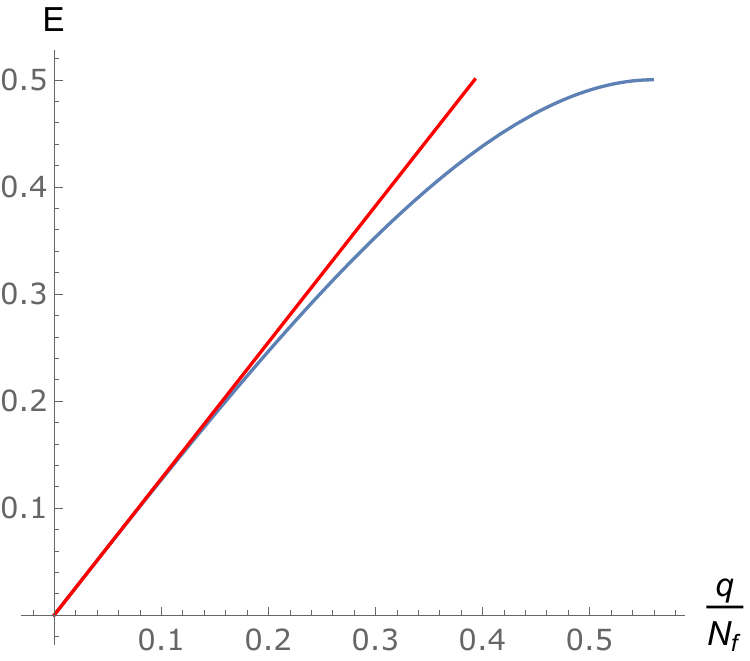}
\caption{Plot of the value of $E$ (in blue) in \eqref{eq_QED3_Ftr} as a function of $q/N_f$ (for $q>0$) as determined from \eqref{eq_QED3_SaddleEq}. The red line corresponds to the linearized result in \eqref{eq_QED3_lin}; as expected, the two curves perfectly agree for small $q/N_f$.}\label{fig:plot_QED3}
\end{figure}

To this aim, we have to compute the fluctuation determinant in \eqref{eq_QED3_A} for arbitrary values of $E$. This can be conveniently done by exploiting Weyl invariance to map the theory to AdS$_2\times S^1$. We provide details on the calculation in  appendix \ref{app_largeN}.  The result for $E=E(q/N_f)$ is shown in blue in figure \ref{fig:plot_QED3}, where we also compare with the linearization \eqref{eq_QED3_lin} (in red).  We find that the critical value for the instability is found at:
\begin{equation}\label{eq_QED3_result}
\left|E\left(\frac{q_c}{N_f}\right)\right|=\frac12\quad\implies\quad\frac{|q_c|}{N_f}\simeq 0.56\,.
\end{equation}
We also comment that the functional determinant develops an imaginary part for $E>1/2$, in agreement with the existence of an instability.

Note that the result for the critical charge in \eqref{eq_QED3_result} is larger than the one obtained by naively extrapolating the linear approximation \eqref{eq_QED3_lin}.  In particular, there are more than $N_f$ independent stable lines (counting both positive and negative values of $q$). An intuitive justification for this fact is as follows. Imagine adding a mass term to the model \eqref{eq_QED3} in a maximally parity breaking form.  As well known, integrating out all the fermions in this setup results in a $U(1)_{N_f}$ Chern-Simons theory in the IR. The latter is a topological theory which admits $N_f$ independent Wilson lines. Therefore it is natural to expect that the number of independent stable lines in the UV theory should also be at least $N_f$.  Interestingly, the linear extrapolation \eqref{eq_QED3_lin} would give less than $N_f$ stable lines (as easily read from the red line on figure~\ref{fig:plot_QED3}), which would lead to a tension with the above RG argument.

The most physically interesting value of $N_f$ for the model at hand is $N_f=2$  \cite{2007PhRvL..98k7205R,2008PhRvB..77v4413H}. In this case the result~\eqref{eq_QED3_result} suggests the existence of two nontrivial Wilson lines. It would be interesting to analyze the effect of subleading corrections in $1/N_f$ for this prediction.

Finally, we comment that for $0<|E|<1/2$ the $j=\pm1/2$ modes admit alternate boundary conditions on the line, and lead to several ultraviolet fixed points on the defect. Since the fluctuation determinant in \eqref{eq_QED3_SaddleEq} implicitly depends on the boundary conditions of the Dirac field at the defect, the relation between the electric field and $q/N_f$ at these fixed points is different than the one at the infrared fixed point shown in figure~\ref{fig:plot_QED3}. It would be interesting to determine the new curve, which is constrained to have the same endpoint at $E=1/2$ as the one in figure~\ref{fig:plot_QED3}.

\subsubsection{Comments on scalar \texorpdfstring{QED$_3$}{QED3}}
\label{ScalarQEDTwoPlusOne}

We now consider theories with $N_f\gg 1$ charged scalars $\Phi_a$ coupled to an Abelian gauge field:
\begin{equation}\label{eq_QED3_scalar}
\mL=\sum_{a=1}^{N_f}|D_\mu\Phi_a|^2-V(|\Phi_a|^2)\,,
\end{equation}
where $D_\mu=\pd_\mu-i A_\mu$ and we omitted again the kinetic term for the gauge field. The theory \eqref{eq_QED3_scalar} admits several multicritical fixed points in the large $N_f$ limit depending on the potential $V(|\Phi_a|^2)$,  which may break the $SU(N_f)$ symmetry, see e.g. \cite{Benvenuti:2019ujm}. Parity forbids a Chern-Simons term as in \eqref{eq_QED3}.

As before, to leading order in $N_f$ the dynamics of the gauge field follows from integrating out the scalar fields in \eqref{eq_QED3_scalar}, leading to a propagator for the gauge field proportional to \eqref{eq_QED3_2ptA}.  Below we briefly comment about the fate of Wilson lines in this class of theories.

Let us consider first the tricritical theory, which is defined by $V(|\Phi_a|^2)=0$ at large $N_f$ \cite{Benvenuti:2019ujm}. It is easy to see that all Wilson lines are unstable in the double-scaling limit $q\rightarrow\infty$, $N_f\rightarrow\infty$ with $q/N_f=\text{fixed}$.  By conformal invariance, inserting a Wilson line results in a Coulomb field of the form \eqref{eq_QED3_Ftr}.  For $q=0$ the singlet scalar bilinear $\sum_a|\Phi_a|^2$ has dimension $\Delta=1$ and provides a marginal deformation of the trivial line defect. Therefore the AdS$_2$ mass of the $\ell=0$ mode of the scalar sits exactly at the BF bound for $q=0$,
and for any value of $q\neq 0$ the electric field leads to an instability.\footnote{Equivalently, this means that the one-loop determinant of the scalar fields is complex for arbitrary (real) values of $E$ \cite{Pioline:2005pf}.} 

The instability does not imply necessarily that Wilson lines are trivial in the infrared. Indeed, as in subsection~\ref{subsec_multiflavor}, for $q\neq 0 \ {\rm mod} \ N_f$ the endpoint operators transform in a nontrivial projective representation of $PSU(N_f)$. Therefore, even if the electric and scalar fields were fully screened,  there must be a 0+1 dimensional system on the line furnishing a representation of $SU(N_f)$ with  $q \ {\rm mod} \ N_f$ boxes. Since the operator in the adjoint of $SU(N_f)$ $\Phi_a\Phi^\dagger_b-{1\over N_f}\delta^b_a|\Phi|^2$ has scaling dimension 1 in the large $N_f$ limit, it can couple to the 0+1 dimensional system via a marginal coupling and one is required to consider $1/N_f$ corrections to understand the true infrared limit, and whether there is a fixed point akin to the spin impurity fixed points discussed in \cite{vojta2000quantum,Cuomo:2022xgw,Beccaria:2022bcr}.

The situation is different in the presence of a non trivial potential. Consider in particular the critical theory which is obtained including a $SU(N_f)$ invariant deformation $V=\lambda(\sum_a|\Phi_a|^2)^2$. Via a standard Hubbard-Stratonovich transformation,  the singlet scalar bilinear $\sum_a|\Phi_a|^2$ is seen to have dimension $\Delta=2$ for $N_f\rightarrow\infty$ at the IR fixed point. For small $|q|/N_f$ the dimension of the singlet bilinear defect operator can be obtained perturbatively, i.e. $\Delta=2+O(q^2/N_f^2)$, and no instability is expected until a critical value $|q|=q_c\sim N_f $.  We thus expect $\sim N_f$ stable Wilson lines at the critical fixed point.

As in the tricritical theory, 
for $q\neq 0 \ {\rm mod} \ N_f$  we have a projective representation of $PSU(N_f)$ living on the line, and again  (around the trivial saddle-point $\Phi=A_\mu=0$) there is a marginal coupling due to the adjoint bilinear, which might be important to take into account at the next order in $1/N_f$.

In the future it would be interesting to compute the value of $q_c/N_f$ similarly to the previous section in the critical theory with $V=\lambda(\sum_a|\Phi_a|^2)^2$.\footnote{To this aim, one would also need to compute the one-point function of the Hubbard-Stratonovich field $\sigma\sim \sum_a |\Phi_a|^2$ for $q\neq 0$,  since this contributes to the AdS$_2$ mass of the fundamental fields similarly to \cite{Cuomo:2021kfm}.} This analysis might provide hints about the fate of Wilson lines in the $N_f=1$ theory, i.e.  the Abelian-Higgs model. Because of particle-vortex duality, Wilson lines in the critical Abelian-Higgs model should correspond to defects in the $O(2)$ model. Some implications of the duality for Wilson lines were discussed in \cite{Metlitski:2007fu}.

\subsubsection{\texorpdfstring{$U(1)_k$}{U(1)k}  with \texorpdfstring{$2N_f$}{2Nf} fermions}\label{subsec_U1k_Nf}

As a final example, we consider the theory that we obtain upon adding a level $k$ Chern-Simons term to the action \eqref{eq_QED3}.  In the presence of a charge $q$ Wilson line the action reads
\begin{equation}\label{eq_QED3k}
S_q=\int d^3x\left[i\sum_{a=1}^{2N_f}\bar{\Psi}_a\left(\slashed{\pd}-i\slashed{A}\right)\Psi_a+\frac{k}{4\pi}\varepsilon^{\mu\nu\rho}A_\mu\pd_\nu A_\rho\right]- q \int dt A_0\,.
\end{equation}
The theory \eqref{eq_QED3k} admits a natural triple scaling limit for $N_f\sim k\gg 1$ with $q/k$ (and hence $q/N_f$) fixed.
We would like to determine for which values of $q$ there is an instability of the matter fields in this limit. To this aim we analyze below the response of the gauge field to the Wilson line.  

Despite the similarity with the action \eqref{eq_QED3}, the Chern-Simons term has a remarkable consequence as explained in section \ref{sec5CS}: Wilson lines with different charges are identified modulo $k$.  We therefore need to analyze only Wilson lines with charges $-\frac{|k|}{2}< q\leq \frac{|k|}{2}$. 

Proceeding as in the previous section, the equation of motion for the gauge field in Euclidean signature can be written as
\begin{equation}\label{eq_QED3k_SaddleEq}
2\frac{\delta \text{Tr}\left[\log\left(\slashed{\pd}-i\slashed{A}\right)\right]}{\delta A_\mu(x)}-i \frac{k}{4\pi N_f}\varepsilon^{\mu\nu\rho}F_{\nu\rho}=i\delta^\mu_0\frac{q}{N_f}\delta^2(x_\bot)\,.
\end{equation}
The most general solution consistent with conformal invariance is given by a Coulomb field with a holonomy in the angular direction (which is allowed since the Chern-Simons term breaks parity):
\begin{equation}\label{eq_QED3k_Ansatz}
F_{\tau r}=i\frac{E}{r^2}\qquad\text{and}\qquad A_\theta=b=\text{const}.\,,
\end{equation}
where both $E$ and $b$ are functions of $k/N_f$ and $q/N_f$.  Note that the fluctuation determinant $\text{Tr}\left[\log\left(\slashed{\pd}-i\slashed{A}\right)\right]$ is a periodic function of $b$ according to the former discussion.

In the presence of the electromagnetic field \eqref{eq_QED3k_Ansatz}, the scaling dimension of the defect operator corresponding  to the spin $j$ KK mode of the fermion is given by
\begin{equation}\label{eq_QED3_Delta2}
\Delta_j=\frac12+\sqrt{\left(j+b \right)^2-E^2}\,,\qquad
j\in\frac12+\mathds{Z}\,.
\end{equation}
The theory develops an instability towards charge screening when $\left(j+b \right)^2-E^2<0$ for some value of $j$.

When $|q|\ll |k|, N_f$ we can linearize the fluctuation determinant and find explicitly the values of $E$ and $b$ which solve \eqref{eq_QED3k_SaddleEq}:
\begin{equation}
E\simeq
\frac{4 \pi  N_f q}{16 k^2+\pi ^2 N_f^2}\,,\qquad
b\simeq \frac{16 k q}{16 k^2+\pi ^2 N_f^2}
\,.
\end{equation}
It is less trivial to solve \eqref{eq_QED3k_SaddleEq} for general values of $q$. In practice, rather than solving \eqref{eq_QED3k_SaddleEq} for fixed values of $k_R \equiv k/N_f$ and $q_R \equiv q/N_f$, it is easier to do the opposite. Namely, given a certain value of $E$ and $b$ for which $\left(j+b \right)^2-E^2>0$ for all $j\in \mathbb{Z}+\frac{1}{2}$,  we determine the values of $k$ and $q$ that solve \eqref{eq_QED3k_SaddleEq}.  Since, as we explained before, an integral holonomy is unphysical, it is enough to determine the region $R$ in the $(k,q)$ plane where $b\in(-1/2,1/2)$ and $ |E|<1/2-|b|$, so that no mode is tachyonic.  

In appendix~\ref{app_largeN} we compute numerically the functional determinant and determine the region $R$.  The result is perhaps surprising. We find that the region $R$ spanned by the possible values of $(k_R,q_R)$ strictly includes the one specified by the inequality $-|k|/2\leq q \leq |k|/2$, which sets the number of independent Wilson lines. This implies that there exists at least one real stable saddle-point solution to \eqref{eq_QED3k_SaddleEq} for all Wilson lines (remember that $q\sim q+k$).  Additionally, the region $R$ includes points where $|q|>|k|/2$.  Since the region $R$ is obtained by restricting the value of the holonomy to $|b|<1/2$, via shifts of the form $b\rightarrow b\pm n$ and $q\rightarrow q\mp k n$, with $n\in\mathds{N}$,  the points in $R$ for which $|q|>|k|/2$ correspond to additional saddle-points in the physical region $|q|\leq |k|/2$. In other words, for certain values of $q$ and $k$ there are multiple saddle-point solutions for the gauge field. 

\begin{figure}[t!]
\centering
\includegraphics[scale=0.5]{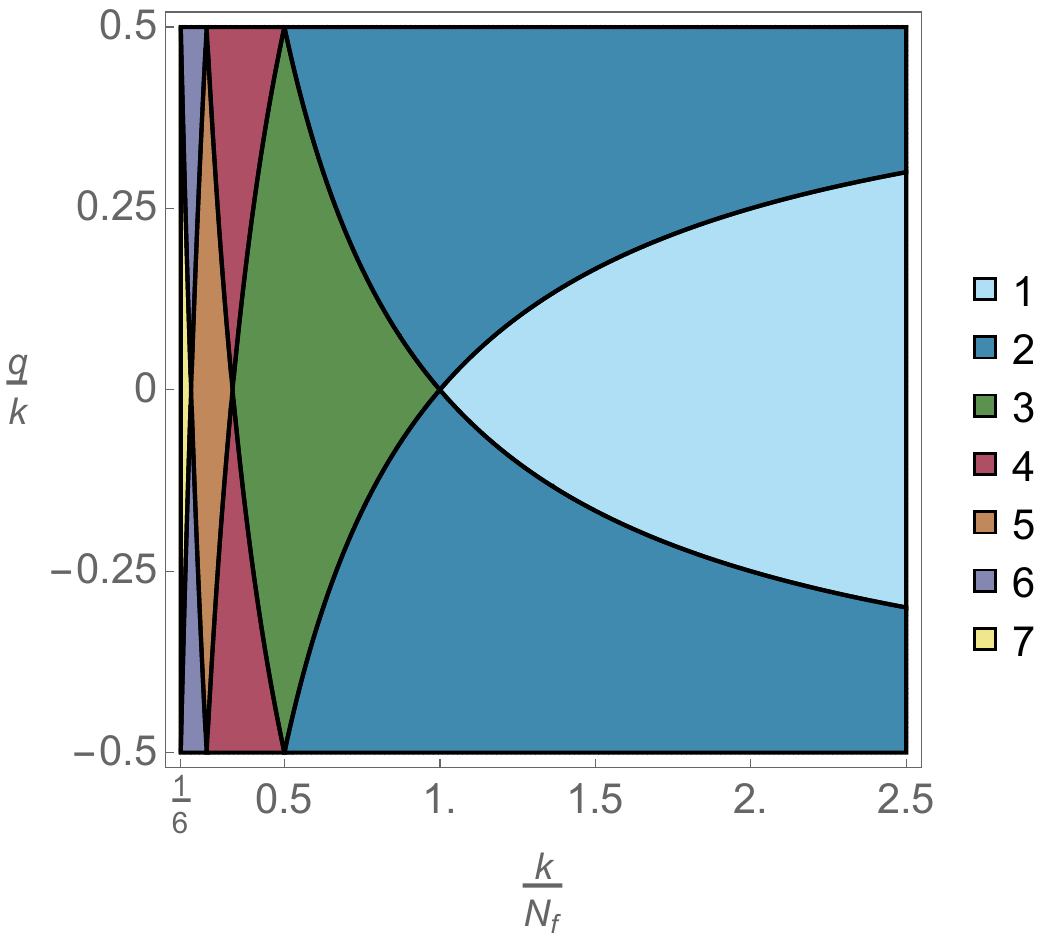}
\caption{In this plot we separate the $(k/N_f,q/k)$ plane into different regions according to the number of saddle-points; the legend on the right associates the color to the number of solutions. The plot is restricted to the physical regions $|q|\leq |k|/2$, and to $1/6<k/N_f<2.5$; a specular plot can be drawn for negative $k$.}\label{fig:plot_QED3k}
\end{figure}

Our results are summarized in figure~\ref{fig:plot_QED3k}, where we separate the physical region $|q|\leq |k|/2$ into smaller subregions according to the number of saddle-points found. Note that the number of solutions corresponding to a given charge $q$ increases as we lower $|k|/N_f$.  We did not analyze the question of stability of these saddle-points; we expect that the only stable saddle-points at a nonperturbative level are those with the minimal absolute value for the holonomy $b$ (when restricting to $-|k|/2\leq  q\leq |k|/2$). It would be interesting to confirm or disprove this expectation.

In conclusion, in the theory \eqref{eq_QED3k}  Wilson lines with different charges are identified modulo $k$, because of the Chern-Simons term.  We find that all Wilson lines are stable.

Wilson lines in various other Chern-Simons-Matter theories are of great interest as well (e.g. due to the their connection with boson-fermion duality, holography etc). We do not study them here. For some recent results see~\cite{Gabai:2022vri,Castiglioni:2022yes,Gabai:2022mya,Castiglioni:2023uus}.

\subsection{Graphene}

It is well known that in a layer of graphene, due to its peculiar lattice structure, the quasi-particles at the Fermi energy are described in terms of an effective Lorentz-invariant theory consisting of four three-dimensional Dirac fermions 
moving at an effective speed $v_f\approx 1/300$ \cite{kotov2012electron}, in the usual relativistic units where the speed of light is set to one.  Since $v_f\ll 1$, these quasiparticles experience an enhanced coupling to the $3+1$-dimensional Coulomb field \cite{kotov2012electron}:
\begin{equation}\label{graphene_alpha_eff}
e_{eff}^2=\frac{e^2}{v_f}\gg e^2\,.
\end{equation}
Naively using the modified coupling in the formula for the Coulomb potential sourced by a Wilson line, $A_0=\frac{e_{eff}^2 q}{4\pi  r}$, the general analysis in section  \ref{sec_Fer} implies that there should be an instability towards charge screening for $|q|\geq|q_c|=  2\pi /e^2_{eff}\simeq 0.2$.  This observation motivated several works in the condensed matter literature,  see \cite{kotov2012electron} for a review.  Of particular relevance for us is the analysis of the screening cloud and the related resonances in \cite{shytov2007atomic,shytov2007vacuum,Nishida:2014zfa}, which largely inspired our analysis in section \ref{subsec_fermions_supercritical}. Remarkably, this instability and the corresponding screening cloud were experimentally observed in \cite{wang2013observing} by introducing an external ion close to the material layer.

In practice, the formula~\eqref{graphene_alpha_eff} neglects important polarization effects that arise because of the strong interaction. To model these effects in a controlled setup it was proposed in \cite{Son:2007ja} to study a model of $2N_f\gg 1$ Dirac fields living on an interface coupled to the Coulomb field $A_0$:
\begin{equation}\label{graphene}
S=-\frac{1}{e^2}\int dzd^3x{1\over 4} (\pd_i A_0)^2+i\sum_{a=1}^{2N_f}\int_{z=0}d^3x
\bar{\Psi}_a\left[\frac{1}{v_f}\gamma^0\left(\pd_0-i A_0\right)+\gamma^i\pd_i\right]\Psi_a \,,
\end{equation} 
where $\Psi_a$ are Dirac fields as in section \ref{subsec_QED3_N}.  The coupling to $A_0$ is fixed by gauge-invariance and breaks the emergent Lorentz-invariance on the interface.   Note that the model~\eqref{graphene} neglects the spatial components $A_i$ of the gauge-field since their interaction with the Dirac quasiparticles is not enhanced by $v_f$. For $N_f=2$ \eqref{graphene} describes the low energy limit of graphene, but following \cite{Son:2007ja} we allowed for an arbitrary number of fermions. See also \cite{Herzog:2017xha,Herzog:2018lqz} for discussions of related models.

In the model~\eqref{graphene} the bulk coupling $e^2$ is given by the QED value and cannot be renormalized by interactions with the fields on the interface. Due to the lack of Lorentz invariance, the value of $v_f$ may instead be renormalized by interactions. It was shown in \cite{Son:2007ja} that the velocity $v_f$, and thus the effective strength of the coupling \eqref{graphene_alpha_eff}, undergo a nontrivial RG flow at order $1/N_f$.  The RG admits an IR relativistic fixed-point at $v_f\rightarrow\infty $ and a UV quantum critical point (corresponding to $v_f=0$) with Lifshitz scaling. Due to the smallness of the measured value of $v_f$,  this suggests that the physical theory might display approximate Lifshitz scaling.

In the following we will study the model~\eqref{graphene} in the presence of a Wilson line of charge $q$ at $x=y=z=0$.  We will work in the triple-scaling limit defined by
\begin{equation}
N_f \sim \frac{1}{e_{eff}^2}\sim q\rightarrow \infty\quad\text{with}\quad
e_{eff}^2 N_f\sim e_{eff}^2 q=\text{fixed}\,.
\end{equation}
In this limit the running of the velocity $v_f$ can be neglected. We will use below the technology that we developed in the analysis of QED$_3$ in the large $N_f$ limit to compute the critical charge in this approximation.\footnote{A former analysis of the Coulomb impurity problem in this model appeared in \cite{Biswas:2007vh}; that work however focused on charges $q\ll N_f\sim 1/e_{eff}^2$, in which case the fluctuation determinant in \eqref{graphene_saddle} can be linearized. This is not possible for nearly critical electric fields, as figure \ref{fig:plot_QED3} shows. }

Wick rotating to Euclidean signature, we integrate out the fermions as in section \ref{subsec_QED3_N} to obtain the effective action for the gauge field
\begin{equation}
S[A]=\frac{1}{e_{eff}^2 v_f^2}\int dzd^3x{1\over 2} (\pd_i A_0)^2
-2N_f\text{Tr}\left[\log\left(\tilde{\slashed{\pd}}-i\slashed{A}/v_f\right)\right]+iq \int d\tau A_0\,,
\end{equation}
where we defined $\tilde{\pd}_\mu=\{\pd_0/v_f,\pd_i\}$.  The gauge field sourced by the line takes the form
\begin{equation}
A_0=v_f\bar{A}_0\,,\qquad
\bar{A_0}=\frac{iE}{\sqrt{x_{\bot}^2+z^2}}\,,
\end{equation}
which becomes critical  for $|E|=1/2$.   The value of $E$ is determined by the saddle-point equation
\begin{equation}\label{graphene_saddle}
\pd_i^2 \bar{A}_0+2
e_{eff}^2N_f\frac{\delta \text{Tr}\left[\log\left(\tilde{\slashed{\pd}}-i\slashed{\bar{A}}\right)\right]}{\delta \bar{A}_0(x)}=
i \,e_{eff}^2q\,\delta^2(x_\bot)\delta(z)\,.
\end{equation}
Since the rescaling $t\rightarrow t/v_f$ leaves invariant the fermion one-loop determinant, we can use the result\eqref{eq_QED3_result} to find the critical value for the charge
\begin{equation}\label{graphene_results}
|q_c|=\frac12\frac{4\pi}{e_{eff}^2 }+ 0.56\,N_f\,.
\end{equation}
The (unjustified) extrapolation of the result \eqref{graphene_results} to the physical theory, for which $N_f=2$ and $e_{eff}^2/(4\pi)= \alpha_{EM}/v_f\simeq 2.2$,  gives $|q_c|\simeq 1.35$, which is not too far from the experimentally observed $|q_c|\approx 2\divisionsymbol 3$ \cite{wang2013observing}.

We finally mention that it is possible to consider other instances of charged matter fields on a $3d$ interface or boundary coupled to a four-dimensional Abelian gauge field. This setup often gives rise to a continuous family of BCFTs,  parametrized by the gauge coupling $e$ and the $\theta$ angle of the theory \cite{DiPietro:2019hqe}. We leave the analysis of Wilson lines in these theories for future work.

\section{'t Hooft line operators} \label{sectHooft}

\subsection{'t Hooft lines in Abelian gauge theories}

We discuss the case of $4$ space-time dimensions, and take the gauge group to be $U(1)$, with matter fields that are some massless fermions and scalars with $U(1)$ charges that we will specify later. Such a theory admits a magnetic $U(1)$ one-form symmetry, since the current $(\star F)_{\mu\nu}$ is conserved due to the Bianchi identity $dF=0$. Furthermore, the one-form symmetry operator $e^{i\alpha \int_{\Sigma_2} F}$  can be cut open in a straightforward fashion, by just allowing $\Sigma_2$ to have a boundary $\partial \Sigma_2$. The non-genuine line operator on $\partial \Sigma_2$ can be viewed as a Wilson line with fractional charge.

Therefore, our considerations in subsection~\ref{subsec_1form} imply that 't Hooft lines in such theories cannot be trivial, or topological. This is simply because the magnetic field cannot be screened. Let us therefore make some remarks about the defect conformal theories arising from 't Hooft lines.  

Inserting a 't Hooft line representing the worldline of a monopole of magnetic charge $n$ leads to a boundary condition for the gauge field on a small $S^2$ surrounding the 't Hooft line, which can be written (up to gauge transformations) as:\footnote{To see that $n$ is an integer we consider the region near the south pole, where we have $A\sim n d\phi$, which can be interpreted as being due to a transparent solenoid if $n$ is an integer. We adopted spherical coordinates $ds^2=dt^2-\left[dr^2+r^2\left( d\theta^2+r^2\sin^2(\theta) d\phi^2\right)\right]$. 
}
\begin{equation}\label{tHooftgauge}
A={n\over 2}(1-\cos(\theta))d\phi.
\end{equation}
In fact, setting all the matter fields to vanish and adopting~\eqref{tHooftgauge} everywhere in space leads to a solution of the equations of motion, and as before, we can now investigate the fluctuations around this background.

We start from a scalar field $\Phi$ of charge $q$  and we consider it in the background~\eqref{tHooftgauge}. We separate the variables as usual $\Phi=e^{i\omega t}Y(\theta,\phi){1\over r}R(r)$, and find that the function $Y(\theta,\phi)$ has to be a monopole spherical harmonic~\cite{Wu:1976ge,Wu:1977qk,Shnir:2005vvi}:
\begin{equation}
\left[\Delta_{S^2}      -\frac{qn/2}{ \cos^2(\theta/2)}\left(i\partial_\phi+{qn\over 2}\right) \right]Y_{qn/2,\ell m}=-\ell(\ell+1) \,Y_{qn/2,\ell m}\,.
\end{equation}
The $Y_{qn/2,\ell m}$ transform in the spin $\ell$ representation of $so(3)$. Most importantly, 
\begin{equation}
\ell\in |qn/2|+\mathbb{Z}^+\,,
\end{equation}
 which famously leads to ground state degeneracy in the presence of a monopole as it removes the $s$-wave modes for nonzero $qn$.
We can now turn to the radial equation, assuming the angular part of the wave function is in a state with spin $\ell=|qn/2|$.
We obtain the radial wave equation:
\begin{equation}\label{magneticradial}
- \partial^2_r R+{|qn|\over2r^2}R=\omega^2 R~.
\end{equation}
The potential is effectively always repulsive and there is no instability for any $qn$. The dimension of the defect operator $\Phi^\dagger\Phi$ is inferred from the behavior of $R(r)$ near the origin, as before
\begin{equation}\label{scalingtHooft}
\Delta=1+2\sqrt{{|qn|\over 2}+\frac14}~.
\end{equation}
We see that the already for the minimal 't Hooft line with $qn=1$ the dimension of the defect operator is $1+\sqrt 3$ which is larger than the bulk scaling dimension of $\Phi^\dagger\Phi$, which is 2. The charged bosons therefore never furnish relevant or marginal operators on the 't Hooft line, unlike the case of the Wilson lines. \eqref{scalingtHooft} is a good approximation as long as $e^2,\lambda\ll1$, where $\lambda$ is the scalar quartic coupling. Note that unlike for Wilson lines, here no double scaling limit is necessary.

The analysis of fermions around a 't Hooft line is analogous. The Dirac equation for a 4 component fermion in spherical coordinates can be written as
\begin{equation}
\left[\gamma^0{\partial\over \partial t}+\gamma^r{\partial\over \partial r}+{1\over r}\gamma^\theta{\partial\over \partial \theta}+{1\over r\sin(\theta)}\gamma^\phi({\partial\over \partial \phi}-iqA_\phi)\right]\Psi=0\,,
\end{equation}
acting on a four-component spinor with the following matrices: 
\begin{equation}
\begin{gathered}
\gamma^0=\left(\begin{matrix}1 & 0 & 0 & 0\cr  0& 1 & 0 & 0\cr 0 & 0 & -1 & 0\cr 0 & 0 & 0 & -1 \end{matrix}\right)~,\quad \gamma^r=\left(\begin{matrix}0 & 0 &  \cos(\theta) & \sin(\theta) e^{-i\phi} \cr 0 & 0 & \sin(\theta) e^{i\phi} &  -\cos(\theta) \cr -\cos(\theta) & -\sin(\theta) e^{-i\phi} & 0 & 0 \cr -\sin(\theta) e^{i\phi} & \cos(\theta) & 0 & 0 \end{matrix}\right)~,
\\[1em]
 \gamma^\theta=\left(\begin{matrix}0 & 0 & -\sin(\theta) & \cos(\theta) e^{-i\phi} \cr 0 & 0 & \cos(\theta) e^{i\phi} & \sin(\theta)\cr \sin(\theta) & -\cos(\theta) e^{-i\phi} & 0 & 0 \cr -\cos(\theta) e^{i\phi} & -\sin(\theta) & 0 & 0\end{matrix}\right),\quad\gamma^\phi=\left(\begin{matrix}0 & 0 & 0 & -i e^{-i\phi} \cr 0 & 0 & i e^{i\phi} & 0 \cr 0 & i e^{-i\phi} & 0 & 0\cr -i  e^{i\phi} & 0 & 0 & 0 \end{matrix}\right). 
\end{gathered}
\end{equation}
 An important novelty compared to the boson, is that now there are modes with total angular momentum $|qn|/2-1/2$ if $|qn|>0$. This is the minimal achievable angular momentum. These modes are often called  ``spinor spherical harmonics of the third type'' and the explicit formula in terms of the ordinary monopole harmonics with azimuthal angular momentum $m$ is (denoting $\mu=|qn|/2$):
\begin{equation}
\Omega^{(3)}_{\mu,\mu-1/2,m} =\left(\begin{matrix}-\sqrt{{\mu-m+1/2\over 2\mu+1 }}Y_{\mu,\mu,m-1/2}\cr \sqrt{{\mu+m+1/2\over 2\mu+1 }}Y_{\mu,\mu,m+1/2}\end{matrix}\right)~.
\end{equation}
See~\cite{Shnir:2005vvi} for an exposition to this subject.
An ansatz for the solution of the Dirac equation is $\Psi=e^{-iEt}{1\over r}\left(\begin{matrix} F(r)\Omega^{(3)}_{\mu,\mu-1/2,m}\cr iG(r)\Omega^{(3)}_{\mu,\mu-1/2,m}\end{matrix}\right)$. The Dirac equation then reduces to 
${dG\over dr} = E F~,{dF\over dr} = -E G$. There are two independent solutions here, $F=e^{iEr},G=-i e^{iEr}$ and 
$F=e^{-iEr},G=i e^{-iEr}$. The doublet $\left(\begin{matrix}F \cr   G \end{matrix}\right)$ is acted upon by the Hamiltonian $H=\left(\begin{matrix}0 & {d\over dr} \cr -{d\over dr} & 0\end{matrix}\right)$. Since $r=0$ plays the role of a boundary, one needs to impose a boundary condition for the existence of a well-defined variational problem (equivalently, the existence of a Hermitian Hamiltonian). Therefore we require that 
\begin{equation}
\bigl\langle \left(\begin{matrix}F \cr  G \end{matrix}\right) ,\left(\begin{matrix}0 & {d\over dr} \cr -{d\over dr} & 0\end{matrix}\right)\left(\begin{matrix}F \cr   G \end{matrix}\right)\bigr\rangle=\bigl\langle\left(\begin{matrix}0 & {d\over dr} \cr -{d\over dr} & 0\end{matrix}\right)\left(\begin{matrix}F \cr   G \end{matrix}\right) ,\left(\begin{matrix}F \cr   G \end{matrix}\right)\bigr\rangle
\end{equation}
The Hamiltonian is Hermitian if $F^*(0)G(0)$ is purely real.
The most general admissible solution is thus
\begin{equation}\label{thetaappear}F=A e^{iEr}+Be^{-iEr}~,\quad G=-iAe^{iEr}+iBe^{-iEr} ~,\quad B=e^{i\theta+i\pi/2} A\,,\end{equation}
where $\theta$ is the $\theta$-angle (not to be confused with the azimuthal coordinate). (The $\pi/2$ shift is a convention.)
For instance, assuming $\theta=0$ we get $F\sim \cos(Er-\pi/4)$ and $G\sim \sin(Er-\pi/4)$. The falloff of the wave function near the origin allows us to read the dimensions of defect operators as usual. We can therefore interpret the coefficient $A$ (or $B$) as an operator of dimension $1/2$.\footnote{For $qn=0$, i.e. the trivial defect, the wave functions behave as $\sin(Er)/r$ for small $r$ and the corresponding defect operator has dimension $3/2$, which is nothing but the original bulk fermion on the trivial defect. }

The conclusion is that for any non-zero $n$ and $q$ we have an operator of dimension $1/2$ on the defect and hence gauge-invariant marginal bilinears of dimension 1. In particular there is a marginal bilinear of spin 0, which has a clear interpretation -- it allows to change the $\theta$ angle appearing in the boundary condition for the fermion in~\eqref{thetaappear}. The appearance of a $\theta$ angle for monopoles in an Abelian gauge theory was noted already in~\cite{Yamagishi:1982wp}. We will soon see that this angle has a very simple interpretation following from the symmetries of the theory.  

Before discussing in more detail the $\theta$ angle and the corresponding marginal operator, we would like to generalize our discussion of the boundary conditions corresponding to a 't Hooft line to arbitrary $U(1)$ gauge theories.

In the background of a monopole with $n$ units of magnetic field, a charge $q$ left moving 4d Weyl fermion reduces (in its lowest angular momentum mode on $S^2$) to $nq$ left moving 2d fermions if $nq>0$ and $-nq$ right moving fermions for $nq<0$. 
Either way, they transform in the spin $(|nq|/2-1/2)$ representation of $SU(2)$, which indeed has $|nq|$ components. (For $nq=0$ the special representations of spin $(|nq|/2-1/2)$ do not exist.)
These massless fermions live on a half-infinite line $r\geq 0$. We must therefore choose boundary conditions at $r=0$. 

It is useful to quickly review some facts about boundary conditions in 2d. For a recent discussion of the connections between anomalies and boundary conditions see~\cite{Jensen:2017eof,Thorngren:2020yht}. Unless $c_L=c_R$ no boundary condition which is time-translation-invariant exists. Similarly, the $Tr[U(1)]^2$ anomaly precludes the existence of a $U(1)$-preserving boundary condition and the $Tr[U(1)_AU(1)_B]$ anomaly precludes the existence of a boundary condition preserving both $U(1)_A$ and $U(1)_B$.

Since the 't Hooft line has to be gauge-invariant, we must insist that a $U(1)$-preserving boundary condition exists. The $|nq|$ left moving fermions contribute to the $Tr[U(1)]^2$ anomaly $nq\times q^2=n q^3$ for $q>0$, and the $|nq|$ right moving fermions give $-|nq^3|$ for $q<0$. Summing them all up we have a condition equivalent to $\sum q^3=0$ over all Weyl fermions. Therefore, as long as the original 4d gauge theory is consistent (free of gauge anomalies) we have no obstruction to picking a gauge-invariant boundary condition at the 't Hooft line.

For there to exist time-translational-invariant boundary conditions a necessary condition is that the number of left and right moving fermions coincides, so that a 2d gravitational anomaly is absent. This gives 
\begin{equation}
\sum_{q_i>0} |q_i|-\sum_{q_i<0} |q_i|=0\,.
\end{equation}
This is realized if the four dimensional theory has no anomaly of the form
\begin{equation}\label{mixedgg}
\partial j_{gauge} \sim (\sum_{{\rm weyl \ fermions}} q_i) R\wedge R~.
\end{equation}
Traditionally, the 4d anomaly~\eqref{mixedgg}, which was first described in~\cite{Delbourgo:1972xb}, is interpreted as an obstruction to gauge invariance in curved space. Here we see that upon introducing a 't Hooft line, it leads to an imbalance of left- and right-moving fermions and consequently, obstructs the existence of time-translationally and rotationally invariant 't Hooft lines. 
In modern parlance the anomaly~\eqref{mixedgg} should be described by a two-group symmetry involving the magnetic one-form symmetry and Lorentz symmetry~\cite{Cordova:2018cvg}. 
We conclude that it is not possible to construct 't Hooft lines that preserve rotational invariance and time translational invariance in such theories with a two-group symmetry. (The calculation of the $SU(2)$ rotation symmetry anomalies is in section 5 of~\cite{vanBeest:2023dbu}. The conclusion is that if the theory is free of a gauge anomaly and the two-group symmetry involving the magnetic one-form symmetry and Lorentz symmetry is trivial, then there is no obstruction to choosing a boundary condition at the monopole which is rotationally invariant.)

Next let us consider a global $U(1)_Q$ symmetry with charges $Q_i$ such that $\sum Q_iq_i^2=0$, i.e. it suffers from no ABJ anomaly (and thus it is a true continuous symmetry), but such that $\sum Q_i^2q_i\neq0$, namely, the global symmetry and the magnetic one-form symmetry furnish a two-group~\cite{Cordova:2018cvg}. It is easy to see that the 2d modes in the lowest angular momentum sector have an anomaly $Tr[U(1)_Q]^2\neq0$ and hence no 't Hooft lines can preserve $U(1)_Q$. Any line operator that violates a global symmetry leads to an exactly marginal ``tilt'' operator  and hence there would be  exactly marginal tilt operators corresponding to such $U(1)_Q$ global symmetries. 

Let us now comment on the axial symmetry with charges $Q_i^A$, for which there is an ABJ anomaly in 4d $\sum Q^A_iq_i^2\neq 0$. This anomaly removes the continuous symmetry in 4d. Depending on the monopole charge, a discrete subgroup can be preserved by the monopole boundary conditions.  Furthermore, the $\theta$ angle in~\eqref{thetaappear} couples to an operator that is marginal at tree level, but already at the next order in $e^2$ it becomes marginally irrelevant~\cite{vanBeest:2023dbu} (and references therein). Indeed, since there is no continuous axial symmetry in 4d, there is no reason to expect a tilt operator.

Let us summarize the main highlights about 't Hooft lines in Abelian gauge theories: 
\begin{itemize} 
\item The axial symmetry leads to a marginal operator at tree level, but this operator and the corresponding $\theta$ angle are irrelevant in the full theory.
 
\item Unless $\sum_i q_i=0$, no 't Hooft lines which are time independent and rotationally symmetric exist.
\item Bosons generally receive positive anomalous dimensions and do not lead to low-lying operators on the defect.
\item Exactly marginal defect operators arise from global $U(1)$ symmetries which participate in a nontrivial two-group with the magnetic one-form symmetry.
\end{itemize} 

\subsection{'t Hooft lines in non-Abelian gauge theories and S-duality}

In weakly coupled non-Abelian gauge theories, to specify a 't Hooft line, we can simply fix the magnetic fluxes for the gauge fields in the Cartan sub-algebra~\cite{Goddard:1976qe,Kapustin:2005py}. Then, expanding around this classical solution with all the other fields vanishing, the main novelty that we encounter compared to the Abelian theory is that we also have spin $1$ massless charged fields (the off-diagonal W-bosons).

Let us therefore generalize the discussion in the previous subsection to the problem of charge $q$, spin $s$ particles, with magnetic $g$-factor $g_m$. We need to determine the centrifugal barrier for these particles. 
The total angular momentum is made out of the angular momentum  $\ell$ and the internal spin $s$. The range of $\ell$ is like for the bosonic wave functions in the presence of the unit monopole: $\ell=|nq|/2,|nq|/2+1,...$.

The general result for the centrifugal barrier matrix is given by~\cite{Shnir:2005vvi} 
\begin{equation}
V={\ell(\ell+1)-n^2q^2/4-\frac12 nq g_m \hat r^a  S^a\over r^2}~, 
\end{equation}
where $S^a$ are spin $s$ representation matrices  and $\hat r$ is the unit vector. 
The eigenvalues of $\hat r^a  S^a$ in principle can be between $-s,-s+1,..,s$. However for the short representations with total angular momentum $j=\ell-s$ which are possible for $\ell\geq s$, only a subset of these values is realized, see for instance~\cite{Weinberg:1993sg}: 
$\hat r S$ has to be $+s$ for positive $nq$ and $-s$ for negative $nq$.

Without loss of generality, taking positive $nq$ we find the centrifugal barrier for representations with spin $|nq|/2-s$ is \begin{equation}\label{lowbarr}V=\frac12nq{1-g_m s\over r^2}~.\end{equation}
For scalars we have a repulsive centrifugal force exactly consistent with~\eqref{magneticradial}.
For fermions with the standard magnetic moment $g_m=2$ the numerator vanishes and we have no centrifugal barrier, as we have seen in the previous subsection. 

Now let us discuss charged vector bosons. If these vector bosons are approximately fundamental particles, as they are in weakly coupled gauge theories, then we have $g_m=2$. 
The discussion for vector bosons has to be split between the case of $nq\geq 2$ and $nq=0,1$.
In the latter cases the special representation with spin $nq/2-1$ which gives rise to~\eqref{lowbarr} does not exist and there are no relevant defect operators associated to the vector bosons. In the case that $nq\geq 2$ the formula~\eqref{lowbarr} is valid, and we clearly see that the potential is attractive with coefficient $-\frac12nq{1\over r^2}$, which for $nq\geq 2$ always leads to an instability of the vector bosons  -- the bilinear operators associated to the vector bosons do not have a real scaling dimension, similarly to the super-critical Wilson lines. 
This means that we have to condense vector bosons with $nq\geq 2$ and the infrared limit of such 't Hooft lines remains to be determined. 

Let us now consider some examples -- for instance, the $\mathcal{N}=4$ SYM theory with gauge group $SU(2)$ and the $\mathcal{N}=4$ SYM theory with gauge group $SO(3)$. Those theories have supersymmetric 't Hooft lines which are stable, but we discuss here the non-supersymmetric $SO(6)_R$-invariant 't Hooft lines which couple only to the gauge field. For $SU(2)$ the minimal monopole has $n=1$ and the vector boson charge is $q=2$ (in units of the minimal charge). Therefore, even the minimal 't Hooft line is unstable to W-boson condensation at weak coupling, and deep in the infrared it presumably becomes trivial. (The cloud of W-bosons that forms remains to be computed.)  
In the latter case, with gauge group $SO(3)$, the charge of the W-boson is 1 and hence in the background of the minimal 't Hooft line $n=1$ we have no such instability and the minimal 't Hooft line should furnish a healthy conformal defect. Higher 't Hooft lines are all unstable to W-boson condensation, though.\footnote{For the $PSU(N)$ $\mathcal{N}=4$ SYM theory the statement would be that only 't Hooft lines in the fully anti-symmetric representation are unscreened at weak coupling. When we refer to the $SO(3)$ gauge theory we really have in mind the $SO(3)_+$ gauge theory which admits purely magnetic lines, and similarly, when we refer to $PSU(N)$ gauge theory, we have in mind the one with purely magnetic lines~\cite{Aharony:2013hda}.}

These results are consistent with one-form symmetry. The $SO(3)$ theory has a magnetic $\mathbb{Z}_2$ one-form symmetry protecting the minimal 't Hooft line while the $SU(2)$ theory does not have such a magnetic one-form symmetry and hence the 't Hooft lines are unprotected.

Let us now make some comments about S-duality for the $SO(6)$ R-symmetry invariant lines. Let us start from the $SU(2)$ theory. As we crank up the coupling constant on the conformal manifold, we have seen that fewer and fewer Wilson lines remain as nontrivial infrared DCFTs. Presumably at strong coupling only the unique Wilson line in the fundamental representation remains, as it is protected by the electric $\mathbb{Z}_2$ one-form symmetry. By S-duality this should map to 't Hooft lines in weakly coupled $SO(3)$ gauge theory. Indeed, we have just argued that at weak coupling no 't Hooft lines other than the minimal one exist. For the $SO(3)$ gauge theory, Wilson lines again gradually disappear as the coupling is cranked up, presumably leaving none at strong coupling, consistently with the absence of any conformal 't Hooft lines in weakly coupled $SU(2)$ gauge theory. 
 Therefore, our results for the non-supersymmetric line operators in $\mathcal{N}=4$ SYM theory are consistent with $S$-duality.

\section*{Acknowledgements}

We thank S. Bolognesi, I. Klebanov, J. Maldacena, M. Metlitski, S. Sachdev, N. Seiberg, A. Sever, S.-H. Shao, Y. Wang, and S. Yankielowicz for useful discussions.  The work of OA was supported in part  by an Israel Science Foundation (ISF) center for excellence grant (grant number 2289/18), by ISF grant no. 2159/22, by Simons Foundation grant 994296 (Simons
Collaboration on Confinement and QCD Strings), by grant no. 2018068 from the United States-Israel Binational Science Foundation (BSF), by the Minerva foundation with funding from the Federal German Ministry for Education and Research, by the German Research Foundation through a German-Israeli Project Cooperation (DIP) grant ``Holography and the Swampland'', and by a research grant from Martin Eisenstein. OA is the Samuel Sebba Professorial Chair of Pure and Applied Physics. 
GC was supported by the Simons Foundation grants 488647, 397411 (Simons Collaboration on the Non-perturbative Bootstrap) and 994296 (Simons Collaboration on Confinement and QCD Strings).  ZK, MM and ARM are supported in part by the Simons Foundation grant 488657 (Simons Collaboration on the Non-Perturbative Bootstrap) and the BSF grant no. 2018204.  MM gratefully acknowledges the support and hospitality from the Simons Center for Geometry and Physics during the final stages of this work. ARM is an awardee of the Women’s
Postdoctoral Career Development Award.  

\appendix

\section{Details on scalar \texorpdfstring{QED$_4$}{QED4}}\label{app_scalar}

\subsection{Defect propagator and double-trace deformation for subcritical charge}\label{app_KG_prop}

In this section we consider a charged field in AdS$_2$, with Euclidean action:
\begin{equation}\label{eq_app_scalar_bulk}
S_{bulk}=\int_{\text{AdS}_2} d^2x\sqrt{g}\left[|D_\mu\Phi|^2+m^2|\Phi|^2\right]\,,
\end{equation}
where $D_\mu=\pd_\mu-i A_\mu$ and $A_\mu=-i\delta_\mu^0 g/r$ is a subcritical Coulomb potential, i.e. such that $1+4m^2-4g^2>0$.  We assume $g>0$ with no loss of generality. For such a model the near boundary ($r\to 0$) expansion of the operator $\Phi$ reads:
\begin{equation}\label{eq_app_Phi_modes}
\Phi\sim \alpha r^{1/2-\nu}+\beta r^{1/2+\nu}\,,
\end{equation}
where $\nu=\sqrt{1/4+m^2-g^2}>0$. We will compute the propagator for the mode $\alpha$ at the alternate quantization fixed point $\beta=0$.  We will then use this result to obtain the exact propagator in the presence of a double-trace defect deformation of the form $\sim f\bar{\alpha}\alpha$.  We will finally argue that such a propagator displays a tachyon pole when the coefficient of the deformation is negative. This signals an instability of the trivial vacuum $\Phi=0$, whose end point we analyze in section \ref{subsec_scalar_stability}.

We start by computing the propagator at the alternate quantization fixed-point. 
We use the generating function approach, which is typically employed for AdS/CFT calculations \cite{Aharony:1999ti}. To this aim, differently from in the main text, we consider the theory with the following Dirichlet boundary conditions in terms of the modes \eqref{eq_app_Phi_modes}
\begin{equation}\label{eq_app_toy_J}
\beta(\tau)=\frac{1}{\nu}J(\tau)\,,\qquad
\beta^\dagger(\tau)=\frac{1}{\nu}\bar{J}(\tau)\,,
\end{equation}
where $J$ is a complex fixed function that we will soon interpret as an external source. For the action to be stationary with boundary conditions of the form \eqref{eq_app_toy_J} we add the following boundary term:
\begin{equation}\label{eq_app_scalar_bdry}
S_{bdry}=\lim_{r_0\rightarrow 0}\int_{r=r_0} \hspace*{-1em} d\tau
\left[\left(\Phi^\dagger\pd_r\Phi+c.c.  \right)
-\frac{1-2\nu}{2 r_0}\Phi^\dagger\Phi\right]\,.
\end{equation}
The sum of the bulk action \eqref{eq_app_scalar_bulk} and the boundary term \eqref{eq_app_scalar_bdry} is finite on-shell and can be written as
\begin{equation}
\begin{split}
S_{bulk}+S_{bdry}\vert_{on-shell}&=\nu\int d\tau
\left[\beta^\dagger(\tau)\alpha(\tau)+\alpha^\dagger(\tau)\beta(\tau)\right]
\\
&=\int d\tau\left[\bar{J}(\tau)\alpha(\tau)+\alpha^\dagger(\tau)J(\tau)\right]\,,
\end{split}
\end{equation}
where in the last step we used the Dirichlet conditions \eqref{eq_app_toy_J}.

We now follow the GKPW prescription and interpret the theory with Dirichlet boundary conditions \eqref{eq_app_toy_J} as the deformation of the alternate fixed point by a complex source $J$ for $\alpha$ \cite{Witten:1998qj,Gubser:1998bc}. It follows by Wick's theorem that the propagator for the boundary field $\alpha$ at the alternate quantization fixed point is
\begin{equation}\label{eq_app_prop_pre}
\langle\alpha(\omega)\alpha^\dagger(\omega')\rangle_{f=0}=2\pi\delta(\omega-\omega')G_{\alpha}^{(0)}(\omega)\,,\qquad
G_{\alpha}^{(0)}(\omega)=-\left[\frac{\alpha(\omega)}{J(\omega)}+\frac{\alpha^\dagger(\omega)}{\bar{J}(\omega)}\right]\,,
\end{equation}
where $\alpha$ in $G_{\alpha}^{(0)}$ is obtained from the boundary limit of a regular solution of the bulk equations of motion with boundary condition \eqref{eq_app_toy_J} for $r\rightarrow 0$. The Fourier transform is defined according to
\begin{equation}
\alpha(\tau)=\int \frac{d\omega}{2\pi}e^{-i\omega \tau}\alpha(\omega)\,,\qquad
\alpha^\dagger(\tau)=\int \frac{d\omega}{2\pi}e^{i\omega \tau}\alpha^\dagger(\omega)\,,
\end{equation}
and similarly for $J(\tau)$, $\bar{J}(\tau)$. 

All that is left to do is to solve the Euclidean Klein-Gordon equation in a Coulomb potential:
\begin{equation}
-r^2(\pd_0-i A_0)^2\Phi-r^2\pd_r^2\Phi+m^2\Phi=0\,.
\end{equation}
Using $A_0=-i g/r$ and setting $\Phi(\tau,r)=e^{-i\omega\tau}\Phi(r)$, we find
\begin{equation}\label{eq_app_KG_eucl}
-r^2\pd_r^2\Phi+\left[m^2-(g+i r\omega)^2\right]\Phi=0\,.
\end{equation}
The most general solution of \eqref{eq_app_KG_eucl} can be written as a linear combination of Whittaker's W functions:
\begin{equation}\label{eq_app_scalar_eucl_sol}
\Phi(r)=
c_1 W_{ig,-\nu}(2 r \omega )+c_2W_{-i g,\nu}(-2 r \omega )\,.
\end{equation}
In the following we focus on $\omega>0$. From $W_{x,y}(z)\overset{z\rightarrow\infty}{\propto} e^{-z/2}$, we infer that regularity at $r\to \infty$ implies that we need to set $c_2=0$ in \eqref{eq_app_scalar_eucl_sol}.  We then extract $\alpha$ and $\beta$ from the comparison of \eqref{eq_app_Phi_modes} with the expansion of the Whittaker's function
\begin{equation}\label{eq_app_Wexp}
W_{x,y}(z)\overset{z\rightarrow 0}{\sim}z^{\frac{1}{2}-y}
\frac{ \Gamma (2 y)}{\Gamma \left(\frac{1}{2}-x+y\right)}+z^{\frac{1}{2}+y} \frac{\Gamma (-2 y)}{\Gamma \left(\frac{1}{2}-x-y\right)}\,.
\end{equation}
Similarly solving the equation for $\Phi^\dagger$, we conclude
\begin{equation}
\frac{\alpha(\omega)}{\beta(\omega)}=
\frac{\alpha^\dagger(\omega)}{\beta^\dagger(\omega)}
=(2\omega)^{-2\nu}
\frac{ \Gamma (2\nu ) \Gamma \left(\frac{1}{2}-\nu-i g\right)}{\Gamma (-2\nu ) \Gamma \left(\frac{1}{2}+\nu - i g\right)}
\,,\qquad
\omega>0\,.
\end{equation}
Note that $\alpha$, $\beta$ are not complex conjugates of $\alpha^\dagger$, $\beta^\dagger$ on the solution. The propagator then follows from \eqref{eq_app_prop_pre}:
\begin{equation}\label{eq_app_scalar_prop}
G_\alpha^{(0)}(\omega)=
(2\omega)^{-2\nu}
\frac{4\Gamma (2 \nu ) \Gamma \left(\frac{1}{2}-\nu-i g \right)}{\Gamma (1-2 \nu ) \Gamma \left(\frac{1}{2}+\nu-i g \right)}
\,,\qquad
\omega>0\,.
\end{equation}

An important remark follows. Consider the Euclidean propagator \eqref{eq_app_scalar_prop} analytically continued to complex values of $\omega=|\omega|e^{i\lambda}$. We find that for $g>0$ and $0<\nu< 1/2$, the imaginary part of the propagator vanishes for a value $\lambda=\lambda_*$ between $0$ and $\pi/2$:
\begin{equation}\label{eq_app_Im_G}
\Im\left[G_\alpha^{(0)}(|\omega|e^{i\lambda_*})\right]=0\quad
\text{for }0\leq \lambda_*<\pi/2\,.
\end{equation}
Additionally, such a zero is unique for $-\pi/2\leq \lambda\leq \pi/2$.
The property \eqref{eq_app_Im_G} follows by noticing that the equation $\Im\left[G_\alpha^{(0)}(|\omega|e^{i\lambda_*})\right]=0$ is equivalent to
\begin{equation}\label{eq_app_Im_G_proof}
e^{4 \pi  g} \sin \left(2\left(\pi -2 \lambda _*\right) \nu \right)-2 e^{2 \pi  g} \sin \left(4 \lambda _* \nu \right)-\sin \left(2\left(\pi+2 \lambda _* \right) \nu \right)=0\,.
\end{equation}
\eqref{eq_app_Im_G_proof} is obtained by writing the ratio $G_\alpha^{(0)}/\left[G_\alpha^{(0)}\right]^*$ using the identity $\Gamma(1/2+x)\Gamma(1/2-x)=\pi/\cos(\pi x)$ to simplify the Gamma functions.
The unique solution of \eqref{eq_app_Im_G_proof} for $-\pi/2\leq \lambda\leq \pi/2$ can be written in a simple form in the limits $g\rightarrow 0$ and $g\rightarrow \infty$:
\begin{equation}
\lambda_*=\begin{cases}\displaystyle
0+\frac{\pi  \sin (\pi  \nu )}{2\nu \cos (\pi  \nu )}g
+O\left(g^3\right)
&\text{for } g\rightarrow 0
\\[1em]
\displaystyle\frac{\pi}{2}-\frac{ \sin (2\pi  \nu )}{2\nu }e^{-2 \pi  g}+
O\left(e^{-4 \pi  g}\right)&\text{for } g\rightarrow\infty\,.
\end{cases}
\end{equation}
It can be checked that $\lambda_*$ monotonically grows from $0$ to $\pi/2$ as $g$ increases (with $0<\nu<1/2$). Note that the propagator \eqref{eq_app_scalar_prop} is real on the Euclidean axis for $g=0$: $G_{\alpha}^{(0)}(\omega)=\omega ^{-2 \nu }\frac{2^{2 \nu +1}  \Gamma (\nu )}{\Gamma (1-\nu )}>0$. We also find that the real part of the propagator is positive when the imaginary part vanishes, $\Re\left[G_\alpha^{(0)}(|\omega|e^{i\lambda_*})\right]>0$.

We finally consider a double-trace defect deformation of the form
\begin{equation}
\delta S= f\int d\tau \alpha^\dagger \alpha\,.
\end{equation}
This deformation is relevant for $\nu<1/2$.
The exact Euclidean propagator in this case can be obtained from \eqref{eq_app_prop_pre} by resumming the perturbative series in $f$, leading to the well known result (see e.g.  \cite{Gubser:2002vv,Faulkner:2010tq}):
\begin{equation}\label{eq_app_scalar_propF}
\langle\alpha(\omega)\alpha^\dagger(\omega')\rangle=2\pi\delta(\omega-\omega')G_{\alpha}(\omega)\,,\qquad
G_{\alpha}(\omega)=\frac{G_{\alpha}^{(0)}(\omega)}{1+f G_{\alpha}^{(0)}(\omega)}\,.
\end{equation}
The retarded propagator $G_{\alpha,R}$ is obtained by analytically continuing the Euclidean expression from $\omega>0$ as 
\begin{equation}\label{eq_app_retarded}
G_{\alpha,R}(\omega_L)=\begin{cases}
G_{\alpha}(|\omega_L| e^{-i\frac{\pi}{2}}) & \text{for }\omega_L>0\,\\
G_{\alpha}(|\omega_L| e^{i\frac{\pi}{2}}) & \text{for }\omega_L<0\,.
\end{cases}
\end{equation}

The property \eqref{eq_app_Im_G} implies that, for $f<0$, the retarded propagator analytically continued to the upper half plane has a tachyon pole for $\Im(\omega_L)>0$ and $\Re(\omega_L)<0$.  Such a pole corresponds to a solution of the classical equations of motion with purely outgoing boundary conditions which grows in time. Therefore it signals an instability of the vacuum.
No such pathology occurs for $f>0$, in which case we can safely expand the result at small frequencies
$\omega/|f|^{\frac{1}{2\nu}}\ll 1$ and find a result corresponding to an operator of scaling dimension $\Delta=\frac{1}{2}+\nu$  (up to a contact term).

\subsection{Tachyons for a supercritical Coulomb potential}\label{app_KG_tachyons}

In this section we study the Klein-Gordon equation for a charged field in AdS$_2$ in an external potential,
\begin{equation}\label{eq_app_eq1}
r^2(\pd_0-i A_0)^2\Phi-r^2\pd_r^2\Phi+m^2\Phi=0\,,
\end{equation}
where we work in Lorentzian signature, such that $A_0=g/r$ with $g^2>1/4+m^2$, corresponding to a supercritical Coulomb potential.  We will be particularly interested in the regime $\tilde{\nu}=\sqrt{g^2-m^2-1/4}\ll 1$.
We introduce a cutoff at a small radius $r=r_0$, and impose the most general linear boundary condition on $\Phi$ as in~\eqref{SupCritBC}:
\begin{equation}\label{eq_app_bc1}
\left[r\pd_r\Phi-\left(\frac12+\hat{f}\right)\Phi\right]_{r=r_0}=0\,.
\end{equation}
In the following we show that the problem specified by \eqref{eq_app_eq1} and \eqref{eq_app_bc1} admits infinitely many tachyonic solutions with negative real part of the frequency.

We consider the following solution to \eqref{eq_app_eq1}:
\begin{equation}\label{eq_app_sol_tac1}
\Phi\propto e^{-i\omega t}W_{i g,i \tilde{\nu }}(-2 i r \omega )\,.
\end{equation}
\eqref{eq_app_sol_tac1} behaves as $\Phi\propto e^{-i\omega t}e^{i\omega r}$ for $r\rightarrow \infty$, and thus corresponds to purely outgoing boundary conditions for $\Re(\omega)<0$.  The expansion for $r\ll \omega^{-1}$ of the solution \eqref{eq_app_sol_tac1} takes the general form
\begin{equation}\label{eq_app_sol_AB1}
\Phi\sim \alpha r^{1/2-i\tilde{\nu}}+\beta r^{1/2+i\tilde{\nu}}\,,
\end{equation}
where the ratio between the modes reads
\begin{equation}\label{eq_app_sol_AB2}
\frac{\alpha}{\beta}=(-2i \omega )^{-2 i \tilde{\nu }} \frac{\Gamma \left(2 i \tilde{\nu }\right) \Gamma \left(\frac{1}{2}-i g-i \tilde{\nu }\right)}{\Gamma \left(-2 i \tilde{\nu }\right) \Gamma \left(\frac{1}{2}-i g+i \tilde{\nu }\right)}\,.
\end{equation}
Focusing on $\omega\ll 1/r_0$, we can express the boundary condition \eqref{eq_app_bc1} in terms of the modes \eqref{eq_app_sol_AB1}. Using \eqref{eq_app_sol_AB2} and working at leading order in $\tilde{\nu}\ll 1$, we find the condition:
\begin{equation}\label{eq_app_tac_cond}
\left(-2\,c\,\omega \,r_0 \, 
e^{i\tilde{\gamma}}\right)^{-2i\tilde{\nu}}=1\,,
\end{equation}
where we defined 
\begin{equation}
\tilde{\gamma}=\frac{ \pi }{e^{2 \pi  g}+1}\,,
\end{equation}
and $c$ is an $O(1)$ positive number given by
\begin{equation}
c=\exp\left[
\frac{1}{2} \psi\left(\frac{1}{2}+i g\right)+\frac{1}{2} \psi\left(\frac{1}{2}-i g\right)-\frac{1}{\hat{f}}+2 \gamma_E \right]\,,
\end{equation}
with $\gamma_E$ is the Euler-Mascheroni constant.
\eqref{eq_app_tac_cond} has infinitely many solutions given by
\begin{equation}\label{eq_app_scalar_tachyons}
\omega_n=-\frac{1}{2c r_0}e^{-i\tilde{\gamma}-n\pi/\tilde{\nu}}\,,\qquad
n\in\mathbb{Z}^+ \,,
\end{equation}
where we excluded $n\leq 0$ since our approximations break down for $\omega\gtrsim 1/r_0$. Noticing that $0<\tilde{\gamma}< \pi/2$ for $g>0$ (as we assumed throughout this section),  we see that the frequencies \eqref{eq_app_scalar_tachyons} have $\Re(\omega_n)<0$ and $\Im(\omega_n)>0$. The corresponding solutions \eqref{eq_app_sol_tac1} thus grow in time and signal an instability of the $\Phi=0$ saddle-point. For the physical case of the $\ell=0$ mode of a $4d$ scalar we have $m=0$, thus (from the condition of small $\tilde\nu$) $g\simeq 1/2$ and we find
\begin{equation}
\Im (\omega_n)\simeq -0.13\Re (\omega_n)\quad\text{for } g=1/2\,.
\end{equation}

\subsection{The effective potential from the soliton solution}\label{app:EffectivePot}

In a semiclassical theory the connected generating functional $W[J]$ and one point function $\bar A[J]$ are:
\es{WAbar}{
W[J]&=\le(S+\int dt\, J A\ri)\Big\vert_\text{$A$ saddle}\,,\\
\bar A[J]&={\de W[J]\ov \de J}\,,
}
where in the first line we used the saddle point approximation and set ${\cal W}(A)=0$ for simplicity, cf.~\eqref{WVdef}. (We will restore it later.) The Legendre transform of $W[J]$ is the 1PI effective action:
\es{GamJ}{
\Gamma[\bar A]&=W[J[\bar A]]-\int dt\, J[\bar A]\, \bar A\,,\qquad
J[\bar A]=-{\de \Gamma[\bar A]\ov \de \bar A}\,.
}
Now we specialize to constant $\bar A$ (and hence constant source $J$) and integrate the above equation to get:\footnote{We can also obtain that 
\es{WJcomp}{W(J)=T\int_0^{\bar A(J)}d\bar A'\, \le(J-J(\bar A')\ri)\,;} we can verify that this equation solves the second line of \eqref{WAbar} by taking the $J$ derivative.}
\es{Gammacomp}{
\Gamma(\bar A)\equiv T\,{\cal V}(\bar A)=-T\int_0^{\bar A}d\bar A'\, J(\bar A')\,.
}
Now let us determine $J$ in terms of the quantities we know. The saddle point condition from the first line of \eqref{WAbar} is
\es{JBrel}{
0&={\de S\ov \de A}+J\quad \implies \quad J=-4\nu B(A)\,,
}
where we used \eqref{BulkBdyVar} combined with the fact that the phase of the scalar $\Phi$ is constant for the soliton solution. Plugging this result back into \eqref{Gammacomp} our final formula is:
\es{Gammacomp2}{
{\cal V}(\bar A)&=-4\nu \int_0^{\bar A}d\bar A'\, B(\bar A')\\
&=4\nu \int_0^{\bar A}d\bar A'\, s(g) \le(\bar A'\ri)^{1/2+\nu\ov1/2-\nu}\\
&=4\nu \le(\frac12-\nu\ri)s(g) \le(\bar A\ri)^{1\ov1/2-\nu}\,,
}
which agrees with \eqref{WVdef} with ${\cal W}(A)=0$. We can restore ${\cal W}(A)$ dependence by adding $4\nu{\cal W}(A)$ to $S$ in \eqref{WAbar} and the first line of \eqref{JBrel}, which then shifts $J$ by $4\nu {\cal W}'(A)$, which upon integrating over $A$ as in \eqref{Gammacomp} simply adds $4\nu{\cal W}(A)$ to the result in \eqref{Gammacomp2} as stated in the main text in  \eqref{WVdef}.

\subsection{Quantization of the screening cloud for a light massive scalar}\label{app_quantization}

In section \ref{subsec_1form} we proved that a line charged under a one-form symmetry cannot flow to a topological line in the IR, if the topological operator implementing the one-form symmetry can be cut-open. In this appendix we explore an application of this result to Wilson lines in QED$_4$ in the Coulomb phase with a charge $q_{\phi}>1$ massive scalar.
Because of the one-form symmetry, supercritical Wilson lines with charge $q\neq 0 \mod q_{\phi}$ cannot become topological in the IR. We show how this expectation is borne out by explicitly quantizing the screening soliton. We find that, for sufficiently small mass, the endpoint of the defect RG flow is a Wilson line of charge $q_{IR}\neq 0$.

The action of the model we consider reads:
\begin{equation}\label{eq_appS_quantization}
S=\frac{1}{e^2}\int d^4x\left[\left|\pd_\mu\Phi-i  q_{\phi} A_\mu\Phi\right|^2-m^2|\Phi|^2-\frac14 F_{\mu\nu}^2\right]\,,
\end{equation}
where we take $q_\phi>1$ (integer), $m^2>0$ and we neglected the quartic scalar vertex for simplicity; this will not affect our considerations. 

We now consider a Wilson line of charge $q\gg 1$. As in section \ref{subsec_scalar_supercritical}, we regulate this insertion by cutting off space at a surface $r=r_0$. As we will focus on distances $r\gg r_0$, the detailed form of the boundary conditions at $r=r_0$ will not be important for us.  

In the massless limit, the trivial saddle-point $A_0=\frac{e^2q}{4\pi r}$ is unstable when the Wilson line is supercritical for $q>2\pi/(q_{\phi}e^2)$ or when deforming the alternate quantization fixed point by a double-trace operator with negative coefficient $f$.  In both cases we schematically denote $R_{cloud}$ the radius of the screening cloud. For supercritical lines with $\tilde{\nu} \ll 1$ this scales as $R_{cloud}\sim r_0 e^{\pi/\tilde{\nu}}$, where $\tilde{\nu}=\sqrt{e^4q_{\phi}^2 q^2/(4\pi^2)-1}$. For double-trace deformations it is parametrically set by the coupling, $R_{cloud}\sim |f|^{1/(2{\hat \nu})}$, where ${\hat \nu}=\sqrt{1-e^4q_{\phi}^2 q^2/(4\pi^2)}$.

A sufficiently large mass term $m\gtrsim R_{cloud}^{-1}$ sets an IR cutoff for the screening cloud. The scalar therefore does not fully screen the Wilson line anymore, leaving a remnant Coulomb field at distances $r\gtrsim m^{-1}$ irrespective of the value of $q\mod q_{\phi}$. A quantitative analysis for $q_{\phi}=1$ can be found in \cite{Greiner:1985ce}.  Quantum effects do not qualitatively change the IR limit of the Wilson line (though they might change the value of the charge of the endpoint by an $O(1)$ amount for $q_\phi>1$).

In what follows we focus on the regime $m\ll R_{cloud}^{-1}$.  In this case, a naive extrapolation of the analysis of the massless setup in sections \ref{subsec_scalar_stability} and \ref{subsec_scalar_supercritical} suggests that the Coulomb field of unstable Wilson line is fully screened for every value of $q$ and $q_{\phi}$. We will see below that this is not the case and that for distances larger than the Compton wavelength there is a nontrivial electric flux, in agreement with the one-form symmetry charge.

Consider first the theory in the presence of a charge $q$ Wilson line, such that $q/q_{\phi}\in\mathds{N}$.  In this case we do not expect any subtlety and we can describe the IR limit of the line defect at distances $r\gtrsim R_{cloud}$ via the effective description \eqref{eq_eft_free}. As discussed in section  \ref{subsec_IR_line}, this amounts to expanding the scalar field around a non-trivial solution of the equations of motion in the presence of a source \eqref{eq_eft_Heom}. In the massive case, up to gauge transformations, the profile takes the following form 
\begin{equation}\label{eq_app_qH}
\langle \Phi(r)\rangle=\langle\bar \Phi (r)\rangle=\frac{h_s(r)}{\sqrt{2}}\,,\qquad
h_s(r)=\frac{e^2v}{4\pi} \frac{e^{-m r}}{r}\,.
\end{equation}

We now quantize the zero modes of the theory in the background \eqref{eq_app_qH}. We work in the gauge $A_r=0$.  This completely specifies the gauge up to $r$-independent gauge transformations. We define fluctuations as follows:
\begin{equation}\label{eq_app_h}
\Phi=\frac{h_s(r)}{\sqrt{2}}+e\frac{\delta\phi+i\delta\psi}{\sqrt{2}}\,,\qquad
A_0=0+e \,a_0\,,
\end{equation}
where $\delta\phi$ and $\delta\psi$ are real and we neglected the angular components of the gauge field as these will not play a role in what follows. The quadratic action then reads:
\begin{equation}\label{eq_app_S2}
S\simeq\int d^4x\left[\frac12\left(\pd_\mu\delta\phi\right)^2+\frac12\left(\pd_\mu\delta\psi-q_{\phi} h_s(r) a_0\delta_\mu^0\right)^2-\frac{m^2}{2}\left(\delta\phi^2+\delta\psi^2\right)+\frac12\left(\nabla a_0\right)^2\right]+\ldots\,.
\end{equation}
The linearized scalar $U(1)$ charge density is given by:
\begin{equation}\label{eq_appJ}
j_0=-iq_{\phi}\frac{\pd\mathcal{L}}{\pd\dot{\Phi}^{\dagger}}\Phi+c.c=\frac{q_{\phi}}{e}h_s(r)\left(\delta\dot{\psi}-q_{\phi} h_s(r) a_0\right)+\ldots\,,
\end{equation}
which is normalized so that $Q=\int j_0\in q_{\phi}\mathbb{Z}$, as we will see. \eqref{eq_appJ} measures the charge density at distance $r\gtrsim R_{cloud}$, since we are working in the effective  intermediate energy description of the line, where only the tail of the screening cloud is visible.

From \eqref{eq_app_S2} we derive the equations of motion for $\delta\psi $ and $a_0$:
\es{eq_app_qEOM}{
-\pd^2\delta\psi+q_{\phi} \dot{a}_0 h_s(r)-m^2\delta\psi &=0\,,\\
\nabla^2 a_0+q_{\phi} h_s(r)(\delta\dot{\psi}-q_\phi h_s(r) a_0)&=0\,.
}
In what follows we will need three nontrivial solutions of the equations \eqref{eq_app_qEOM}. The first is given by:
\begin{equation}\label{eq_app_sol1}
\delta\psi \propto h_s(r)\,,\quad a_0=0\,.
\end{equation}
This solution clearly corresponds to an infinitesimal $U(1)$ rotation of the scalar profile \eqref{eq_app_h}.

To find the other, we set $\delta\psi=\dot{a}_0=0$ and consider the radial equation for $a_0$:
\begin{equation}\label{eq_app_a_eq}
\frac{1}{r^2}\pd_r\left(r^2\pd_r a_0\right)=q^2_\phi h_s^2(r) a_0\,.
\end{equation}
This equation can be solved numerically. It admits two solutions, which can be distinguished by their behavior for $r\ll m^{-1}$:
\begin{equation}\label{eq_app_sols}
a_0^{(1)}(r)\stackrel{mr\rightarrow 0}{\sim}\frac{1}{r^{\delta}}\,,\qquad
a_0^{(2)}(r)\stackrel{mr\rightarrow 0}{\sim}\frac{1}{r^{1-\delta}}
\,,
\end{equation}
where 
\begin{equation}
\delta=\frac12+\frac12\sqrt{1+q_{\phi}^2\frac{ e^4v^2}{4\pi^2}}>1\,,
\end{equation}
so that $a^{(1)}_0$ is singular and $a^{(2)}_0$ is regular for $r\rightarrow 0$.
The solutions in \eqref{eq_app_sols} are exact in the massless limit.  For $r m \gg 1$, both solutions take the asymptotic form $a^{(i)}(r)\sim c_1^{(i)}+c_2^{(i)}/r$, where $c_1^{(i)}$ and $c_2^{(i)}$ are constants.\footnote{Note that $c_1^{(i)}$ can be removed by a large gauge transformation involving $\delta\psi$.} We will not need the explicit expressions for $a_0^{(1)}(r)$ and $a_0^{(2)}(r)$ in what follows.

For our purposes it is more convenient to consider the two nontrivial solutions of~\eqref{eq_app_a_eq} in terms of two linear combinations of the modes $a_0^{(1)}(r)$ and $a_0^{(2)}(r)$ in \eqref{eq_app_sols}. The first linear combination is such that the gauge field has no electric flux on the cutoff surface $R_0\gtrsim R_{cloud}$ for the effective defect field theory description. ($R_0$ is not to be confused with the UV cutoff $r_0$.)  We call this the \emph{normalizable} solution. This is formally written as
\begin{equation}\label{eq_app_aNor}
a^{(nor)}_0(r)=
\alpha_1 R_0^{\delta-1} a_0^{(1)}(r)+
\alpha_2 R_0^{-\delta }a_0^{(2)}(r)\quad \text{such\ that}\quad
4\pi r^2\pd_r a^{(nor)}_0(r)\vert_{r=R_0}=0\,,
\end{equation}
where $\alpha_1/\alpha_2=(\delta-1)/\delta$ for $R_0 m\ll 1$. Without loss of generality, we normalize $\alpha_1$ and $\alpha_2$ in \eqref{eq_app_aNor} so that
\begin{equation}\label{eq_app_int0}
\lim_{r\rightarrow\infty}4\pi r^2\pd_r a^{(nor)}_0(r)=-1\,,
\end{equation}
which because of Gauss's law \eqref{eq_app_a_eq} implies
\begin{equation}\label{eq_appJ_int}
4\pi q^2_{\phi} \int_{R_0}^\infty dr\, r^2 h^2_s(r)a^{(nor)}_0(r)=-1\,.
\end{equation}
Importantly for what follows, when we take the massless limit $m\rightarrow 0$ at fixed $r/R_0$, we have $\alpha_2\propto\alpha_1\rightarrow 0$ for the integral in \eqref{eq_appJ_int} to converge. 

For future purposes we also define another linear combination $\tilde{a}_0(r)$ of the solutions \eqref{eq_app_sols} which has the same flux at $r=R_0$ and at infinity:
\begin{equation}\label{eq_app_atilde_def}
4\pi r^2\pd_r \tilde{a}_0(r)\vert_{r=R_0}=\lim_{r\rightarrow\infty}4\pi r^2\pd_r \tilde{a}_0(r)=-1\,,
\end{equation}
that, because of Gauss's law \eqref{eq_app_a_eq}, imply
\begin{equation}
4\pi q^2_{\phi} \int_{R_0}^\infty dr \,r^2 h^2_s(r)\tilde{a}_0(r)=0\,.
\end{equation}
Explicitly this solution reads
\begin{equation}\label{eq_app_atilde}
\tilde{a}_0(r)=
\beta_1 R_0^{\delta-1} a_0^{(1)}(r)+
\beta_2 R_0^{-\delta }a_0^{(2)}(r)\,,
\end{equation}
where for $R_0 m\ll 1$ we have $\beta_1=\alpha_1+1/(4\delta\pi)$ and $\beta_2=\alpha_2$.  Therefore in the massless limit $\tilde{a}_0(r)\propto a^{(1)}(r)$, which is regular at infinity.

\begin{figure}[t]
\centering
\subcaptionbox{\label{fig:app_Enor}}
{\includegraphics[scale=0.35]{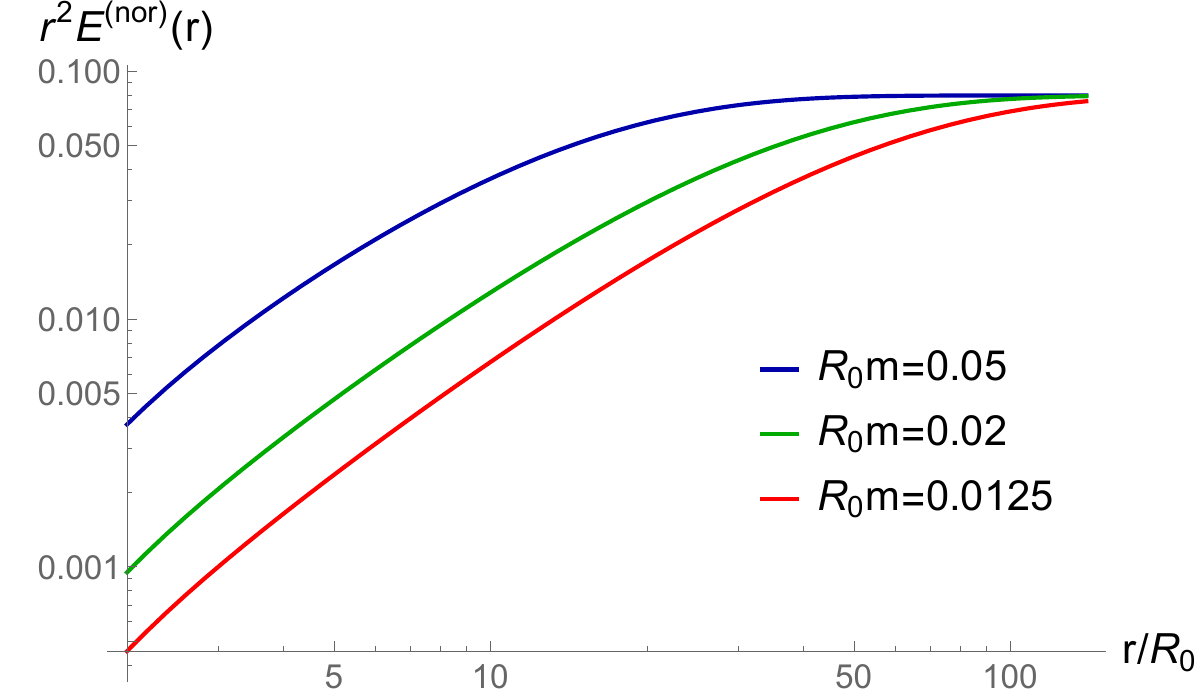}}\hspace*{2em}
\subcaptionbox{ \label{fig:app_Et}}
{\includegraphics[scale=0.35]{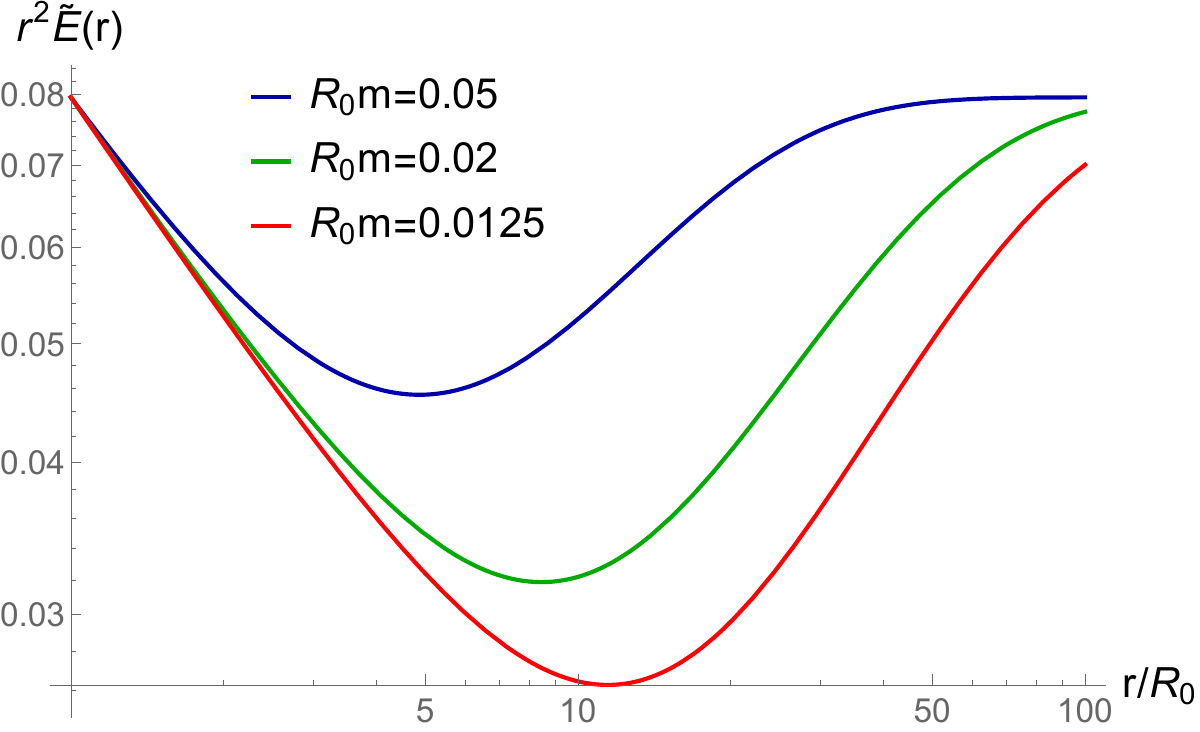}}
\caption{The electric flux $r^2 E= r^2 F_{tr}$ associated with the solutions~$a_0^{(nor)}(r)$ (figure~\ref{fig:app_Enor}) and~$\tilde{a}(r)$ (figure~\ref{fig:app_Et}) for different values of the mass $m$ (in units of the cutoff $R_0$). These plots were obtained by setting $\frac{q_{\phi}e^2v}{4\pi}=1$.}
\end{figure}

We plot the schematic form of the dimensionless electric flux $r^2 F_{tr}$ associated with the solutions above in figures~\ref{fig:app_Enor} and~\ref{fig:app_Et}.  In the normalizable solution $a^{(nor)}(r)$, the flux continuously increases until distances $r\sim 1/m$, at which it becomes constant. For the solution $\tilde{a}(r)$ instead the electric flux first decreases, reaching a minimum, and then it starts rising and  asymptotically approaches a constant value. It is important to stress that the distance at which the electric flux reaches its minimum increases as we make the mass smaller, $r_{min}\sim 1/m$, and it drifts to infinity in the massless limit.  

When quantizing the theory, the two solutions \eqref{eq_app_sol1} and \eqref{eq_app_aNor} provide the zero-modes inside the decomposition of the fields $\delta \psi$ and $a_0$
\begin{equation}\label{eq_app_dec}
\begin{aligned}
\delta\psi(t,x)&=h_s(r)\,\hat{x}+\text{wave-modes}\,,\\
a_0(t,x)&=e q_{\phi}\,a^{(nor)}_0(r)\,\hat{p}+\text{wave-modes}\,.
\end{aligned}
\end{equation}
To bypass the full quantization of the system (which would require solving also for the wave modes) and directly find the commutation relation between $\hat{x}$ and $\hat{p}$, we impose the charge action on $\Phi$:
\begin{equation}\label{eq_app_Qcomm}
[Q,\Phi]=q_{\phi}\Phi\quad\implies\quad [\delta\psi,Q]=i q_{\phi} h_s(r)\,,
\end{equation}
where we linearized around the solution \eqref{eq_app_qH}. From \eqref{eq_appJ} and \eqref{eq_appJ_int} we read the charge operator
\begin{equation}\label{eq_app_Qexpr}
Q=-\frac{4\pi q_{\phi}^2}{e}  \int_{R_0}^\infty dr \,r^2 h^2_s(r)a_0(t,r)
=q_{\phi}\hat{p}\,.
\end{equation}
We conclude that at quantum level the operators $\hat{x}$ and $\hat{p}$ form a canonical pair
\begin{equation}\label{eq_app_CCR}
[\hat{x},\hat{p}]=i.
\end{equation}

Remarkably, the decomposition \eqref{eq_app_dec} and the commutation relation \eqref{eq_app_CCR} are all we need to construct solitonic states with the properly quantized charge. Explicitly, we notice that the phase of $\Phi$ is a compact field. When linearizing around the background \eqref{eq_app_qH}, this implies that $\hat{x}$ is defined only modulo $2\pi$ in \eqref{eq_app_dec}. Therefore, calling $|0\rangle$ the state such that $\hat{p}|0\rangle=0$, we can construct the following quantum states
\begin{equation}
|n\rangle= e^{i n\hat{x}}|0\rangle\,\qquad\text{for }n\in\mathds{N}\,.
\end{equation}
\eqref{eq_app_CCR} implies that
\begin{equation}\label{eq_app_n_def}
\hat{p}|n\rangle=n |n\rangle\,,
\end{equation}
and therefore, using \eqref{eq_app_Qexpr} we conclude that the states $|n\rangle$ have quantized values for the gauge charge in units of $q_{\phi}$
\begin{equation}\label{eq_app_n_q}
Q|n\rangle=n q_{\phi} |n\rangle\,.
\end{equation}
\eqref{eq_app_n_q} implies that the screeneing cloud for the state $|n\rangle$ has an extra $q_{\phi} n$ units of charge with respect to the ground state, which fully screens the Wilson line. The expectation value of the gauge field and the charge density are similarly computed from \eqref{eq_appJ} and \eqref{eq_app_dec}.
Note that to linear order there is no difference in the expectation value of the scalar field profile between a state $|n\rangle$ and $|0\rangle$; this is no longer true at higher orders.

We are finally ready to address the issue of screening of Wilson lines with charge $\notin q_{\phi}\mathds{N}$. To this aim, consider a Wilson line with charge $q+\delta q$, with $q/q_{\phi}\in \mathds{N}$  and $\delta q$ an $O(1)$ correction, $\delta q\ll q$. We model this setup by perturbing the scalar IR defect previously analyzed via the following term
\begin{equation}
\delta S_{D}=-\delta q\int dt \,A_0\,.
\end{equation}
This term induces a classical profile for the gauge field in addition to the quantized part.  Thus \eqref{eq_app_dec} is modified to:
\begin{equation}\label{eq_app_dec2}
\begin{aligned}
\delta\psi(t,x)&=h_s(r)\,\hat{x}+\text{wave-modes}\,,\\
a_0(t,x)&= e q_{\phi} a^{(nor)}_0(r)\,\hat{p}+e\,\delta q \,\tilde{a}_0(r)+\text{wave-modes}\,.
\end{aligned}
\end{equation}
where $\tilde{a}_0(r)$ is the solution \eqref{eq_app_atilde_def} which has the same flux at $r=R_0$ and $r\rightarrow \infty$ (cf. \eqref{eq_app_atilde_def} and figure~\ref{fig:app_Et}); by Gauss's law thus, $\tilde{a}_0(r)$ does not contribute to the total charge of the cloud. It does however contribute to the electric flux, which for $r \gg m^{-1}$ on a state $|n\rangle$ is given by
\begin{equation}\label{eq_app_quant_res}
\lim_{r\rightarrow\infty}4\pi r^2\pd_ra_0(r)|n\rangle=-e\left(q_{\phi} n+\delta q\right)|n\rangle\,.
\end{equation}

\begin{figure}[t]
\centering
\includegraphics[scale=0.45]{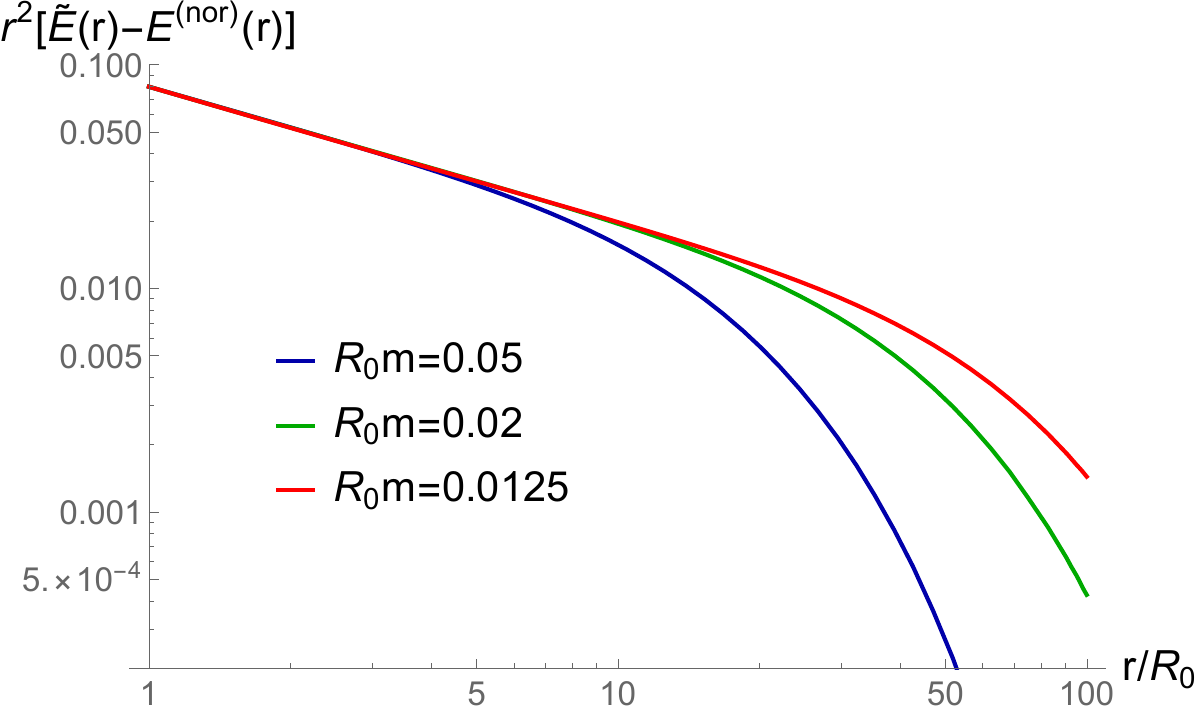}
\caption{A log-log plot of the electric flux $E=r^2 F_{tr}$ associated with the solution~$\tilde{a}_0(r)-a_0^{(nor)}(r)$  which describes the screened electric field for $\delta q \mod q_{\phi}=0$. The plot corresponds to $\frac{q_{\phi}e^2v}{4\pi}=1$.}
\label{fig:app_Ediff}
\end{figure}

From \eqref{eq_app_quant_res} we conclude that when $\delta q=0 \mod q_{\phi}$, the flux is fully screened on the state $|-\delta q/q_{\phi}\rangle$. This state is obviously the energetically favored one. The expectation value of the gauge field on this state  reads
\begin{equation}\label{eq_app_q_qphi_gauge}
\langle a_0(r)\rangle=\delta q\,e \left[\tilde{a}_0(r)-a_0^{(nor)}(r)\right]=\frac{e\,\delta q}{4\pi\delta} R_0^{\delta-1}a_0^{(1)}(r)\,,
\end{equation}
where $a_0^{(1)}(r)$ is the mode which is singular for $r m\ll 1$ in \eqref{eq_app_sols}.  In figure~\ref{fig:app_Ediff} we show the behavior of the resulting electric field (normalized by $r$ to be dimensionless); as expected the field vanishes at large distances. For $rm \ll 1$ the flux decays as a power law according to~\eqref{eq_app_sols}.

For $\delta q\neq 0 \mod q_{\phi}$, instead, \eqref{eq_app_quant_res} implies that the electric flux cannot be fully screened by the scalar field, in agreement with our expectation. In this case we expect the ground state to be given by the state in which the maximal possible amount of charge has been screened at infinity, \footnote{For instance,  we expect a Wilson line of supercritical odd charge $q>0$ interacting with a charge $2$ scalar field to flow to a (positive) charge $1$ Wilson line at distances $r\gg 1/m$.} at least for sufficiently small $m$. The expectation value of the gauge field on this state is a linear combinations of the two modes $a_0^{(1)}(r)$ and $a_0^{(2)}(r)$ in \eqref{eq_app_sols}:
\begin{equation}\label{eq_app_q_qphi_gauge2}
\begin{split}
\langle a_0(r)\rangle 
&= e\left(\delta q-n q_{\phi}\right)\tilde{a}_0(r)+n q_{\phi} \left[\tilde{a}_0(r)-a_0^{(nor)}(r)\right]
\\&
=e\left[\frac{\delta q}{4\pi\delta} +\alpha_1(\delta q-n q_\phi)\right]R_0^{\delta-1}a_0^{(1)}(r)
+e\,\alpha_2 (\delta q-n q_\phi) R_0^{-\delta}a_0^{(2)}(r)\,.
\end{split}
\end{equation}
As evident from the plot~\ref{fig:app_Et}, the term $e\left(\delta q-n q_{\phi}\right)\tilde{a}_0(r)$ leads to a nontrivial electric field which can be measured at large distances $r\gg 1/m$.

We close this section with some comments on the massless limit, taken as $m\rightarrow 0$ for fixed $r/R_0$. For a line such that $q=0\mod q_{\phi}$, this limit is smooth. Indeed in this case the expectation value of the gauge field is given by \eqref{eq_app_q_qphi_gauge}, which for $m=0$ reduces exactly to the expression in \eqref{eq_app_sols},  in agreement with the discussion in sections \ref{subsec_scalar_supercritical} and \ref{subsec_IR_line}. For $q\neq 0\mod q_{\phi}$ the gauge field \eqref{eq_app_q_qphi_gauge2} admits also a contribution proportional to $a^{(2)}_0(r)$. However, according to the discussion below \eqref{eq_app_aNor}, the coefficients $\alpha_1$ and $\alpha_2$ become infinitesimal in the massless limit.  Thus from \eqref{eq_app_q_qphi_gauge} and \eqref{eq_app_q_qphi_gauge2} we conclude that the electric flux, and more in general the screening cloud, do not depend on the value of  $q\mod q_{\phi}$ for $r\ll m^{-1}$.  Indeed, as already commented,  figure~\ref{fig:app_Et} shows that the electric flux constantly decreases for $r\ll 1/m$. Physically this behavior is due to the fact that the scalar wave-function is delocalized over distances of the order of the Compton wavelength. It is thus necessary to make measurements at $r\gtrsim m^{-1}$ to see the effects of the quantization of the $U(1)$ charge of the matter field. In the massless limit, the wave-function can be spread over arbitrary distances and it is possible to store fractional units of charge at $r\rightarrow \infty$.\footnote{This is the physical meaning of \eqref{eq_app_q_qphi_gauge2} in the massless limit, since $\alpha_2$  is infinitesimal while the solution $a^{(2)}(r)\sim r^{\delta}$ grows with the distance.} This is not in contradiction with the general theorem proven in section \ref{subsec_1form}, since the defect remains nontrivial at all scales also in the massless limit due to the conformal one-point function for the scalar field \eqref{eq_app_qH}.

The same general discussion remains true upon including a quartic coupling, which only modifies the expressions for the solutions in \eqref{eq_app_sols}.  In particular analogous conclusions are found for massive scalars, and Wilson lines charged under the one-form symmetry lead a remnant Coulomb field at large distances.  In the massless limit, one finds again the screening cloud does not depend on the value of  $q\mod q_{\phi}$.  As remarked in section \ref{subsec_1form},  this raises a small puzzle, since the analysis in section~\ref{subsec_IR_line} shows that all Wilson lines flow logarithmically to a trivial one in the double-scaling limit. We plan to analyze  this issue further in future work.

\section{Details on fermionic \texorpdfstring{QED$_4$}{QED4}}\label{app_DC_problem}

\subsection{Defect propagator and double-trace deformation for subcritical charge}\label{app_DC_problem_sub}

In this section we compute the propagator of the defect operator $\alpha$ for the theory \eqref{eq_fer_toy_action} at the alternate quantization fixed point. We then use this result to compute the exact propagator in the presence of a double-trace deformation as in \eqref{eq_fer_toy_negF}. We argue that such a propagator does not have a tachyon pole.

We start by computing the propagator at $f=0$. We consider the action in Euclidean signature
\begin{equation}\label{eq_fer_toy_action_EUCL}
S=\int_{\text{AdS}_2} d^2x\sqrt{g}\,\bar{\psi} \left(
\overset{\leftrightarrow}{\slashed{\nabla}}_{\text{AdS}_2}-g\gamma^0+m\right)\psi\,,
\end{equation}
where the Euclidean gamma matrices are given by $\gamma^0=\sigma_1$ and $\gamma^1=\sigma_3$.

To extract the propagator we pursue the generating function approach as in appendix \ref{app_KG_prop}.
We consider the theory with Dirichlet boundary conditions in terms of the modes \eqref{eq_fer_modes},
\begin{equation}\label{eq_app_fer_toy_J}
\beta(\tau)=\frac{m+\nu}{\nu}J(\tau)\,,\qquad
\bar{\beta}(\tau)=\frac{m+\nu}{\nu}\bar{J}(\tau)\,,
\end{equation}
where $J$ is an external Grassmanian function, which is interpreted as an external source. The action is stationary with Dirichlet boundary conditions of the form \eqref{eq_app_fer_toy_J} if we add the following boundary term:\footnote{This is just the Euclidean version of the term \eqref{eq_bdry_fer2} in the limit $r_0\rightarrow 0$; note however that its interpretation is now different, as we are imposing Dirichlet conditions, rather than minimizing the action for arbitrary values of the fluctuations.}
\begin{equation}\label{eq_fer_toy_action_EUCL_bdry}
S_{bdry}=\frac{\nu}{m+\nu}\int d\tau
\left[\bar{\alpha}(\tau)\beta(\tau)+\bar{\beta}(\tau)\alpha(\tau)\right]\,
=\int d\tau\left[\bar{\alpha}(\tau)J(\tau)+\bar{J}(\tau)\alpha(\tau)\right] \, ,
\end{equation}
where in the last step we used the Dirichlet conditions \eqref{eq_app_fer_toy_J}. 

Importantly, the fact that the bulk action is linear in derivatives implies that the on-shell action coincides with the boundary term \eqref{eq_fer_toy_action_EUCL_bdry}.
We may thus follow the GKPW prescription as in appendix \ref{app_KG_prop} and interpret the theory with Dirichlet boundary conditions \eqref{eq_app_fer_toy_J} as the deformation of the alternate fixed point by a complex source $J$ for $\alpha$ \cite{Witten:1998qj,Gubser:1998bc}. 
Then the propagator for the boundary field $\alpha$ at the alternate quantization fixed point is\footnote{$G_{\alpha}^{(0)}(\omega)$ is the sum of $\al(\om)$ with $J(\om)$ stripped off, and similarly for  $\bar\al(\om)$, just like in the scalar case \eqref{eq_app_prop_pre}. Here we have to write the expression a bit differently from the scalar case, since we are dealing with Grassmannian quantities.}
\begin{equation}\label{eq_app_fer_prop_pre}
\langle\alpha(\omega)\bar{\alpha}(\omega')\rangle_{f=0}=2\pi\delta(\omega-\omega')G_{\alpha}^{(0)}(\omega)\,,\qquad
G_{\alpha}^{(0)}(\omega)=-\left[\frac{\pd \alpha(\omega)}{\pd J(\omega)}+\frac{\pd \bar{\alpha}(\omega)}{\pd \bar{J}(\omega)}\right]\,,
\end{equation}
where similarly to \eqref{eq_app_prop_pre}, $\alpha$ is obtained from the boundary limit (cf.~\eqref{eq_fer_modes_basic}) of a regular solution of the bulk equations of motion with boundary condition \eqref{eq_app_fer_toy_J} for $r\rightarrow 0$. The Fourier transform is defined according to
\begin{equation}
\alpha(\tau)=\int \frac{d\omega}{2\pi}e^{-i\omega \tau}\alpha(\omega)\,,\qquad
\bar{\alpha}(\tau)=\int \frac{d\omega}{2\pi}e^{i\omega \tau}\bar{\alpha}(\omega)\,,
\end{equation}
and similarly for $J(\tau)$, $\bar{J}(\tau)$. 

To find the value of $\alpha(\omega)$ to be used in \eqref{eq_app_fer_prop_pre} we thus need to solve the Euclidean Dirac-Coulomb equation for $\psi(\tau,r)=e^{-i\omega\tau}\psi(r)$:
\begin{equation}\label{eq_app_fer_prop_eq1}
\left[\left(-i\omega r-g\right)\gamma^0+\left(r\pd_r-\frac12\right)\gamma^1+m\right]\psi=0\,.
\end{equation}
We set
\begin{equation}\label{eq_app_fer_psi12}
\displaystyle\psi=\left(\begin{array}{c}
\displaystyle\frac12(\psi_1+\psi_2) \\[0.6em]
\displaystyle\frac{i}{2}(\psi_1-\psi_2)
\end{array}\right)\,,
\end{equation}
so that \eqref{eq_app_fer_prop_eq1} reduces to
\begin{align}\label{eq_app_fer_prop_eq2a}
&r \pd_r\psi_1+ \left(r \omega-i g -\frac{1}{2}\right)\psi_1+m\psi_2=0\,,\\
& r \pd_r\psi_2+m \psi_1 +\left(i g-r \omega -\frac{1}{2}\right)\psi_2=0\,.
\label{eq_app_fer_prop_eq2b}
\end{align}
Using \eqref{eq_app_fer_prop_eq2b} to solve for $\psi_1$ in terms of $\psi_2$ and $\pd_r \psi_2$, we find
\begin{equation}
\psi_1=\frac{1}{2 m}\left[\left(1+2 r \omega -2i g\right)\psi_2- 2 r \pd_r \psi_2\right]\,,
\end{equation}
and \eqref{eq_app_fer_prop_eq2a} reduces to
\begin{equation}
-r^2\pd_r^2\psi_2+ \left[m^2-(g+i r \omega )^2+r \omega -\frac{1}{4}\right]\psi_2=0\,.
\end{equation}
The most general solution to this equation is a linear combination of Whittaker $W$ functions
\begin{equation}\label{eq_app_fer_prop_sol1}
\psi_2= c_1 W_{i g-\frac{1}{2},\nu}(2r \omega )+c_2 W_{\frac{1}{2}-i g,\nu}(-2 r\omega )\,.
\end{equation}
In the following we focus on $\omega>0$. From $W_{x,y}(z)\overset{z\rightarrow\infty}{\propto} e^{-z/2}$, we infer that regularity implies that we need to set $c_2=0$ in \eqref{eq_app_fer_prop_sol1}.  Comparing \eqref{eq_app_fer_psi12} with \eqref{eq_fer_modes_basic}, we find that the $r\rightarrow 0$ limit of $\psi_2$ can be written as
\begin{equation}
\psi_2\overset{r\rightarrow 0}{\sim}
\frac{2(g+i \nu)}{\nu+m-i g}\beta
 r^{\frac{1}{2}+\nu}+\frac{2  m}{\nu+m-i g}\alpha 
 r^{\frac{1}{2}-\nu}\,.
\end{equation}
We thus extract $\alpha$ and $\beta$ by comparing with the expansion of the Whittaker's function \eqref{eq_app_Wexp}.
Performing the same steps also for $\bar{\psi}$, we eventually find
\begin{equation}
\frac{\alpha(\omega)}{\beta(\omega)}=\frac{\bar{\alpha}(\omega)}{\bar{\beta}(\omega)}=i  \omega ^{-2\nu }
\frac{ \Gamma (2\nu ) \Gamma \left(1-i g-\nu\right)}{ m 2^{2\nu }\Gamma (-2\nu ) \Gamma \left(-i g+\nu\right)}\,,\qquad
\omega>0\,.
\end{equation}
Note that $\bar{\alpha}/\bar{\beta}\neq \alpha^\dagger/\beta^\dagger$ on the solution. The propagator then follows from \eqref{eq_app_fer_prop_pre}:\footnote{Note that the result \eqref{eq_app_fer_prop} cannot be straightforwardly continued to $\omega<0$, since the fermion propagator is discontinuous at $\omega=0$.}
\begin{equation}\label{eq_app_fer_prop}
G_\alpha^{(0)}(\omega)=-
2i  \omega ^{-2\nu }\frac{\nu /m}{m+\nu}
\frac{ \Gamma (2\nu ) \Gamma \left(1-i g-\nu\right)}{ 2^{2\nu }\Gamma (-2\nu ) \Gamma \left(-i g+\nu\right)}\,,\qquad
\omega>0\,.
\end{equation}

For future purposes, we study the imaginary part of \eqref{eq_app_fer_prop} for $g>0$ and $0<\nu<1/2$.  First note that for $g=0$, we find $G_{\alpha}^{(0)}(\omega)=i\omega^{-2\nu}|c|$ where $|c|>0$ is a constant. More generally, we find that for $g>0$ and $0<\nu<1/2$ the propagator \eqref{eq_app_fer_prop} always has a positive imaginary part for $\Re (\omega)>0$:
\begin{equation}\label{eq_app_fer_propIm}
\Im (\left[G_{\alpha}^{(0)}(|\omega|e^{i\lambda})\right])>0\,\quad
\text{for}\quad -\frac{\pi}{2}\leq \lambda\leq \frac{\pi}{2}\,.
\end{equation}

We now consider a double-trace defect deformation as in \eqref{eq_fer_toy_negF}. The exact Euclidean propagator in this case can be obtained from \eqref{eq_app_fer_prop} by resumming the perturbative series in $f$
as in \eqref{eq_app_scalar_propF}:
\begin{equation}\label{eq_app_fer_propF}
\langle\alpha(\omega)\bar{\alpha}(\omega')\rangle=2\pi\delta(\omega-\omega')G_{\alpha}(\omega)\,,\qquad
G_{\alpha}(\omega)=\frac{G_{\alpha}^{(0)}(\omega)}{1+f G_{\alpha}^{(0)}(\omega)}\,.
\end{equation}
The retarded propagator $G_{\alpha,R}$ is obtained by analytically continuing the Euclidean expression from $\omega>0$ as 
in \eqref{eq_app_retarded}.
Then the property \eqref{eq_app_fer_propIm} implies that the retarded propagator analytically continued on the upper half plane $\Im(\omega)>0$ has no singularities irrespectively of the sign of $f$. In particular, there is no tachyon pole, differently than in the scalar setup analyzed in appendix \ref{app_KG_prop}. The expansion of \eqref{eq_app_fer_propF} for 
$\omega/|f|^{\frac{1}{2\nu}}\ll 1$ takes qualitatively the same form for both signs of $f$, corresponding (up to a contact term) to an operator of scaling dimension $\Delta=\frac{1}{2}+\nu$.\footnote{Note however that the propagator differs by a sign in both cases, hinting at a change in nature between creation and annihilation operators and thus at a screening mechanism; we analyze this mechanism in section~\ref{subsec_fer_negative}.}
In conclusion, the defect fermionic propagator does not display any pathology.

We also comment that, by numerically studying $G_{\alpha}^{(0)}(\omega)$ for $m=1$ and $m=1/2$, and $g$ such that $0<\nu<1/2$, we found that the imaginary part $\Im (\left[G_{\alpha}^{(0)}(|\omega|e^{i\lambda})\right])$ admits zeroes for $\lambda=\frac{\pi}{2}+\delta_{m,g}$ where $\delta_{m,g}>0$ is a numerically small number. For instance for $m=1$ and $g=0.9$ we find $\delta_{m,g}\simeq 0.0016$, while for $g\ll 1$ and arbitrary $m$ we find $\delta_{m,g}\simeq g^2\pi$. Such a zero of the imaginary part implies a pole in the second sheet for the double-trace deformed propagator \eqref{eq_app_fer_propF} at $f<0$, for $\Re(\omega)<0$ and $\Im(\omega)<0$. As commented in footnote \ref{footnote_decay_fer}, we expect the imaginary value of $\omega$ on this pole to be associated with the lifetime of the unstable vacuum after a negative double-trace defect deformation is suddenly turned on. It would be interesting to investigate this connection further.

\subsection{Massive Dirac-Coulomb equation for subcritical charge}\label{app_DC_problem_M_sub}

In this section we study the AdS$_2$ Dirac-Coulomb equation for the model \eqref{eq_fer_toy_action} in the presence of the deformation \eqref{eq_fer_M_def}:
\begin{equation}\label{eq_app_DC_M}
\left[i\left(\slashed{\pd}-i\slashed{A}\right)-m\pm i\,r\gamma^3 M\right]\psi^{(\pm)}(t,r)=0\,,
\end{equation}
where $m>0$, $M>0$, $A_0=g/r$ with $g>0$ and we work in Lorentzian signature with the gamma matrices given by \eqref{eq_gamma_matrices_2d} and \eqref{eq_gamma3def}. Note that the mass term $M$ does not modify the near-boundary behavior \eqref{eq_fer_modes_basic}.

We are interested in finding bound states, i.e.~solutions of \eqref{eq_app_DC_M} with energy $\omega$ such that $M^2>\omega^2$. To this aim we follow \cite{Greiner:1985ce} and set
\begin{equation}\label{eq_app_DC_M_ansatz}
\psi^{(\pm)}(t,r)=\sqrt{r}e^{-i\omega t}\left(\begin{array}{c}
e^{-\frac{\rho}{2}}\sqrt{M\mp\omega}\left(\Phi_2^{(\pm)}(\rho)
\pm\Phi_1^{(\pm)}(\rho)\right)\\[0.4em]
e^{-\frac{\rho}{2}}\sqrt{M\pm\omega}\left(\Phi_1^{(\pm)}(\rho)
\mp\Phi_2^{(\pm)}(\rho)\right)
\end{array}\right)\,,
\end{equation}
where we defined
\begin{equation}
\rho=2\lambda r\,, \qquad
\lambda\equiv\sqrt{M^2-\omega^2}\,.
\end{equation}
Using \eqref{eq_app_DC_M_ansatz},  \eqref{eq_app_DC_M} reduces to
\begin{align} \label{eq_app_DC_M_eq1}
&
\pd_\rho\Phi_1^{(\pm)}+\left(\frac{g \omega }{\rho  \sqrt{M^2-\omega ^2}}-1\right)\Phi_1^{(\pm)}
+\left(\frac{g M }{\lambda}\pm m\right)\Phi_2^{(\pm)}=0\,,\\
& \pd_\rho\Phi_2^{(\pm)}
-\left(\frac{g M }{\lambda}\mp m\right)\Phi_1^{(\pm)}
-\frac{g \omega }{\rho \lambda}\Phi_2^{(\pm)}=0\,.
\label{eq_app_DC_M_eq2}
\end{align}
Solving \eqref{eq_app_DC_M_eq2} for $\Phi_1^{(\pm)}$ as
\begin{equation}\label{eq_app_DC_M_U_pre}
\Phi_1^{(\pm)}=\frac{ \lambda \rho\pd_\rho\Phi_2^{(\pm)}-g \Phi_2^{(\pm)} \omega }{g M\mp m \lambda}\,,
\end{equation}
we recast \eqref{eq_app_DC_M_eq1} in the form
\begin{equation}\label{eq_app_DC_M_eq3}
\rho\pd_\rho^2 \Phi_2^{(\pm)}+(1-\rho) \pd_\rho \Phi_2^{(\pm)}+
\left(\frac{g\omega}{\lambda}-\frac{\nu}{\rho}\right)\Phi_2^{(\pm)}=0\,,
\end{equation}
where $\nu=\sqrt{m^2-g^2}$.  The solution of \eqref{eq_app_DC_M_eq3} that is regular for $\rho\rightarrow\infty$ (with $\rho>0$) is written in terms of a confluent hypergeometric function
\begin{equation}\label{eq_app_DC_M_U}
 \Phi_2^{(\pm)}(\rho)\propto\rho ^{\nu} U\left(\nu-\frac{g \omega }{\lambda},1+2\nu ;\rho \right)\,.
\end{equation}

We want to find the quantization condition on $\omega$ for the  most general linear boundary condition on the modes \eqref{eq_fer_modes_basic}:
\begin{equation}\label{eq_app_fer_AB_bc_M}
\beta /\alpha =\frac{m+\nu}{\nu}f=\text{sgn}(f)\mu^{2\nu}\,,
\end{equation}
where we defined $\mu=\left(\frac{m+\nu}{\nu}|f|\right)^{1/(2\nu)}>0$ as the mass scale associated with the double-trace perturbation. We are particularly interested in the consequences of a negative $f$ in \eqref{eq_app_fer_AB_bc_M}, but we will study both signs for generality. 

To proceed, we compare the solution \eqref{eq_app_DC_M_ansatz} and \eqref{eq_app_fer_AB_bc_M} to rewrite the small $\rho$ expansion of $\Phi_2^{(\pm)}$ in terms of $\alpha$ and $\beta$
\begin{equation}\label{eq_app_DC_ABexp}
\begin{split}
\Phi_2^{(\pm)} \overset{\rho\rightarrow 0}{\sim}
&\beta  \left(\frac{\rho }{2\sqrt{M^2-\omega ^2}}\right)^{\nu} \left[\frac{ g}{2( m+\nu) \sqrt{M\pm\omega }}\mp \frac{1}{2\sqrt{M\mp \omega }}\right]\\
+&\alpha   \left(\frac{\rho }{2\sqrt{M^2-\omega ^2}}\right)^{-\nu} \left[\frac{1}{2\sqrt{M\pm\omega }}\mp \frac{g}{2\left(m+\nu\right) \sqrt{M\mp\omega }}\right]\,.
\end{split}
\end{equation}
Using the expansion of the confluent hypergeometric function,
\begin{equation}\label{eq_app_DC_Uexp}
U(x,1+y;z)\overset{z\rightarrow 0}{\sim}z^{-y} \frac{\Gamma (y)}{\Gamma (x)}+\frac{\Gamma (-y)}{\Gamma (x-y)}\,,
\end{equation}
we extract the ratio $\alpha/\beta$ from the comparison of \eqref{eq_app_DC_M_U} and \eqref{eq_app_DC_ABexp}:
\begin{equation}\label{eq_app_DC_M_ratio}
\frac{\beta^{(\pm)}}{\alpha^{(\pm)}}=
\frac{4^{\nu } \lambda ^{2\nu }(M\mp\omega )[ g (M\pm\omega )\mp \lambda  (\nu + m)] \Gamma (-2\nu )  \Gamma \left(1+\nu -\frac{g \omega }{\lambda }\right) }{\Gamma (2\nu ) \Gamma \left(-\nu-\frac{g \omega }{\lambda }\right) \left[\left(\nu  m +\nu^2\right) M^2\pm M \left(g^2 \omega -\nu  g \lambda \right)- m \omega  (\nu  \omega + g \lambda )- m^2 \omega ^2\right]}\,.
\end{equation}

Using \eqref{eq_app_DC_M_ratio}, the boundary condition \eqref{eq_app_fer_AB_bc_M} provides a condition on $\omega$ from which we infer the energies $\omega_n$ of the bound states. Let us state the results for $f=0$ and $f\rightarrow+\infty$:
\begin{itemize}
\item $f\rightarrow +\infty$: this sets $\alpha=0$, corresponding to standard quantization.  We find
\begin{equation}\label{eq_app_DC_M_q1}
\omega_n=\frac{M}{\sqrt{1+\frac{g^2}{(n+\nu)^2}}}\,,
\end{equation}
where $n=1,2,\ldots$ for $(\delta)=+$ and $n=0,1,2,\ldots $ for $(\delta)=-$.  \eqref{eq_app_DC_M_q1} agrees with the well known result for the relativistic Hydrogen atom \cite{Greiner:1985ce}.
\item $f=0$: this sets $\beta=0$, corresponding to alternate quantization.  We find
\begin{equation}\label{eq_app_DC_M_q2}
\omega_n=\frac{M(n-\nu)}{\sqrt{(n-\nu)^2+g^2}}\,,
\end{equation}
where again $n=1,2,\ldots$ for $(\delta)=+$ and $n=0,1,2,\ldots $ for $(\delta)=-$.  Note that $\omega_0=-\frac{\nu}{m}M$ is negative. \eqref{eq_app_DC_M_q2} is a new result to the best of our knowledge.
\end{itemize}
Increasing the coupling $f$ from $0$ to $\infty$ smoothly transforms the spectrum \eqref{eq_app_DC_M_q2} into \eqref{eq_app_DC_M_q1}. For a sufficiently negative $f<0$ we instead encounter an interesting phenomenon: as we increase $\mu$ in \eqref{eq_app_fer_AB_bc_M} the lowest bound state eventually reaches $\omega= -M$ and then \emph{dives} into the continuum part of the spectrum. To see this, we look for a solution of \eqref{eq_app_fer_AB_bc_M} with $\omega\simeq -M$. We expand the ratio \eqref{eq_app_DC_M_ratio} as:
\begin{equation}\label{eq_app_fer_DC_M_exp_ratio}
\frac{\beta^{(\pm)}}{\alpha^{(\pm)}}=(g M)^{2\nu }\left[-c_1^{(\pm)}+c_2^{(\pm)}\frac{\omega+M}{M g^2}+
O\left(\frac{(\omega+M)^2}{M^2 g^4}\right)\right]\,,
\end{equation}
where we defined the following coefficients
\begin{align}
c_1^{(\pm)}&= 
\frac{\pi  \,2^{2\nu }( m\pm\nu )}{g\,\sin\left(2\pi  \nu \right)  \Gamma (2\nu ) \Gamma (1+2\nu )}\,,
 \\[0.6em]
c_2^{(\pm)}&=c_1^{(\pm)} \frac{\nu}{3}   \left(1-4\nu^2+6 m^2\mp 3 m\right)\,.
\end{align}
Importantly, they are both positive for $0<\nu<1/2$:\footnote{To see this one needs to use $\nu=\sqrt{m^2-g^2}$ and remember that we assumed $m>0$ and $g>0$ everywhere.} $c_1^{(\pm)}>0$ and $c_2^{(\pm)}>0$. Note also that $c_1^{(+)}>c_1^{(-)}$. We define the critical value for $\mu$ as:
\begin{equation}
\mu_c^{(\pm)}\equiv g M\left[c_1^{(\pm)}\right]^{\frac{1}{2\nu}}\,.
\end{equation}
For $f<0$ and $\mu=\mu_c^{(\pm)}-\delta\mu$ with $0<\delta\mu\ll g M$ we can use the expansion \eqref{eq_app_fer_DC_M_exp_ratio} to solve for the energy of the lowest bound state:
\begin{equation}\label{eq_app_DC_M_result}
\omega\simeq-M\left\{1-\frac{2\delta\mu}{M}\frac{   \nu\,g  }{c_2^{(\pm)}}\left[c_1^{(\pm)}\right]^{1-\frac{1}{2\nu}}\right\}\,.
\end{equation}
\eqref{eq_app_DC_M_result} clearly shows that as we lower $\delta\mu$ to $0$ the energy eventually becomes $-M$, at which point we have a completely delocalized bound state solution.  This solution is sometimes referred to as a \emph{diving} state \cite{Greiner:1985ce}. As we discuss in subsection \ref{subsec_fer_negative}, this phenomenon implies that one unit of charge is screened in the vacuum for $\mu>\mu_c$. Note that $\mu_c^{(+)}>\mu_c^{(-)}$, as it could have been intuitively expected since the lowest energy mode, $n=0$, is absent from \eqref{eq_app_DC_M_q2} for $(\delta)=+$; therefore it takes a stronger perturbation for $(\delta)=+$ than for $(\delta)=-$ to make the lowest bound state join the negative continuum part of the spectrum.

\subsection{Massive Dirac-Coulomb equation for supercritical charge}\label{app_DC_problem_Sup_M}

In this section we solve for the diving states of the massive Dirac-Coulomb equation \eqref{eq_app_DC_M} in the presence of the gauge field
\begin{equation}\label{eq_app_fer_A0_super}
A_0=\begin{cases}\displaystyle
\frac{g}{r_0}& \text{for }r<r_0\\[0.7em]
\displaystyle\frac{g}{r} & \text{for }r\geq r_0\,,
\end{cases}
\end{equation} 
where $g>m>0$.  We will be interested in the limit $1/r_0\gg  M$ and $\nt=\sqrt{g^2-m^2}\ll 1$.

The equation for $\psi^{(\pm)}(t,r)=e^{-i\omega t}\psi_{<}^{(\pm)}(r)$ for $r<r_0$ reads
\begin{equation}
\left[r\left(\omega+\frac{g}{r_0}\right)\gamma^0+i\left(r\pd_r-\frac12\right)\gamma^1-m\pm i r\gamma^3 M\right]\psi^{(\pm)}_{<}(r)=0\,.
\end{equation}
In the limit of interest $-\omega\simeq M\ll g/r_0$, the above reduces to
\begin{equation}
\left[i\left(r\pd_r-\frac12\right)\gamma^1-m+r\frac{g}{r_0}\gamma^0\right]\psi^{(\pm)}_{<}(r)\simeq 0\,,
\end{equation}
whose solutions satisfying standard boundary conditions \eqref{eq_fer_standard_in} for $r\rightarrow 0$ are
\begin{equation}\label{eq_App_DCinside}
\psi_{<}^{(\pm)}(r)\propto\left(\begin{array}{c}
\displaystyle r J_{m+\frac{1}{2}}\left(\frac{g r}{r_0}\right)\\[0.9em]
\displaystyle r J_{m-\frac{1}{2}}\left(\frac{g r}{r_0}\right)
\end{array} \right)\,.
\end{equation} 
The important conclusion for us is that the ratio between the two components at $r=r_0$ is independent of $\omega$ and $M$ (cf. \eqref{fer_two_comp_def} for the notation):
\begin{equation}
\frac{\chi_{<}^{(\pm)}(r)}{\xi_{<}^{(\pm)}(r)}\simeq
\frac{J_{m+\frac{1}{2}}\left(g\right)}{J_{m-\frac{1}{2}}\left(g\right)}\equiv R_g\,.
\end{equation}
A different potential for $r<r_0$ might change the value of $R_g$, which would however remain approximately independent of $\omega$ and $M$. For $m=1$ we have $R_g=1/g-\cot (g)$.

The solution $\psi^{(\pm)}(t,r)=e^{-i\omega t}\psi_{>}^{(\pm)}(r)$ for $r>r_0$ is obtained as in appendix \ref{app_DC_problem_M_sub} and can be found by replacing $\nu\rightarrow i\nt$ in \eqref{eq_app_DC_M_U_pre} and \eqref{eq_app_DC_M_U}.  The boundary condition arises from the requirement of continuity at $r=r_0$, which leads to
\begin{equation}\label{eq_app_m_bc_sup}
\frac{\chi_{>}^{(\pm)}(r_0)}{\xi_{>}^{(\pm)}(r_0)}=\frac{\chi_{<}^{(\pm)}(r_0)}{\xi_{<}^{(\pm)}(r_0)}\simeq R_g\,.
\end{equation}
For\footnote{This requirement arises since the expansion \eqref{eq_fer_modes_super} holds for $\omega  r\ll 1$. } $\omega g\ll 1/r_0$, we can express the outer solutions $\chi_{>}^{(\pm)}(r_0)$ and $\xi_{>}^{(\pm)}(r_0)$ using the small $r$ mode expansion \eqref{eq_fer_modes_super} and write the boundary condition \eqref{eq_app_m_bc_sup} as
\begin{equation}\label{eq_app_m_bc2_sup}
\frac{\beta}{\alpha}=\frac{ 
g R_g-m-i \nt}{g-m R_g- i \nt R_g}r_0^{-i 2\nt }\,,
\end{equation}
which  in particular implies $|\alpha|=|\beta|$.
Proceeding as in the previous section, we find that the ratio $\beta^{(\pm)}/\alpha^{(\pm)}$ for the solution $\psi_{>}^{(\pm)}(r)$ at $\omega=-M$ reads
\begin{equation}
\frac{\beta^{(\pm)}}{\alpha^{(\pm)}}=\frac{ \left(m\pm i \nt \right) \Gamma (-2i \nt ) }{g \Gamma (2i \nt )}
(2M g)^{2i \nt }\,.
\end{equation}
We conclude that the condition to have a bound state with energy $\omega=-M$ reads
\begin{equation}\label{eq_app_DC_sup_M_condition}
(2M g r_0)^{2i \nt }=e^{2i \nt \eta^{(\pm)}}\,,
\end{equation}
where we defined the following real quantity for convenience
\begin{equation}
\begin{split}
\eta^{(\pm)}&=\frac{1}{2i\nt}\log\left[\frac{g\left(m- 
g R_g+i \nt\right)\Gamma (2i \nt )}{\left(m\pm i \nt \right)\left(g-m R_g- i \nt R_g\right) \Gamma (-2i \nt )}\right]\\[0.6em]
&=\frac{R_g+1}{2 m( R_g-1)}\mp\frac{1}{2m}-2 \gamma_E+O\left(\nt^2\right)\,,
\end{split}
\end{equation}
where $\gamma_E$ is the Euler-Mascheroni constant.
For us it is only relevant that $\eta^{(\pm)}$ does not depend on $\nt$ for $\nt \ll 1/2$.  The condition \eqref{eq_app_DC_sup_M_condition} can be conveniently written as
\begin{equation}\label{eq_app_DC_sup_M_result}
\nt\log (2M g r_0)=\nt \eta^{(\pm)}-\pi n\,,\qquad
n=1,2,\ldots\,.
\end{equation}
where we excluded $n\leq 0$ since the approximations leading to \eqref{eq_app_m_bc2_sup} break down for $M r_0\gtrsim O(1)$.\footnote{In practice, an additional solution for $M\simeq 1/r_0$, intuitively corresponding to $n=0$ in \eqref{eq_app_DC_sup_M_result}, may exist for different potentials at $r<r_0$, somewhat similarly to the negative double-trace deformation discussed in section \ref{subsec_fer_negative}.} The result \eqref{eq_app_DC_sup_M_result} agrees with the classic analysis by Pomeranchuk and Smorodinsky \cite{pomeranchuk1945energy}.

\subsection{Massless Dirac equation for supercritical charge}\label{app_DC_problem_Sup}

In this appendix we study the massless Dirac equation in the presence of a supercritical Coulomb potential \eqref{eq_app_fer_A0_super}:
 \begin{equation}\label{eq_app_DC_Massless}
\left[i\left(\slashed{\pd}-i\slashed{A}\right)-m\right]\psi(t,r)=0\,,
\end{equation}
where $\psi(t,r)=e^{-i\omega t}\psi(r)$.
The solution for $r<r_0$ coincides with \eqref{eq_App_DCinside} (at small frequencies). For $r>r_0$ the equation in Fourier space coincides with the Euclidean equation \eqref{eq_app_fer_prop_eq1} up to the replacement $\omega\rightarrow -i\omega$. Therefore, writing the spinor as in \eqref{eq_app_fer_psi12}, we obtain the following equation for $\psi_2$:
\begin{equation}\label{eq_app_DC_sup_eq2}
r^2\pd_r^2\psi_2+ \left[\frac{1}{4}-m^2+i r\omega+(g+ r \omega )\right]\psi_2=0\,,
\end{equation}
while $\psi_1$ is given by
\begin{equation}\label{eq_app_DC_sup_eq1}
\psi_1=\frac{1}{2 m}\left[\left(1-2 i g-2i r \omega \right)\psi_2- 2 r \pd_r \psi_2\right]\,.
\end{equation}

We first show that \eqref{eq_app_DC_Massless} admits infinitely many resonances in the negative energy continuum. To this aim, we use the fact that the solution with purely outgoing boundary conditions for $\Re(\omega)<0$ reads
\begin{equation}\label{eq_app_DC_psiB}
\psi(t,r)\propto  e^{-i\omega t}
\left(\begin{array}{c}
\frac{1}{2 m}W_{\frac{1}{2}+i g,i \nt}(-2 i r \omega )+\frac{1}{2} W_{-\frac{1}{2}+i g,i \nt}(-2 i r \omega )
\\
\frac{i }{2 m}W_{\frac{1}{2}+i g,i \nt}(-2 i r \omega )-\frac{i}{2} W_{-\frac{1}{2}+i g,i \nt}(-2 i r \omega )
\end{array}\right)\,\qquad
\text{for }r>r_0\,,
\end{equation}
which behaves as $\psi\sim e^{-i\omega t+i \omega r}$ for $|r\omega|\gg 1$.  From the expansion \eqref{eq_app_Wexp}, we find that the ratio between the small $\omega r$ modes \eqref{eq_fer_modes_super} corresponding to the solution \eqref{eq_app_DC_psiB} is given by
\begin{equation}
\frac{\beta}{\alpha}=(-2i \omega )^{2i \nt}\frac{i \Gamma \left(-2i\nt\right) \Gamma \left(1-i g+i \nt\right)}{m \Gamma \left(2i \nt\right) \Gamma \left(-i g-i\nt\right)}\,.
\end{equation}
Using this expression in the boundary condition \eqref{eq_app_m_bc2_sup} and working at leading order in $\nt\ll 1$, we obtain the following condition on the frequency $\omega$:
\begin{equation}\label{eq_app_res_cond}
\left(-2c\,\omega r_0 
e^{-i\tilde{\gamma}}\right)^{2i\nt}=1\,,
\end{equation}
where we defined 
\begin{equation}
\tilde{\gamma}=\frac{ \pi }{e^{2 \pi  m}-1}\,,
\end{equation}
and $c$ is an $O(1)$ positive number given by
\begin{equation}
c=\exp\left[
\frac{1+R_g}{2m (R_g-1)}+\frac{\psi (1+i m)+
\psi (1-i m)}{2}+ 2\gamma_E \right]\,.
\end{equation}
For $m>\frac{\log (2)}{2\pi}\approx 0.11 $,\footnote{This restriction applies in all the physical cases, since $m_0\geq 1/2 $ for $d\geq 3$. } we have $0<\tilde{\gamma}<\pi$ and \eqref{eq_app_res_cond} admits the following infinite family of solutions with $\Re(\omega)<0$:
\begin{equation}\label{eq_app_res_result}
\omega_n=-\frac{1}{2c r_0}e^{i\tilde{\gamma}-\pi n/\nt}\,,\qquad
n=1,2,\ldots\,,
\end{equation}
where the restriction on $n$ arises from the requirement $|\omega_n r_0|\ll 1$, which is needed in order to be able to use the expansion \eqref{eq_fer_modes_super} at $r=r_0$.  The solutions \eqref{eq_app_res_result} are resonances with $\Re(\omega_n)\sim \Im(\omega_n)=-\exp\left(-\pi n/\nt\right)/r_0$ and correspond to poles of the retarded Green's function analytically continued to the second sheet. 
Note that by increasing $m$ the imaginary part becomes smaller. For $m=1/2$ and $m=1$, as appropriate for the $\ell=0$ modes in $d=3$ and $d=4$, we find a numerically small imaginary part
\begin{equation}\label{eq_app_res_result_num}
\Im (\omega_n)\simeq\Re(\omega_n)\times\begin{cases}
0.14 &\text{for }m=\frac12 \\
 0.006 &\text{for }m=1\,.
\end{cases}
\end{equation}
The result \eqref{eq_app_res_result} was previously obtained in \cite{shytov2007atomic}.

We now find the scattering wave-functions. The most general solution of \eqref{eq_app_DC_sup_eq2} is
\begin{equation}\label{eq_app_DC_psi2}
\begin{split}
\psi_2&=\beta \frac{ 2i  \left( g-\nt\right)}{ g+ i m-\nt }
 (-2i \omega )^{-\frac{1}{2}-i \nt} M_{-\frac{1}{2}+i g,i \nt}(-2 i r \omega )\\
 &+
\alpha \frac{2 i  m}{ g+ i m-\nt}(2 i \omega )^{-\frac{1}{2}+i \nt} M_{\frac{1}{2}-i g,-i\nt }(2 i r \omega )\,,
\end{split}
\end{equation}
where $M_{x,y}(z)$ is the Whittaker $M$ function and $\psi_1$ is found from \eqref{eq_app_DC_sup_eq1}:
\begin{equation}\label{eq_app_DC_psi1}
\begin{split}
\psi_1 &=\beta\frac{2  m  }{ g+ i m-\nt}
(-2i \omega )^{-\frac{1}{2}-i \nt} M_{\frac{1}{2}+i g,i \nt}(-2 i r \omega )
\\
 &+
\alpha \frac{  2g-2\nt}{ g+ i m-\nt}(2 i \omega )^{-\frac{1}{2}+i \nt}
M_{-\frac{1}{2}-i g,-i\nt} (2 i r \omega )\,.
\end{split}
\end{equation}
The coefficients $\alpha$ and $\beta$ are chosen such that they precisely coincide with those in \eqref{eq_fer_modes_super}, as can be seen using
\begin{equation}\label{eq_app_Mexp}
M_{x,y}(z)\overset{z\rightarrow 0}{\sim}z^{\frac{1}{2}+y}\,.
\end{equation}
The ratio $\beta/\alpha $ is thus determined by the boundary condition \eqref{eq_app_m_bc2_sup}. In the following we determine  the absolute value $|\alpha|=|\beta|$ as well by demanding the orthonormality condition
\begin{equation}\label{eq_app_ortho}
\int dr\sqrt{g g^{rr}}\psi^\dagger_\omega(r)\psi_{\omega'}(r)=(2\pi)\delta(\omega-\omega')\,,
\end{equation}
where $\psi_\omega$ denotes the wave-function at frequency $\omega$.

To compute the integral \eqref{eq_app_ortho}, we use that the Dirac equation \eqref{eq_app_DC_Massless} implies
\begin{equation}\label{eq_app_DC_property}
(\omega-\omega')\psi^\dagger_{\omega }(r)\psi_{\omega'}(r)=\frac{i}{\sqrt{g g^{rr}}}
\pd_r\left[\sqrt{g g^{rr}}\,\bar{\psi}_{\omega}(r)\gamma^1 \psi_{\omega'}(r)\right]\,.
\end{equation}
\eqref{eq_app_DC_property} and the continuity of the wave-functions let us express the integral \eqref{eq_app_ortho} as a boundary term
\begin{equation}\label{eq_app_DC_norm_limit}
\begin{split}
\int dr\sqrt{g g^{rr}}\psi^\dagger_{\omega}(r)\psi_{\omega'}(r)&=i\lim_{r\rightarrow\infty}
\frac{\bar{\psi}_{\omega}(r)\gamma^1 \psi_{\omega'}(r)}{r(\omega-\omega')}\\
&=\lim_{r\rightarrow\infty}
\frac{\psi^{\dagger}_1(r)\psi'_1(r)-\psi^\dagger_2(r)\psi_2'(r) }{2ir(\omega'-\omega)}\,,
\,
\end{split}
\end{equation}
where we used \eqref{eq_app_fer_psi12} and we have to use $\omega'$ in the primed spinors. To evaluate the limit, we use the following expansion of the Whittaker function
\begin{equation}
M_{x,y}(z)\overset{z\rightarrow\infty}{\sim}\frac{e^{z/2} z^{-x} \Gamma (2 y+1)}{\Gamma \left(1/2-x+y\right)}\left[1+O\left(\frac{1}{z}\right)\right]+\frac{e^{-\frac{z}{2}} (-1)^{x-y+\frac{3}{2}} z^x \Gamma (2 y+1)}{\Gamma \left(1/2+x+y\right)}\left[1+O\left(\frac{1}{z}\right)\right]\,,
\end{equation}
from which we obtain
\begin{align}
\psi_1(r)&\overset{r\rightarrow\infty}{\sim}
\alpha \, r^{1/2+i g} e^{i r \omega } (2 i \omega )^{i g-i \nt}A(\omega)\left[1+O\left(\frac{1}{r}\right)\right]\,,\\
\psi_2(r)&\overset{r\rightarrow\infty}{\sim} i
\beta \, r^{1/2-i g} e^{-i r \omega } (-2i \omega )^{- i g+i \nt}B(\omega)\left[1+O\left(\frac{1}{r}\right)\right]\,,
\end{align}
where
\es{}{
A(\omega)&=\frac{\beta}{\alpha}
\frac{2   m \Gamma \left(1+2i \nt\right)}{  \Gamma \left(1+i g+i\nt\right) \left( g+ i m-\nt\right)}
-i\frac{ 2 (2i \omega )^{2i\nt} \Gamma \left(1-2i \nt\right)}{\Gamma \left(i g-i\nt\right) \left( g+i m-\nt\right)}\,,\\
B(\omega)&=
\frac{\alpha}{\beta}\frac{2 m\Gamma \left(1-2i\nt\right)}{\Gamma \left(1-i g-i\nt\right) \left( g+i m-\nt\right)}+i
\frac{ 2  (-2 i \omega )^{-2i\nt}   \Gamma \left(1+2i \nt\right) }{  \Gamma \left(i\nt-i g\right) \left( g+ i m-\nt\right)}\,.
}
Recall that the ratio $\beta/\alpha$ is an $\omega$-independent phase given by \eqref{eq_app_m_bc2_sup}. We notice that $A(\omega)$ and $B(\omega)$ are log-periodic functions of $\omega$
\begin{equation}\label{eq_app_DC_AB_prop}
A(\omega)=A(\omega e^{\pm \pi/\nt})\,,\qquad
B(\omega)=B(\omega e^{\pm \pi/\nt})\,,
\end{equation}
and that $|A(\omega)|^2=|B(\omega)|^2$.
We therefore find
\begin{multline}
\lim_{r\rightarrow\infty}
\frac{\psi^{\dagger}_1(r)\psi'_1(r)-\psi^\dagger_2(r)\psi_2'(r) }{2ir(\omega'-\omega)}\\
=\lim_{r\rightarrow\infty}
e^{\text{sgn}(\omega)\pi  \left(\nt- g\right)}
\frac{|A(\omega)|^2|\alpha|^2 e^{ir(\omega'-\omega)}-|B(\omega)|^2|\beta|^2 e^{-ir(\omega'-\omega)}}{2i(\omega'-\omega)}\,,
\end{multline}
where we kept only the leading terms in the expansion for $\omega\rightarrow\omega'$, as the limit clearly vanishes (in the distributional sense) when $\omega\neq \omega'$. Finally, using $|A(\omega)|^2=|B(\omega)|^2$ and $|\alpha|^2=|\beta|^2$, we get
\begin{equation}
\begin{split}
\lim_{r\rightarrow\infty}
\frac{\psi^{\dagger}_1(r)\psi'_1(r)-\psi^\dagger_2(r)\psi_2'(r) }{2ir(\omega'-\omega)}&=
e^{\text{sgn}(\omega)\pi  \left(\nt- g\right)}|A(\omega)|^2|\beta|^2
\lim_{r\rightarrow\infty}\frac{\sin{[r(\omega'-\omega)]}}{(\omega'-\omega)}\\
&=
 e^{\text{sgn}(\omega)\pi  \left(\nt- g\right)}|A(\omega)|^2|\beta|^2\pi\delta(\omega-\omega')\,,
\end{split}
\end{equation}
and, from \eqref{eq_app_DC_norm_limit},
\begin{equation}
\int dr\sqrt{g g^{rr}}\psi^\dagger_{\omega}(r)\psi_{\omega'}(r)=\pi
e^{\text{sgn}(\omega)\pi  \left(\nt- g\right)}|A(\omega)|^2|\beta|^2\delta(\omega-\omega')\,.
\end{equation}
The normalization condition \eqref{eq_app_ortho} is thus satisfied by setting
\begin{equation}\label{eq_app_DC_norm}
|\beta|^2=\frac{2 e^{-\text{sgn}(\omega)\pi  \left(\nt- g\right)}}{|A(\omega)|^2}\,.
\end{equation}

Our result has an important consequence. Plugging \eqref{eq_app_DC_norm} into the solutions \eqref{eq_app_DC_psi2} and \eqref{eq_app_DC_psi1} and recalling the property \eqref{eq_app_DC_AB_prop}, we infer
\begin{equation}\label{eq_app_DC_period}
\psi_{\omega}(r e^{-\pi n/\nt})\
\simeq e^{-\pi n/(2\nt)}\psi_{\omega e^{-\pi n/\nt}}(r)\,, \qquad
n\in \mathds{Z}\,.
\end{equation}
In practice, in our calculations we assumed $\omega r_0\ll 1$, and thus \eqref{eq_app_DC_period} holds only as long as $\omega r_0,\,\omega  e^{-\pi n/\nt}r_0\ll 1$.  As we explain in section \ref{subsec_fermions_supercritical}, \eqref{eq_app_DC_period} has important implications for the screening of supercritical lines when $\nt\ll 1$.

\section{Vector boson instability}\label{app_vectors}

In this section, we study perturbative instabilities for the vector bosons in a $SU(2)$ gauge theory, in the presence of a Wilson-line in a $(2s+1)$-dimensional representation.  The result was originally derived in \cite{Mandula:1976uh,Mandula:1976xf}.

According to the discussion in section \ref{secNonAbelian}, we study the fluctuations for the gauge field in the presence a background Coulomb potential in the third direction:
\begin{equation}\label{eq_appD_Coulomb}
A^3_0=\frac{g_{YM}^2 s}{4\pi r}\,.
\end{equation}
The equations of motion deriving from the action \eqref{eq_actionSU2} are
\begin{equation}\label{eq_appD_EOM1}
\nabla^\mu F_{\mu\nu}^a+\varepsilon_{abc}A^b_\mu g^{\mu\sigma} F_{\sigma\nu}^c=0\,,
\end{equation}
where as in sections \ref{sec_ScalarQED} and \ref{sec_Fer} we work on AdS$_2\times S^2$, with metric
\begin{equation}
ds^2=\frac{dt^2-dr^2}{r^2}-d \Omega_2^2\,.
\end{equation}
We are interested in the equation for the fluctuations $A^1_\mu$ and $A^2_\mu$. It is convenient to define a charged $W$-boson as
\begin{equation}
W_\mu= A^1_\mu+i A^2_\mu\,.
\end{equation}
The linearization of the equation of motion \eqref{eq_appD_EOM1} reads
\begin{equation}\label{eq_appD_EOM2}
D^\mu \left(D_\mu W_\nu-D_\nu W_\mu\right)-iW^\mu F_{\mu\nu}=0\,,
\end{equation}
where we defined an Abelian covariant derivative as
\begin{equation}
D_\mu W_\nu=(\nabla_\mu +i A^3_\mu) W_\nu\,,\qquad
D_\mu D_\nu W_\rho=(\nabla_\mu +i A^3_\mu) D_\nu W_\rho\,,
\end{equation}
and $F_{\mu\nu}=\pd_\mu A^3_\nu-\pd_\nu A^3_\mu$ is the Abelian field strength associated with the Coulomb potential \eqref{eq_appD_Coulomb}. \eqref{eq_appD_EOM2} is invariant under the linearized gauge transformations\footnote{To check this, one needs to use $\nabla_\mu F^{\mu\nu}=0$.}
\begin{equation}
\delta W_\mu= D_\mu\lambda-\lambda_3 W_\mu\,,\qquad
\delta A^3_\mu=-\pd_\mu\lambda_3\,,
\end{equation}
where $\lambda=\lambda_1+i\lambda_2$ and $\lambda_3$ are infinitesimal parameters. $\lam$ carries the same charge as $W$, while $\lam_3$ is neutral. 

To proceed, we decompose $W_\mu=(W_a,W_i)$, where $a,b,c,\ldots$ denote AdS$_2$ indices and $i,j,k,\ldots$ the $S^2$ indices. \eqref{eq_appD_EOM2} then explicitly reads
\begin{align}\label{eq_appD_EOM3a}
&D_a D^a W^i+\nabla_k\nabla^k W^i-\mathcal{R}^i_{\;j}W^j-\nabla^i\left(D_a W^a+\nabla_j W^j\right)=0\,,\\
& D_b (D^b W^a-D^a W^b) +\nabla_i\nabla^i W^a-iW^b F_b\,^{a}
-D^a\nabla_i W^i=0\,,
\label{eq_appD_EOM3b}
\end{align}
where $\mathcal{R}^\mu_{\;\nu}$ is the Ricci tensor. In the following we choose the gauge
\begin{equation}\label{eq_appD_gauge1}
\nabla_i W^i=0\,,
\end{equation}
which leaves a residual gauge freedom $\delta W_a=D_a\lambda$ with $\nabla_i\lambda=0$. Using \eqref{eq_appD_gauge1} in \eqref{eq_appD_EOM3a} we obtain $\nabla_i(D_a W^a)=0$. We thus can use the residual freedom to further impose
\begin{equation}\label{eq_appD_gauge2}
D_a W^a=0\,.
\end{equation}
The conditions \eqref{eq_appD_gauge1} and \eqref{eq_appD_gauge2} ensure that \eqref{eq_appD_EOM3a} and \eqref{eq_appD_EOM3b} decouple. Note that \eqref{eq_appD_gauge2} still leaves a residual gauge freedom of the form
\begin{equation}\label{eq_appD_gauge_residual}
\delta W_a=D_a \lambda\quad
\text{for }\lambda \text{ such\ that }
D_a D^a \lambda=0\,.
\end{equation}
This will be important in what follows.

The analysis of the first equation \eqref{eq_appD_EOM3a} is straightforward.  The condition \eqref{eq_appD_gauge1} is compatible with setting
\begin{equation}\label{eq_appD_Wi}
W_i= \sqrt{g_{S^2}}\, \varepsilon_{ij}\nabla^j W^T\,,
\end{equation}
where $W^T$ is a scalar. Then the $W^T$ equation reduces to the Klein-Gordon equation in a Coulomb field
\begin{equation}
(D_aD^a+\nabla_i\nabla^i)W^T=0\,.
\end{equation}
The analysis of section \ref{subsec_2DCFTs} lets us conclude that the defect scaling dimensions are given by
\begin{equation}\label{eq_appD_result1}
\Delta_{\ell}=\frac12\pm\frac12\sqrt{1+4\ell(\ell+1)-\frac{g_{YM}^4 s^2}{4\pi^2}}\qquad\text{for }\ell=1,2,\ldots\,,
\end{equation}
where the $\ell=0$ mode is excluded since it does not contribute to \eqref{eq_appD_Wi}.

To analyze \eqref{eq_appD_EOM3b} we set $W_a(t,r)=e^{-i\omega t} w_a(r)$ and solve the condition \eqref{eq_appD_gauge2} in terms of the components $w_a=(w_0,w_r)$
\begin{equation}
w_0=i\frac{w_r}{\omega-A^3_0}\,.
\end{equation}
Decomposing $W_r$ into spherical harmonics
\begin{equation}
w_r=e^{-i\omega t}\sum_{\ell,m}Y_{\ell,m}(\hat{n})w_{\ell,m}(r)\,,
\end{equation}
we obtain
\begin{equation}
r^2 \pd_r^2 w_{\ell,m}+\frac{2 A^3_0 }{A^3_0 - \omega }r\pd_rw_{\ell,m}+
\left[r^2\left(\omega-A^3_0\right)^2-\ell(\ell+1)\right]w_{\ell,m}=0\,.
\end{equation}
Looking for solutions in the form $w_{\ell,m}\sim r^{\Delta_{\ell}-1}$, we  again find the same $\Delta_\ell$ as in \eqref{eq_appD_result1}.
In this polarization (in the AdS$_2$ directions) the $\ell =0$ mode solution is excluded because it is equivalent to a shift of the form \eqref{eq_appD_gauge_residual}.

From the result \eqref{eq_appD_result1} we conclude that the first instability is found for the $\ell=1$ modes at
\begin{equation}
s=\frac{6\pi}{g_{YM}^2}\,.
\end{equation}

\section{Details on Wilson lines in large \texorpdfstring{$N_f$}{Nf} \texorpdfstring{QED$_3$}{QED3}}\label{app_largeN}

\subsection{The saddle-point equations}

In this section we provide details on the calculation of the gauge field sourced by a Wilson line in large $N_f$ QED$_3$, with and without a Chern-Simons term.  For the sake of generality we consider right away the action with a Chern-Simons term \eqref{eq_QED3k}. We thus want to extremize the following Euclidean effective action for the gauge field
\begin{equation}\label{eq_app_QED3k}
S_q[A]=-2N_f\text{Tr}\left[\log\left(\slashed{\pd}-i\slashed{A}\right)\right]-i\frac{k}{4\pi}\int d^3x\varepsilon^{\mu\nu\rho}A_\mu\pd_\nu A_\rho+i q \int d\tau A_0\,.
\end{equation}

To proceed we consider the ansatz
\begin{equation}\label{eq_app_QED3k_Ansatz}
F_{\tau r}=i\frac{E}{r^2}\qquad\text{and}\qquad A_\theta=b=\text{const}.\,.
\end{equation}
The ansatz \eqref{eq_app_QED3k_Ansatz} is dictated by conformal invariance; for $k=0$ parity demands $b=0$. It is further convenient to exploit Weyl invariance to map the theory to AdS$_2\times S^1$.  The one-loop fermion determinant naturally decomposes into a sum over the contributions of the AdS$_2$ KK modes, labeled by the angular momentum $j\in\frac12+\mathds{Z}$, as in section \ref{subsec_fermions_general_con}:
\begin{equation}\label{eq_app_dets}
\text{Tr}\left[\log\left(\slashed{\pd}-i\slashed{A}\right)\right]_{
\text{AdS}_2\times S^1}=\text{Vol}(\text{AdS}_2)\sum_{j\in\frac12+\mathds{Z}}\Sigma_{j}(E,b)\,.
\end{equation}
We defined $\Sigma_{j}$ to be proportional to the one-loop determinant for the AdS$_2$ Dirac operator in a constant electric field
\begin{equation}\label{eq_app_sigma}
\text{Vol}(\text{AdS}_2)\Sigma_{j}(E,b)=
\text{Tr}\left[\log\left(\slashed{\pd}-i\slashed{A}+\tilde{m}_{j}\right)\right]_{\text{AdS}_2}\,,
\end{equation}
where the KK masses receive a contribution from the holonomy
\begin{equation}\label{eq_app_mKK}
\tilde{m}_{j}=j+ b\,.
\end{equation} 
We factored out explicitly the AdS$_2$ volume in \eqref{eq_app_dets} for future convenience. We will discuss how to explicitly evaluate  $\Sigma_{\ell}$ and the infinite sum in \eqref{eq_app_dets} in the next section.

To conveniently write the last two terms in \eqref{eq_app_QED3k} we parametrize AdS$_2$ using global coordinates
\begin{equation}
ds^2_{\text{AdS}_2}=d\sigma^2+\sinh^2\sigma d\phi^2\,,
\end{equation}
and we choose the following gauge for the AdS$_2$ gauge field
\begin{equation}\label{eq_app_ansatzAds2}
A_\sigma=0\,,\quad A_\phi=i(\cosh\sigma-1)E
\quad\implies\quad F_{\sigma\phi}=iE\sinh \sigma \,.
\end{equation}
It can be checked that \eqref{eq_app_ansatzAds2} is indeed equivalent to the electric field in \eqref{eq_app_QED3k_Ansatz}. Then using \eqref{eq_app_ansatzAds2} we obtain
\begin{equation}\label{eq_app_QED3k_2terms}
\begin{split}
-i\frac{k}{4\pi}\int d^3x\varepsilon^{\mu\nu\rho}A_\mu\pd_\nu A_\rho+i q \int dx^\mu A_\mu &=-i \frac{k}{2\pi}\int_{\text{AdS}_2\times S^1
\hspace*{-2.3em}}d^3x\, F_{\sigma\phi}b
+i q \int_{\pd \text{AdS}_2 \hspace*{-1.5em}} d\phi\, A_\phi \\[1em]
&=\left(k b E-q E\right)\text{Vol}(\text{AdS}_2)\,.
\end{split}
\end{equation}
We used that the AdS$_2$ volume in global coordinates is given by
\begin{equation}\label{eq_app_VOL}
\text{Vol}(\text{AdS}_2)=2\pi\int_0^{\sigma_c}\sinh\sigma=2\pi\left[\frac{e^{\sigma_c}}{2}-1+O\left(e^{-\sigma_c}\right)\right]\,,
\end{equation}
where we introduced a (large) radial cutoff $\sigma_c$.  As explained in \cite{Cuomo:2021kfm},  since $\text{Vol}(\pd\text{AdS}_2)=\pi e^{\sigma_c}+O\left(e^{-\sigma_c}\right)$, the terms proportional to $e^{\sigma_c}$ in the action can be absorbed into a defect cosmological constant counterterm and thus we can replace $\text{Vol}(\text{AdS}_2)$ with its well known regulated expression $\text{Vol}(\text{AdS}_2)\vert_{\text{reg}}=-2\pi$ \cite{Casini:2011kv,Klebanov:2011uf}.  

Overall, from \eqref{eq_app_dets} and \eqref{eq_app_QED3k_2terms} we obtain
\begin{equation}
\frac{S_q[A]}{\text{Vol}(\text{AdS}_2)}=-2N_f\sum_{j\in\frac12+\mathds{Z}}\Sigma_{j}(E,b)+k\, b E-q\, E\,,
\end{equation}
from which we obtain the saddle-point equations
\begin{align}\label{eq_app_saddle_1}
&2  \sum_{j\in\frac12+\mathds{Z}}\frac{\pd \Sigma_{j}(E,b)}{\pd E}=\frac{k}{N_f}b -\frac{q}{N_f}\,,\\
&2  \sum_{j\in\frac12+\mathds{Z}}\frac{\pd \Sigma_{j}(E,b)}{\pd b}=\frac{k}{N_f} E\,.
\label{eq_app_saddle_2}
\end{align}

\subsection{The fluctuation determinant via zeta function regularization}

In this appendix we explain how to evaluate the fluctuation determinant \eqref{eq_app_sigma}, as well as its derivatives in  \eqref{eq_app_saddle_1} and \eqref{eq_app_saddle_2}. 

We start by commenting on two important properties of the sum in \eqref{eq_app_saddle_1} as a function of the holonomy  $b$.  From the definitions \eqref{eq_app_sigma} and \eqref{eq_app_mKK} it follows that
\begin{equation}
\Sigma_{j}(E,b\pm n)=\Sigma_{j\pm n}(E,b)\quad
\text{for }n\in\mathds{Z}\,.
\end{equation}
This implies that the sum
\begin{equation}
\sum_{j\in\frac12+\mathds{Z}}\Sigma_{j}(E,b)\,,
\end{equation}
is a periodic function of $b$ with unit period, in agreement with the discussion on integral holonomies in section  \ref{Sec3dCFTs}. Additionally,  we will soon see that $\Sigma(E,b)$ is an even function of both $E$ and $b$. This implies in particular
\begin{equation}\label{eq_app_check}
\sum_{j\in\frac12+\mathds{Z}}\left.\frac{\pd \Sigma_{j}(E,b)}{\pd b}\right\vert_{b=0}=\sum_{j\in\frac12+\mathds{Z}}\left.\frac{\pd \Sigma_{j}(E,b)}{\pd E}\right\vert_{E=0}=0\,.
\end{equation}
This ensures that \eqref{eq_app_saddle_2} is solved by $b=0$ for $k=0$, as expected from parity invariance. In general,  given a solution $(E,b)$ of \eqref{eq_app_saddle_1} and \eqref{eq_app_saddle_2} for certain values $(k,q)$,  this implies that $(-E,b)$ is a solution for $(-k,-q)$,  $(E,-b)$ is a solution for $(-k,q)$ and  $(-E,-b)$ solves the equations for $(k,-q)$.

To proceed, we express the determinant of the Dirac operator in a constant electric field \cite{Comtet:1984mm,Pioline:2005pf} as an integral,
\begin{equation}\label{eq_app_Sigma_mu}
\Sigma_j(E,b)=\int_{-\infty}^\infty d\nu\,\mu_{E}(\nu)\log\left(\nu^2-E^2+\tilde{m}_j^2\right)\,,
\end{equation}
where $\mu_E(\nu)$ is the appropriate hyperbolic spectral density\footnote{In \cite{Comtet:1984mm} the determinant is given for an Euclidean electric field, in which case the spectral density receives an additional discrete contribution; it can be checked that upon Wick rotating the electric field to be real in Lorentzian signature, as in \eqref{eq_app_QED3k_Ansatz}, the only effect of the discrete contribution is to introduce the principal value prescription in \eqref{eq_app_muE}.}
\begin{equation}\label{eq_app_muE}
\mu_E(\nu)=P\left(\frac{\nu\sinh(2\pi\nu)}{4\pi\left[\cosh(2\pi\nu)-\cosh(2\pi E)\right]}\right)\,;
\end{equation}
the prefix P specifies that the pole at $\nu=E$ in the spectral density should be integrated according to the principal value prescription. \eqref{eq_app_Sigma_mu} makes it clear that the effective action develops an imaginary part for supercritical electric fields, i.e. when $E^2>\tilde{m}_j^2$ for some $j$.  Below we will focus on subcritical fields, for which \eqref{eq_app_Sigma_mu} is real.

Even upon taking derivatives of \eqref{eq_app_Sigma_mu} with respect to $E$ and $b$, both the integration over $\nu$ and the sums in \eqref{eq_app_saddle_1} and \eqref{eq_app_saddle_2} do not converge. We therefore need to regulate the calculation. We decided to use zeta function regularization.  This amounts to rewriting the fluctuation determinant in \eqref{eq_app_dets} as
\begin{equation}\label{eq_app_detS}
\frac{\text{Tr}\left[\left(\slashed{\pd}-i\slashed{A}\right)\right]_{
\text{AdS}_2\times S^1}}{\text{Vol}(\text{AdS}_2)}=-\lim_{s\rightarrow 0}\frac{d}{ds}\sum_{j\in\frac12+\mathds{Z}}\Sigma_{j}^{(s)}(E,b)\,,
\end{equation}
where
\begin{equation}\label{eq_app_SigmaS}
\Sigma_j^{(s)}(E,b)=\int_{-\infty}^\infty d\nu\,\frac{\mu_{E}(\nu)}{\left(\nu^2-E^2+\tilde{m}_j^2\right)^{s}}\,.
\end{equation}
The idea then is to compute the sum on the right hand-side of \eqref{eq_app_detS} for sufficiently large $s$, so that both the integration and the sum converge, and then analytically continue the result. 

In practice \eqref{eq_app_detS} and its derivatives can only be evaluated numerically.  We sketch below the strategy to evaluate the fluctuation determinant itself; the derivatives are computed analogously.

First,  we deal with the integral over $\nu$. To this aim, we notice that the spectral density in \eqref{eq_app_muE} admits the following asymptotic expansion for large $\nu$
\begin{equation}
\mu_E(\nu)\sim \frac{|\nu|}{4\pi}+O\left(e^{-2\pi|\nu|}\right)\quad\text{for }|\nu|\rightarrow\infty\,.
\end{equation}
We therefore define a subtracted spectral density
\begin{equation}
\tilde{\mu}_E(\nu)=\mu_E(\nu)-\frac{|\nu|}{4\pi}\,,
\end{equation}
which decays exponentially for $\nu\rightarrow\infty$. We then separate \eqref{eq_app_SigmaS} into two contributions
\begin{equation}
\Sigma_j^{(s)}(E,b)=\Sigma_j^{(s,1)}(E,b)+\Sigma_j^{(s,2)}(E,b)\,,
\end{equation}
where 
\begin{align}\label{eq_app_Sigma1}
\Sigma_j^{(s,1)}(E,b)&=2\int_0^\infty d\nu\frac{\nu}{4\pi \left(\nu^2-E^2+\tilde{m}_j^2\right)^{s}}=-\frac{\left(\tilde{m}_j^2-E^2\right)^{1-s}}{4 \pi(1- s)}\,, \\ \label{eq_app_Sigma2}
\Sigma_j^{(s,2)}(E,b)&=2\int_0^\infty d\nu
\frac{\tilde{\mu}_{E}(\nu)}{\left(\nu^2-E^2+\tilde{m}_j^2\right)^{s}}\,.
\end{align}
We evaluated the integral in \eqref{eq_app_Sigma1} by analytically continuing the result for $s>1$.  Even though we were not able to perform the integration in \eqref{eq_app_Sigma2} in closed form, the integral converges for arbitrary $s$ since $\tilde{\mu}_{E}(\nu)=O( e^{-2\pi|\nu|})$ for $\nu\rightarrow \infty$.  

Next we isolate the divergent contributions to the sum from $\Sigma_j^{(s,1)}(E,b)$ and $\Sigma_j^{(s,2)}(E,b)$. Grouping terms with opposite spin we have
\es{eq_app_Sigma1Div}{
\left[\Sigma_{j}^{(s,1)}(E,b)+\Sigma_{-j}^{(s,1)}(E,b)\right]_{div}&=
j^{-2 s} \left[\frac{j^2}{2 \pi  (s-1)}+\frac{-(1-2 s) b^2-E^2}{2 \pi }\right]\,,\\
\left[\Sigma_{j}^{(s,2)}(E,b)+\Sigma_{-j}^{(s,2)}(E,b)\right]_{div}&=
\frac{4}{j^{2s}}\int_0^{\infty}d\nu\,\tilde{\mu}_E(\nu) \,,
}
where the last integral can be evaluated numerically for arbitrary values of $E$.  All we have to do then is to write the sum we are interested in as
\begin{equation}\label{eq_app_final}
\begin{split}
\sum_{j\in\frac12+\mathds{Z}}\Sigma_{j}^{(s)}(E,b)&=
\sum_{j>0}\left\{
\left[\Sigma_{j}^{(s,1)}(E,b)+\Sigma_{-j}^{(s,1)}(E,b)\right]_{div}
+\left[\Sigma_{j}^{(s,2)}(E,b)+\Sigma_{-j}^{(s,2)}(E,b)\right]_{div}
\right\}\\
&+\sum_{j>0}\left\{\Sigma_{j}^{(s,1)}(E,b)+\Sigma_{-j}^{(s,1)}(E,b)
-\left[\Sigma_{j}^{(s,1)}(E,b)+\Sigma_{-j}^{(s,1)}(E,b)\right]_{div}\right\}\\
&+\sum_{j>0}\left\{\Sigma_{j}^{(s,2)}(E,b)+\Sigma_{-j}^{(s,2)}(E,b)-\left[\Sigma_{j}^{(s,2)}(E,b)+\Sigma_{-j}^{(s,2)}(E,b)\right]_{div}\right\}\,.
\end{split}
\end{equation}
The sum in the first line can be evaluated analytically in terms of generalized zeta functions.  The sums in the last two lines instead converge for $s\rightarrow 0$.  Therefore their contributions to \eqref{eq_app_detS} can be straightforwardly evaluated numerically.  We do not report further details.

Finally we comment that for $k=b=0$ it is also simple to compute (numerically) the determinant of the Dirac-Coulomb operator \eqref{eq_app_dets} in dimensional regularization.  As a crosscheck, we verified that the results of zeta function regularization and dimensional regularization agree in the overlapping regime. We also checked that the numerical result is periodic as a function of $b$ and satisfies \eqref{eq_app_check}.

\subsection{Solving the saddle-point equations}

We proved in the previous section that \eqref{eq_app_saddle_2} is always satisfied for $k=b=0$. In this case, by computing numerically the left-hand side of \eqref{eq_app_saddle_1} for $0<E<1/2$, we obtain the curve $q(E)/N_f$; figure \ref{fig:plot_QED3} is obtained plotting $\{E,q(E)/N_f\}$. By studying the limit of $q(E)$ for $E\rightarrow 1/2$ from below we find the result \eqref{eq_QED3_result}.

For nonzero $k$ it is harder to solve \eqref{eq_app_saddle_1} and \eqref{eq_app_saddle_2}. We instead compute the sums on the left-hand side of \eqref{eq_app_saddle_1} and \eqref{eq_app_saddle_2} as a function of $E$ and $b$ and use the result to read off the corresponding values of $k$ and $q$. Note that different values of $E$ and $b$ may correspond to the same pair $(k,q)$,  i.e. multiple saddle-points may exist for the same value of the charge and the Chern-Simons level. This is indeed what we find.

As explained in section  \ref{subsec_U1k_Nf},  in order to decide what the stable Wilson lines are in the theory at hand, we carve out the region $R$ of  solutions $(k_R,q_R)$ to the equations \eqref{eq_app_saddle_1} and \eqref{eq_app_saddle_2} for
\begin{equation}
-1/2<b<1/2\quad\cap\quad |E|<1/2-|b|\,,
\end{equation}
where the latter condition comes from the requirement of stability.
By symmetry, it is enough to focus on $E>0$.  To determine the boundary of the region it turns out to be convenient to span the $(E,b)$ plane using the curves defined by
\begin{equation}
c_{\pm}(x)=\left\{(E,b)\;\text{such\ that}\;\left(\pm1/2+b\right)^2-E^2=x^2\;\cap\; \mp b>0\right\}\,.
\end{equation}
In some sense, the curve $c_{\pm}(x)$ specifies all points in the $(E,b)$ plane which are equidistant from an instability of the $j=\pm 1/2$ mode.  The restriction on the sign of $b$ ensures that the two curves do not intersect.

\begin{figure}[t]
   \centering
		\subcaptionbox{  \label{fig_QED3kapp1a}}
		{\includegraphics[scale=0.33]{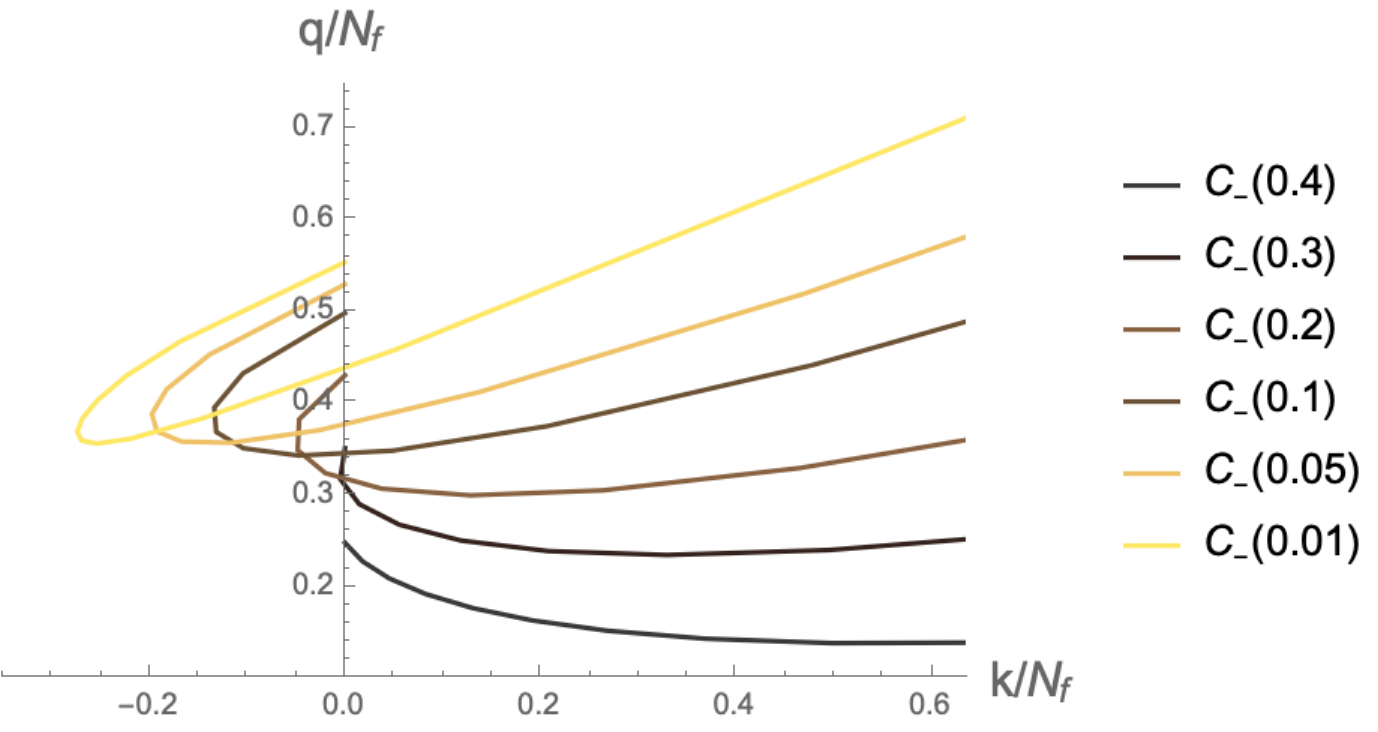}}
	\quad 
		\subcaptionbox{ \label{fig_QED3kapp1b}}
		{\includegraphics[scale=0.36]{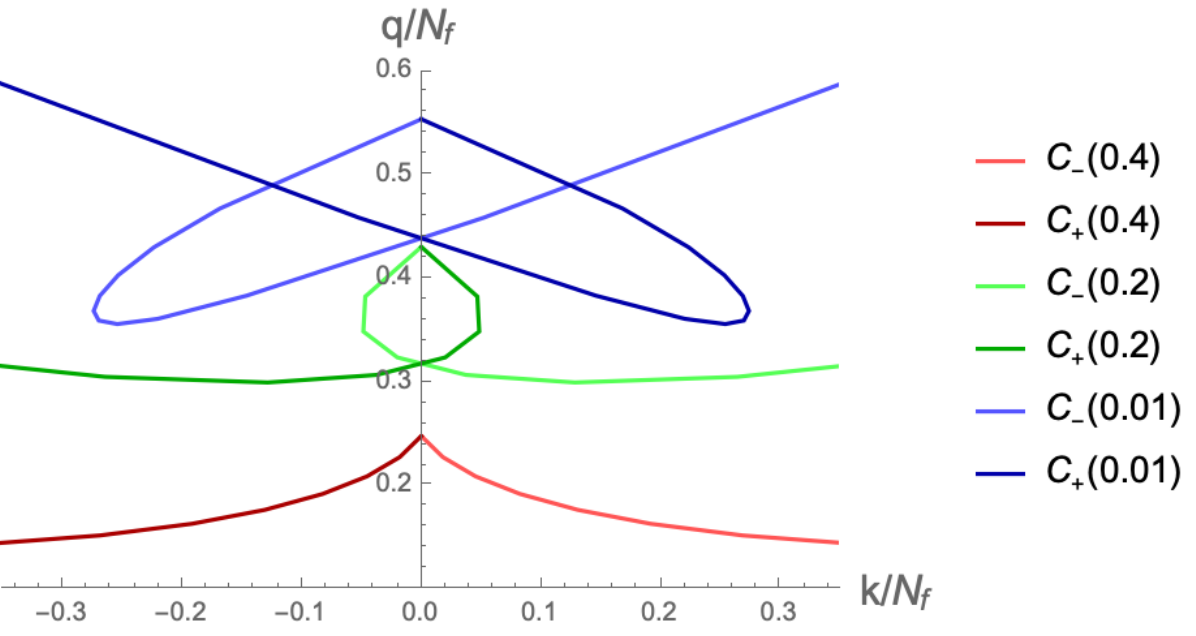}}
        \caption{In figure \ref{fig_QED3kapp1a} we plot the curve $C_-(x)$ for $x=0.4,\,0.3,\,0.2,\,0.1,\,0.05,\,0.01$, while in figure \ref{fig_QED3kapp1b} we show both the curves $C_-(x)$ and $C_+(x)$ for $x=0.4,\,0.2,\,0.01$.   }
\label{fig_QED3kapp1}
\end{figure}

Let us call $C_{\pm}(x)$ the set of solutions $(k,q)$ to \eqref{eq_app_saddle_1} and \eqref{eq_app_saddle_2} for the values of $(E,b)$ which lie on the curve $c_{\pm}(x)$. The $C_{\pm}(x)$ are thus curves on the $(k,q)$ plane.   Examples of the curve $C_-(x)$ for different values of $x$ are shown in figure \ref{fig_QED3kapp1a}.  The symmetry properties of the equations imply that $C_+(x)$ is obtained by mirroring $C_-(x)$ around the $k$ axis.  Interestingly for $x$ sufficiently small the curve $C_-(x)$ intersects the $k=0$ axis at two different points. For $x\rightarrow 0$ the first intersection point approaches $(k/N_f,q/N_f)=(0,1/2)$,\footnote{We determined the value of the intersection point analytically by studying the fluctuation determinant for $b\rightarrow \pm 1/2$.}  while the second intersection point approaches the critical value determined in \eqref{eq_QED3_result}, namely $(k/N_f,q/N_f)\simeq(0,0.56)$.  We also notice that the curves $C_-(x)$ and $C_+(x)$ intersect each other at two other points, see figure~\ref{fig_QED3kapp1b}. 

\begin{figure}[t]
   \centering
		\subcaptionbox{  \label{fig_QED3kapp2a}}
		{\includegraphics[scale=0.34]{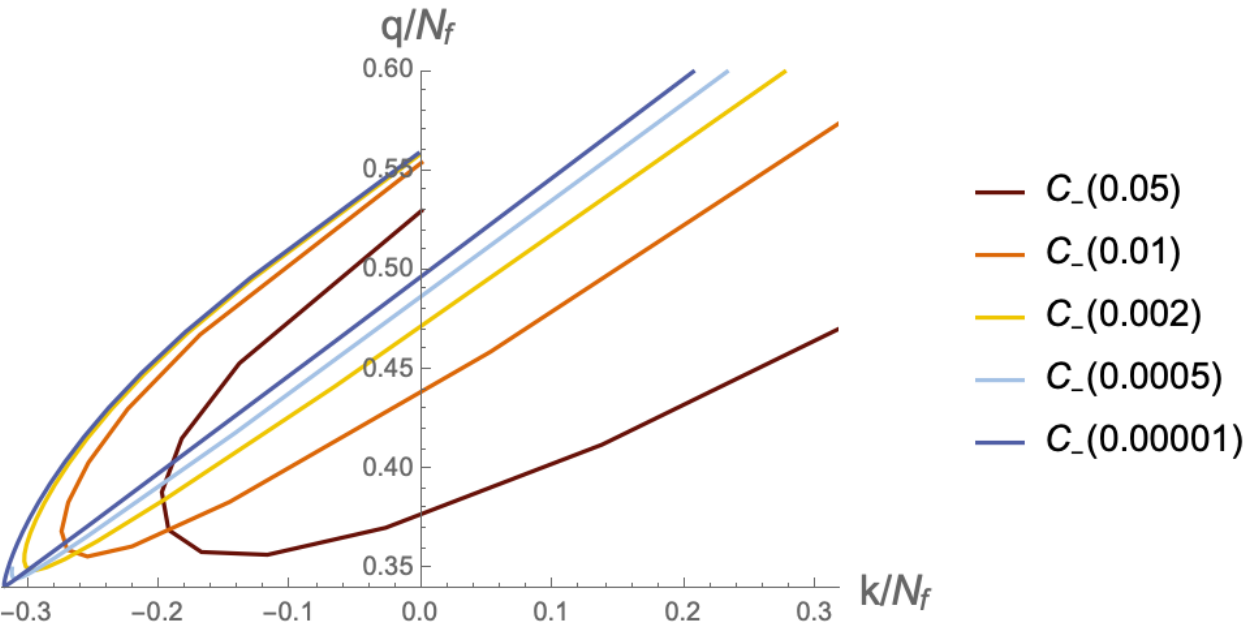}}
	\qquad 
		\subcaptionbox{ \label{fig_QED3kapp2b}}
		{\includegraphics[scale=0.34]{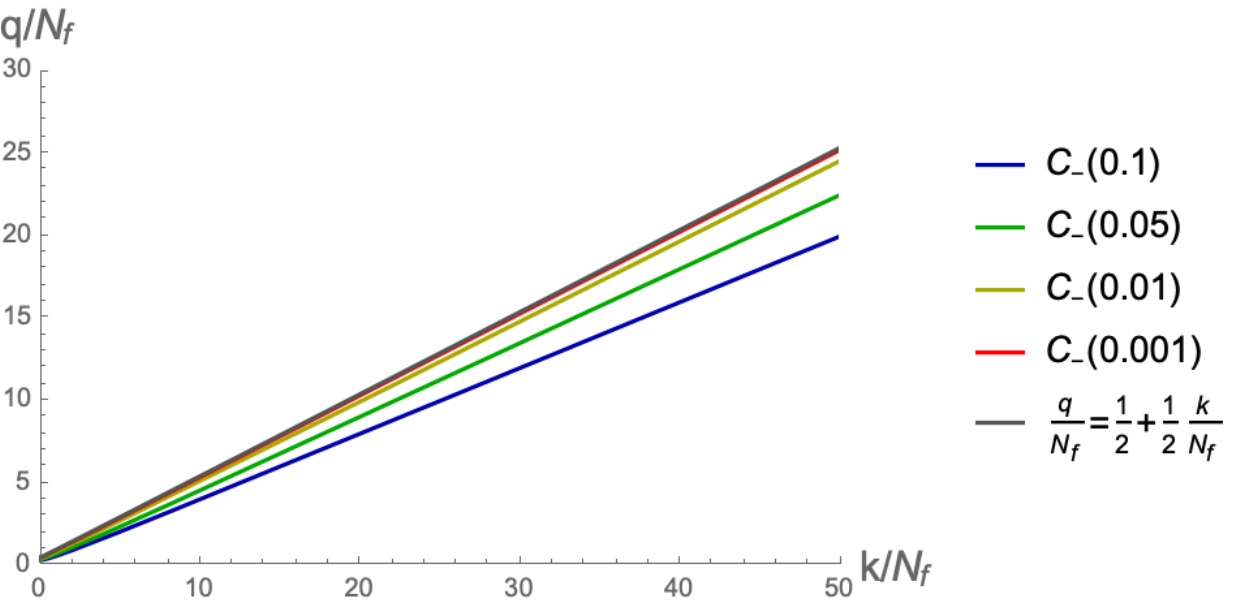}}
        \caption{In figure \ref{fig_QED3kapp2a} we plot the curve $C_-(x)$ as $x$ approaches $0$. In figure \ref{fig_QED3kapp2b} we compare the curve $C_-(x)$ to the line $q/N_f=\frac12 |k|/N_f+\frac12$ for $x\rightarrow 0$ and $k/N_f> 0$.    }
\label{fig_QED3kapp2}
\end{figure}

It can be seen by changing $x$ from $1/2$ to $0$ that the curves $C_\pm(x)$ cover the full region below the curve specified by the union of $\lim_{x\rightarrow 0} C_-(x)$ and $\lim_{x\rightarrow 0} C_+(x)$.  As figure~\ref{fig_QED3kapp2a} shows, as $x\to 0$, the curve $C_-(x)$ approaches a limit that is composed of two parts: a generic-looking curve that starts at $(k/N_f,q/N_f)\simeq(0,0.56)$ and ends at the point $(k^*/N_f,q^*/N_f)\simeq (1/\pi,0.34)$, and a second part that is a straight line that we conjecture to be $q/N_f=\frac12 |k|/N_f+\frac12$, see plot \ref{fig_QED3kapp2b}.  

\begin{figure}[h]
\centering
\includegraphics[scale=0.45]{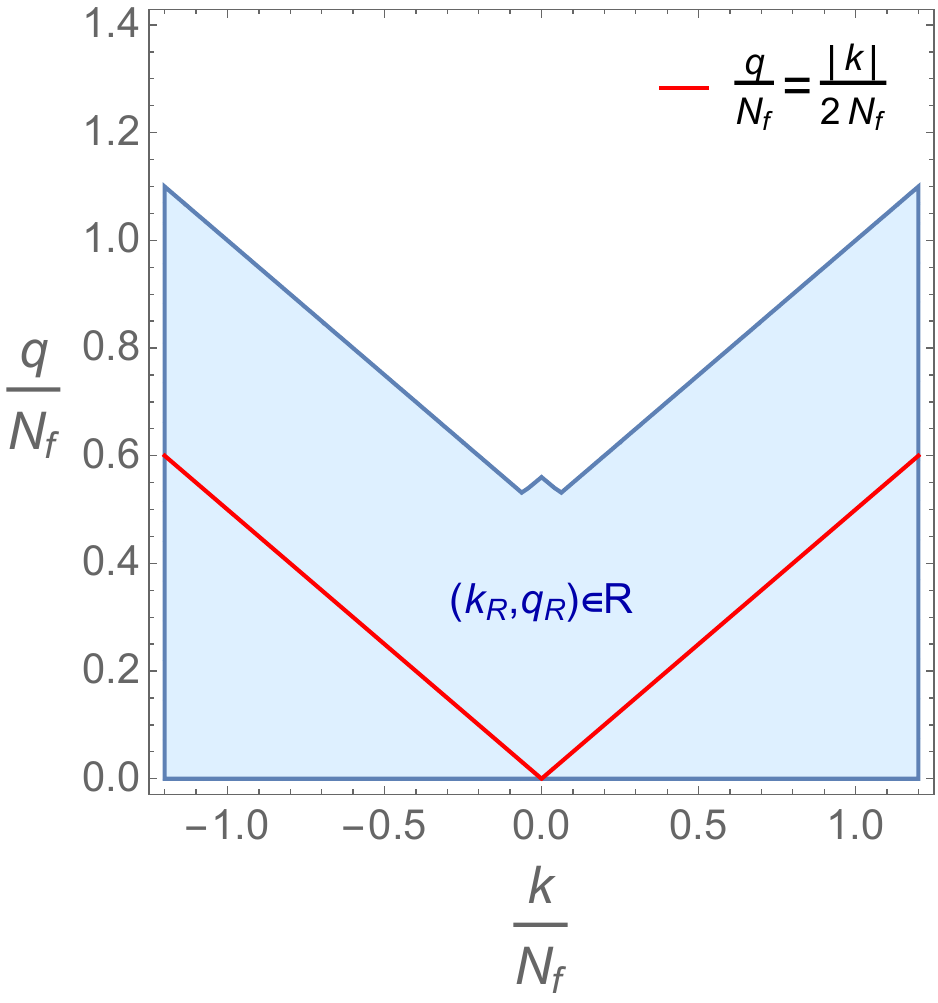}
\caption{Plot of the region $R$ specified by the solutions $(k_R,q_R)$  of \eqref{eq_QED3k_SaddleEq} for $b\in(-1/2,1/2)$ and $E<1/2-|b|$; the plot is restricted to $q>0$ since the region is symmetric for $q\leftrightarrow -q$.  For $|k|/N_f\gtrsim 0.061$ the boundary of $R$ is a straight line $q/N_f=\frac12|k|/N_f+\frac12$, while for $|k|/N_f\lesssim 0.061$ the boundary is curved (despite appearances). In red we plot the line $q=|k|/2$.}\label{fig:plot_QED3k_app}
\end{figure}

By considering the union of the two curves discussed above we obtain the region $R$ in figure \ref{fig:plot_QED3k_app}.  Note that the region $R$ strictly includes the one specified by $|q|\leq |k|/2$ and marked by red on the plot. Hence all physical Wilson lines correspond to at least one perturbatively stable saddle-point. 

As explained in the main text,  the points in $R$ for which $|q|>|k|/2$ correspond to additional saddle-points in the physical region $|q|\leq |k|/2$. To determine the value of $q$ to which a saddle-point $(q^*,k)$ with $|q^*|>|k|/2$ corresponds to, we simply need to perform a shift $q^*\rightarrow q^*- k n\equiv q$ (which is accompanied by the shift $b^*\rightarrow b^*+n$), with $n\in\mathds{Z}$,  such that $|q|<|k|/2$. Note that for each point $(q^*,k)$ in $R$ there is a single value of $n \in\mathds{Z}$ such that $|q^*- k n|<|k|/2$ is in the physical region. 

In practice, we consider the upper and lower boundary curves $q_{\pm}(k)$ of the region $R$. We then draw the shifted curves $q_{\pm}(k)\mp k$, $q_{\pm}(k)\mp 2 k$, $q_{\pm}(k)\mp 3 k$; the intersection of these curves separates the physical region $|q|<|k|/2$ of the $(q,k)$ plane into subregions according to the number of saddle-points. The result of this geometrical procedure is shown in the main text in figure~\ref{fig:plot_QED3k}.

\bibliography{Biblio}
	\bibliographystyle{JHEP.bst}

\end{document}